\documentclass[astrosymb,twocolumn]{aastex631}

\usepackage{gensymb}
\usepackage{lmodern}
\usepackage{amsmath}
\usepackage{natbib}
\usepackage{amssymb}
\usepackage{xfrac}
\usepackage{bm}
\usepackage{placeins}
\usepackage{xspace}
\usepackage{hyperref}
\usepackage[nameinlink]{cleveref}

\usepackage[T1]{fontenc} 
\usepackage[utf8]{inputenc} 

\setcitestyle{citesep={;}}
\setcitestyle{aysep={}}
\creflabelformat{equation}{#2\textup{#1}#3}

\newcommand{\RNum}[1]{\uppercase\expandafter{{\scshape\romannumeral #1\relax}}}

\newcommand{\API}{Anton Pannekoek Institute for Astronomy, University of Amsterdam, Science Park 904, 1098 XH Amsterdam, The Netherlands}

\newcommand{\HJOR}{$1.5^{+2.2}_{-1.1}$\%\xspace}
\newcommand{\WJOR}{$1.9^{+2.6}_{-1.4}$\%\xspace}
\newcommand{\WJORonly}{1.9}
\newcommand{\BDOR}{$<$3.6\%\xspace}

\newcommand{\HJcomp}{0.79\xspace}
\newcommand{\WJcomp}{0.63\xspace}
\newcommand{\BDcomp}{0.98\xspace}

\shorttitle{The Occurrence Rate of Young Giant Planets}
\shortauthors{Tran et al.}
\graphicspath{{Figures/}}

\accepted{June 4, 2025}
\submitjournal{AJ}

\begin{document}

\title{The Epoch of Giant Planet Migration Planet Search Program. III. \\ The Occurrence Rate of Young Giant Planets Inside the Water Ice Line\footnote{Based on observations obtained with the Hobby-Eberly Telescope (HET), which is a joint project of the University of Texas at Austin, the Pennsylvania State University, Ludwig-Maximillians-Universitaet Muenchen, and Georg-August Universitaet Goettingen. The HET is named in honor of its principal benefactors, William P. Hobby and Robert E. Eberly.}}

\correspondingauthor{Quang H. Tran}
\author[0000-0001-6532-6755]{Quang H. Tran}
\email{quang.tran@yale.edu}
\altaffiliation{51 Pegasi b Fellow.}
\affiliation{Department of Astronomy, Yale University, New Haven, CT 06511, USA}
\affiliation{Department of Astronomy, The University of Texas at Austin, 2515 Speedway, Stop C1400, Austin, TX 78712, USA}

\author[0000-0003-2649-2288]{Brendan P. Bowler}
\affiliation{Department of Physics, University of California, Santa Barbara, Santa Barbara, CA 93106, USA}
\affiliation{Department of Astronomy, The University of Texas at Austin, 2515 Speedway, Stop C1400, Austin, TX 78712, USA}

\author[0000-0001-9662-3496]{William D. Cochran}
\affiliation{McDonald Observatory and Center for Planetary Systems Habitability}
\affiliation{Department of Astronomy, The University of Texas at Austin, 2515 Speedway, Stop C1400, Austin, TX 78712, USA}

\author[0000-0003-4384-7220]{Chad F. Bender}
\affiliation{Steward Observatory, University of Arizona, 933 N Cherry Ave, Tucson, AZ 85721, USA}

\author[0000-0003-1312-9391]{Samuel Halverson}
\affiliation{Jet Propulsion Laboratory, California Institute of Technology, 4800 Oak Grove Drive, Pasadena, CA 91109, USA}

\author[0000-0001-9596-7983]{Suvrath Mahadevan}
\affiliation{Department of Astronomy \& Astrophysics, The Pennsylvania State University, 525 Davey Lab, University Park, PA 16802, USA}
\affiliation{Center for Exoplanets \& Habitable Worlds, University Park, PA 16802, USA}
\affiliation{Penn State Astrobiology Research Center, University Park, PA 16802, USA}

\author[0000-0001-8720-5612]{Joe P.\ Ninan}
\affil{Department of Astronomy and Astrophysics, Tata Institute of Fundamental Research, Homi Bhabha Road, Colaba, Mumbai 400005, India}

\author[0000-0003-0149-9678]{Paul Robertson}
\affiliation{Department of Physics \& Astronomy, The University of California, Irvine, Irvine, CA 92697, USA}

\author[0000-0001-8127-5775]{Arpita Roy}
\affiliation{Astrophysics \& Space Institute, Schmidt Sciences, New York, NY 10011, USA}

\author[0000-0001-7409-5688]{Guðmundur Stefánsson}
\affil{\API}

\author[0000-0002-4788-8858]{Ryan C. Terrien}
\affil{Carleton College, One North College St., Northfield, MN 55057, USA}

\begin{abstract}
    We present statistical results from the Epoch of Giant Planet Migration RV planet search program. This survey was designed to measure the occurrence rate of giant planets interior to the water ice line of young Sun-like stars, compare this to the prevalence of giant planets at older ages, and provide constraints on the timescale and dominant inward migration mechanism of giant planets. Our final sample amounts to 85 single young (20--200~Myr) G and K dwarfs which we target across a 4-year time baseline with the near-infrared Habitable-zone Planet Finder spectrograph at McDonald Observatory's Hobby-Eberly Telescope. As part of this survey, we discovered the young hot Jupiter HS Psc b. We characterize survey detection completeness with realistic injection-recovery tests and measure an occurrence rate of \WJOR for intermediate-age giant planets ($0.3 < m \sin i < 13$~$M_\mathrm{Jup}$) within 2.5~AU. This is lower than the field age occurrence rate for the same planet masses and separations and favors an increase in the prevalence of giant planets over time from $\sim$100~Myr to several Gyr, although our results cannot rule out a constant rate. A decaying planet occurrence rate is, however, strongly excluded. This suggests that giant planets located inside the water ice line originate from a combination of in situ formation or early migration coupled with longer-term inward scattering. The completeness-corrected prevalence of young hot Jupiters in our sample is \HJOR---similar to the rate for field stars---and the 95\% upper limit for young brown dwarfs within 5000~d is \BDOR.
\end{abstract}

\section{Introduction}

Young planets provide unique opportunities to test theories of orbital migration \citep[e.g.,][]{Ward1997, Kley2012}, atmospheric mass loss \citep[e.g.,][]{Lopez2013, Owen2017}, tidal decay and realignment \citep[e.g.,][]{Albrecht2012}, and thermal contraction \citep[e.g.,][]{David2019, Suarez2022}. Detecting and characterizing planets at ages when these evolutionary processes are expected to take place provide clues about how these populations evolve over time. In particular, determining the demographics of young planetary systems and contrasting this with well-constrained statistics for field-age stars offers an opportunity to identify trends over a broad range of stellar ages and peer into the evolutionary pathways of planets.

This task has been hindered by the difficulty of detecting young planets. Young Sun-like and low-mass stars rotate quickly and are more magnetically active, producing surface features such as starspots and plages that can manifest as quasi-periodic signals in time series measurements \citep{Jahn1984, Lagrange2010, Dumusque2011b, Boisse2012, Vanderburg2016}. This correlated noise, or jitter, can mask and even mimic the dynamical signals of planets in radial velocities (RVs) and photometry \citep[e.g.,][]{Queloz2001, Boisse2011, Aigrain2012, Haywood2014}. Traditionally, RV programs designed to search for planets have predominately targeted old, quiescent stars to minimize stellar jitter. As a result, knowledge of planet demographics has been primarily limited to ages of $\sim$1--10 Gyr.

The Epoch of Giant Planet Migration (EGPM) planet search program was designed to constrain the dominant migration pathway of giant planets \citep{Tran2021}. Giant planets are expected to migrate inward through two general mechanisms: an early ($\lesssim$10~Myr) and non-disruptive process where the planet and gaseous protoplanetary disk exchange angular momenta, and long-term ($\sim$10$^7$--10$^{10}$~yr) dynamical processes caused by gravitational interactions with other bodies in the system \citep{Wu2003, Fabrycky2007, Chatterjee2008, Triaud2010}. The EGPM survey launched in 2018 and was carried out with the near-infrared (NIR) Habitable-zone Planet Finder (HPF) spectrograph at McDonald Observatory's Hobby-Eberly Telescope (HET). The long time baseline ($\sim$4~yr) of the survey enables detection sensitivities out to orbital separations near the water ice line of Sun-like stars \citep[$\sim$2.5~AU;][]{Martin2012}. By operating in the NIR, the EGPM program leverages the wavelength dependence of RV jitter caused by starspot-driven activity signals \citep[e.g.,][]{Crockett2012, Bailey2012, Carleo2020, Tran2021, Cale2021, Carmona2023}. Moving to longer wavelengths compared to traditional visible-light RV programs reduces RV contributions from stellar activity, facilitating the discovery and statistical characterization of planets at young ages.

The goal of this survey is to measure the frequency of giant planets orbiting Sun-like stars (G and K dwarfs) within 2.5~AU at intermediate ages (20--200~Myr) and compare that occurrence rate with measurements at older ages ($\gtrsim$1~Gyr). Determining when giant planets arrive inside the water ice line provides direct constraints on the mechanism of inward migration.

In this work, we present the EGPM survey results. In \autoref{sec:survey_details}, we provide an overview of the EGPM program, including a summary of the observations and target characterization. The observed NIR RV variability of our targets is described in \autoref{sec:jitter}. The statistical analysis, including detection sensitivities, detected companions, and measured giant planet frequency, is reported in \autoref{sec:occurrence_rate}. In \autoref{sec:discussion}, we discuss how our findings compare to previous RV surveys and the future of observing young stars and planets. Finally, we summarize our results and conclusions in \autoref{sec:summary}.

\begin{deluxetable*}{lcc}
    \setlength{\tabcolsep}{3em}
    \tablecaption{\label{tab:ymg_assoc} Young Moving Groups of EGPM Survey Targets}
    \tablehead{\colhead{Young Moving Group} & \colhead{Age (Myr)} & \colhead{Age Ref.}}
    \startdata
    AB Doradus (AB Dor) & $133_{-20}^{+15}$ & \citet{Gagne2018c} \\
    $\beta$ Pic & $24 \pm 3$ & \citet{Bell2015} \\
    Carina & $45^{+11}_{-7}$ & \citet{Bell2015} \\
    Carina-Near & $200 \pm 50$ & \citet{Zuckerman2006} \\
    Columba & $42^{+6}_{-4}$ & \citet{Bell2015} \\
    Octans-Near (Oct-Near) & 30--100 & \citet{Zuckerman2013} \\
    Tucana-Horologium (Tuc-Hor) & $45\pm4$ & \citet{Bell2015} \\
    Pleiades & $110\pm15$ & \citet{Gaia2018} \\
    32 Ori & $24 \pm4 $ & \citet{Bell2017} \\
    Argus & 40--50 & \citet{Zuckerman2019} \\
    Pisces & 30--50 & \citet{Binks2018}
    \enddata
\end{deluxetable*}

\section{The Epoch of Giant Planet Migration Planet Search Program\label{sec:survey_details}}

\subsection{Stellar Sample}

Our survey targets comprise bona fide and high-probability candidate members of eleven young moving groups (YMGs; \autoref{tab:ymg_assoc}). Previously known spectroscopic binaries, close visual binaries (projected physical separations $<$1000~AU), and fast rotating stars ($v \sin i_* > 35$~km~s$^{-1}$) are removed to produce a sample consistent with giant planet RV search programs at field ages \citep[e.g.,][]{Johnson2010}. For a detailed description of the target selection and summary of target properties, see \citet{Tran2021}.

In the final statistical sample and analysis, 19 systems from our original sample of 104 young stars are removed. Some binary systems ultimately evaded scrutiny and were present in our original sample, and several recently identified binaries that were not known at the start of the survey were announced after our program began. Altogether, 18 targets are close binaries. In addition, one system that is now considered a field-age star is removed. \autoref{sec:appendix_excluded} summarizes the properties of these 19 excluded targets. One star (HD 24194) exhibits a low-amplitude, long-term RV trend (see \autoref{sec:appendix_long-term_trend}). As there is no additional evidence indicating that this trend is caused by a stellar binary companion, HD 24194 is retained in the subsequent analysis as a young, single star.

Our final statistical sample therefore consists of 85 young single Sun-like stars with G and K spectral types. Properties of the final targets, including spectral types, moving group associations, and summary statistics of HPF RV observations are reported in \autoref{tab:stellar_sample} in \autoref{sec:appendix_targ_info}.

\subsection{HPF Observations\label{sec:hpf_obs}}

The EGPM survey operated between October 2018 and March 2024 using the HPF spectrograph \citep{Mahadevan2012, Mahadevan2014, Mahadevan2018}. HPF is a near-infrared (0.81--1.27 $\mu$m) high-resolution \citep[$\overline{R} \sim$ 55,000;][]{Ninan2019} spectrograph located on the 9.2-meter HET \citep{Ramsey1998, Hill2021}. It achieves temperature stability at the 0.6~mK level \citep{Stefansson2016} and utilizes three fibers for simultaneous observations of the science target, sky background, and calibration data \citep{Kanodia2018b}, enabling precise RVs at the m s$^{-1}$ level \citep{Metcalf2019a}. Observations were obtained following the strategies described in \citet{Tran2021} and \citet{Tran2024}. For each epoch, three contiguous observations are acquired to assess potential high-frequency variations. All observations are obtained using HET's queue-mode scheduling system \citep{Shetrone2007}. Following \citet{Ninan2018}, \citet{Kaplan2019}, and \citet{Metcalf2019a}, 1D spectra are optimally extracted with the standard HPF reduction pipeline.

Relative RVs, differential line widths (dLWs), and chromatic indices (CRXs) are measured from each 1D spectra with a custom least-squares matching pipeline based on the \texttt{SERVAL} algorithm \citep{Zechmeister2018, Tran2021}. Similarly, the \ion{Calcium}{2} infrared triplet (\ion{Ca}{2} IRT) emission line indices are computed following \citet{Stefansson2020b}. During this process, only 8 \'echelle orders, corresponding to wavelength ranges in the $z$-band (8535--8895~\AA) and $Y$-band (9933--10767~\AA), are used for the RV extraction to avoid contamination of strong telluric absorption \citep[orders 4, 5, 6, 14, 15, 16, 17, 18;][]{Tran2021}. Hereafter, an epoch refers to a visit, which consists of three contiguous single exposure observations, and reported RVs refer to individual observations, or spectra.

\begin{figure*}[!ht]
    \centering
    \includegraphics[width=1.0\linewidth]{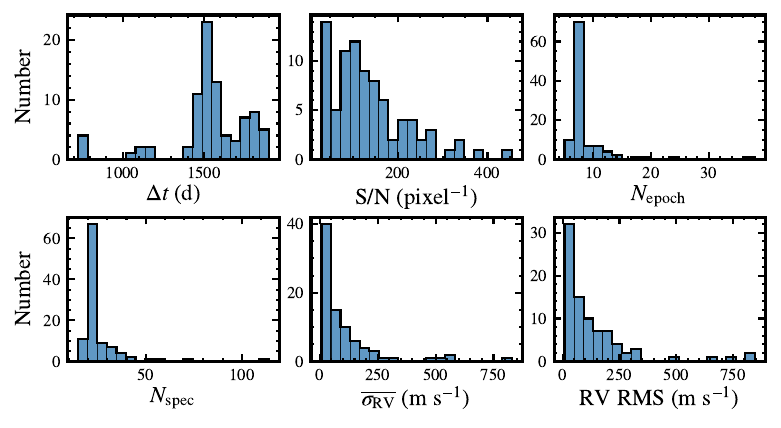}
    \caption{Summary of RV observations for the EGPM statistical sample. 91\% our targets have at least 7 epochs of RVs and a baseline of at least 4 years.}
    \label{fig:rv_dists}
\end{figure*}

In total, we obtained 2654 spectra across 884 epochs for 104 stars. The average number of epochs is 9 per star, average number of spectra is 26 per star, average time baseline is 1536~d (4.2~yr), and average signal-to-noise (S/N) at 1.07~$\mu$m is 136 pixel$^{-1}$. \autoref{fig:rv_dists} displays the distributions of time baseline ($\Delta t$), S/N, number of epochs, number of spectra, average RV uncertainty ($\overline{\sigma_\mathrm{RV}}$), and RV RMS for our sample, which are also detailed in \autoref{tab:stellar_sample}. \autoref{tab:hpf_measurements} in \autoref{sec:appendix_measurements} reports relative RVs and spectroscopic indicators of each target.

Between 2022 May to June, maintenance was carried out on HPF, which included a vacuum warm-up and cool-down cycle that introduces a velocity offset in the RV measurements. We choose to empirically model and correct for this offset using RV standard stars. This ensures that we retain sensitivity to the full survey time baseline and spectral template quality for targets with fewer observations, and it enables us to extract RVs from both observing seasons simultaneously. Based on three RV standard stars, the empirical velocity offset between post- and pre-maintenance HPF RVs (HPF$_\mathrm{post}$ -- HPF$_\mathrm{pre}$) is $-53.1\pm2.7$~m~s$^{-1}$. For the subsequent analysis, we treat the two time series as one single instrument by applying this offset and adding the associated error in quadrature to all post-maintenance HPF RVs and uncertainties, respectively. For more details on this empirical offset, see \autoref{sec:appendix_offset}.

\subsection{Projected Rotational Velocities\label{sec:characterization}}

We determine the projected rotational velocity ($v \sin i$) of each target from our HPF spectra following the procedures described in \citet{Tran2024}. Briefly, we adopt the empirical spectral matching framework from \citet{Stefansson2020a}, which uses the \texttt{SpecMatch-Emp} algorithm \citep{Yee2017}. The matching algorithm determines a best-fit $v \sin i$ for each of the eight HPF \'echelle orders used for RV extraction by comparing the science target's highest-S/N spectrum to a library of high-S/N spectra of slowly rotating stars convolved with a rotational kernel \citep{Gray1992}. The adopted $v \sin i$ and uncertainty are the median and standard deviation of the eight \'echelle order measurements. All measured $v \sin i$ values are reported in \autoref{tab:stellar_sample}.

\section{NIR RV Jitter at Intermediate Ages\label{sec:jitter}}

RV jitter in the optical has been studied extensively \citep[e.g.,][]{Saar1998, Santos2000, Lovis2011, Lagrange2013, Suarez2017, Luhn2020}, but less in known about trends of RV jitter in the NIR. Measuring how NIR RV jitter behaves across a broad range of ages, spectral types, and rotational velocities may provide clues about how the physical origin of these signals, such as starspots, behaves as a function of those parameters \citep[e.g.,][]{Luhn2020}. Here, we examine activity levels of targets in our sample as measured from the RMS of RV observations from this survey. To our knowledge, this represents the largest analysis of NIR activity levels of young stars to date.

\begin{figure*}[!t]
    \centering
    \includegraphics[width=1.0\linewidth]{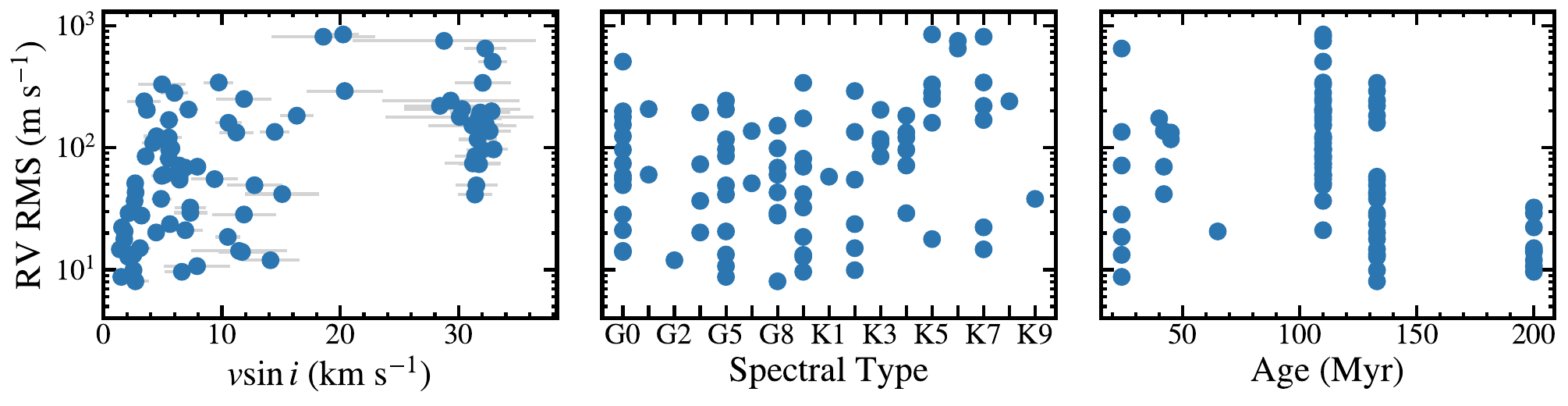}
    \caption{RV RMS, or intrinsic RV variability, as a function of projected rotational velocity, stellar spectral type, and stellar age for all stars in our sample.}
    \label{fig:rv_rms_relations}
\end{figure*}

\autoref{fig:rv_rms_relations} displays the average RV RMS of all targets as a function of their projected rotational velocities, stellar spectral type, and stellar age. The average RV uncertainty and RMS for our sample are also listed in \autoref{tab:stellar_sample}. Furthermore, the RMS of the five activity indicators that we measure from the HPF spectra (dLW, CRX, and the \ion{Ca}{2} emission line indices) as a function of projected rotational velocity, stellar spectral type, and stellar age are displayed in \autoref{fig:act_rms_vs_prop} in \autoref{sec:appendix_activity}.

There are broad trends between the RMS of the different measurements and projected rotational velocities and stellar spectral types. Over this range of $v \sin i$ values, there appears to be a slight rise in the upper and lower envelopes of RV RMS. RV measurement scatter is expected to increase for more rapidly rotating stars due to enhanced Doppler broadening of absorption lines \citep{Gray1999}. Across stellar spectral types, RV RMS is relatively flat with a modest increase for later-type stars. This feature likely corresponds to an observed enhancement in starspot covering fractions for cooler stars that results in larger RV scatter \citep{Saar1998, JacksonRJ2009, Morris2019}. The dLW and CRX RMS behave similarly to the RV RMS, which is consistent with the expectation that stellar jitter should manifest similarly in RVs, dLWs, and CRXs \citep{Zechmeister2018}. On the other hand, the RMS of the \ion{Ca}{2} IRT indices decreases as a function of $v \sin i$, but this negative correlation can be attributed to reduced measurement scatter for those indicators caused by broadened line profiles.

If starspots are the dominant origin of RV jitter in young Sun-like stars, then the observed spread in RV RMS suggests that starspots can vary significantly across this range of $v \sin i$ and stellar spectral type. Furthermore, this wide distribution in RV scatter is likely enhanced by the conflicting contributions between rotational Doppler broadening and starspots, as earlier-type stars rotate faster \citep{Saar1997}.

Following \citet{Tran2021}, we investigate the ratio between RV measurement precision and RV RMS, a metric that can be used to identify anomalously high RV dispersion that may be caused by the dynamical signature of a planet. If this ratio is substantially less than one, then the observed dispersion is not primarily driven by random measurement uncertainty but instead points to a astrophysical or dynamical origin. While it is not possible to distinguish between these two signals using this metric alone, systems with both RV RMS values greater than the typical level of our sample and uncertainty ratios less than 1.0 represent promising targets that may host giant planets.

\begin{figure}[!t]
    \centering
    \includegraphics[width=1.0\linewidth]{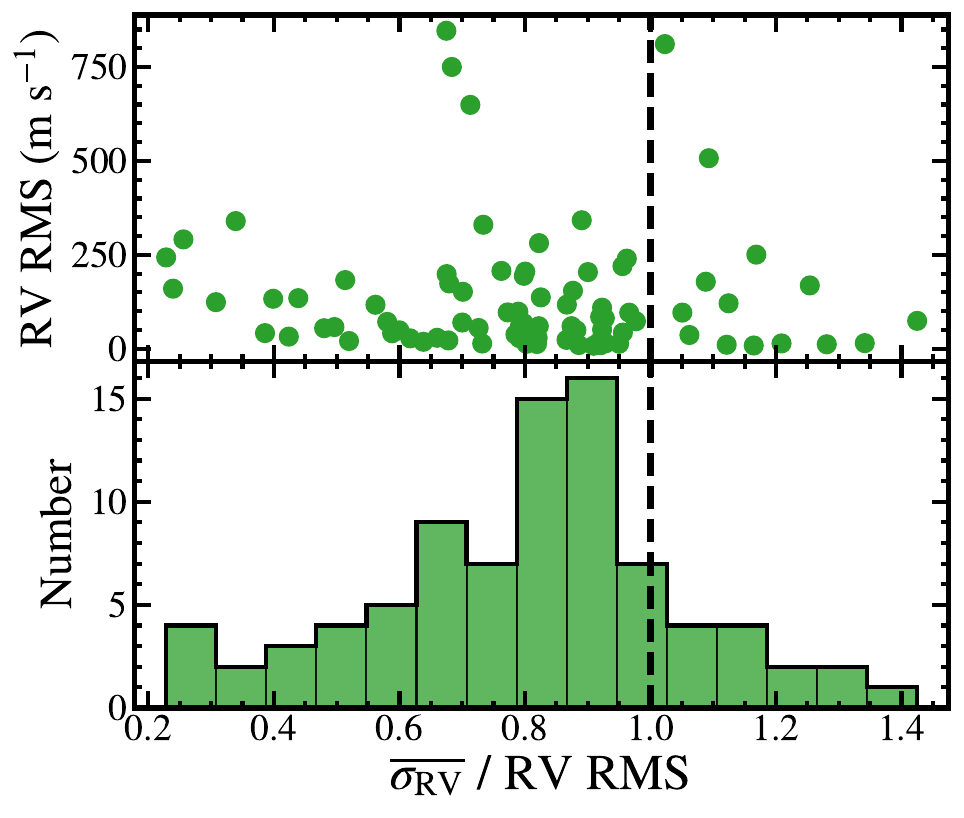}
    \caption{The ratio of average RV measurement error ($\overline{\sigma_\mathrm{RV}}$) to the RV RMS measurement of targets in our sample. An RV error-to-RMS ratio less than one indicates that the observed RV scatter is large and that activity or dynamical signals from a companion dominate over scatter associated with individual RV measurement precision. Objects with a low uncertainty ratio but high RV RMS value may be promising planet candidates.}
    \label{fig:error_rms_ratio}
\end{figure}

The distribution of this uncertainty ratio as a function of RV RMS is shown in \autoref{fig:error_rms_ratio}. The majority of our sample (84\%) does not appear to be dominated by RV measurement error ($\overline{\sigma_\mathrm{RV}}$/RV RMS $< 1$), which suggests that RV scatter in our sample is primarily driven by stellar activity, perhaps with additional contribution from planets. There are three stars with low uncertainty ratios and prominent RV RMS that rise above the rest of the sample (RV RMS $\gtrsim 500$~m~s$^{-1}$)---2MASS J05234246+0651581, V623 Tau, and V677 Tau. Further observations are needed in order to determine if these RV observations are a result of abnormally strong stellar activity, a stellar companion, or a planet. Note, however, that in the framework of our statistical analysis, these are not considered planet candidates; \autoref{sec:identify_candidates} describes our procedure for identifying and following up promising planet-hosting stars.

\section{The Occurrence Rate of Young Giant Planets Inside the Water Ice Line\label{sec:occurrence_rate}}

\subsection{Identifying Planet Candidates\label{sec:identify_candidates}}

The objective of the EGPM program is to robustly measure the occurrence rate of giant planets located within the water ice line at young ages. To minimize potential biases in this measurement introduced by human decision making, careful considerations must be made when adopting specific observing strategies as part of the main survey and for follow-up planet validation efforts \citep[e.g.,][]{Teske2021}. This is especially important for a program targeting young stars where activity levels are high and span a wide dynamic range (tens to hundreds of m~s$^{-1}$).

Below, we describe our approach to searching for periodic signals, identifying compelling targets for increased RV cadence, and, when applicable, determining the nature of each observed signal for targets without known stellar companions. Systems that meet the following criteria are designated as ``objects of interest,'' or targets with significant periodicities that require further characterization to be confirmed or refuted as originating from planets.
\begin{enumerate}
    \item RV time series must yield a significant peak in their Generalized Lomb-Scargle periodograms \citep[GLS;][]{Zechmeister2009} above a false alarm probability (FAP) threshold of 0.01\%.
    \item The power of the strongest peak rises above the median of all the peak powers by at least four times the standard deviation of the peak distribution. This ensures that the periodic signals of interest are distinguished from the statistical noise distribution of peak powers for each target.
    \item No prominent features can be present in the spectral window function at or near the observed peak of the RV time series, and the period of the significant peak cannot be consistent with an alias or harmonic of the stellar rotation period \citep{VanderPlas2018}.
\end{enumerate}
When possible, objects of interest that meet these conditions will have their RV cadence increased to weekly observations to assess if the periodic signal remains robust over time. For these systems, the planetary nature of each signal is then confirmed or refuted using a comprehensive characterization with all available data, including follow-up observations or previous measurements from other studies.

\subsubsection{Generalized Lomb-Scargle Periodogram Search\label{sec:gls_periodogram}}

Using the HPF RVs, we search for any periodic signals that could indicate the presence of a companion in the remaining systems. We compute the GLS periodogram over the frequency range 0.0007--1.0526~d$^{-1}$ (0.95--1500~d) using the open-source Python package \texttt{astropy.timeseries} \citep{Astropy2013, Astropy2018, Astropy2022} for each target. FAPs, which describe the probability that observations with no periodic signal would exhibit a periodogram peak with this power, are calculated using the bootstrap method. Three systems (HD 236717, HS Psc, and PW And) display significant peaks that simultaneously rise above a FAP threshold of 0.01\% and the surrounding periodogram noise. The periodograms of all three objects of interest are displayed in \autoref{fig:app_periodogram0} in \autoref{sec:appendix_periodograms}.

Periodic signals with significant periodogram peaks can originate from sampling features, rotationally modulated stellar activity signals, or periodic Keplerian motion from a dynamical companion. To test whether the observed periodogram peaks in the HPF RVs are caused by a dynamically-induced signal, we investigate if these peaks are driven by strong features in the spectral window function, are aliases or harmonics of the stellar rotation period, or are correlated with spectroscopic indicators.

\subsubsection{Stellar Rotation Periods of Objects with Significant Periodic Signals}

We measure the stellar rotation periods of each object of interest using available Transiting Exoplanet Survey Satellite (TESS) photometry following the procedure described in \citet{Tran2024}.\footnote{We do not measure stellar rotation periods for HD 236717. The TESS light curves of HD 236717 does not demonstrate clear and persistent modulations, which prevents an accurate inference of the stellar rotation period.} For each target, photometry from available TESS sectors \citep{Smith2012, Stumpe2012, Stumpe2014} is downloaded from the MAST data archive.\footnote{\href{https://archive.stsci.edu/missions-and-data/tess}{https://archive.stsci.edu/missions-and-data/tess.} All TESS data used is available at MAST:\dataset[10.17909/t9-nmc8-f686]{http://dx.doi.org/10.17909/t9-nmc8-f686} \citep{doi.org10.17909/t9-nmc8-f686}.} Each sector is then median-normalized. Outlier points outside of three standard deviations are removed after flattening the light curve using a Savitzky-Golay filter \citep{Savitzky1964}. Light curves are then binned to a cadence of 60 minutes to improve computational efficiency while retaining rotationally modulated structure.

Rotation periods are inferred by fitting a quasi-periodic Gaussian process to each separate TESS sector, or continuous group of TESS Sectors. This strategy is adopted as there are often large observational gaps between non-contiguous sectors, during which the observed variability can evolve significantly leading to inaccurate rotation period estimates. We use a quasi-periodic kernel with the functional form:

\begin{equation} \label{eq:quasi_per_kernel_lc}
    k_\mathrm{QP}(\tau) = A^2 \: \mathrm{exp}\left( -\frac{\tau^2}{2 l^2} - \frac{2 \: \mathrm{sin}^2\left(\frac{\pi \tau}{P_\mathrm{rot}}\right)}{\theta^2} \right),
\end{equation}

\noindent{}where $\tau$ is the difference between any two points in time, $A$ is the amplitude, $l$ is the localized correlation timescale, $\theta$ is the smoothness of the periodic component, and $P_\mathrm{rot}$ is stellar rotation period.

Posterior distributions of the Gaussian process kernel hyperparameters and a white-noise term ($\sigma_\mathrm{jit}$) are sampled using the \texttt{emcee} open Python package \citep{Foreman-Mackey2013, Foreman-Mackey2019}. We adopt uniform priors for $\sigma_\mathrm{jit}$, $A$, $l$, $\theta$, and $P_\mathrm{rot}$: $\mathcal{P}(\sigma_\mathrm{jit}) = \mathcal{U}[\log_e(10^{-12}), \log_e(1)]$, $\mathcal{P}(A) = \mathcal{U}[0.001, 5]$, $\mathcal{P}(l) = \mathcal{U}[0.001, 100]$, $\mathcal{P}(\theta) = \mathcal{U}[0.001, 100]$, and $\mathcal{P}(P_\mathrm{rot}) = \mathcal{U}[0.1, 30.0]$~d. If the object only has a single TESS sector or the TESS sectors are contiguous, then we adopt the average and standard deviation of the $P_\mathrm{rot}$ posterior distribution as the rotation period and uncertainty. If there are multiple, isolated TESS sectors, then the weighted average and standard deviation of all posterior distributions are adopted as the rotation period and uncertainty, respectively. Convergence of chains is confirmed using an integrated autocorrelation time threshold of 50 \citep{Goodman2010}.

Light curves and posterior distributions of the stellar rotation periods are shown in \autoref{fig:app_prot0} and adopted stellar rotation periods are reported in \autoref{tab:app_prot_vals} in \autoref{sec:appendix_lc_prot}. A Gaussian process regression yields a well-constrained stellar rotation period for each system. For systems where rotation periods are inferred using two independent sets of TESS sectors separated by an observational gap, the posterior distributions of $P_\mathrm{rot}$ are in good agreement.

\subsubsection{Correlations between RVs and Spectral Indices of Stars with Significant Periodic Signals\label{sec:ooi_corre}}

To identify RV signals that may originate from stellar activity, we search for significant correlations between the HPF RVs and various spectroscopic indices. This test is performed as spectral line profile distortions driven by active regions manifest similarly in both spectral indicators and RVs \citep{Jahn1984, Donati1992, Schuessler1996}. Thus, strong linear correlations between RVs and activity indices indicate that observed RV signals are not dynamical in nature \citep[e.g.,][]{Queloz2001, Boisse2011, Meunier2013, Robertson2014, Zechmeister2018}.

We calculate the Pearson's correlation coefficient (Pearson's $r$), associated $p$ values, and best-fit slope value ($m$) between the RVs and each indicator for all objects of interest. Pearson's $r$ values correspond to the strength of correlation or anti-correlation between two time series. Corresponding $p$ values describe the probability of detecting an $r$ value at least that extreme by chance if the two time series are not correlated such that a lower $p$ corresponds to stronger evidence of correlation.

Correlations between the differential line width, chromatic index, and the \ion{Ca}{2} emission line indices for all objects of interest are shown in \autoref{fig:app_corre0} in 
\autoref{sec:appendix_correlations}.

\subsection{Objects of Interest Designations}

Of the three targets that have a significant peak in their periodograms, one (HD 236717) exhibits significant correlation between the HPF RVs and all spectral indicators (\autoref{fig:app_corre0}). The remaining two targets (HS Psc and PW And) are designated as objects of interest. For each object of interest, we search the literature to assess whether previous observations might aid in the interpretation of the periodogram signals we identify. Furthermore, we increase the observing cadence with HPF to further investigate each signal.

As a result of these efforts, we are able to refine the nature of signals from two stars: HS Psc and PW And. HS Psc has a young planetary candidate, HS Psc b, detected as part of the EPGM survey \citep{Tran2024}. PW And is considered a non-detection as further HPF observations revealed that the periodic signal was caused by stellar activity.

\subsection{
Refuted Objects of Interest}

\subsubsection{HD 236717}

HD 236717 is a K0 dwarf \citep{Yoss1961} member of the $\approx$133~Myr AB Dor moving group \citep{Gagne2018b}. It has a Gaia DR3 distance of $38.63_{-0.02}^{+0.03}$~pc \citep{Bailer-Jones2021}. In total, 57 spectra were obtained over 4.6~yr. A significant peak in the GLS periodogram of all HPF RVs emerges at 7.3~d. This peak is likely a harmonic of the second-strongest peak at 29.4~d, which is consistent with a monthly-sampling alias. Furthermore, the HPF RVs exhibits strong correlations ($p \leq 0.01$) between its RVs and every spectral indicator, which supports the hypothesis that any other observed periodicity will be dominated by stellar activity signals. We therefore remove this system from further consideration as harboring a planet candidate.

\subsubsection{PW And}

PW And is an AB Dor moving group member \citep{Zuckerman2004}. It is a K2 dwarf \citep{Montes2001a} with a Gaia DR3 parallactic distance of $28.3\pm0.01$~pc \citep{GaiaDR3_2022} and cluster membership age of $\approx$133~Myr. A total of 114 spectra were obtained of PW And between December 2018 and July 2023. A GLS periodogram applied to all HPF RVs reveals a significant peak at 1.76~d, which corresponds to the stellar rotation period as determined from the TESS light curve ($P_\mathrm{rot} = 1.765 \pm 0.003$~d). The other significant peaks are consistent with harmonics of this period and coincide with strong peaks in the dLW and the \ion{Ca}{2} IRT triplet periodograms. As a result, we attribute the signal to rotationally modulated activity and designate PW And as a non-detection.

\subsection{Giant Planet Candidates}

\subsubsection{HS Psc b}

HS Psc b is a young giant planet candidate detected as part of the EGPM survey. The host star is a K7V member of the $133_{-20}^{+15}$~Myr AB Dor Moving group \citep{Schlieder2010, Malo2013, Gagne2018b, Gagne2018c}. It has a mass of $0.69 \pm 0.07$~$M_\odot$ \citep{Tran2024} and a Gaia DR3 parallactic distance of $37.67 \pm 0.04$~pc \citep{GaiaDR3_2022}. This star was selected for intensive follow-up observations based on its high RV RMS of $>$200~m~s$^{-1}$ from eight initial RV epochs \citep{Tran2021}.

We increased the frequency of observations for this system over the next 3 years to determine if this large RV scatter could be attributed to a planetary companion. This led to the discovery of the candidate young hot Jupiter HS Psc b, which was presented in \citet{Tran2024}. A periodogram analysis of 83 HPF spectra over 4 years highlighted a strong periodic signal at $P = 3.99$~d, which differed from any integer harmonics of the stellar rotation period as measured by two TESS Sectors ($P_\mathrm{rot} = 1.086 \pm 0.003$~d). Applying a joint Keplerian and Gaussian process stellar activity model to the HPF RVs produced a minimum planetary mass of $m \sin i = 1.46_{-0.44}^{+0.57}$~$M_\mathrm{Jup}$, an orbital period of $P_b = 3.986_{-0.003}^{+0.044}$~d, and a broad eccentricity with a preference for low values ($e_b = 0.27_{-0.27}^{+0.21}$). We include HS Psc b as a detection in the final statistical analysis.

\subsection{Survey Completeness and Detection Limits\label{sec:completeness}}

To accurately determine the occurrence rate of giant planets from our survey, we need to assess the sensitivity of our observations to potential detections and missed planets. These detection limits are described by the survey search completeness, or the probability that a planet with particular physical and orbital characteristics would be detected in our survey. We characterize the search completeness map as a function of orbital period and velocity semi-amplitude, or $C(P, K)$, for the 85 young, single stars comprising our statistical sample following the methodology of \citet{Howard2010}. For each target, we calculate the maximum Keplerian semi-amplitude, $K$, that best match the observed RV observations at a given orbital period, $P$. We use the open source Python package \texttt{radvel} \citep{Fulton2018} to fit circular orbits to the HPF RVs over a range of 100 logarithmically-spaced orbital periods between 1.5 and 5000~d for each target. In each fit, $P$ is allowed to vary within $\pm$1 day.

We also convert RV semi-amplitude to planetary minimum mass to transform $C(P, K)$ into $C(P, m \sin i)$, which requires knowledge of the stellar mass. We interpolate the evolutionary models of \citet{Baraffe1998} and use the absolute $J$-band magnitudes and stellar ages reported in \autoref{tab:stellar_sample} of each star to estimate their mass. The distribution of these inferred masses is centered at 0.9 $M_\odot$ and spans about 0.6--1.2 $M_\odot$. Changing these masses by as much as $\pm$20\% does not significantly impact $C(P, m \sin i)$ nor the final occurrence rate we measure from this survey. 

\begin{figure}[!t]
     \centering
     \includegraphics[width=1.0\linewidth]{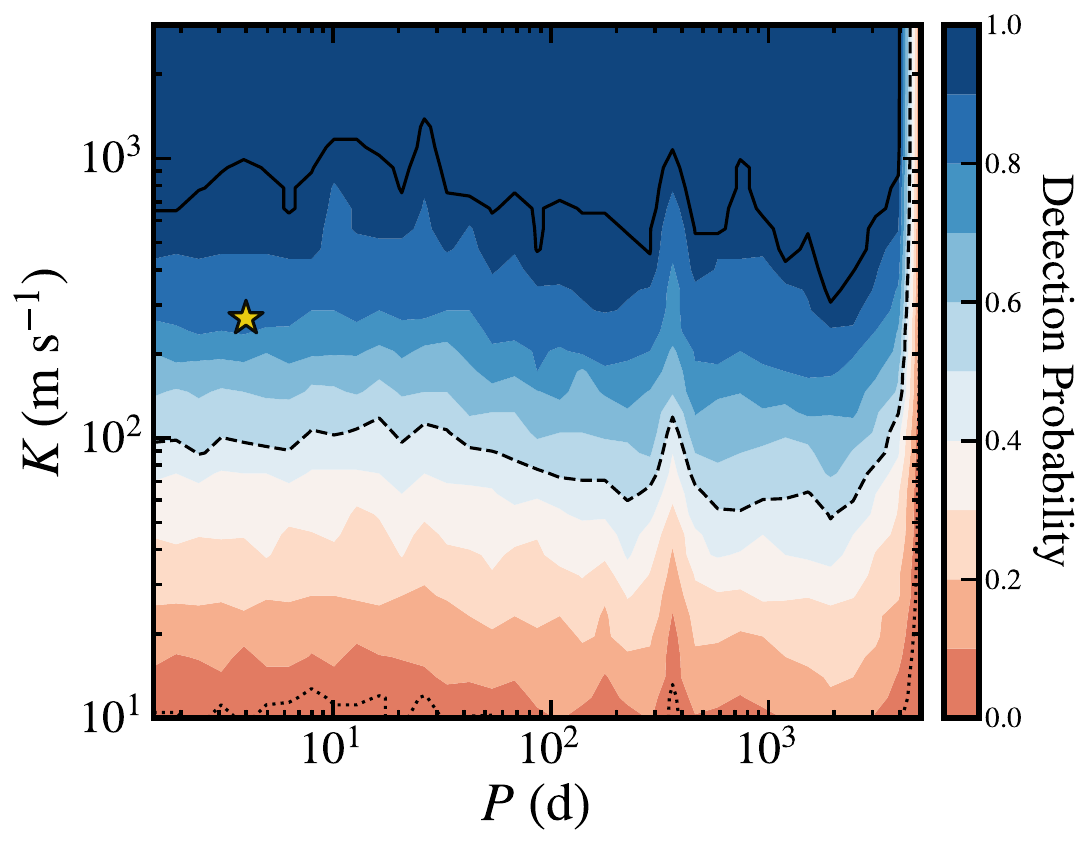}
     \hspace*{-0.4em}\includegraphics[width=1.0\linewidth]{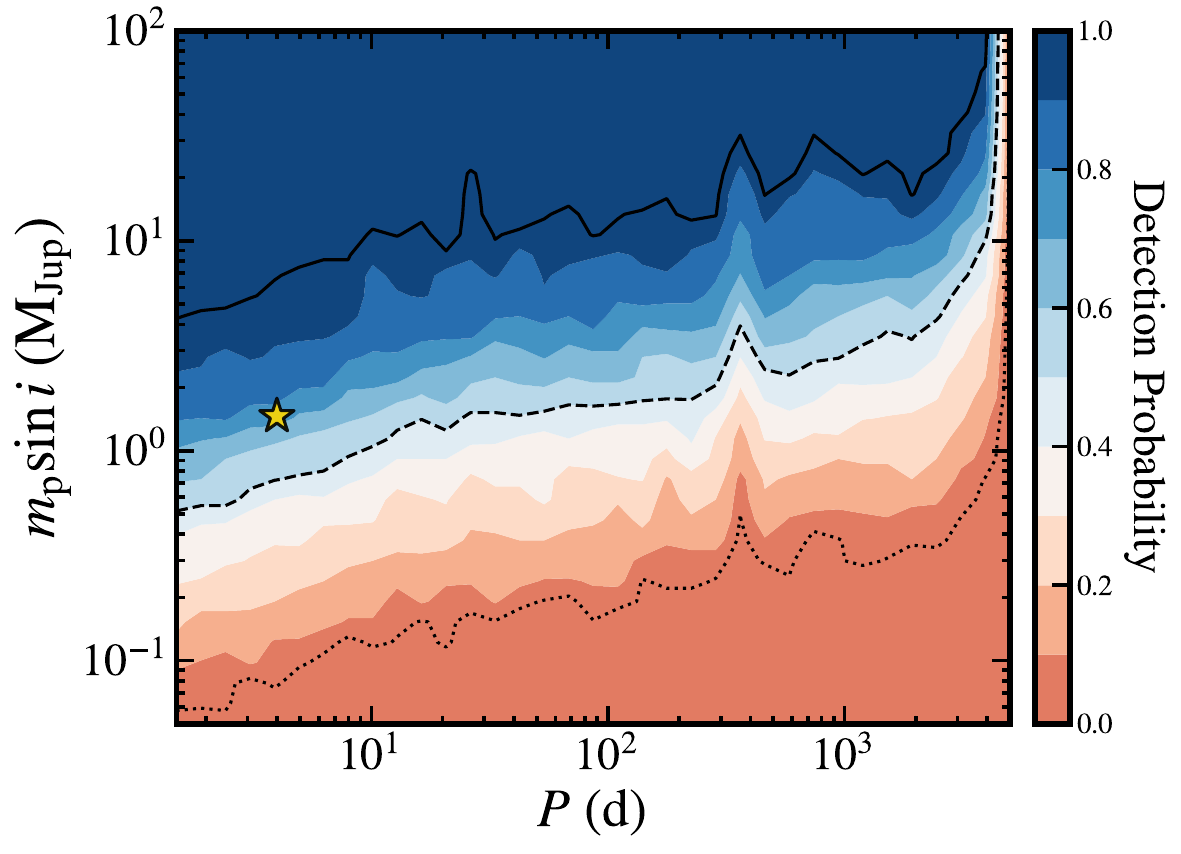}
     \caption{Search completeness function of the EGPM survey as a function of RV semi-amplitude (top) and minimum planetary mass (bottom). The 5\%, 50\%, and 95\% completeness contour lines are shown as dotted, dashed, and solid black lines, respectively. The yellow star represents the detected planet candidate HS Psc b \citep{Tran2024}.}
    \label{fig:survey_completeness}
\end{figure}

The search completeness function of our survey represents the fraction of stars with detection limits lower than some $K$ or $m \sin i$ threshold for a given $P$. When calculating an occurrence rate over a parameter domain of $P_1 < P < P_2$ and $K_1 < K < K_2$ (or $m \sin i_1 < m \sin i < m \sin i_2$), the survey completeness $C$ is the average search completeness function value over that corresponding physical domain. Following \citet{Howard2010}, the densely sampled completeness function is recomputed to the upper limit envelope over 35 logarithmically-spaced bins. This averages over rapid variations in the completeness function at short orbital periods, which is a result of increased precision at shorter orbital periods from an observational baseline of several years. \autoref{fig:survey_completeness} displays the average search completeness function from the statistical sample of our survey as a function of both RV semi-amplitude and minimum planetary mass. \Crefrange{fig:rv_comp0}{fig:rv_comp16} in \autoref{sec:appendix_completeness_function} display all RV time series and corresponding search completeness function in both RV semi-amplitude and minimum planetary mass.

Our survey is almost entirely sensitive to massive giant planets and brown dwarfs within orbital periods of 10~d, but detection limits are reduced to about 60\% for hot Jupiters with masses of 1 $M_\mathrm{Jup}$ and even lower for 51 Peg b analogs with masses of 0.5 $M_\mathrm{Jup}$. We are insensitive to Saturn-mass and sub-Saturn mass planets beyond about 100~d. There is structure at $\sim$27~d and $\sim$365~d, corresponding to monthly and annual observing cycles, and a sharp upward trend beyond the time baseline of the survey ($\sim$4~year or $\approx$1460~d).

\subsection{Young Giant Planet Occurrence Rate\label{sec:survey_or_results}}

We follow the methodology described in \citet{Bowler2015} to calculate the occurrence rate of young giant planets interior to the water ice line based on our survey sensitivity map. In surveys with perfect detection efficiency, each target star represents a Bernoulli trial and the probability distribution of planet frequency, $f$, is represented with the binomial distribution. However, when surveys have some level of incompleteness, the number of Bernoulli trials is not the sample size $N$. Instead, the number of trials must be corrected by the survey sensitivity. Using the search completeness function computed in \autoref{sec:completeness}, we ``correct'' for our detection sensitivities by multiplying our sample size by the search completeness function averaged over the relevant physical domain. This correction allows us to account for mass-period phase space that is not probed due to limited sensitivity, providing us with an effective number of trials, $n$.

In this instance, the effective number of trials will be a non-integer, and the planet occurrence rate is described by the generalized binomial distribution \citep{Artin1964, Bowler2015},

\begin{equation}
    P(f | n, k) = \frac{\Gamma(n + 1)}{\Gamma(k + 1) \Gamma(n - k + 1)} f^k (1 - f)^{n - k} (n + 1),
\end{equation}
where the binomial coefficient uses Gamma functions in place of factorials, $n$ is the effective number of trials, $k$ is the number of successes, and the $(n + 1)$ factor is a normalization constant. Using the effective number of trials and generalized binomial distribution is similar to the ``missing planet correction'' approach of \citet{Howard2010} and is also is comparable to the methods of \citet{Nielsen2008}, who use the Poisson distribution.

We calculate the survey completeness and planet frequency over a $P$--$K$ domain of $20 < K < 1500$~m~s$^{-1}$ and $0 < P < 1461$~d. This corresponds to giant planets orbiting within $\sim$2.5 AU and is constructed to closely match the sample considered in \citet{Johnson2010}. In this regime, we have a completeness of \WJcomp, an effective number of trials of 56.0, and 1 detection. Based on those values, we determine an occurrence rate of \WJOR for young giant planets inside the water ice line.

As this survey is sensitive to most hot Jupiters (HJs) and brown dwarfs (BDs) within a few AU, we also determine their occurrence rates at young ages. To remain consistent with other studies, we define HJs as giant planets (0.3~$M_\mathrm{Jup} < m \sin i <$ 13~$M_\mathrm{Jup}$) on short orbits ($P < 10$ d) and BDs as being between the hydrogen- and deuterium-burning limits (13~$M_\mathrm{Jup} < m \sin i <$ 80~$M_\mathrm{Jup}$). BDs could have been detected at much wider separations, so we consider those with orbital periods within 5000~d. We calculate survey completeness functions over the corresponding $m \sin i$--$P$ domains and find an average sensitivity of \HJcomp and \BDcomp for HJs and BDs, respectively. Repeating the methodology outlined above, we then determine occurrence rates for each population. As there are no detections of BDs in the EGPM survey, we only place an upper limit on their prevalence. For young HJs and BDs, we infer an occurrence rate of \HJOR and a 95\% upper limit of \BDOR, respectively. \autoref{tab:ors} summarizes our inferred occurrence rates for each subsample of our survey. We discuss the implications of these results in \autoref{sec:discussion}.

\begin{figure*}[!t]
     \centering
     \includegraphics[width=1.0\linewidth]{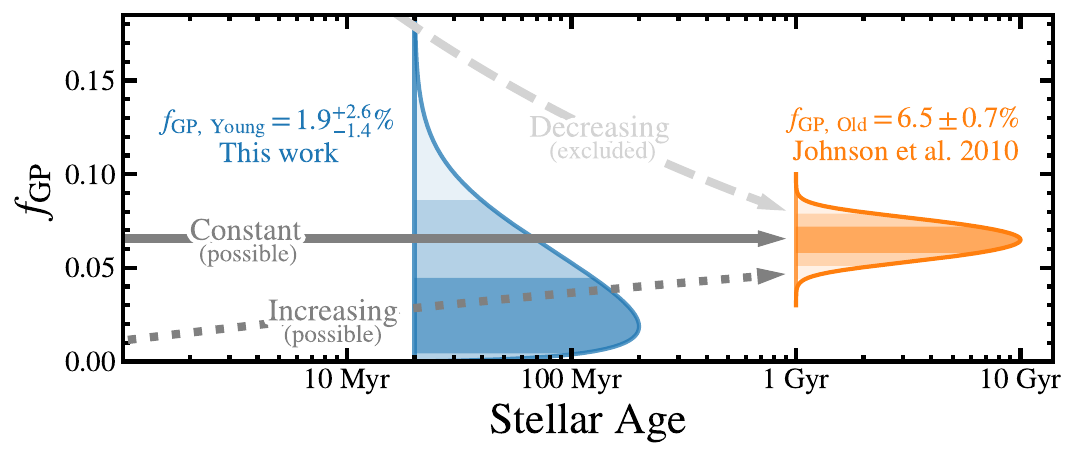}
     \caption{The inferred occurrence rates of giant planets orbiting interior to the water ice line of Sun-like stars as a function of age. Schematic decreasing, consistent, and increasing giant planet frequency over time trends, resulting from different migration pathways, are also plotted as grey dashed, solid, and dotted lines, respectively. Our young measurement of \WJOR and the field-age measurement of $6.5 \pm 0.7$\% from \citet{Johnson2010} are displayed in the blue and orange distributions, respectively. The shaded regions under each distribution show the 1$\sigma$, 2$\sigma$, and 3$\sigma$ confidence intervals. If the occurrence rate of giant planets increases over time, then dynamical processes are the dominant migration pathway of giant planets. If the giant planet frequency is consistent between young and old systems, then giant planets likely arrived at their present-day locations through disk migration. A decreasing trend suggests that giant planets are efficiently engulfed by their host stars at early ages.}
    \label{fig:results_comparison}
\end{figure*}

\begin{figure*}[!t]
     \centering
     \includegraphics[width=1.0\linewidth]{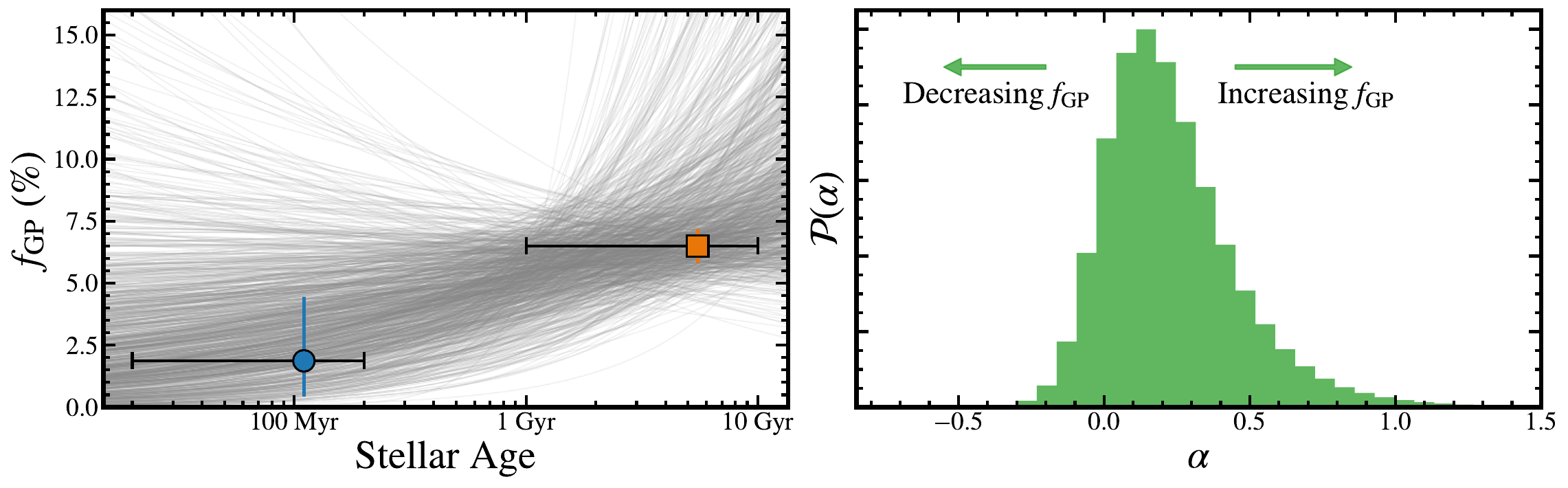}
     \caption{Left: Constraints on the giant planet frequency inside the water ice line of Sun-like stars over time, assuming this evolution follows a power law functional form. Overplotted is the inferred occurrence rate of giant planets at young (blue circle) and old (orange square) ages. The spread of power law trends represents the possible evolutionary pathways that the occurrence rate of giant planets may undergo. Right: The distribution of the power law coefficient, $\alpha$, generated from the young sample in this work compared with the field-age sample from \citet{Johnson2010}. A positive $\alpha$ corresponds to an increase in the frequency of giant planets over time. A negative $\alpha$ means that the giant planet occurrence rate decays over Gyr timescales. If $\alpha$ is 0, then the frequency remains constant.}
    \label{fig:gp_powerlaw_alphas}
\end{figure*}

\section{Discussion\label{sec:discussion}} 

\subsection{The Frequency of Giant Planets Inside the Water Ice Line Over Time\label{sec:discuss_frequency}}

To place constraints on the migration pathway of giant planets, we compare our measurement of the occurrence rate of giant planets interior to the water ice line at young ages to similar measurements at older ages and assess how this frequency evolves over time.

\citet{Johnson2010} used a sample of 1266 stars from the California Planet Survey to measure the prevalence of planets with
$K > 20$~m~s$^{-1}$ orbiting within 2.5~AU of old (several Gyr) Sun-like ($0.7 < M_*/M_{\odot} < 1.3$ and [Fe/H] = 0~dex) stars. Their field-age stellar sample is an older analog of our targets, allowing us to directly compare results from both surveys. However, their large sample size translates to a higher precision relative to our results inferred with a sample of 85 stars. Moreover, substantially higher jitter levels in our program translate into reduced sensitivity, which further lowers the effective number of targets in the statistical analysis (\autoref{sec:survey_or_results}). Our occurrence rate constraint, which peaks at \WJORonly\%, is lower but statistically consistent within 2$\sigma$ of the $6.5 \pm 0.7$\% frequency measured by \citet{Johnson2010}.

\autoref{fig:results_comparison} displays two measurements of the prevalence of giant planets inside the water ice line as a function of stellar age and how observed trends in this frequency over time broadly trace different migration mechanisms. These results point to an increasing occurrence rate over time and suggest that the inner regions of planetary systems are being increasingly populated with giant planets over Gyr timescales that formed beyond the water ice line and migrated inward. This implies that a dynamically disruptive process, such as planet-planet scattering, is an important inward migration mechanism shaping this population of giant planets \citep[e.g.,][]{Mustill2017, Anderson2017, Marzari2019, Frelikh2019}. These dynamical processes operate on timescales longer than the ages of the systems in this study ($\sim$20--200~Myr).

Although an increasing occurrence rate is preferred, we cannot confidently exclude the possibility of a constant frequency over time. In situ formation, inspiraling disk migration, or early planet scattering is also plausible \citep[e.g.,][]{Ward1997, Kley2012, Albrecht2012}. In this case, the occurrence rate we observed was established earlier than the ages of our targets ($\sim$$10^7$--$10^8$~yr). Larger sample sizes of young giant planet systems are needed to confidently rule out statistically distinct measurements over this broad range of stellar ages. Instead, we can more securely rule out a decaying frequency over time.

We further attempt to model the evolution of the giant planet frequency. We can approximate this trend with a power law relationship using the giant planet frequency we measure at young ages together with the value for field stars from \citet{Johnson2010}. We use a power law with the functional form: 
\begin{equation}
    f_\mathrm{GP}(t) = C t^\alpha,
\end{equation}
where $f_\mathrm{GP}$ is the frequency of giant planets inside the water ice line at a particular stellar age $t$, $C$ is a constant, and $\alpha$ is the power law coefficient that represents the strength of growth or decay in the giant planet frequency over time. This coefficient is equivalent to the slope of a linear relationship between the two measurements in logarithmic space. Thus, we can evaluate the power law coefficient using the differences between giant planet frequencies at young and old ages ($\log \Delta f_\mathrm{GP}$) and the differences between the stellar ages ($\log \Delta t$). More simply, $\alpha$ is equivalent to the ratio of these differences ($\log \Delta f_\mathrm{GP}$ / $\log \Delta t$). If $\alpha = 0$, that would mean that there is no change in the giant planet frequency over Gyr timescales. A positive $\alpha$ means that the prevalence of giant planets increases over time, while a negative $\alpha$ represents a decay of the giant planet frequency over time.

As $\alpha$ is computed using the logarithm of the difference between simulated giant planet occurrence rates at young ($f_\mathrm{GP, \; Young}$) and old ($f_\mathrm{GP, \; Old}$) ages, each $\alpha$ value can be translated to a ratio between the two frequencies, or the fractional change of the frequency over time. As a result, the distribution of $\alpha$ values can be mapped to a distribution of $f_\mathrm{GP, \; Young}$/$f_\mathrm{GP, \; Old}$ ratios, which quantifies the relative change of $f_\mathrm{GP}$ over time. For instance, the characteristic $\alpha$ value of $0.23^{+0.14}_{-0.26}$ corresponds to a +147$^{+175}_{-158}$\% fractional change in the frequency of giant planets over time.

From the $\alpha$ and corresponding $f_\mathrm{GP, \; Young}$/$f_\mathrm{GP, \; Old}$ distributions, we exclude the possibility that the young age rate is 1.3$\times$ and 1.9$\times$ the field age rate at 95\% and 99\% confidence, respectively. All fractional changes larger than these values are excluded at even higher confidence levels, strongly ruling out the most severe cases of decaying occurrence rates over time. This rejects a scenario where planets are being significantly lost over time on timescales of $10^8$--$10^9$~yr, through processes such as engulfment by their host stars via tidal decay.

\begin{deluxetable*}{ccccccc}[!t]
    \centerwidetable
    \setlength{\tabcolsep}{4pt}
    \tablecaption{Inferred occurrence rates of giant planets and brown dwarfs from RV surveys targeting young stars. \label{tab:ors}}
    \tablehead{\colhead{$K$ or $m \sin i$ Range} & \colhead{$P$ Range} & \colhead{Sample Age} & \colhead{Frequency} & \colhead{$1\sigma$ HDI\tablenotemark{a}} & \colhead{$2\sigma$ HDI} & \colhead{Reference}}
    \startdata
    \multicolumn{7}{c}{\textbf{Hot Jupiters}} \\
    $0.3 < m \sin i < 13$~$M_\mathrm{Jup}$ & $P < 10$~d & 20--200~Myr & 1.5\% & 0.5--3.7\% & 0.06--6.99\% & This work \\
    $5 < m \sin i < 13$~$M_\mathrm{Jup}$ & $P < 3$~d & 30--300~Myr & $<$2.9\%\tablenotemark{b} & 0.0--2.9\% & 0.0--4.1\% & \citet{Takarada2020} \\
    $m \sin i < 13$~$M_\mathrm{Jup}$ & $P \leq 10$~d & 10--400~Myr & $<$0.9\%\tablenotemark{c} & 0.0--0.9\% & $\cdots$ & \citet{Zakhozhay2022} \\
    $1 < m \sin i < 13$~$M_\mathrm{Jup}$ & $1 < P < 10$~d & $<$600~Myr & $<$1.2\%\tablenotemark{d} & 0.0--3.9\% & 0.0--6.6\% & \citet{Grandjean2023} \\
    $m \sin i < 13$~$M_\mathrm{Jup}$ & $P < 10$~d & 500--625~Myr & 0.93\%\tablenotemark{e} & 0.40--1.83\% & $\cdots$ & \citet{Quinn2016} \\
    \hline
    \multicolumn{7}{c}{\textbf{Giant Planets Inside the Water Ice Line}} \\
    $K > 20$~m~s$^{-1}$ & $P < 1461$~d & 20--200~Myr & 1.9\% & 0.4--4.5\% & 0.1--8.6\% & This work \\
    $1 < m \sin i < 13$~$M_\mathrm{Jup}$ & $1 < P < 1000$~d & $<$600~Myr & $<1.3$\%\tablenotemark{d} & 0.0--4.2\% & 0.0--7.1\% & \citet{Grandjean2023} \\
    \hline
    \multicolumn{7}{c}{\textbf{Brown Dwarfs}} \\
    $13 < m \sin i < 80$~$M_\mathrm{Jup}$ & $P < 5000$ d & 20--200~Myr & $<$3.6\% & 0.0--1.4\% & 0.0--3.6\% & This work \\
    $13 < m \sin i < 80$~$M_\mathrm{Jup}$ & $P < 3$~d & 30--300~Myr & $<$1.9\%\tablenotemark{c} & 0.0--1.9\% & 0.0--2.7\% & \citet{Takarada2020} \\
    $13 < m \sin i < 80$~$M_\mathrm{Jup}$ & $1 < P < 1000$~d & $<$600~Myr & $<$1.2\%\tablenotemark{d} & 0.0--4.0\% & 0.0--6.7\% & \citet{Grandjean2023} \\
    \enddata
    \tablenotetext{a}{HDI refers to the highest density interval, the minimum range encompassing a given fraction of a distribution.}
    \tablenotetext{b}{Estimated by combining results with \citet{Paulson2006} and \citet{Bailey2018}.}
    \tablenotetext{c}{Estimated by combining results with \citet{Grandjean2021}.}
    \tablenotetext{d}{Value as reported by original study.}
    \tablenotetext{e}{Estimated by combining results with \citet{Paulson2004}.}
\end{deluxetable*}

\subsection{Comparison to Young Giant Planet Occurrence Rates from Literature \label{sec:lit_young_ors}}

Here, we contextualize our results in the larger landscape of young giant planet frequency measurements. The demographics of giant planets has also been the focus of a number of RV programs targeting young stars \citep[e.g.,][]{Cochran2002, Paulson2004, Paulson2006, Prato2008, Crockett2012, Bailey2012, Lagrange2013, Donati2014, Gagne2016, Johns-Krull2016, Yu2017, Bailey2018, Tang2023}. However, due to the difficulty of detecting and characterizing planets around more magnetically active young stars, only a handful of RV campaigns have reported planetary occurrence rates at younger ages. \autoref{tab:ors} compiles measurements of the giant planet and brown dwarf occurrence rates from RV surveys targeting stars younger than 1 Gyr.

Our inferred frequencies of young HJs, giant planets interior to the water ice line, and BDs are consistent with values determined by other RV surveys. Altogether, it is clear that the frequency of giant planets and brown dwarfs orbiting within several AU are intrinsically low at intermediate ages. However, these constraints are fairly broad relative to measurements of field-age stars due to smaller sample sizes and more modest survey sensitivities. Further RV observations and a larger sample of young stars are needed to increase the precision of these occurrence rate measurements. This would enable a more robust constraint on how the giant planet frequency evolves over time, opening the possibility of confidently distinguishing between a constant and increasing frequency over time.

\subsection{Tracing Evolutionary Pathway\\ versus Birth Environment}

Our sample of young stars was assembled to match the stellar parameters of targets from \citet{Johnson2010}. This enables us to directly compare these two results in a fair way without inadvertently being impacted by known correlations in the occurrence rate of giant planets, such as with stellar metallicity and mass \citep{Fischer2005, Johnson2010, Buchhave2012, Nielsen2019}.

However, one of the less explored properties of these samples is stellar birth environment. The majority of field stars are expected to have formed in clusters embedded in giant molecular clouds \citep{Lada2003}. Differences in initial cluster properties, surrounding environment, and evolutionary histories, such as star formation histories and rates \citep[e.g.,][]{Pecaut2016}, proximity to other, more massive stars \citep{Hester2005}, and previous dynamical encounters \citep{Portegies2016} will all subsequently affect the sizes and lifetimes of protoplanetary disks \citep{Lynden-Bell1974, Haisch2001}. The formation and evolution of giant planets is expected to heavily depend on these fundamental disk properties.

If the birth environments of YMG stars observed in this study are distinct from those of the field stars observed in \citet{Johnson2010}, then observed trends in the prevalence of giant planets could reflect differences in stellar birth environment instead of (or in addition to) time evolution. Precise measurements of the giant planet frequency across many distinct well-characterized environments (e.g., open clusters and young associations) may enable us to identify this environmental bias.

\section{Summary\label{sec:summary}}

We have presented the results of the Epoch of Giant Planet Migration planet search program, a precise RV survey targeting intermediate-age Sun-like stars with the NIR HPF spectrograph located on McDonald Observatory's HET. Below, we summarize our main conclusions:

\begin{itemize}
    \item We use the HPF spectrograph to obtain 2666 spectra for 104 young stars. 18 systems were identified as binaries and 1 system was later determined to be older than 200 Myr, leaving 85 for our main young planet search program. Each target in our sample has an average of 9 epochs, 26 individual RVs, and a time baseline of $\sim$4~yr. From these RV observations, we estimate survey completeness of \HJcomp, \WJcomp, and \BDcomp for giant planets ($0.3 < m \sin i < 13$ $M_\mathrm{Jup}$) within 10 d, giant planets located within 2.5 AU, and brown dwarfs ($13 < m \sin i < 80$ $M_\mathrm{Jup}$) within 5000 d, respectively.
    \item We detect one young hot Jupiter candidate, HS Psc b. Taking into account completeness corrections, we estimate an occurrence rate of \WJOR for giant planets located interior to the water ice line of Sun-like stars. By assuming a power law relationship between our constraint at young ages and the observed frequency of giant planets at older ages, $6.5\pm0.7$\% \citep{Johnson2010}, we find a power law coefficient of $\alpha = 0.23^{+0.14}_{-0.26}$. This constraint favors an increasing giant planet occurrence rate over time. Furthermore, we are able to reject the possibility that the young frequency is 1.3$\times$ and 1.9$\times$ larger than the old age rate at 95\% and 99\% confidence, respectively.
    \item As we are sensitive to giant planets within 10~d and more massive companions, we also measure the prevalence of HJs and BDs at young ages. We determine occurrence rates of \HJOR and \BDOR, respectively, which agree with values from other young star RV surveys. Furthermore, this young HJ frequency is similar to measurements at field ages \citep{Wright2012, Fressin2013}.
    \item Altogether, these results indicate that young giant planets and brown dwarfs orbiting within several AU of their host stars are intrinsically rare, which is broadly similar to their older, field-age counterparts. Our measurement favors a picture in which the close-in giant planet population is a mixture of planets that are present early---perhaps through in situ formation or early inward migration---combined with planets that migrate in from beyond the snow line over longer timescales of $10^8$--$10^9$~yr. Larger samples are needed to more precisely constrain the giant planet occurrence rate at young ages and in turn place more stringent constraints on the timescale of giant planet migration.
\end{itemize}

\section{Acknowledgements}

We thank the referee for the constructive feedback which improved this manuscript, as well as Kyle Franson, Marvin Morgan, and Adam Kraus for insightful conversations on the evolution of giant planets and young stars. This manuscript benefited from discussions with Lily Zhao and Andrea Lin on radial velocity extraction, and Keith Hawkins and Caroline Morley on stellar characterization. The authors are also grateful to Steven Janowiecki, Greg Zeimann, and all resident astronomers and telescope operators at the HET for supporting these observations and data processing.

Q.H.T. and B.P.B. acknowledge the support from a NASA FINESST grant (80NSSC20K1554). B.P.B. acknowledges support from the National Science Foundation grant AST-1909209, NASA Exoplanet Research Program grant 20-XRP20$\_$2-0119, and the Alfred P. Sloan Foundation.

These results are based on observations obtained with the Habitable-zone Planet Finder Spectrograph on the HET. The HPF team acknowledges support from NSF grants AST-1006676, AST-1126413, AST-1310885, AST-1517592, AST-1310875, ATI 2009889, ATI-2009982, AST-2108512, AST-2108569, and the NASA Astrobiology Institute (NNA09DA76A) in the pursuit of precision radial velocities in the NIR. The HPF team also acknowledges support from the Heising-Simons Foundation via grant 2017-0494. The Hobby-Eberly Telescope is a joint project of the University of Texas at Austin, the Pennsylvania State University, Ludwig-Maximilians-Universität München, and Georg-August Universität Gottingen. The HET is named in honor of its principal benefactors, William P. Hobby and Robert E. Eberly. 
We acknowledge the Texas Advanced Computing Center (TACC) at The University of Texas at Austin for providing high performance computing, visualization, and storage resources that have contributed to the results reported within this paper. Computations for this research were also performed on the Pennsylvania State University’s Institute for Computational and Data Sciences Advanced CyberInfrastructure (ICDS-ACI, now known as Roar), including the CyberLAMP cluster supported by NSF grant MRI1626251.

We would like to acknowledge that the HET is built on Indigenous land. Moreover, we would like to acknowledge and pay our respects to the Carrizo \& Comecrudo, Coahuiltecan, Caddo, Tonkawa, Comanche, Lipan Apache, Alabama-Coushatta, Kickapoo, Tigua Pueblo, and all the American Indian and Indigenous Peoples and communities who have been or have become a part of these lands and territories in Texas, here on Turtle Island.

This paper includes data collected by the TESS mission. Funding for the TESS mission is provided by the NASA's Science Mission Directorate. This work presents results from the European Space Agency (ESA) space mission Gaia. Gaia data are being processed by the Gaia Data Processing and Analysis Consortium (DPAC). Funding for the DPAC is provided by national institutions, in particular the institutions participating in the Gaia MultiLateral Agreement (MLA). This research has made use of the Washington Double Star Catalog maintained at the U.S. Naval Observatory.

This research has made use of the VizieR catalogue access tool, CDS, Strasbourg, France (DOI: 10.26093/cds/vizier). The original description of the VizieR service was published in 2000, A\&AS 143, 23. This publication makes use of data products from the Two Micron All Sky Survey, which is a joint project of the University of Massachusetts and the Infrared Processing and Analysis Center/California Institute of Technology, funded by the National Aeronautics and Space Administration and the National Science Foundation. This publication makes use of data products from the Wide-field Infrared Survey Explorer, which is a joint project of the University of California, Los Angeles, and the Jet Propulsion Laboratory/California Institute of Technology, funded by the National Aeronautics and Space Administration.

\facilities{HET (HPF)}

\software{\texttt{astropy} \citep{Astropy2018},
    \texttt{astroquery} \citep{Ginsburg2019},
    \texttt{barycorrpy} \citep{Kanodia2018a},
    \texttt{emcee} \citep{Foreman-Mackey2013, Foreman-Mackey2019},
    \texttt{HxRGproc} \citep{Ninan2018},
    \texttt{lightkurve} \citep{Lightkurve2018},
    \texttt{matplotlib} \citep{Hunter4160265},
    \texttt{MultiNest} \citep{Feroz2009, Feroz2019},
    \texttt{numpy} \citep{vanderWalt2011},
    \texttt{pandas} \citep{mckinney-proc-scipy-2010},
    \texttt{pyaneti} \citep{Barragan2019, Barragan2022},
    \texttt{PyMultiNest} \citep{Buchner2014},
    \texttt{radvel} \citep{Fulton2018}, 
    \texttt{scipy} \citep{Virtanen2020},
    \texttt{SpecMatch-Emp} \citep{Yee2017},
    \texttt{SERVAL} \citep{Zechmeister2018}.
    }

\appendix

\section{Systems Excluded from Statistical Sample\label{sec:appendix_excluded}}

\begin{figure}[!]
    \centering
    \includegraphics[width=\linewidth]{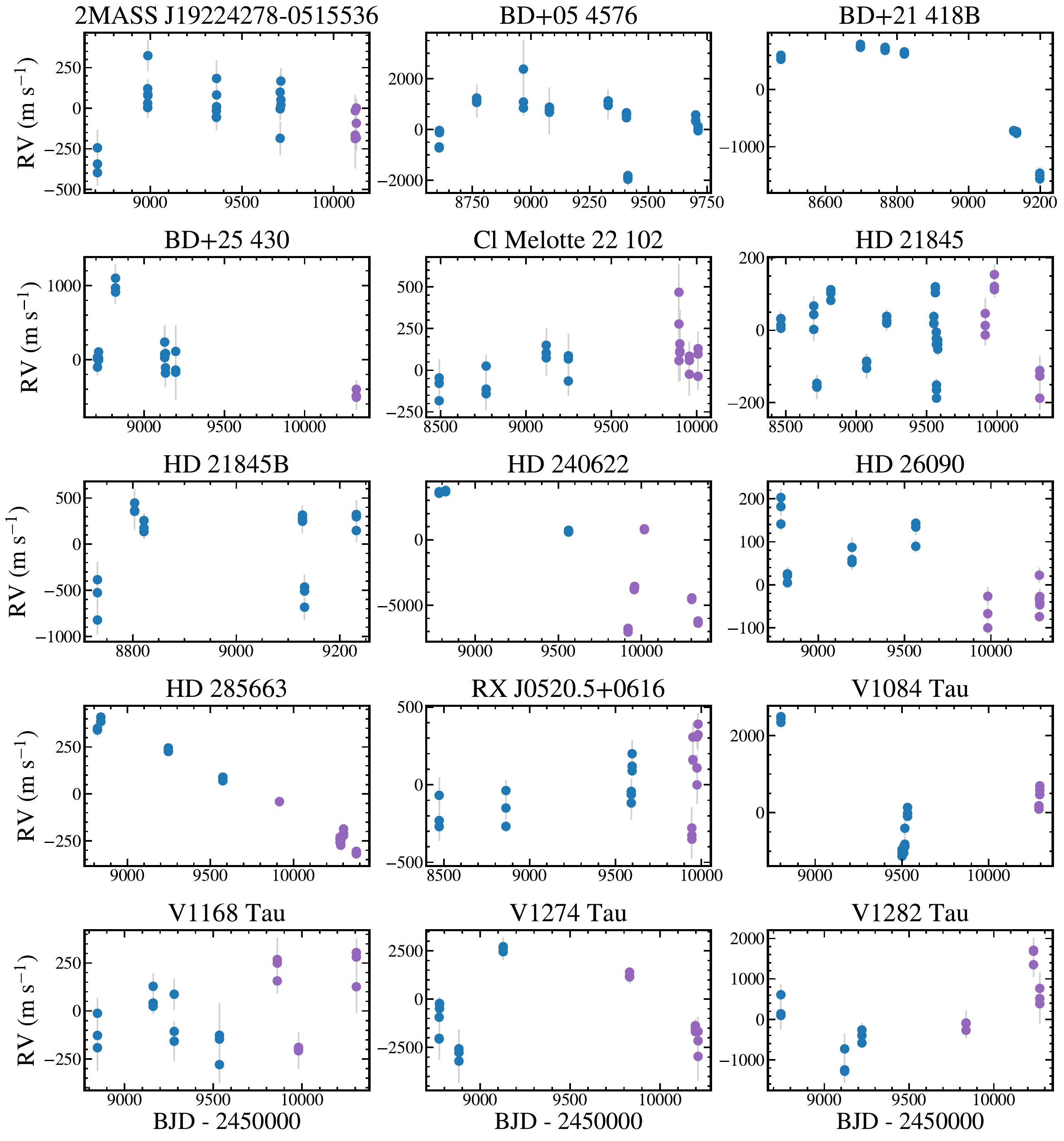}
    \caption{HPF RVs of excluded binary systems (2MASS J19224278-0515536, BD+05 4576, BD+21 418B, BD+25 430, Cl Melotte 22 102, HD 21845, HD 21845B, HD 240622, HD 26090, HD 285663, RX J0520.5+0616, V1084 Tau, V1168 Tau, V1274 Tau, and V1282 Tau). Pre- and post-maintenance HPF RVs are plotted as blue and purple points, respectively. Post-maintenance RVs are corrected with the empirically measured velocity offset reported in \autoref{sec:appendix_offset}. Long-period orbital motion is evident for BD+21 418B and HD 285663.}
    \label{fig:excluded_rvs0}
\end{figure}

18 binaries and 1 field-age target were removed from the final statistical sample. We briefly discuss the properties of these systems below and summarize them in \autoref{tab:excluded_sample}. \Cref{fig:excluded_rvs0,fig:excluded_rvs1} display the HPF RV measurements of all 19 systems excluded from our statistical analysis. \autoref{tab:excluded_hpf_measurements} in \autoref{sec:appendix_measurements} reports the measurements and uncertainties for the relative HPF RVs and spectroscopic indicators for these 19 systems.

\subsection{Stellar Binaries}

Five systems were recently discovered to be long-period spectroscopic binaries in the Pleiades cluster (Cl Melotte 22 102, V1084 Tau, V1168 Tau, V1282 Tau, V366 Tau) identified by \citet{Torres2021}. \citet{Torres2021} used nearly 40 years of RV data to reveal a long-term trend indicating that Cl Melotte 22 102 is a binary in a hierarchical triple system with a period longer than the baseline of their RV observations. V1084 Tau is a double-lined spectroscopic binary on a 757-d orbit. The primary and secondary components have RV semi-amplitudes of 12.25 and 14.30 km~s$^{-1}$, respectively. \citet{Torres2021} determined that V1168 Tau is a binary with an orbital period that is substantially longer than the duration of their observational baseline (40~yr). V1282 Tau was characterized in detail by \citet{Torres2020} using a combination of very long baseline interferometry and spectroscopic observations. It has an orbital period of $18.18\pm0.11$~yr and the components were resolved with an angular separation of $62.33^{+0.45}_{-0.42}$~mas. V366 Tau is a double-lined spectroscopic binary with a 542~d orbit that is in a hierarchical triple system.

BD+21 418 B, HD 21845 B, HD 21845, HD 240622, and HD 285663 have nearby faint sources reported in the Washington Double Star catalog \citep{Mason2001}. BD+21 418 B and its primary component (BD+21 418) are separated by $34\farcs8$ (1760 AU), but BD+21 418 B is itself a close binary. Both HD 240622 and HD 285663 have a faint visual companion ($\Delta V$ = 3.2~mag and 3.4~mag, respectively) listed in the Washington Double Star catalog \citep{Mason2001} with angular and projected physical separations of $\rho = 0\farcs2$ (19~AU) and $\rho = 0\farcs3$ (15~AU), respectively, which exclude them from our statistical sample. Long-term trends, which we attribute to the companions, are evident in the HPF RVs of both systems. Similarly, HD 21845 B and HD 21845 form a wide binary with a projected physical separation of 334~AU ($\rho = 9\farcs2$).

Furthermore, five systems are previously identified binaries or later identified as non-single stars by Gaia. HD 26090 is a known double-lined spectroscopic binary on a 60~yr orbit \citep{Griffin2001} and V1874 Ori is a known single-line spectroscopic binary on a 463-d orbit \citep{Torres2002}. BD+05 4576 and V1274 Tau were both compatible with the single lined spectroscopic binary model in the Gaia DR3 non-single star catalog \citep{GaiaDR3_2022, Gaia2022}. They have best-fit orbital periods of 61.2~d and 5.6~d, respectively. BD+25 430 is most compatible with the variable acceleration solution, which indicates an astrometric acceleration due to an intermediate-period companion.

Finally, we assess the Gaia re-normalized unit weight error (RUWE) and the Gaia Image Parameters Determination fraction of multiple peaks (IPDFMP) for the 104 stars in our sample.\footnote{RUWE and IPDFMP values are from Gaia DR3 if available. 2MASS J19224278-0515536 and BD+21 418 B RUWE values are obtained from Gaia DR2.} The Gaia RUWE is an astrometric goodness-of-fit statistic, where $\mathrm{RUWE} > 1.4$ is an approximate threshold that generally signifies a poor single-star model fit \citep{Lindegren2018, Belokurov2020}. The IPDFMP parameter is the percentage of astrometric scans with multiple peaks identified, indicating the possible detection of a visually resolved double star. Known single stars with $\mathrm{RUWE} < 1.4$ in our sample exhibit IPDFMPs below 2\%, with the majority having values of 0\%. Thus, stars with RUWE values significantly exceeding 1.4 and high IPDFMPs are likely binaries.

We find that 19 targets have RUWE values greater than $1.4$. Of these 19 stars, 12 are among the known or newly identified binaries detailed above. The RUWE values of the remaining 7 targets are either moderate ($\leq1.5$; 2MASS J05200029+0613036, 2MASS J05234246+0651581, and HW Cet) or large ($>2.2$; 2MASS J19224278-0515536, RX J0520.5+0616, V395 Peg, and V700 Tau). Notably, the RUWE value for V395 Peg is 19.0. The three stars with RUWE excess between 1.4--1.5 have IPDFMPs of 0\%. On the other hand, 2MASS J19224278-0515536, RX J0520.5+0616, and V700 Tau with RUWE excesses greater than 2.2 and IPDFMPs greater than 20\%. The combination of high IPDFMP and/or RUWE suggests that 2MASS J19224278-0515536, RX J0520.5+0616, V395 Peg, and V700 Tau are likely binary systems.

\subsection{The Field-age System (HD 130322)\label{sec:appendix_HD_130322}}

HD 130322 is a K0 dwarf originally identified as a member of the $110\pm15$~Myr Pleiades cluster \citep{Montes2001a, Gaia2018}. It has a Gaia DR3 parallactic distance of $31.89^{+0.07}_{-0.03}$~pc \citep{GaiaDR3_2022} and a stellar isochrone-inferred mass of $0.79$~$M_\odot$ \citep{Udry2000}.

Multiple approaches have resulted in inconsistent age estimates for HD 130322. \citet{Udry2000} noted a chromospheric activity indicator value of log$\:R^\prime_\mathrm{HK} = -4.39$ that suggested an age of $\sim$350 Myr based on empirical relations \citep{Donahue1993, Garcia2001} and kinematic properties comparable with young Pleiades cluster members. Similarly, \citet{Montes2001a} classified HD 130322 as a member of the Pleiades cluster, which is inconsistent with the distance of 31~pc. The BANYAN $\Sigma$ young association classification tool predicts an 84\% and 16\% probability that HD 130322 is a member of the AB Dor moving group or a field star, respectively \citep{Gagne2018a}. However, subsequent log$\:R^\prime_\mathrm{HK}$ measurements are lower than $-4.6$, supporting an age older than several Gyr for HD 130322 (\citealp[$-4.78$,][]{Wright2004}; \citealp[$-4.64$,][]{Brewer2016}; \citealp[$-4.70$,][]{Meunier2017}). Inferences from lithium abundances similarly predict ages greater than several Gyr \citep{Ramirez2012, Aguilera-Gomez2018, LlorentedeAndres2021}. Finally, The Bayesian age-dating framework BAFFLES predicts an age of $\sim$3.5~Gyr \citep{Stanford-Moore2020}. Based on these various lines of evidence, we consider HD 130322 to be a field-age star. Consequently, we exclude the system from our sample as it no longer meets our target selection criteria.

\begin{longrotatetable}
\begin{deluxetable}{ccccccccccccc}
    \centerwidetable
    \tabletypesize{\footnotesize}
    \setlength{\tabcolsep}{6pt}
    \tablecaption{Properties of Excluded Systems. \label{tab:excluded_sample}}
    \tablehead{
    \colhead{Object} &
    \colhead{$J$\tablenotemark{a}} & \colhead{$N_\mathrm{epoch}$\tablenotemark{b}} &
    \colhead{$N_\mathrm{spec}$\tablenotemark{c}} &
    \colhead{$\Delta t$} & 
    \colhead{$\overline{\mathrm{S/N}}$\tablenotemark{d}} &
    \colhead{$\overline{\sigma_\mathrm{RV}}$} &
    \colhead{RV RMS} &
    \colhead{$v \sin i$\tablenotemark{e}} &
    \colhead{SpT} &
    \colhead{SpT} &
    \colhead{YMG} &
    \colhead{YMG} \\
    \colhead{Name} &
    \colhead{(mag)} &
    \colhead{} &
    \colhead{} &
    \colhead{(d)} & 
    \colhead{(pixel$^{-1}$)} & 
    \colhead{(m s$^{-1}$)} &
    \colhead{(m s$^{-1}$)} &
    \colhead{(km s$^{-1}$)} &
    \colhead{} &
    \colhead{Ref.} &
    \colhead{} &
    \colhead{Ref.}
    }
    \startdata
2MASS J19224278-0515536 & 9.9 & 9 & 27 & 1413.1 & 66 & 83.8 & 161.5 & 10.2$\pm$1.0 & K5 & 1 & Arg & 1 \\
BD+05 4576 & 7.9 & 10 & 30 & 1103.0 & 166 & 362.7 & 986.3 & 4.5$\pm$0.7 & K7 & 2 & AB Dor & 2 \\
BD+21 418B & 8.4 & 7 & 21 & 723.0 & 150 & 80.3 & 880.5 & 5.9$\pm$0.7 & K6 & 3 & AB Dor & 3 \\
BD+25 430 & 7.9 & 7 & 21 & 1613.8 & 180 & 182.5 & 431.3 & 3.4$\pm$1.1 & G0 & 4 & $\beta$ Pic & 5 \\
Cl Melotte 22 102 & 9.1 & 8 & 24 & 1516.8 & 115 & 109.7 & 138.2 & 31.9$\pm$2.5 & G1 & 6 & Pleiades & 7 \\
HD 130322 & 6.7 & 10 & 33 & 1527.9 & 288 & 9.7 & 71.9 & 2.0$\pm$0.1 & K0 & 8 & Field & $\cdots$ \\
HD 21845 & 6.8 & 14 & 43 & 1837.2 & 284 & 21.7 & 95.6 & 9.7$\pm$0.6 & K2 & 9 & AB Dor & 10 \\
HD 21845B & 8.4 & 6 & 18 & 499.8 & 128 & 138.0 & 424.9 & 23.0$\pm$1.5 & K7 & 11 & AB Dor & 10 \\
HD 240622 & 7.6 & 9 & 27 & 1557.8 & 196 & 142.4 & 3906.0 & 10.2$\pm$2.5 & G0 & 9 & Carina-Near & 12 \\
HD 26090 & 7.1 & 7 & 21 & 1498.9 & 247 & 16.3 & 88.4 & 10.5$\pm$2.5 & G0 & 13 & Carina-Near & 12 \\
HD 285663 & 7.8 & 9 & 25 & 1558.9 & 197 & 20.3 & 275.1 & 3.0$\pm$0.7 & K2 & 4 & Carina-Near & 12 \\
RX J0520.5+0616 & 9.2 & 8 & 24 & 1507.9 & 94 & 82.6 & 230.8 & 13.3$\pm$2.3 & K3 & 14 & 32 Ori & 12 \\
V1084 Tau & 9.5 & 7 & 21 & 1491.9 & 84 & 103.0 & 1145.1 & 30.7$\pm$1.9 & G5 & 15 & Pleiades & 7 \\
V1168 Tau & 9.9 & 7 & 21 & 1459.8 & 71 & 84.0 & 186.6 & 32.2$\pm$1.9 & G8 & 15 & Pleiades & 7 \\
V1274 Tau & 8.6 & 7 & 21 & 1440.1 & 118 & 644.5 & 1907.9 & 31.9$\pm$1.7 & K5 & 4 & AB Dor & 5 \\
V1282 Tau & 9.2 & 6 & 18 & 1527.1 & 106 & 283.4 & 877.1 & 32.4$\pm$1.5 & G7 & 6 & Pleiades & 7 \\
V1874 Ori & 10.1 & 8 & 24 & 1482.1 & 61 & 522.8 & 4244.6 & 20.3$\pm$1.8 & K7 & 16 & 32 Ori & 12 \\
V366 Tau & 10.6 & 7 & 21 & 1401.2 & 54 & 294.1 & 1812.6 & 15.7$\pm$1.7 & K7 & 17 & Pleiades & 7 \\
V395 Peg & 9.5 & 13 & 39 & 1506.9 & 90 & 210.4 & 246.1 & 32.3$\pm$1.6 & G8 & 18 & Pisces & 18 \\
    \enddata
    \tablenotetext{a}{\citet{Cutri2003}}
    \tablenotetext{b}{Number of visits, which are comprised of three contiguous observations, for each target. See \autoref{sec:hpf_obs} for more details.}
    \tablenotetext{c}{Number of individual observations, or RVs, for each target. See \autoref{sec:hpf_obs} for more details.}
    \tablenotetext{d}{Signal-to-noise estimated at 1.07~$\mu$m.}
    \tablenotetext{e}{Projected rotational velocity measured in this work (see \autoref{sec:characterization}).}
    \tablerefs{(1) \citet{Malo2013}; 
(2) \citet{Malo2014}; 
(3) \citet{Zuckerman2004}; 
(4) \citet{Nesterov1995}; 
(5) \citet{Gagne2018b}; 
(6) \citet{White2007}; 
(7) \citet{Stauffer2007}; 
(8) \citet{Montes2001b};  
(9) \citet{Cannon1993}; 
(10) \citet{Torres2008}; 
(11) \citet{Mann2013}; 
(12) \citet{Gagne2018a}; 
(13) \citet{Stephenson1986}; 
(14) \citet{Alcala1996}; 
(15) \citet{Soderblom1993}; 
(16) \citet{Shvonski2016}; 
(17) \citet{Prosser1991}; 
(18) \citet{Binks2018}.}
\end{deluxetable}
\end{longrotatetable}

\begin{figure}
    \centering
    \includegraphics[width=\linewidth]{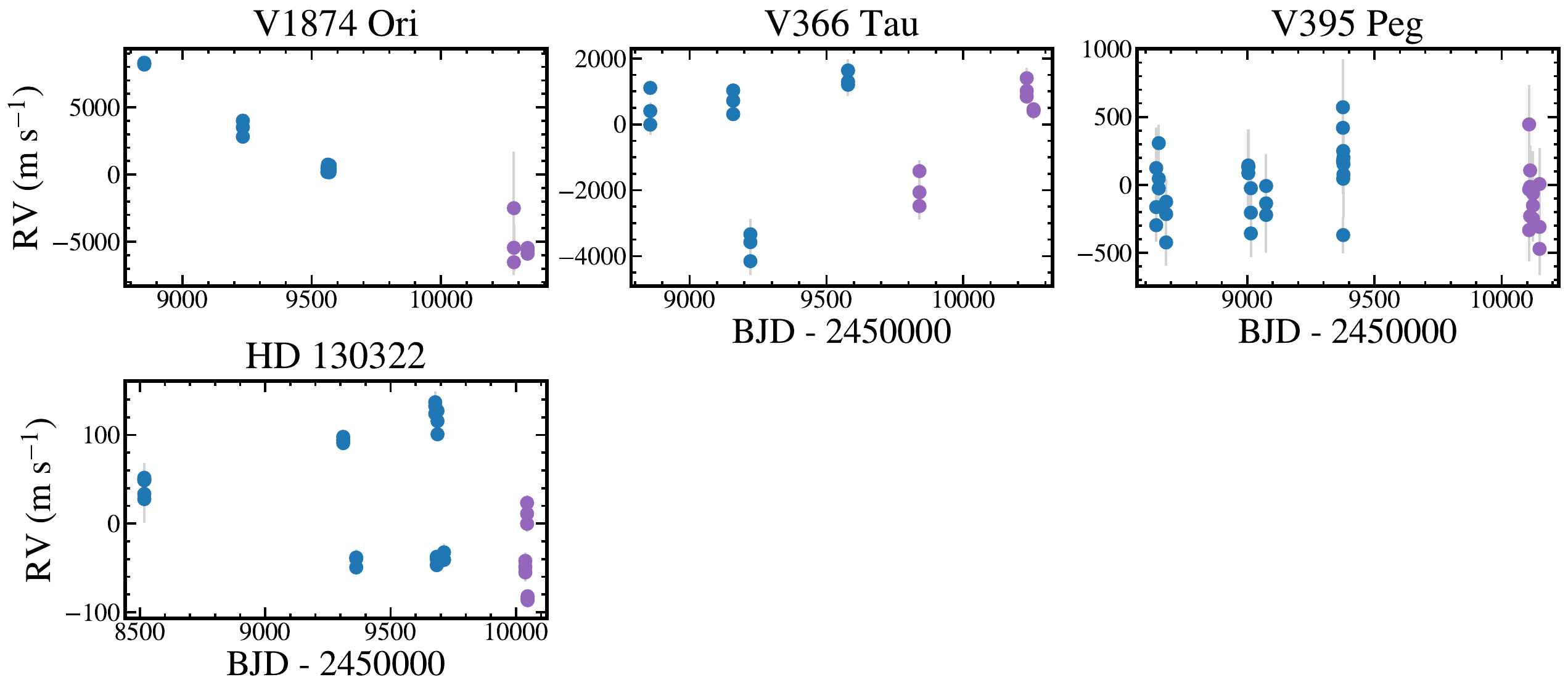}
    \caption{HPF RVs of excluded binary (V1874 Ori, V366 Tau, and V395 Peg) and field-age (HD 130322) systems. Long-period orbital motion is evident for V1874 Ori.}
    \label{fig:excluded_rvs1}
\end{figure}

\clearpage

\subsubsection{Radial Velocity Modeling of HD 130322  b\label{sec:appendix_HD_130322_modeling}}

HD 130322 can be used as a diagnostic to test the detection limits of our HPF RVs. HD 130322 was first reported to host a Jupiter-mass planet on an $\sim$11~d orbit by \citet{Udry2000}. The properties of HD 130322 b were further refined by \citet{Hinkel2015}, who combined new observations with previous measurements from \citet{Udry2000}, \citet{Butler2006}, and \citet{Wittenmyer2009b} to measure a precise RV semi-amplitude of $112.5 \pm 2.4$~m~s$^{-1}$, a minimum mass of $m \sin \; i = 1.15 \pm 0.04$~$M_\mathrm{Jup}$, orbital period of $P = 10.7087 \pm 0.0002$~d, and an eccentricity of $e = 0.03 \pm 0.02$.

Using 33 HPF RVs over 4.2~yr, we detect a significant periodicity at 10.71~d, coinciding with the known orbital period of HD 130322 b. The HPF RVs have an RMS of 70.8~m~s$^{-1}$, peak-to-peak amplitude of 223.4~m~s$^{-1}$, and an average measurement uncertainty of 9.7~m~s$^{-1}$ (\autoref{fig:excluded_rvs1}). We perform the first multi-wavelength characterization of HD 130322 b using a combination of NIR HPF RVs and previously reported optical RVs \citep{Udry2000, Butler2006, Wittenmyer2009b, Hinkel2015} in \autoref{sec:appendix_HD_130322_modeling}.

We use the \texttt{pyaneti} modeling suite \citep{Barragan2019, Barragan2022} to perform a multi-wavelength Keplerian model fit on both new NIR HPF RVs and previously reported optical RVs of HD 130322 for 5 orbital parameters: orbital period ($P_b$), time of inferior conjunction ($T_{0, b}$), RV semi-amplitude ($K_b$), and parameterized forms of eccentricity and argument of periastron ($\sqrt{e_b} \sin \omega_*$ and $\sqrt{e_b} \cos \omega_*$). We also fit for zero-point velocity offsets ($\gamma_i$) and ``jitter'' ($\sigma_i$) terms for each set of RV observations. Based on the well-characterized nature of HD 130322 b, we adopt informed priors from the inferred parameters of \citet{Hinkel2015}. We impose uniform priors on the eccentricity parameters, $\mathcal{P}(\sqrt{e_b} \sin \omega_*) = \mathcal{U}[-1, 1]$ and $\mathcal{P}(\sqrt{e_b} \cos \omega_*) = \mathcal{U}[-1, 1]$, and RV semi-amplitude, $\mathcal{P}(K_b) = \mathcal{U}[50.0, 200.0]$~m~s$^{-1}$. For the orbital period and time of inferior conjunction, we adopt Gaussian priors where the mean and standard deviation are equal to the inferred values and uncertainties inflated by a factor of 10 from \citet{Hinkel2015}, $\mathcal{P}(P_b) = \mathcal{N}[10.7087, 0.0018]$~d and $\mathcal{P}(T_{0, b}) = \mathcal{U}[2456745.594, 0.85]$~d, respectively. Parameter posterior distributions are sampled using a Markov chain Monte Carlo Metropolis-Hasting algorithm \citep{Sharma2017}. Each distribution comprises 50 independent chains of 100,000 iterations with a thinning factor of 10. A Gelman-Rubin statistic threshold of 1.02 is used to determine convergence \citep{Gelman1992}.

\begin{figure}[!t]
    \centering
    \includegraphics[width=\linewidth]{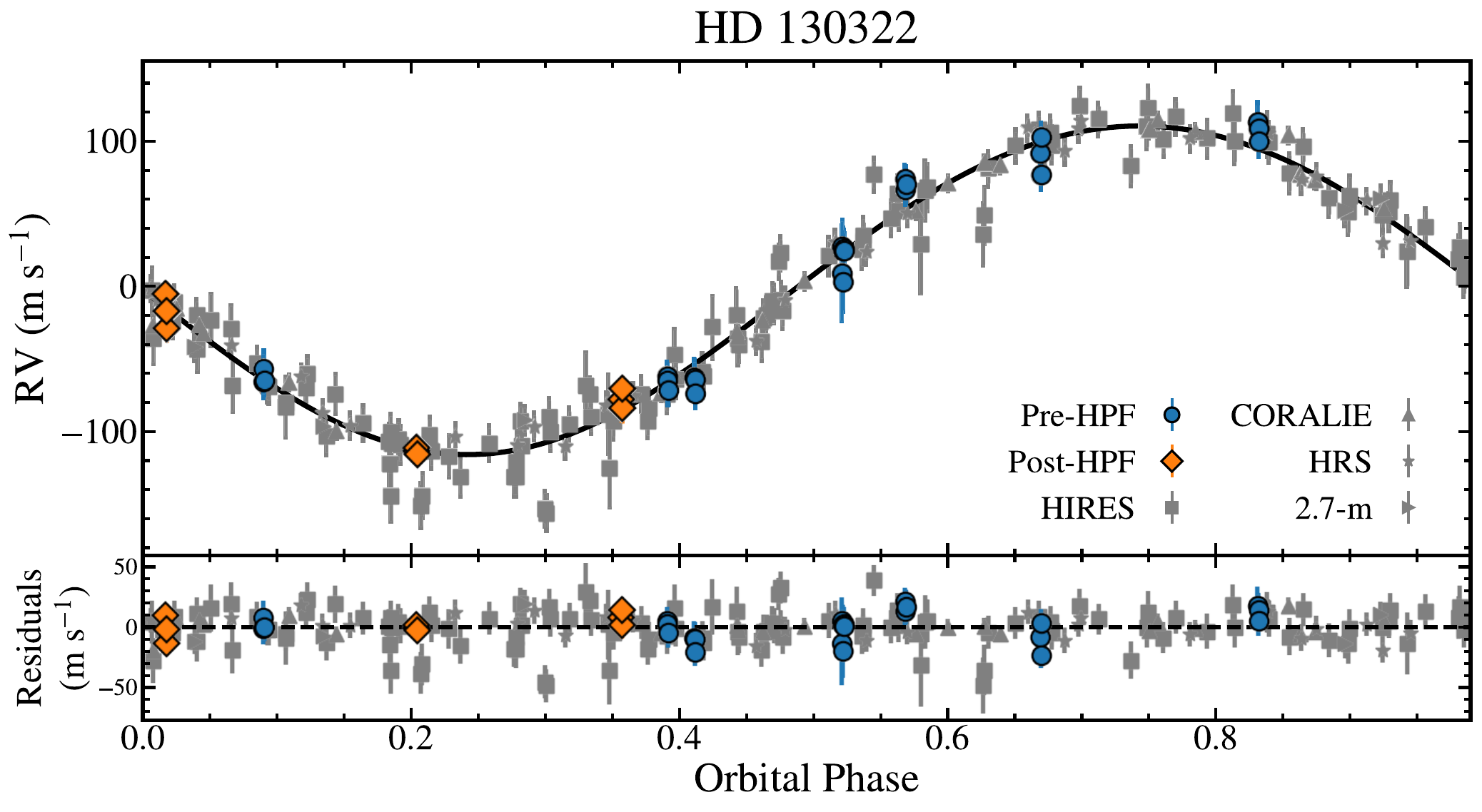}
    \caption{RV curve of HD 130322 phased folded to the best-fit orbital period. The blue and orange points denote the pre- and post-maintenance HPF RVs, respectively, and the best-fit Keplerian model is shown as the solid black line. The grey points represent RV measurements from other instruments. The fit residuals are plotted in the lower panels.}
    \label{fig:app_HD_130322_phase}
\end{figure}

\autoref{tab:app_HD_130322_params} summarizes the results of the Keplerian model fit and \autoref{fig:app_HD_130322_phase} displays the best-fit RV curve phased to the orbital period of HD 130322 b. Our inferred planetary properties are in excellent agreement with the values reported by \citet{Hinkel2015}. The addition of the HPF RVs improve the time baseline of the observations to 24~yr, enabling us to further refine the properties of HD 130322 b.

\begin{deluxetable}{lcc}[!p]
    \setlength{\tabcolsep}{5em}
    \tablecaption{\label{tab:app_HD_130322_params} Parameter Priors and Posteriors from Keplerian Model fit of HD 130322 RVs.}
    \tablehead{\colhead{Parameter} & 
    \colhead{Prior\tablenotemark{a}} & \colhead{MAP\tablenotemark{b} $\pm$ 1$\sigma$}}
    \startdata
    \multicolumn{3}{c}{Fitted Parameters} \\
    \hline
    $P_b$ (d) & $\mathcal{N}(10.7087, 0.0018)$ & $10.7085 \pm 0.0001$ \\
    $T_{0, b}$ (d) & $\mathcal{N}(2456745.594, 0.85)$ & $2456745.49 \pm 0.05$ \\
    $K_b$ (m s$^{-1}$) & $\mathcal{U}[50.0, 200.0]$ & $112.8 \pm 1.3$ \\
    $\sqrt{e_b} \sin{\omega_*}$ & $\mathcal{U}[-1, 1]$ & $-0.001_{-0.061}^{+0.062}$ \\
    $\sqrt{e_b} \cos{\omega_*}$ & $\mathcal{U}[-1, 1]$ & $-0.136_{-0.040}^{+0.061}$ \\
    $\sigma_{\mathrm{pre-HPF}}$ (m s$^{-1}$) & $\mathcal{J}(1, 100)$ & $10.3_{-2.0}^{+2.5}$ \\
    $\sigma_{\mathrm{post-HPF}}$ (m s$^{-1}$) & $\mathcal{J}(1, 100)$ & $2.9_{-2.3}^{+4.9}$ \\
    $\sigma_{\mathrm{HIRES}}$ (m s$^{-1}$) & $\mathcal{J}(1, 100)$ & $7.3_{-1.0}^{+1.3}$ \\
    $\sigma_{\mathrm{CORALIE}}$ (m s$^{-1}$) & $\mathcal{J}(1, 100)$ & $9.8_{-1.5}^{+1.6}$ \\
    $\sigma_{\mathrm{HRS}}$ (m s$^{-1}$) & $\mathcal{J}(1, 100)$ & $6.7_{-1.8}^{+1.9}$ \\
    $\sigma_{\mathrm{2.7m}}$ (m s$^{-1}$) & $\mathcal{J}(1, 100)$ & $11.2_{-6.8}^{+9.3}$ \\
    $\gamma_{\mathrm{pre-HPF}}$ (km s$^{-1}$) & $\mathcal{U}[-0.5495, 0.6370]$ & $0.025 \pm 0.003$ \\
    $\gamma_{\mathrm{post-HPF}}$ (km s$^{-1}$) & $\mathcal{U}[-0.5865, 0.5234]$ & $0.029 \pm 0.003$ \\
    $\gamma_{\mathrm{HIRES}}$ (km s$^{-1}$) & $\mathcal{U}[-0.6278, 0.5860]$ & $-0.028 \pm 0.002$ \\
    $\gamma_{\mathrm{CORALIE}}$ (km s$^{-1}$) & $\mathcal{U}[-0.6340, 0.6460]$ & $0.022 \pm 0.001$ \\
    $\gamma_{\mathrm{HRS}}$ (km s$^{-1}$) & $\mathcal{U}[-0.6132, 0.6104]$ & $-0.002 \pm 0.002$ \\
    $\gamma_{\mathrm{2.7m}}$ (km s$^{-1}$) & $\mathcal{U}[-0.5728, 0.5837]$ & $0.02 \pm 0.01$ \\
    \hline
    \multicolumn{3}{c}{Derived Parameters} \\
    \hline
    $m_b \sin i$\tablenotemark{c} $(M_\mathrm{Jup})$ & $\cdots$ & $1.16 \pm 0.03$ \\
    $T_\mathrm{peri, b}$ (d) & $\cdots$ & $2456748.1_{-0.9}^{+0.8}$ \\
    $e_b$ & $\cdots$ & $0.022 \pm 0.012$ \\
    $\omega_*$ $(\degree)$ & $\cdots$ & $-36.7_{-132.6}^{+205.7}$ \\
    \enddata
    \tablenotetext{a}{$\mathcal{U}[a, b]$ refers to the uniform distribution bounded by $a$ and $b$. $\mathcal{N}(a, b)$ describes a Gaussian distribution with average $a$ and standard deviation $b$. $\mathcal{J}(a, b)$ refers to the modified Jeffreys prior as defined in Equation 6 of  \citet{Gregory2005}, $\mathcal{P}(x) = 1/(a + x) \cdot 1/\ln \left[\left(a + b\right) / a\right]$.}
    \tablenotetext{b}{MAP refers to the \textit{maximum a posteriori} value.}
    \tablenotetext{c}{Planetary mass is derived assuming a stellar mass of $M_* = 0.92 \pm 0.03$ $M_\odot$.}
\end{deluxetable}

\clearpage

\section{Long-term RV Trends\label{sec:appendix_long-term_trend}}

Aside from the binaries discussed in \autoref{sec:appendix_excluded}, we identify one long-term RV trend in the HPF RV time series of HD 24194 (\autoref{fig:rv_comp8}). We fit a linear model using the \texttt{emcee} open Python package \citep{Foreman-Mackey2013, Foreman-Mackey2019} and measure a slope of $65.1 \pm 4.6$~m~s$^{-1}$~yr$^{-1}$ ($0.178 \pm 0.013$~m~s$^{-1}$~d$^{-1}$) over a baseline of 1758~d. The peak-to-peak amplitude of the HPF RVs is 365~m~s$^{-1}$. Following the framework from \citet{Torres1999} and the Monte Carlo approach from \citet{Bowler2021}, we use the RV trend and the Gaia DR3 distance ($131.37_{-0.55}^{+0.36}$~pc) to constrain the mass and orbital separation of a companion that would produce a trend of this amplitude. The observed slope is most consistent with a giant planet with a mass between $\sim$5--15~$M_\mathrm{Jup}$ orbiting beyond the water ice line and within $\sim$4~AU, a brown dwarf orbiting between $\sim$4--9~AU, or a star beyond $\sim$9~AU. We are unable to determine the true nature of the companion with the current RV data, but future observations will be able to identify the true nature of this companion. Gaia DR4, which will be sensitive to massive companions beyond several AU \citep{Perryman2014}, is an especially promising pathway.

\section{Properties of Targets in the EGPM Survey\label{sec:appendix_targ_info}}

\autoref{tab:stellar_sample} reports properties of our sample, including information on RV observations and physical characteristics.

\begin{longrotatetable}
\begin{deluxetable}{ccccccccccccc}
    \centerwidetable
    \tabletypesize{\small}
    \setlength{\tabcolsep}{6pt}
    \tablecaption{Properties of the EGPM Sample of Young Stars.\label{tab:stellar_sample}}
    \tablehead{
    \colhead{Object} &
    \colhead{$J$\tablenotemark{a}} & \colhead{$N_\mathrm{epoch}$\tablenotemark{b}} &
    \colhead{$N_\mathrm{spec}$\tablenotemark{c}} &
    \colhead{$\Delta t$} & 
    \colhead{$\overline{\mathrm{S/N}}$\tablenotemark{d}} &
    \colhead{$\overline{\sigma_\mathrm{RV}}$} &
    \colhead{RV RMS} &
    \colhead{$v \sin i$\tablenotemark{e}} &
    \colhead{SpT} &
    \colhead{SpT} &
    \colhead{YMG} &
    \colhead{YMG} \\
    \colhead{Name} &
    \colhead{(mag)} &
    \colhead{} &
    \colhead{} &
    \colhead{(d)} & 
    \colhead{(pixel$^{-1}$)} & 
    \colhead{(m s$^{-1}$)} &
    \colhead{(m s$^{-1}$)} &
    \colhead{(km s$^{-1}$)} &
    \colhead{} &
    \colhead{Ref.} &
    \colhead{} &
    \colhead{Ref.}
    }
    \startdata
2MASS J03402958+2333040 & 11.7 & 7 & 21 & 1501.9 & 30 & 242.1 & 330.1 & 4.9$\pm$2.0 & K5 & 1 & Pleiades & 2 \\
2MASS J03433440+2345429 & 9.9 & 7 & 21 & 1818.1 & 75 & 49.4 & 60.1 & 5.1$\pm$0.5 & G8 & 3 & Pleiades & 2 \\
2MASS J03444398+2413523 & 9.9 & 7 & 21 & 1479.2 & 79 & 54.8 & 68.7 & 6.9$\pm$0.8 & G8 & 3 & Pleiades & 2 \\
2MASS J03445383+2355165 & 9.3 & 7 & 21 & 1586.9 & 93 & 158.0 & 207.3 & 30.3$\pm$4.9 & G1 & 4 & Pleiades & 2 \\
2MASS J03474811+2313053 & 9.8 & 7 & 21 & 1601.8 & 74 & 47.1 & 51.1 & 2.7$\pm$0.5 & G7 & 5 & Pleiades & 2 \\
2MASS J03515733+2320219 & 11.0 & 10 & 30 & 1502.9 & 42 & 136.1 & 121.1 & 5.5$\pm$1.1 & K4 & 3 & Pleiades & 2 \\
2MASS J05234246+0651581 & 9.9 & 7 & 21 & 1516.0 & 73 & 462.8 & 649.2 & 32.3$\pm$1.8 & K6 & 6 & 32 Ori & 5 \\
ASAS J232157+0721.3 & 9.4 & 11 & 33 & 1500.9 & 89 & 118.1 & 173.9 & 31.7$\pm$1.9 & K0 & 7 & Pisces & 8 \\
BD+11 1690 & 7.5 & 8 & 24 & 1505.7 & 218 & 12.7 & 13.3 & 2.3$\pm$0.9 & G5 & 9 & AB Dor & 10 \\
BD+17 455 & 7.6 & 7 & 21 & 1594.8 & 240 & 8.9 & 9.6 & 6.6$\pm$1.4 & K0 & 5 & Carina-Near & 5 \\
BD+17 641 & 8.2 & 8 & 24 & 1594.6 & 176 & 17.3 & 14.3 & 11.5$\pm$4.1 & G0 & 5 & Carina-Near & 5 \\
BD+20 1790 & 7.6 & 6 & 18 & 1852.7 & 225 & 38.3 & 160.0 & 10.6$\pm$1.1 & K5 & 11 & AB Dor & 12 \\
BD+21 418 & 7.3 & 7 & 21 & 1832.0 & 262 & 24.3 & 41.3 & 31.4$\pm$1.5 & G5 & 13 & AB Dor & 13 \\
BD+21 504 & 8.8 & 7 & 21 & 1503.9 & 124 & 101.1 & 96.2 & 32.0$\pm$1.3 & G0 & 9 & Pleiades & 2 \\
BD+22 548 & 9.0 & 7 & 21 & 1444.9 & 120 & 105.8 & 74.2 & 31.2$\pm$2.4 & G0 & 14 & Pleiades & 2 \\
BD+23 527 & 9.3 & 7 & 21 & 1579.7 & 102 & 46.6 & 58.8 & 4.9$\pm$0.9 & G0 & 15 & Pleiades & 2 \\
BD+25 610 & 9.0 & 7 & 21 & 1475.0 & 111 & 43.5 & 49.3 & 12.8$\pm$2.3 & G0 & 9 & Pleiades & 5 \\
BD+26 592 & 9.0 & 8 & 24 & 1599.6 & 103 & 40.1 & 55.3 & 9.4$\pm$2.0 & G0 & 9 & Pleiades & 2 \\
BD+41 4749 & 7.6 & 13 & 39 & 1651.7 & 187 & 16.4 & 20.2 & 4.4$\pm$0.4 & G4 & 13 & AB Dor & 13 \\
BD+49 646 & 8.0 & 8 & 24 & 1754.0 & 168 & 115.3 & 339.9 & 32.0$\pm$2.4 & K0 & 16 & AB Dor & 10 \\
BD-03 5579 & 8.6 & 10 & 30 & 758.0 & 126 & 53.1 & 133.1 & 11.2$\pm$1.4 & K4 & 7 & Car & 7 \\
BD-05 1229 & 7.4 & 9 & 27 & 1552.7 & 217 & 10.2 & 8.7 & 1.5$\pm$0.6 & G5 & 17 & $\beta$ Pic & 10 \\
BD-08 1195 & 8.5 & 8 & 24 & 1499.9 & 132 & 112.9 & 136.9 & 32.6$\pm$1.8 & G7 & 18 & Columba & 7 \\
BD-08 995 & 8.8 & 9 & 27 & 1865.9 & 100 & 48.9 & 69.8 & 7.9$\pm$0.7 & K0 & 19 & Columba & 19 \\
BD-09 1108 & 8.5 & 8 & 24 & 1440.1 & 127 & 65.8 & 117.1 & 31.6$\pm$1.6 & G5 & 18 & Tuc-Hor & 19 \\
Cl* Melotte 22 DH 352 & 9.5 & 6 & 18 & 1082.0 & 94 & 38.9 & 36.7 & 2.6$\pm$0.6 & G4 & 10 & Pleiades & 2 \\
Cl* Melotte 22 DH 875 & 11.5 & 7 & 21 & 1565.9 & 35 & 292.7 & 250.5 & 11.9$\pm$2.3 & K5 & 20 & Pleiades & 2 \\
Cl Melotte 22 2126 & 10.1 & 7 & 21 & 1522.0 & 66 & 75.3 & 81.2 & 5.5$\pm$0.8 & K0 & 3 & Pleiades & 2 \\
Cl Melotte 22 248 & 9.7 & 7 & 21 & 1494.7 & 92 & 71.6 & 73.3 & 31.7$\pm$1.8 & G4 & 5 & Pleiades & 2 \\
Cl Melotte 22 513 & 11.3 & 7 & 21 & 1468.0 & 35 & 211.4 & 168.6 & 5.5$\pm$0.8 & K7 & 5 & Pleiades & 2 \\
Cl Melotte 22 659 & 10.3 & 7 & 21 & 1460.0 & 67 & 155.0 & 194.2 & 31.8$\pm$2.6 & G4 & 21 & Pleiades & 2 \\
EX Cet & 6.2 & 8 & 24 & 1561.7 & 386 & 10.6 & 13.2 & 2.5$\pm$0.4 & K0 & 22 & $\beta$ Pic & 10 \\
HD 147512 & 6.1 & 17 & 51 & 1380.2 & 454 & 7.3 & 8.0 & 2.7$\pm$1.2 & G8 & 23 & AB Dor & 10 \\
HD 16760B & 8.4 & 5 & 15 & 1846.0 & 171 & 20.5 & 23.7 & 5.6$\pm$0.6 & K2 & 24 & AB Dor & 5 \\
HD 189285 & 8.3 & 11 & 33 & 728.0 & 166 & 29.5 & 49.2 & 31.5$\pm$1.8 & G5 & 25 & AB Dor & 19 \\
HD 20439 & 6.6 & 7 & 21 & 1512.0 & 341 & 10.2 & 14.0 & 11.7$\pm$2.0 & G0 & 26 & Carina-Near & 5 \\
HD 221239 & 6.7 & 12 & 36 & 1657.7 & 276 & 8.8 & 9.9 & 2.5$\pm$0.8 & K2 & 27 & AB Dor & 10 \\
HD 22680 & 8.9 & 8 & 24 & 1900.0 & 109 & 194.2 & 178.4 & 30.1$\pm$6.3 & G0 & 28 & Pleiades & 2 \\
HD 23464 & 7.6 & 8 & 24 & 1651.7 & 201 & 17.0 & 21.1 & 6.9$\pm$1.5 & G0 & 28 & Pleiades & 2 \\
HD 236717 & 7.2 & 19 & 57 & 1699.1 & 261 & 10.5 & 12.8 & 2.1$\pm$0.2 & K0 & 29 & AB Dor & 10 \\
HD 23975 & 8.6 & 6 & 18 & 1477.9 & 121 & 134.2 & 198.7 & 32.8$\pm$0.7 & G0 & 9 & Pleiades & 5 \\
HD 24194 & 9.0 & 6 & 18 & 1758.2 & 115 & 38.2 & 124.0 & 4.5$\pm$1.1 & G0 & 3 & Pleiades & 2 \\
HD 24463 & 8.7 & 7 & 21 & 1746.2 & 126 & 554.4 & 507.2 & 32.9$\pm$1.2 & G0 & 9 & Pleiades & 5 \\
HD 244945 & 8.5 & 8 & 27 & 1542.0 & 143 & 22.4 & 28.3 & 11.9$\pm$2.7 & G0 & 28 & 32 Ori & 10 \\
HD 24681 & 7.7 & 8 & 24 & 1761.3 & 197 & 55.5 & 243.2 & 29.3$\pm$5.8 & G5 & 25 & AB Dor & 10 \\
HD 26257 & 6.7 & 9 & 27 & 1518.7 & 328 & 15.3 & 12.0 & 14.1$\pm$2.4 & G2 & 25 & Carina-Near & 5 \\
HD 282954 & 9.1 & 7 & 21 & 1540.0 & 107 & 135.3 & 154.4 & 32.3$\pm$1.4 & G0 & 17 & Pleiades & 2 \\
HD 282958 & 9.6 & 7 & 21 & 1448.0 & 87 & 78.9 & 85.3 & 31.4$\pm$1.7 & G5 & 17 & Pleiades & 2 \\
HD 283869 & 8.4 & 8 & 24 & 1517.1 & 150 & 15.1 & 22.2 & 1.6$\pm$0.6 & K7 & 17 & Carina-Near & 5 \\
HD 285367 & 7.7 & 8 & 23 & 1498.9 & 213 & 13.7 & 32.3 & 7.3$\pm$1.3 & K0 & 17 & Carina-Near & 5 \\
HD 286693 & 8.1 & 7 & 21 & 1469.2 & 163 & 20.2 & 15.0 & 3.1$\pm$0.9 & K2 & 17 & Carina-Near & 5 \\
HD 287167 & 8.3 & 7 & 21 & 1463.2 & 148 & 11.9 & 18.6 & 10.5$\pm$1.1 & K0 & 17 & $\beta$ Pic & 10 \\
HD 29621 & 7.6 & 10 & 30 & 1523.6 & 229 & 12.0 & 10.7 & 7.9$\pm$2.8 & G5 & 30 & Carina-Near & 5 \\
HD 48370 & 6.7 & 8 & 24 & 1594.7 & 325 & 16.1 & 41.7 & 15.1$\pm$3.1 & K0 & 25 & Columba & 19 \\
HS Psc & 8.4 & 24 & 72 & 1701.1 & 135 & 210.2 & 220.0\tablenotemark{f} & 28.4$\pm$3.1 & K7 & 31 & AB Dor & 32 \\
HW Cet & 8.6 & 9 & 27 & 1832.8 & 145 & 26.3 & 54.7 & 6.4$\pm$0.5 & K2 & 10 & AB Dor & 10 \\
IS Eri & 7.2 & 7 & 21 & 1836.0 & 221 & 19.3 & 29.3 & 7.3$\pm$1.4 & G8 & 25 & AB Dor & 12 \\
LP 745-70 & 8.4 & 14 & 42 & 757.9 & 114 & 29.9 & 38.1 & 4.9$\pm$0.9 & K9 & 33 & AB Dor & 13 \\
MR Tau & 11.5 & 7 & 21 & 1491.9 & 32 & 304.7 & 342.1 & 9.7$\pm$1.3 & K7 & 3 & Pleiades & 2 \\
NX Aqr & 6.4 & 8 & 24 & 735.0 & 268 & 10.7 & 20.6 & 1.8$\pm$0.6 & G5 & 7 & Oct-Near & 34 \\
OT Tau & 11.1 & 6 & 18 & 1782.1 & 40 & 183.8 & 204.1 & 3.7$\pm$0.7 & K3 & 3 & Pleiades & 2 \\
PR Tau & 11.7 & 7 & 18 & 1543.8 & 30 & 829.6 & 811.0 & 18.6$\pm$4.4 & K7 & 35 & Pleiades & 2 \\
PW And & 7.0 & 38 & 114 & 1679.2 & 271 & 74.5 & 291.2 & 20.4$\pm$3.2 & K2 & 36 & AB Dor & 13 \\
Parenago 2752 & 9.6 & 7 & 21 & 1159.8 & 85 & 41.1 & 43.0 & 2.7$\pm$0.4 & G8 & 37 & AB Dor & 19 \\
RX J0520.0+0612 & 9.3 & 7 & 22 & 1103.1 & 97 & 41.5 & 71.5 & 6.4$\pm$1.2 & K4 & 6 & 32 Ori & 5 \\
StKM 1-382 & 9.3 & 8 & 24 & 1469.0 & 105 & 23.8 & 29.1 & 2.1$\pm$0.3 & K4 & 22 & Carina-Near & 5 \\
StKM 1-543 & 8.4 & 8 & 24 & 1524.8 & 137 & 16.4 & 17.8 & 1.8$\pm$0.4 & K5 & 22 & AB Dor & 10 \\
TYC 1090-543-1 & 9.5 & 13 & 39 & 1060.1 & 95 & 93.9 & 182.9 & 16.3$\pm$1.4 & K4 & 37 & AB Dor & 19 \\
TYC 1853-1452-1 & 8.5 & 8 & 24 & 1551.8 & 144 & 13.6 & 14.7 & 1.4$\pm$0.3 & K7 & 9 & AB Dor & 10 \\
TYC 3385-23-1 & 8.1 & 7 & 21 & 1497.9 & 172 & 17.1 & 27.8 & 3.2$\pm$0.4 & G8 & 10 & AB Dor & 10 \\
V1169 Tau & 9.5 & 7 & 20 & 1789.9 & 94 & 52.6 & 60.2 & 5.2$\pm$0.5 & G1 & 3 & Pleiades & 2 \\
V1173 Tau & 11.4 & 7 & 21 & 1523.6 & 33 & 231.3 & 281.2 & 6.0$\pm$1.2 & K5 & 3 & Pleiades & 2 \\
V1174 Tau & 10.7 & 7 & 21 & 1412.9 & 47 & 164.4 & 205.4 & 7.2$\pm$0.8 & G5 & 38 & Pleiades & 2 \\
V1841 Ori & 8.4 & 8 & 25 & 1900.0 & 139 & 59.1 & 134.7 & 14.5$\pm$1.2 & K2 & 7 & $\beta$ Pic & 10 \\
V370 Tau & 11.0 & 7 & 21 & 1515.1 & 51 & 100.9 & 109.3 & 4.2$\pm$1.0 & K3 & 3 & Pleiades & 2 \\
V446 Tau & 11.5 & 7 & 21 & 1799.1 & 32 & 230.9 & 240.0 & 3.4$\pm$1.4 & K8 & 1 & Pleiades & 2 \\
V623 Tau & 11.5 & 7 & 21 & 1829.2 & 34 & 571.2 & 846.5 & 20.2$\pm$1.3 & K5 & 3 & Pleiades & 2 \\
V677 Tau & 11.1 & 7 & 21 & 1589.7 & 41 & 512.7 & 750.1 & 28.8$\pm$7.8 & K6 & 39 & Pleiades & 2 \\
V700 Tau & 10.5 & 7 & 21 & 1500.9 & 53 & 101.8 & 117.5 & 4.9$\pm$1.1 & K3 & 3 & Pleiades & 2 \\
V810 Tau & 10.3 & 6 & 18 & 1151.1 & 59 & 77.9 & 98.7 & 5.7$\pm$0.7 & G8 & 5 & Pleiades & 2 \\
V814 Tau & 10.5 & 6 & 18 & 1602.8 & 54 & 78.1 & 84.9 & 3.6$\pm$0.6 & K3 & 3 & Pleiades & 2 \\
V815 Tau & 10.6 & 7 & 21 & 1622.8 & 50 & 92.8 & 96.1 & 5.6$\pm$0.9 & K4 & 5 & Pleiades & 2 \\
V963 Tau & 9.5 & 8 & 24 & 1816.0 & 93 & 74.8 & 96.8 & 33.0$\pm$1.2 & G5 & 5 & Pleiades & 2 \\
V966 Tau & 10.0 & 7 & 21 & 1764.2 & 80 & 106.4 & 151.7 & 31.2$\pm$3.7 & G8 & 26 & Pleiades & 2 \\
Wolf 1259 & 8.7 & 7 & 21 & 1748.2 & 137 & 28.7 & 57.8 & 6.4$\pm$0.8 & K1 & 10 & AB Dor & 10
\enddata
\tablenotetext{a}{\citet{Cutri2003}}
\tablenotetext{b}{Number of visits, which are comprised of three contiguous observations, for each target. See \autoref{sec:hpf_obs} for more details.}
\tablenotetext{c}{Number of individual observations, or RVs, for each target. See \autoref{sec:hpf_obs} for more details.}
\tablenotetext{d}{Signal-to-noise estimated at 1.07~$\mu$m.}
\tablenotetext{e}{Projected rotational velocity measured in this work (see \autoref{sec:characterization}).}
\tablenotetext{f}{RMS of RV residuals after subtracting the best-fit planetary model reported by \citet{Tran2024}.}
\tablerefs{(1) \citet{Prosser1991}; (2) \citet{Stauffer2007}; (3) \citet{Soderblom1993}; (4) \citet{Wilson1963}; (5) \citet{Gagne2018a}; (6) \citet{Shvonski2016}; (7) \citet{Torres2006}; (8) \citet{Binks2018}; (9) \citet{Roeser1988}; (10) \citet{Gagne2018b}; (11) \citet{Reid2004}; (12) \citet{Zuckerman2011}; (13) \citet{Zuckerman2004}; (14) \citet{Kraft1967}; (15) \citet{Mendoza1956}; (16) \citet{Hill1952}; (17) \citet{Nesterov1995}; (18) \citet{Alcala1996}; (19) \citet{Torres2008}; (20) \citet{Findeisen2010}; (21) \citet{Breger1984}; (22) \citet{Stephenson1986}; (23) \citet{White2007}; (24) \citet{Abt1988}; (25) \citet{Houk1999}; (26) \citet{CayreldeStrobel2001}; (27) \citet{Gray2003}; (28) \citet{Cannon1993}; (29) \citet{Yoss1961}; (30) \citet{Adolfsson1954}; (31) \citet{Bowler2019}; (32) \citet{Schlieder2010}; (33) \citet{Gray2006}; (34) \citet{Zuckerman2013}; (35) \citet{Haro1982}; (36) \citet{Montes2001b}; (37) \citet{Elliott2016}; (38) \citet{Skiff2014}; (39) \citet{Soderblom2005}.}
\end{deluxetable}
\end{longrotatetable}

\section{Instrumental Velocity Offset for HPF\label{sec:appendix_offset}}

Throughout the EPGM survey, we continued regular observation of three RV standard stars, HD 221354 \citep{Bouchy2013}, HD 116442 \citep{Chubak2012}, and HD 3765 \citep{Chubak2012}, to characterize instrumental stability over time. These RV standards were originally established in the optical as slowly rotating stars with constant RVs and, initially, no apparent signs of long-period companions. However, after our survey began, \citet{Rosenthal2021} reported a giant planet candidate with an orbital period of $P = 1226^{+19}_{-23}$~d, RV semi-amplitude of $K = 3.9^{+0.5}_{-0.4}$~m~s$^{-1}$, and an eccentricity of $e = 0.4^{+0.1}_{-0.1}$ around HD 3765. We nevertheless continued to monitor this star throughout the duration of our campaign. \autoref{tab:app_rv_stand_info} and  \autoref{tab:app_rv_stand_hpf_rv_info} summarize the properties and pre-maintenance and post-maintenance HPF RV observations of each RV standard star, respectively.

Having continued monitoring these stars over the instrument maintenance period allows us to estimate the velocity offset between pre- and post-maintenance HPF RV measurements. We fit a flat trend to both datasets and compare the intercepts of each fit. The observed differences between post- and pre-maintenance RVs (HPF$_\mathrm{post}$ -- HPF$_\mathrm{pre}$) are $-53.5\pm1.3$, $-42.4\pm4.0$, and $-54.3\pm1.7$~m~s$^{-1}$ for HD 221354, HD 116442, and HD 3765, respectively. The measured offsets are consistent within 3$\sigma$. The higher offset of HD 116442 may be caused by the lower number of post-maintenance RVs, higher intrinsic measurement error relative to the other two standard stars, and difference in spectral type. The weighted average difference for the three standards is $-53.1\pm2.7$~m~s$^{-1}$.

Furthermore, we note that HD 130322 b provides an additional opportunity to estimate the difference of the velocity zero-points between pre- and post-maintenance HPF RVs. From the model fit of HD 130322 b (\autoref{sec:appendix_HD_130322_modeling}), we find that the difference between velocity offsets of pre- and post-maintenance HPF RVs is 3.8~m~s$^{-1}$, comparable to the uncertainties of the individual offsets (3~m~s$^{-1}$). Furthermore, we perform a similar Keplerian model fit without applying this HPF velocity correction. In this uncorrected fit, the inferred velocity offsets of the pre- and post-maintenance HPF RVs are $25.43 \pm 3$~m~s$^{-1}$ and $-23.9 \pm 3$~m~s$^{-1}$, respectively. The difference between these offsets is $-49.3 \pm 4.2$~m~s$^{-1}$, which is consistent with the $-53.1 \pm 2.7$~m~s$^{-1}$ estimated from three RV standard stars. All other inferred planetary and instrumental parameters are in excellent agreement between the corrected and uncorrected model fits.

\begin{figure}[!]
    \centering
    \includegraphics[width=1.0\linewidth]{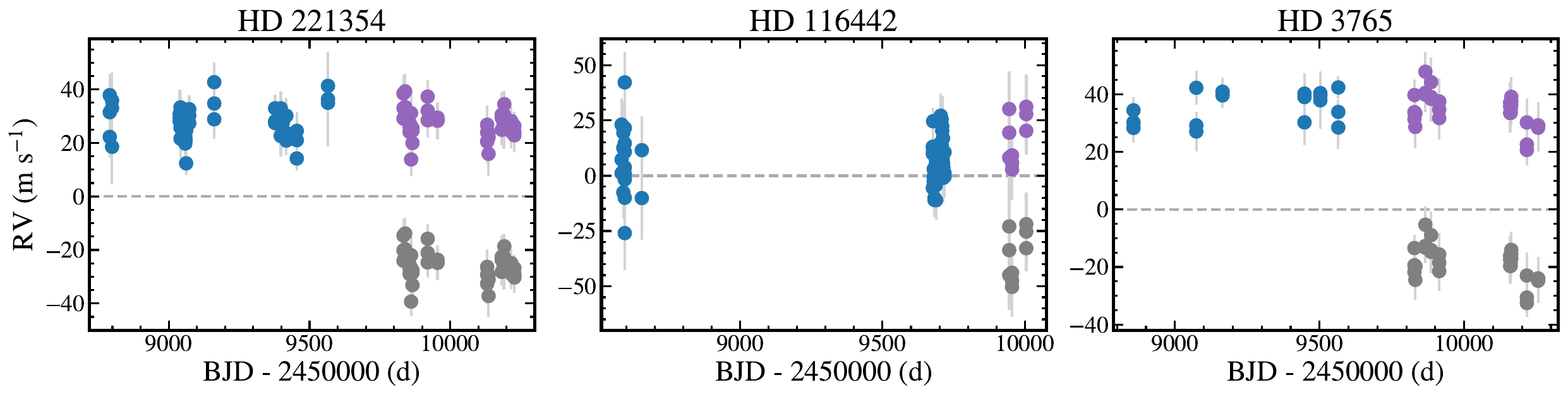}
    \caption{HPF RV time series for the three standard stars, HD 221354, HD 116442, and HD 3765. Pre- and uncorrected post-maintenance RVs are shown as blue and grey circles, respectively. Post-maintenance RVs corrected with the empirically measured offset, $-53.1\pm2.7$~m~s$^{-1}$, are plotted as purple points.}
    \label{fig:rv_standards_break}
\end{figure}

Altogether, these tests support the adoption of the empirically-measured HPF velocity offset. \autoref{fig:rv_standards_break} displays all RV measurements, including the velocity offset correction, for the three standard stars. \autoref{tab:standard_measurements} in \autoref{sec:appendix_measurements} reports measurements for each standard star.

\begin{deluxetable}{lcccccccc}[!b]
    \tabletypesize{\footnotesize}
    \setlength{\tabcolsep}{4pt}
    \tablecaption{\label{tab:app_rv_stand_info} Properties of RV Standards}
    \tablehead{\colhead{RV Standard} & \colhead{$J$\tablenotemark{a}} & \colhead{SpT} & \colhead{SpT Ref.} & \colhead{Optical RV} & \colhead{Optical} & \colhead{Optical} & \colhead{Optical} & \colhead{Optical} \\
    \colhead{Name} & \colhead{(mag)} & \colhead{} & \colhead{} & \colhead{RMS (m s$^{-1}$)} & \colhead{$N_\mathrm{RV}$} & \colhead{Baseline (d)} & \colhead{Inst.} & \colhead{Ref.}}
    \startdata
    HD 221354 & 5.3 & K0 & \citet{Gray2003} & 2.2 & 43 & 174 & SOPHIE+ & \citet{Bouchy2013} \\
    & & & & 2.1 & 406 & 3375 & HIRES & \citet{Butler2017} \\
    HD 116442 & 5.6 & G5 & \citet{Cannon1993} & 4.2 & 74 & 6366 & HIRES & \citet{Butler2017} \\
    HD 3765 & 5.7 & K2 & \citet{Keenan1989} & 4.8 & 198 & 6366 & HIRES & \citet{Butler2017}
    \enddata
    \tablenotetext{a}{\citet{Cutri2003}}
\end{deluxetable}

\clearpage

\begin{deluxetable}{lcccccc}
    \setlength{\tabcolsep}{10pt}
    \tablecaption{\label{tab:app_rv_stand_hpf_rv_info} Summary of pre- and post-maintenance HPF observations for RV Standards}
    \tablehead{\colhead{RV Standard} & \colhead{Pre HPF} & \colhead{Pre HPF} & \colhead{Pre HPF} & \colhead{Post HPF} & \colhead{Post HPF} & \colhead{Post HPF} \\
    \colhead{Name} & \colhead{RV RMS (m s$^{-1}$)} & \colhead{$N_\mathrm{RV}$} & \colhead{Baseline (d)} & \colhead{RV RMS (m s$^{-1}$)} & \colhead{$N_\mathrm{RV}$} & \colhead{Baseline (d)}} 
    \startdata
    HD 221354 & 7.0 & 45 & 775 & 5.9 & 42 & 394 \\
    HD 116442 & 9.3 & 56 & 1134 & 13.5 & 9 & 60 \\
    HD 3765 & 6.0\tablenotemark{a} & 18 & 708 & 6.1\tablenotemark{a} & 30 & 428
    \enddata
    \tablenotetext{a}{For HD 3765, RV RMS is calculated after removing the signal of the giant planet candidate using the best-fit parameters from \citet{Rosenthal2021}.}
\end{deluxetable}

\section{HPF RVs and Activity Indicators Measurements\label{sec:appendix_measurements}}

\autoref{tab:hpf_measurements} reports the measurements and uncertainties of the relative HPF RVs and associated spectral indicators for all science targets in our sample. \autoref{tab:standard_measurements} reports the same values for the three RV standards (HD 116442, HD 221354, and HD 3765) observed as part of the EGPM program. All post-maintenance RV observations have been corrected following the procedure described in \autoref{sec:appendix_offset}. \autoref{tab:excluded_hpf_measurements} reports the same measurements for the 19 excluded systems (see \autoref{sec:appendix_excluded}). See \autoref{sec:hpf_obs} for more details on these observations.

\movetabledown=2in
\begin{rotatetable}
\begin{deluxetable}{ccccccccc}
    \tablecaption{Relative HPF RVs, activity indicators (dLW and CRX), and indices of the \ion{Ca}{2} IRT lines for the 85 science targets. \label{tab:hpf_measurements}}
    \tablehead{\colhead{BJD\textsubscript{TDB}} & \colhead{RV} & \colhead{dLW} & \colhead{CRX} & \colhead{\ion{Ca}{2} IRT 1} & \colhead{\ion{Ca}{2} IRT 2} & \colhead{\ion{Ca}{2} IRT 3} & \colhead{Maintenance} & \colhead{Object} \\
    \colhead{(d)} & \colhead{(m s$^{-1}$)} & \colhead{(m$^2$ s$^{-2}$)} & \colhead{(m s$^{-1}$ Np$^{-1}$)\tablenotemark{a}} & \colhead{} & \colhead{} & \colhead{} & \colhead{Period\tablenotemark{b}} & \colhead{}}
    \startdata
2458779.9590 & 604.2 $\pm$ 163.4 & 498.7 $\pm$ 580.5 & 0.10 $\pm$ 0.25 & 0.809 $\pm$ 0.028 & 0.753 $\pm$ 0.042 & 0.738 $\pm$ 0.033 & Pre & 2MASS J03402958+2333040 \\
2458779.9629 & 56.1 $\pm$ 140.5 & -618.9 $\pm$ 621.2 & 0.07 $\pm$ 0.21 & 0.796 $\pm$ 0.032 & 0.689 $\pm$ 0.050 & 0.846 $\pm$ 0.037 & Pre & 2MASS J03402958+2333040 \\
2458779.9669 & -39.8 $\pm$ 192.6 & 156.8 $\pm$ 322.6 & 0.27 $\pm$ 0.28 & 0.799 $\pm$ 0.057 & 0.750 $\pm$ 0.073 & 0.520 $\pm$ 0.067 & Pre & 2MASS J03402958+2333040 \\
2458784.9519 & -79.5 $\pm$ 99.3 & -1330.8 $\pm$ 433.6 & 0.09 $\pm$ 0.14 & 0.773 $\pm$ 0.040 & 0.582 $\pm$ 0.061 & 0.671 $\pm$ 0.050 & Pre & 2MASS J03402958+2333040 \\
\enddata
\tablenotetext{a}{Np refers to Neper, or logarithmic ($e$) wavelength ratio \citep{Zechmeister2018}.}
\tablenotetext{b}{Denotes whether observation was obtained pre- or post-maintenance period of HPF.}
\tablecomments{This table is published in its entirety in the machine-readable format. A portion is shown here for guidance regarding its form and content.}
\end{deluxetable}
\end{rotatetable}
\movetabledown=1.5in
\begin{rotatetable}
\begin{deluxetable}{ccccccccc}
    \tablecaption{Relative HPF RVs, activity indicators (dLW and CRX), and indices of the Ca II IRT lines for the 3 RV standards. \label{tab:standard_measurements}}
    \tablehead{\colhead{BJD\textsubscript{TDB}} & \colhead{RV} & \colhead{dLW} & \colhead{CRX} & \colhead{\ion{Ca}{2} IRT 1} & \colhead{\ion{Ca}{2} IRT 2} & \colhead{\ion{Ca}{2} IRT 3} & \colhead{Maintenance} & \colhead{Object} \\
    \colhead{(d)} & \colhead{(m s$^{-1}$)} & \colhead{(m$^2$ s$^{-2}$)} & \colhead{(m s$^{-1}$ Np$^{-1}$)\tablenotemark{a}} & \colhead{} & \colhead{} & \colhead{} & \colhead{Period\tablenotemark{b}} & \colhead{}}
    \startdata
2458583.7431 & 1.2 $\pm$ 9.8 & 61.2 $\pm$ 15.0 & -0.01 $\pm$ 0.01 & 0.4988 $\pm$ 0.0009 & 0.3659 $\pm$ 0.0011 & 0.3573 $\pm$ 0.0009 & Pre & HD 116442 \\
2458583.7443 & 7.3 $\pm$ 14.6 & 76.5 $\pm$ 13.6 & -0.00 $\pm$ 0.02 & 0.4972 $\pm$ 0.0007 & 0.3708 $\pm$ 0.0008 & 0.3694 $\pm$ 0.0007 & Pre & HD 116442 \\
2458583.7454 & 23.1 $\pm$ 11.8 & 72.4 $\pm$ 23.7 & 0.02 $\pm$ 0.01 & 0.4884 $\pm$ 0.0016 & 0.3743 $\pm$ 0.0020 & 0.3699 $\pm$ 0.0016 & Pre & HD 116442 \\
2458589.7134 & -7.6 $\pm$ 11.9 & 88.1 $\pm$ 20.4 & -0.00 $\pm$ 0.01 & 0.4964 $\pm$ 0.0007 & 0.3663 $\pm$ 0.0008 & 0.3675 $\pm$ 0.0007 & Pre & HD 116442 \\
    \enddata
\tablenotetext{a}{Np refers to Neper, or logarithmic ($e$) wavelength ratio \citep{Zechmeister2018}.}
\tablenotetext{b}{Denotes whether observation was obtained pre- or post-maintenance period of HPF.}
\tablecomments{This table is published in its entirety in the machine-readable format. A portion is shown here for guidance regarding its form and content.}
\end{deluxetable}
\end{rotatetable}
\movetabledown=2in
\begin{rotatetable}
\begin{deluxetable}{ccccccccc}
    \tablecaption{Relative HPF RVs, activity indicators (dLW and CRX), and indices of the \ion{Ca}{2} IRT lines for the 19 excluded systems. \label{tab:excluded_hpf_measurements}}
    \tablehead{\colhead{BJD\textsubscript{TDB}} & \colhead{RV} & \colhead{dLW} & \colhead{CRX} & \colhead{\ion{Ca}{2} IRT 1} & \colhead{\ion{Ca}{2} IRT 2} & \colhead{\ion{Ca}{2} IRT 3} & \colhead{Maintenance} & \colhead{Object} \\
    \colhead{(d)} & \colhead{(m s$^{-1}$)} & \colhead{(m$^2$ s$^{-2}$)} & \colhead{(m s$^{-1}$ Np$^{-1}$)\tablenotemark{a}} & \colhead{} & \colhead{} & \colhead{} & \colhead{Period\tablenotemark{b}} & \colhead{}}
    \startdata
2458711.6886 & -396.9 $\pm$ 80.1 & -1277.4 $\pm$ 378.5 & -0.07 $\pm$ 0.14 & 0.943 $\pm$ 0.019 & 0.863 $\pm$ 0.025 & 0.737 $\pm$ 0.019 & Pre & 2MASS J19224278-0515536 \\
2458711.6925 & -243.9 $\pm$ 109.7 & -909.5 $\pm$ 333.7 & 0.08 $\pm$ 0.19 & 1.015 $\pm$ 0.015 & 0.874 $\pm$ 0.020 & 0.847 $\pm$ 0.015 & Pre & 2MASS J19224278-0515536 \\
2458711.6966 & -343.7 $\pm$ 108.7 & -922.3 $\pm$ 261.7 & 0.03 $\pm$ 0.19 & 0.950 $\pm$ 0.015 & 0.897 $\pm$ 0.020 & 0.833 $\pm$ 0.016 & Pre & 2MASS J19224278-0515536 \\
    \enddata
\tablenotetext{a}{Np refers to Neper, or logarithmic ($e$) wavelength ratio \citep{Zechmeister2018}.}
\tablenotetext{b}{Denotes whether observation was obtained pre- or post-maintenance period of HPF.}
\tablecomments{This table is published in its entirety in the machine-readable format. A portion is shown here for guidance regarding its form and content.}
\end{deluxetable}
\end{rotatetable}

\clearpage

\section{Relationship Between Activity Indicators and Physical Parameters\label{sec:appendix_activity}}

\begin{figure}[!p]
     \centering
     \includegraphics[width=\linewidth]{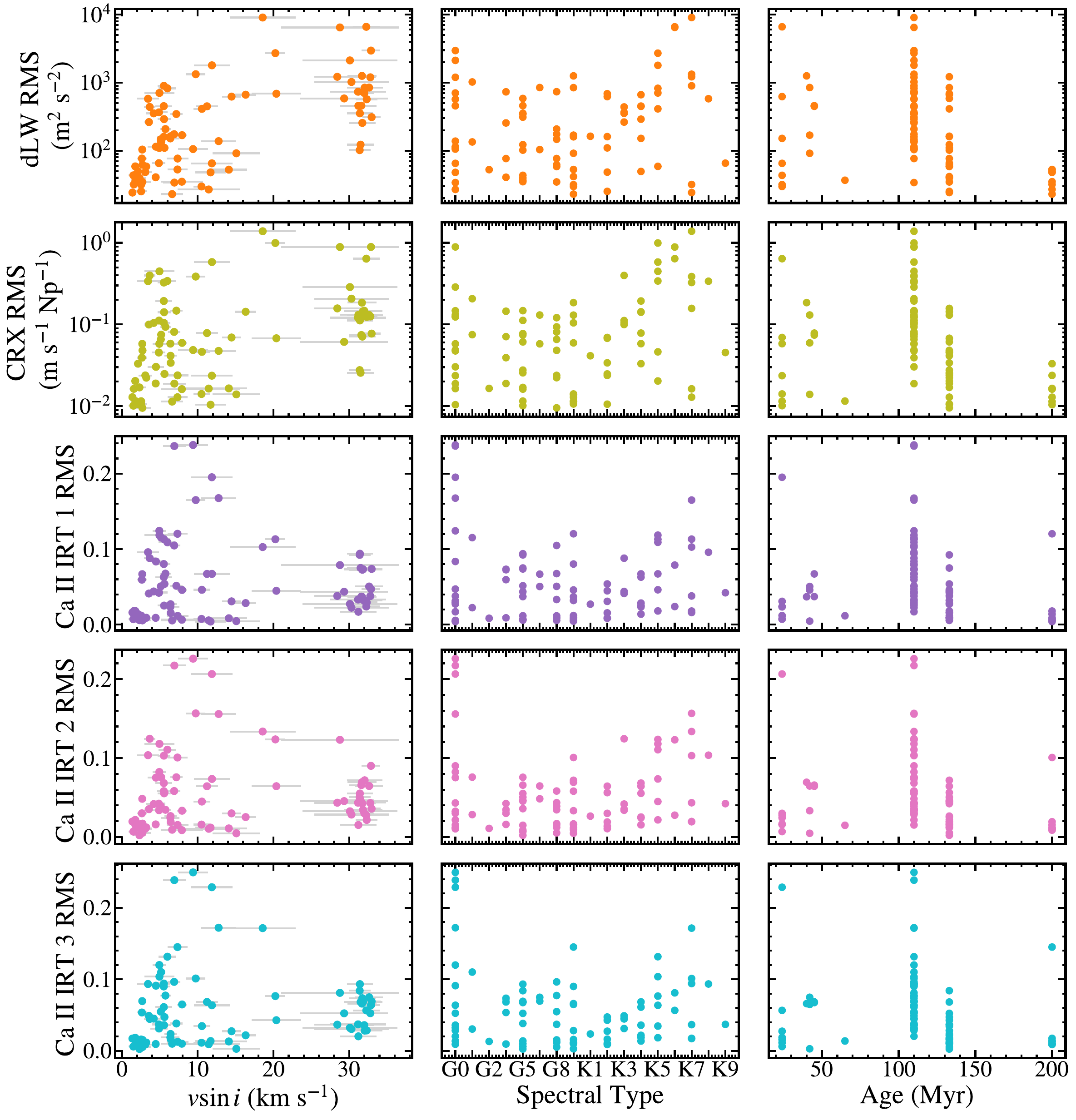}
     \caption{RMS of various activity indicators as a function of projected rotational velocity, stellar spectral type, and stellar age for all stars in our sample without stellar companions. Beginning from the top, each row displays the differential line width (dLW, orange), the chromatic index (CRX, yellow), and the \ion{Ca}{2} IRT indices for the \ion{Ca}{2} 1, 2, and 3 IRT emission lines (purple, pink, and cyan, respectively).}
    \label{fig:act_rms_vs_prop}
\end{figure}

\autoref{fig:act_rms_vs_prop} displays RMS of measured activity indicators (dLW, CRX, \ion{Ca}{2} IRT 1, 2, and 3; for more information on these indicators, see \autoref{sec:hpf_obs}) as a function of each target's projected rotational velocities, spectral types, and ages.

\clearpage

\section{Generalized Lomb-Scargle Periodograms of HPF RVs for Systems with Significant Periodic RV Signals\label{sec:appendix_periodograms}}

\begin{figure}[!h]
     \centering
     \includegraphics[width=\linewidth]
     {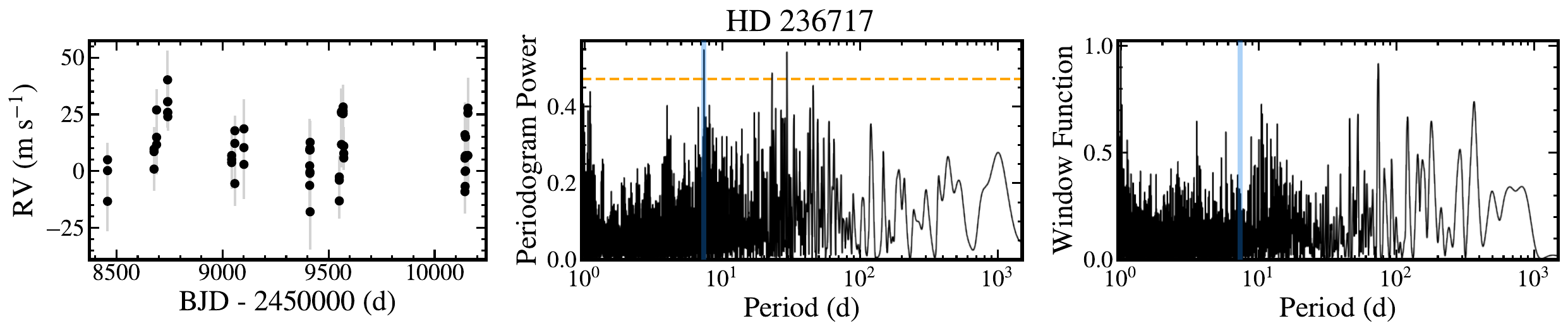}\\
     \vspace{2ex}
     \includegraphics[width=\linewidth]{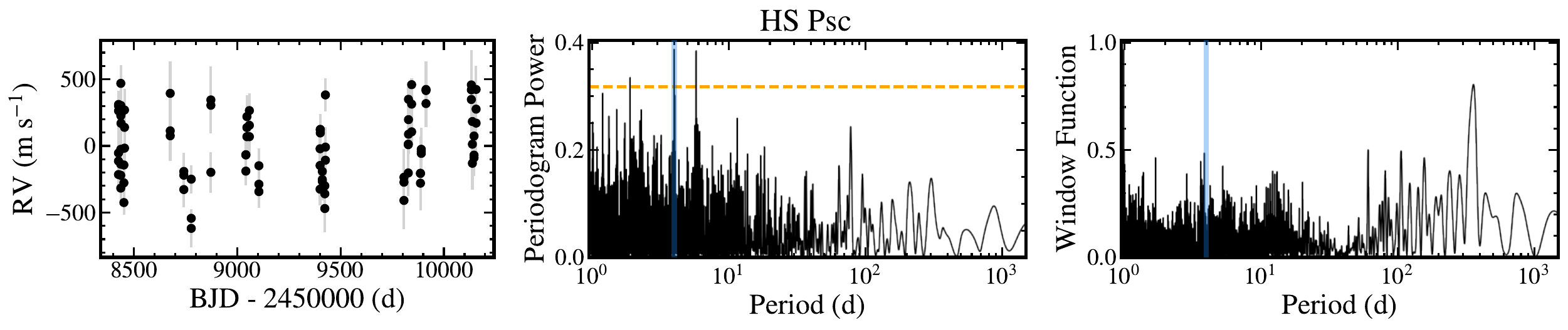}\\
     \vspace{2ex}
     \includegraphics[width=\linewidth]{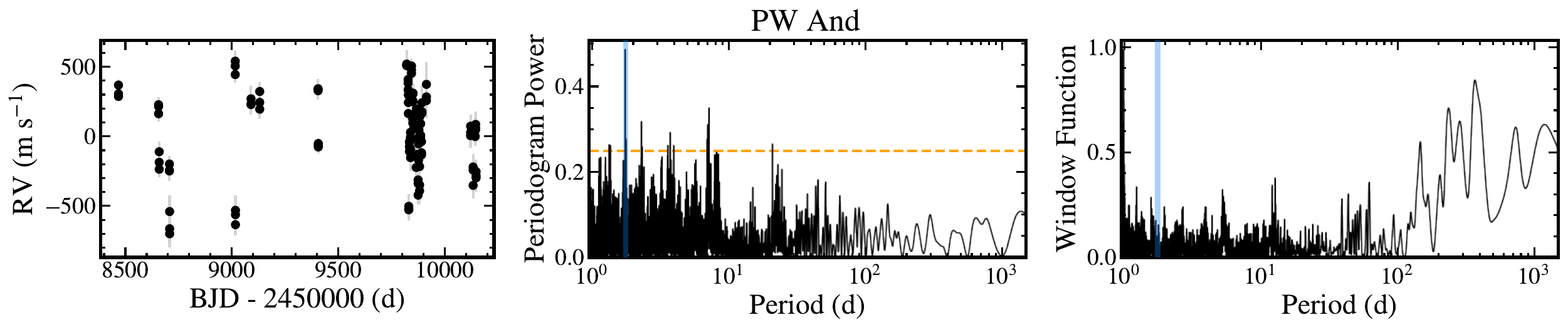}
     \caption{HPF RV measurements, GLS periodograms, and corresponding spectral window functions for targets HD 236717, HS Psc, and PW And. The orange dashed horizontal line represents the FAP = 0.01\% threshold. The most significant peak in the periodogram is highlighted with a blue vertical line. The GLS periodogram is computed up to a period of 1500~d, but only 50~d is shown for each target as all significant peaks have periods less than 50~d.}
    \label{fig:app_periodogram0}
\end{figure}

\autoref{fig:app_periodogram0} shows the HPF RVs, GLS periodograms, and window functions of the three stars that have at least one significant periodic signal identified in \autoref{sec:gls_periodogram} (HD 236717, HS Psc, and PW And). A description of the HPF RV measurements is provided in \autoref{sec:hpf_obs}. The methodology of applying a GLS periodogram to each star is described in \autoref{sec:gls_periodogram}.

\clearpage

\section{Light Curves and Inferred Stellar Rotation Periods for Systems with Significant Periodic RV Signals\label{sec:appendix_lc_prot}}

\begin{figure}[!b]
     \centering
     \includegraphics[width=\linewidth]{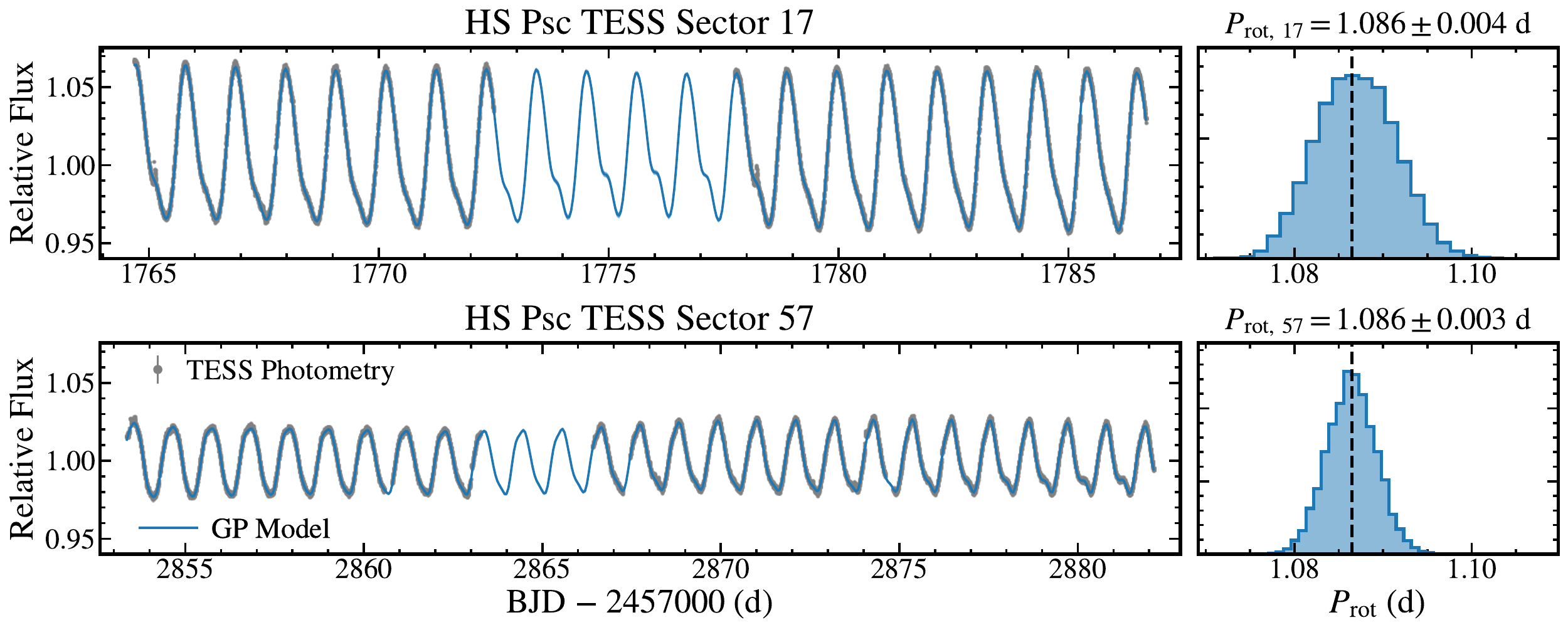}\\
     \vspace{5ex}
     \includegraphics[width=\linewidth]{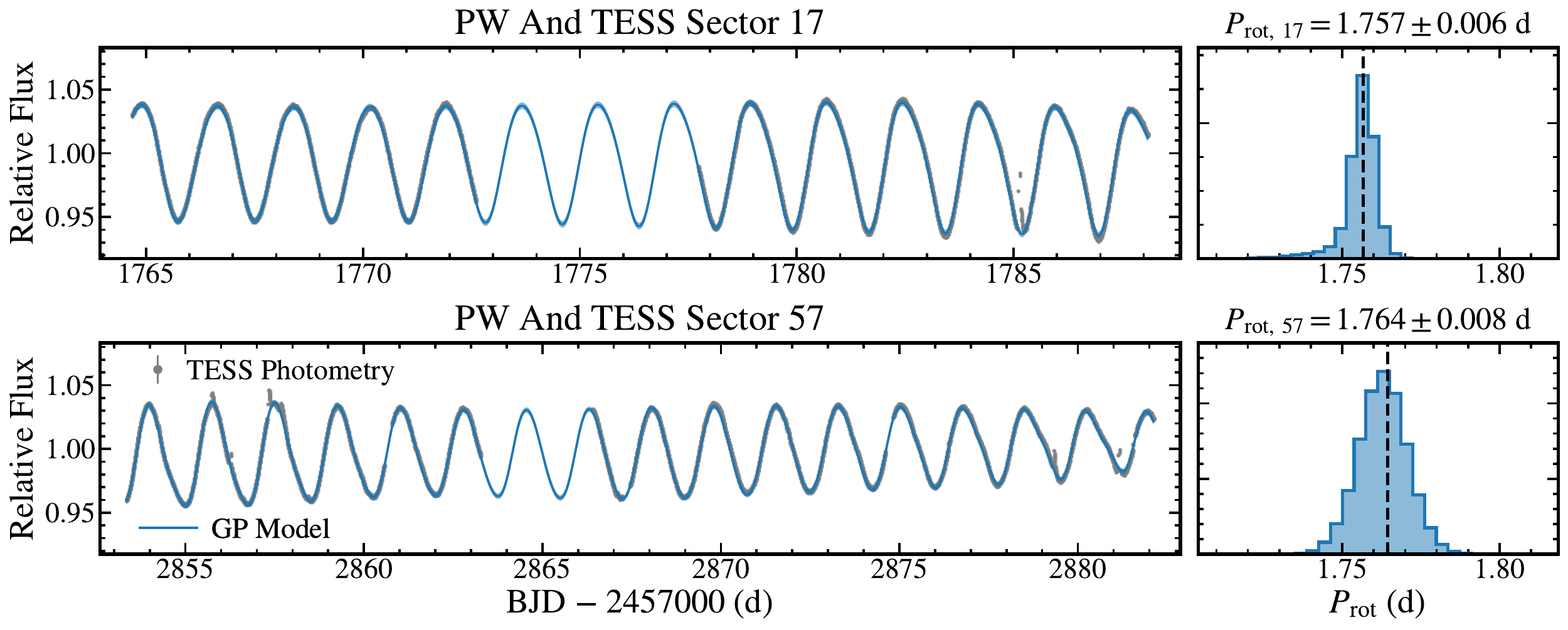}\\

     \caption{TESS light curves for systems that have significant peaks in their periodograms. Best-fit Gaussian process model and 1$\sigma$ variance are plotted as solid blue lines and shaded regions, respectively. The posterior distributions and MAP values (dashed vertical line) for the quasi-periodic Gaussian process $P_\mathrm{rot}$ hyperparameter are displayed in the right panels for targets HS Psc and PW And.}
    \label{fig:app_prot0}
\end{figure}

\autoref{fig:app_prot0} displays the TESS light curve, best-fit quasi-periodic Gaussian process model, and inferred posterior distributions of the stellar rotation period for the two stars that have a significant peak in their GLS periodograms and TESS light curves with persistent and strong modulations (HS Psc and PW And). The inferred stellar rotation periods for all objects are summarized in \autoref{tab:app_prot_vals}. The procedure to obtaining TESS light curves, fitting Gaussian process models, and inferring stellar rotation periods is discussed in \autoref{sec:gls_periodogram}. Multiple TESS sectors with large observational gaps are separated and independent stellar rotation periods are inferred for each isolated set of light curves. If there is only one TESS sector or one continuous group of TESS sectors, then the average and standard deviation of the $P_\mathrm{rot}$ posterior distribution is adopted as the rotation period and uncertainty. If there are multiple contiguous groups of TESS sectors, then the stellar rotation period and uncertainty is the weighted average and standard deviation of each independent $P_\mathrm{rot}$ posterior distributions for each target.

\begin{deluxetable}{lcc}[!h]
    \setlength{\tabcolsep}{3em}
    \tablecaption{Inferred stellar rotation periods of objects with significant periodic signals in HPF RV time series.\label{tab:app_prot_vals}}
    \tablehead{\colhead{Object} & \colhead{TESS} & \colhead{Inferred $P_\mathrm{rot}$} \\
    \colhead{Name} & \colhead{Sector(s)} & \colhead{(d)}}
    \startdata
    HS Psc & 17 & $1.086 \pm 0.004$ \\
    & 57 & $1.086 \pm 0.003$ \\
    & \textbf{Adopted} & $\mathbf{1.086 \pm 0.003}$ \\
    \hline
    PW And & 17 & $1.757 \pm 0.006$ \\
    & 57 & $1.764 \pm 0.008$ \\
    & \textbf{Adopted} & $\mathbf{1.760 \pm 0.003}$ \\
    \enddata
\end{deluxetable}

\section{Correlations between HPF RVs and Spectroscopic Activity Indicators for Systems with Significant Periodic RV Signals\label{sec:appendix_correlations}}

\begin{figure}[!p]
     \centering
     \includegraphics[width=\linewidth]{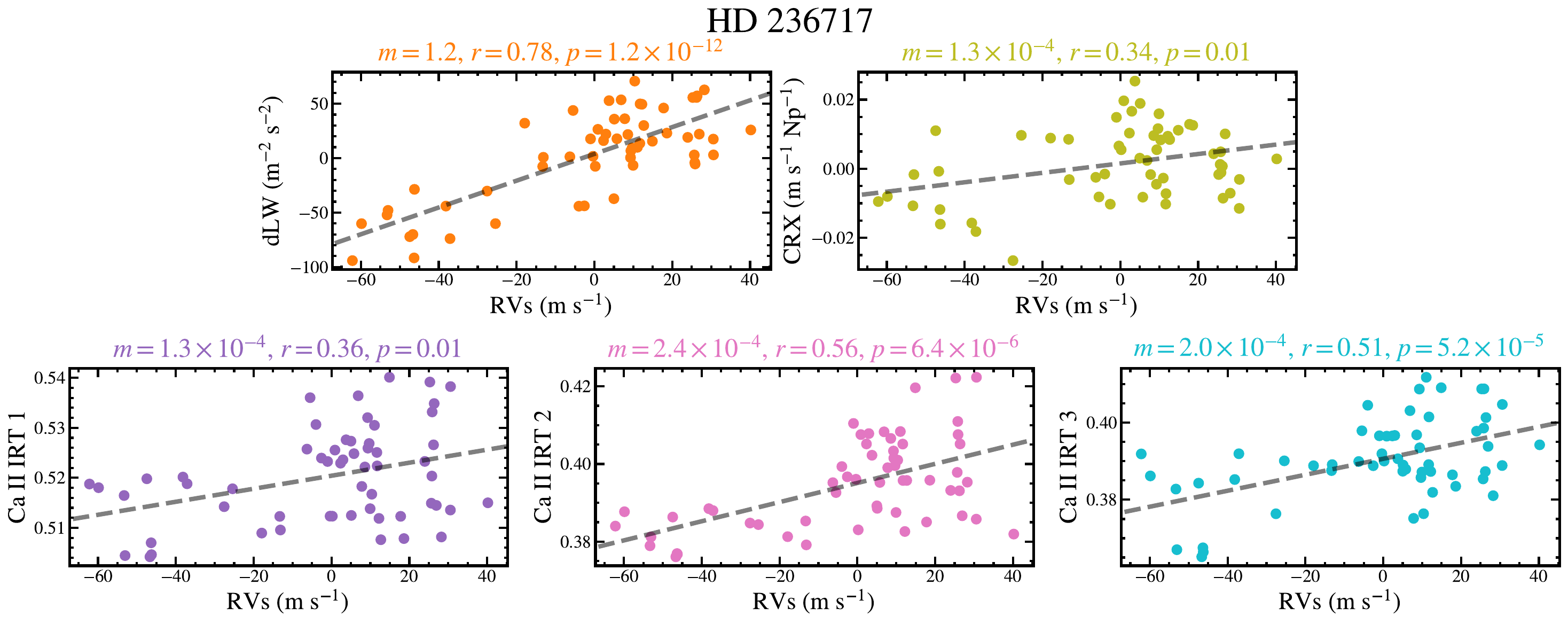}
     \\
     \includegraphics[width=\linewidth]{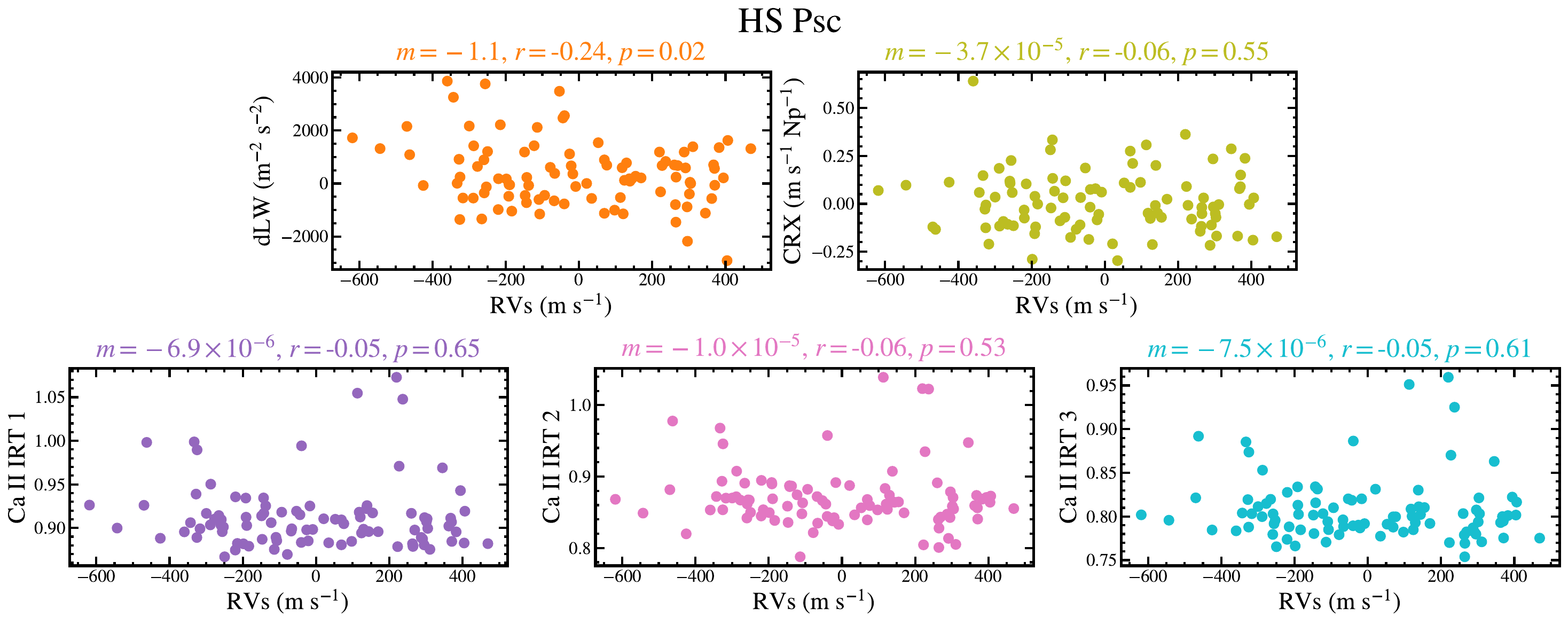}
     \\
     \includegraphics[width=\linewidth]{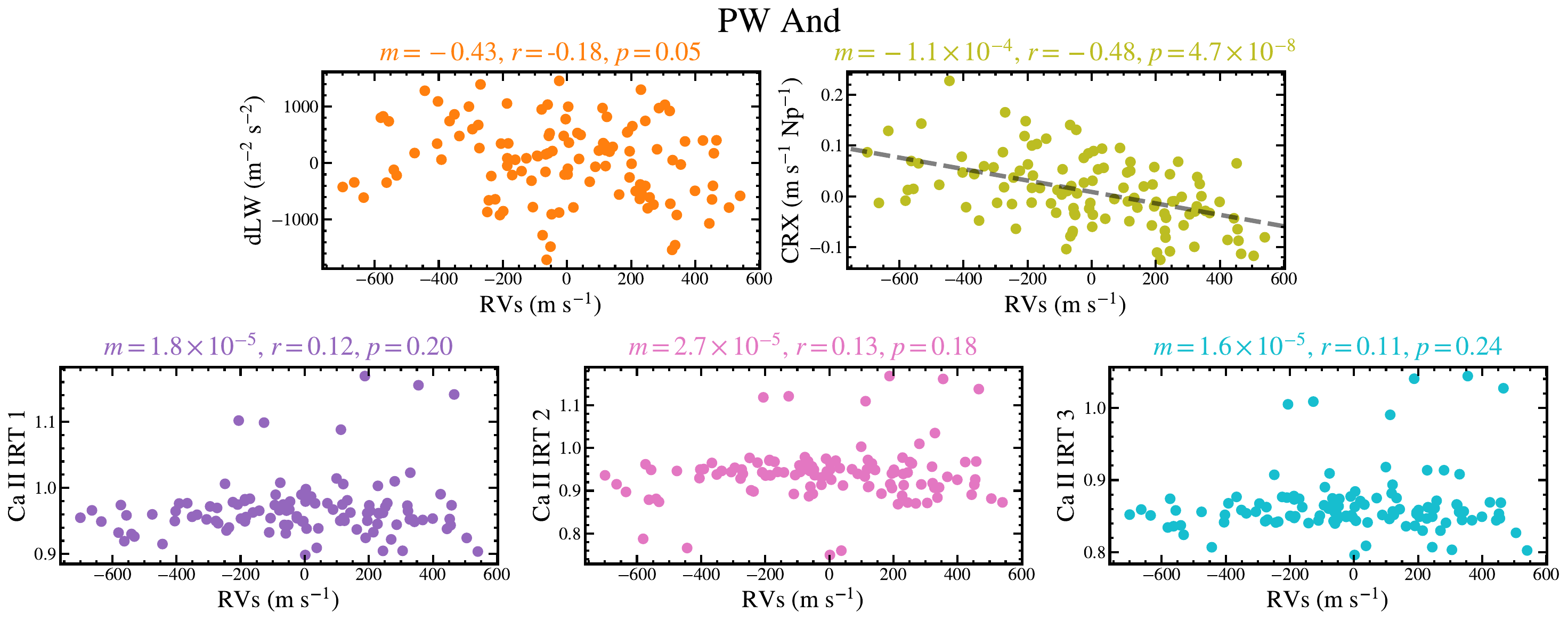} \\
     \caption{Correlations between HPF RVs and various activity indicators for targets HD 236717, HS Psc, and PW And. From the top left to bottom right, each panel plots the correlations for the differential line width (dLW, orange), the chromatic index (CRX, yellow), and the \ion{Ca}{2} IRT indices for the \ion{Ca}{2} 1, 2, and 3 IRT emission lines (purple, pink, and cyan, respectively). The best-fit slope ($m$), Pearson's correlation coefficient ($r$), and corresponding $p$ value are reported at the top of each panel. Significant correlations ($p \leq 0.01$) are denoted in panels where the best-fit linear correlation is overplotted as a dashed black line. The activity indicators for HD 236717 all exhibit significant correlations with the HPF RVs, indicating that observed RV signals are stellar activity-driven.}
    \label{fig:app_corre0}
\end{figure}

\autoref{fig:app_corre0} displays the correlations between the HPF RV time series and various spectral indicators (differential line width, chromatic index, and \ion{Ca}{2} emission line indices) for the three stars that exhibit a significant peak in the periodograms of their RVs. Each panels corresponds to a different index. A description of each time series is presented in \autoref{sec:hpf_obs}. The best-fit slope ($m$), Pearson's correlation coefficient ($r$), and corresponding $p$-values are labeled at the top of each panel. These metrics are described in detail in \autoref{sec:ooi_corre}. Objects with a significant correlation between the HPF RVs and a spectral indicator ($p \leq 0.01$) have the best-fit linear trend overplotted as a black dashed line. Measurements for all time series are reported in \autoref{tab:hpf_measurements} in \autoref{sec:appendix_measurements}.

\clearpage

\section{RV Time Series and RV Semi-amplitude, Planet Minimum Mass, and Orbital Period Completeness Functions\label{sec:appendix_completeness_function}}

\begin{figure}[p]
     \centering
     \includegraphics[width=\linewidth]{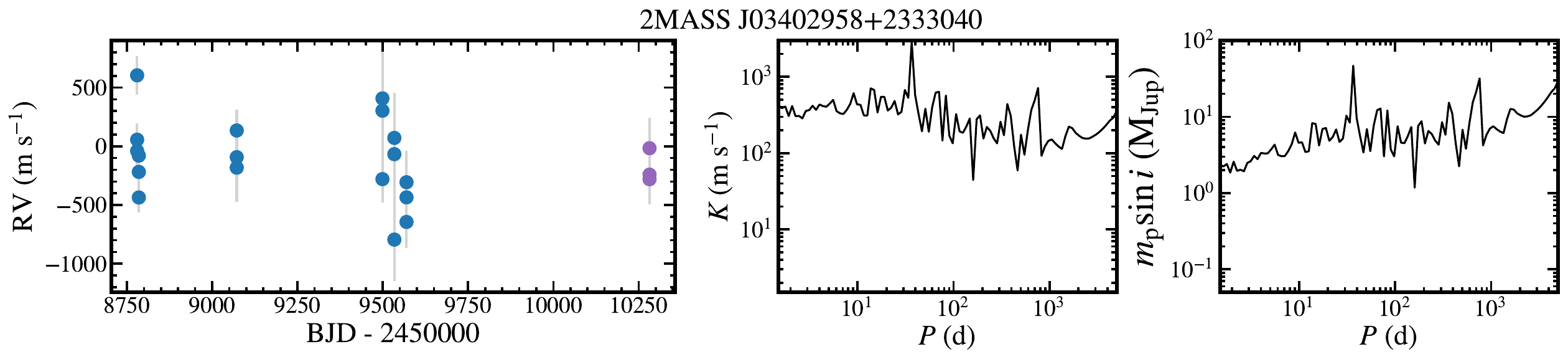}
     \includegraphics[width=\linewidth]{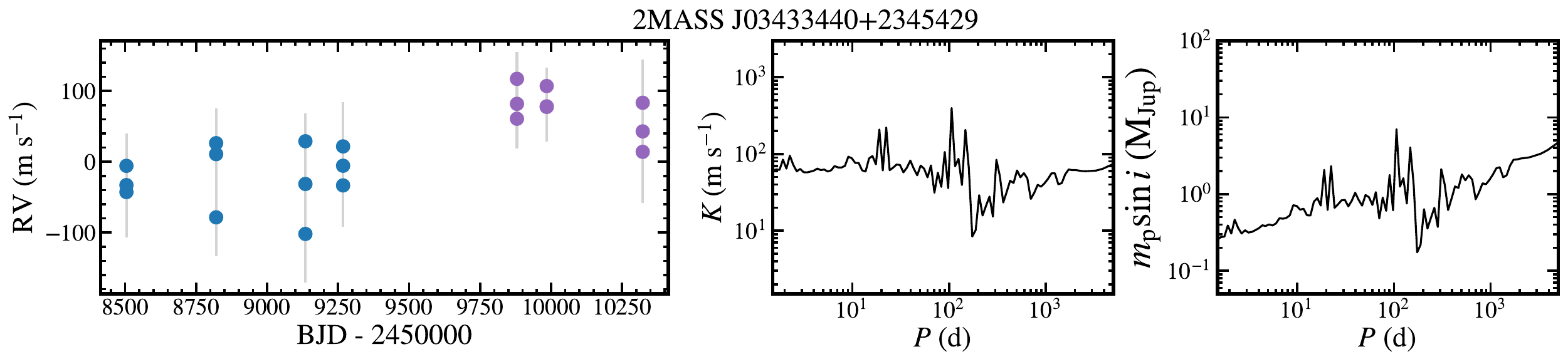}
     \includegraphics[width=\linewidth]{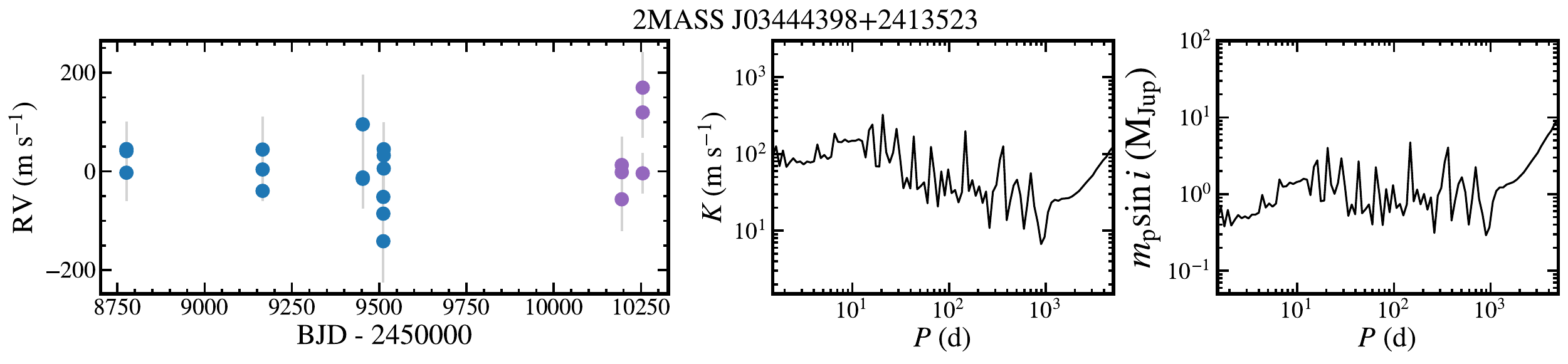}
     \includegraphics[width=\linewidth]{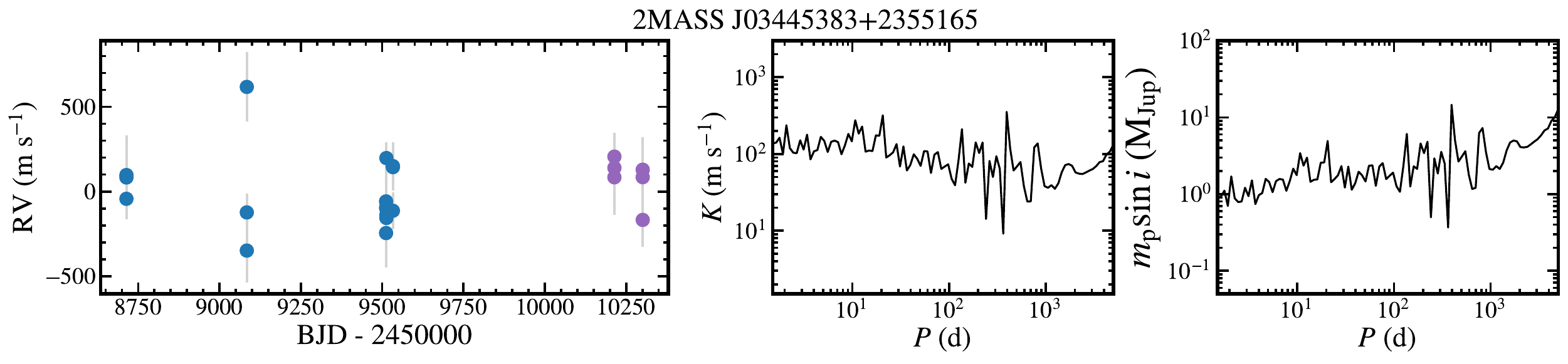}
     \includegraphics[width=\linewidth]{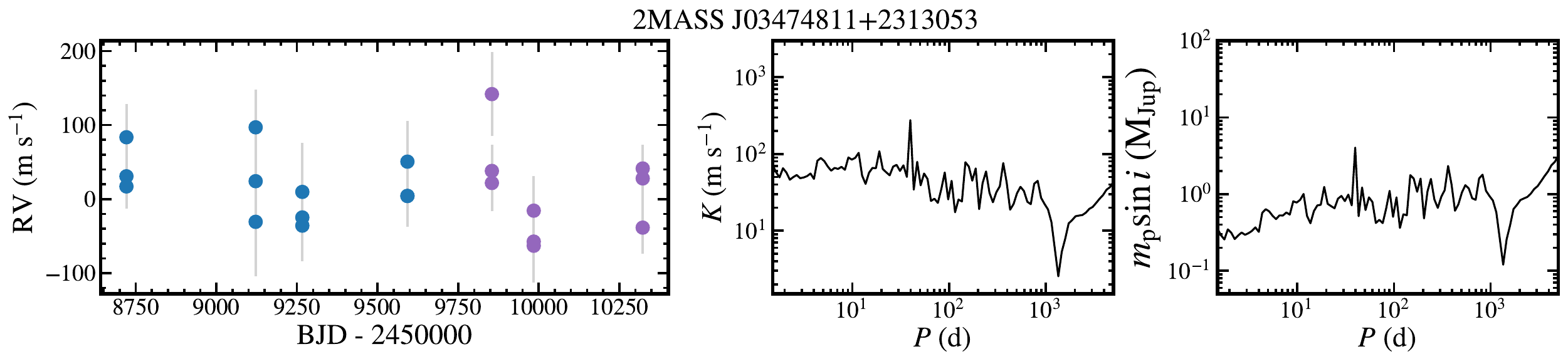}
    \caption{HPF time series and corresponding completeness functions for targets 2MASS J03402958+2333040, 2MASS J03433440+2345429, 2MASS J03444398+2413523, 2MASS J03445383+2355165, and 2MASS J03474811+2313053.}
    \label{fig:rv_comp0}
\end{figure}

\begin{figure}[p]
     \centering
     \includegraphics[width=\linewidth]{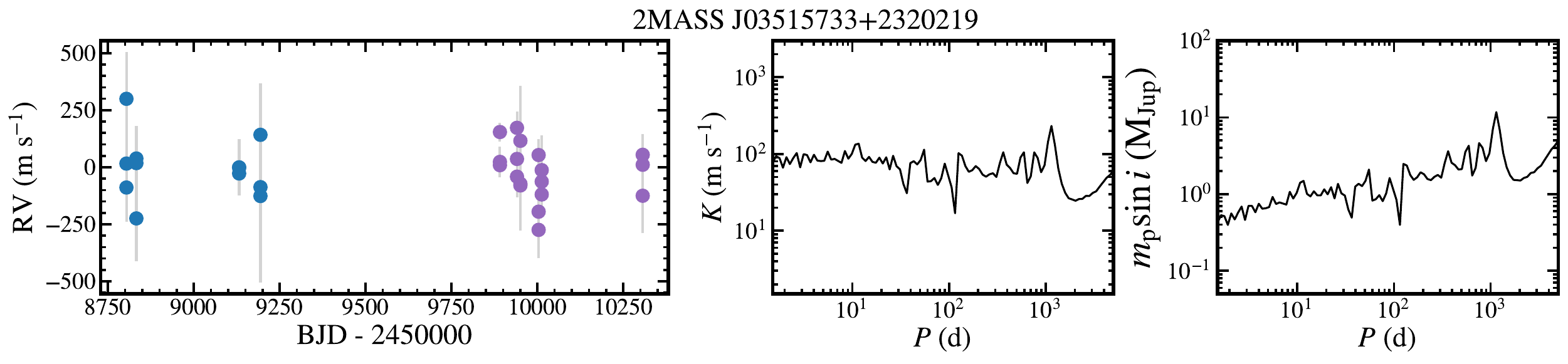}
     \includegraphics[width=\linewidth]{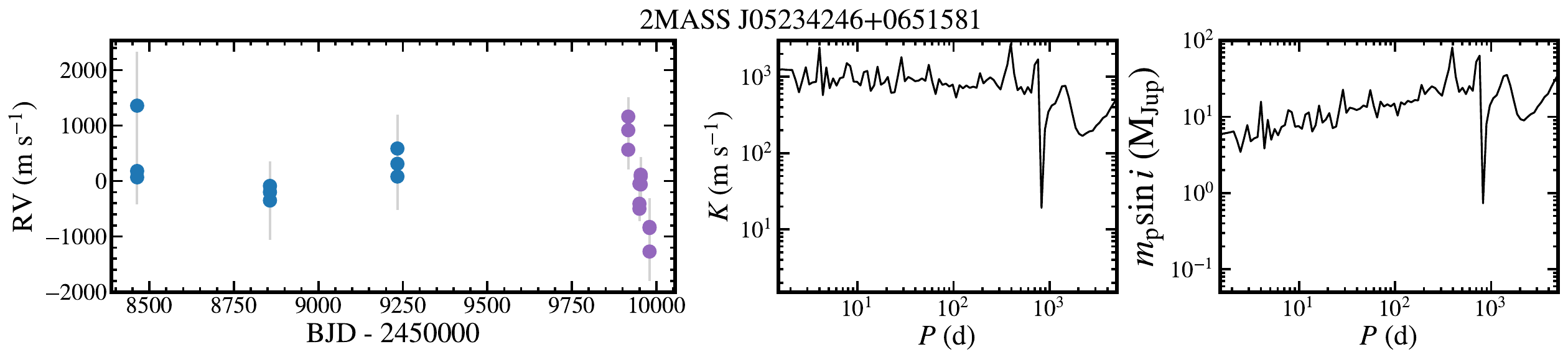}
     \includegraphics[width=\linewidth]{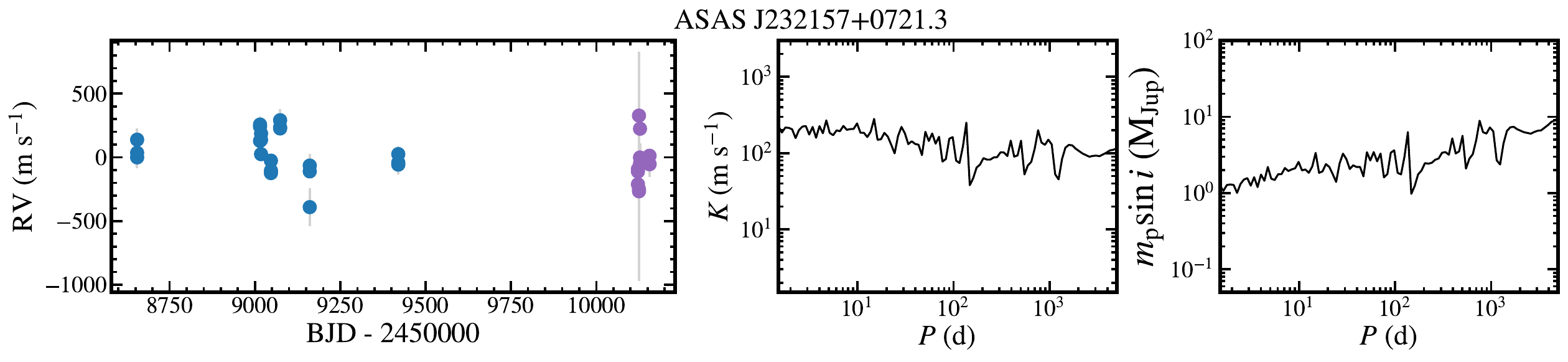}
     \includegraphics[width=\linewidth]{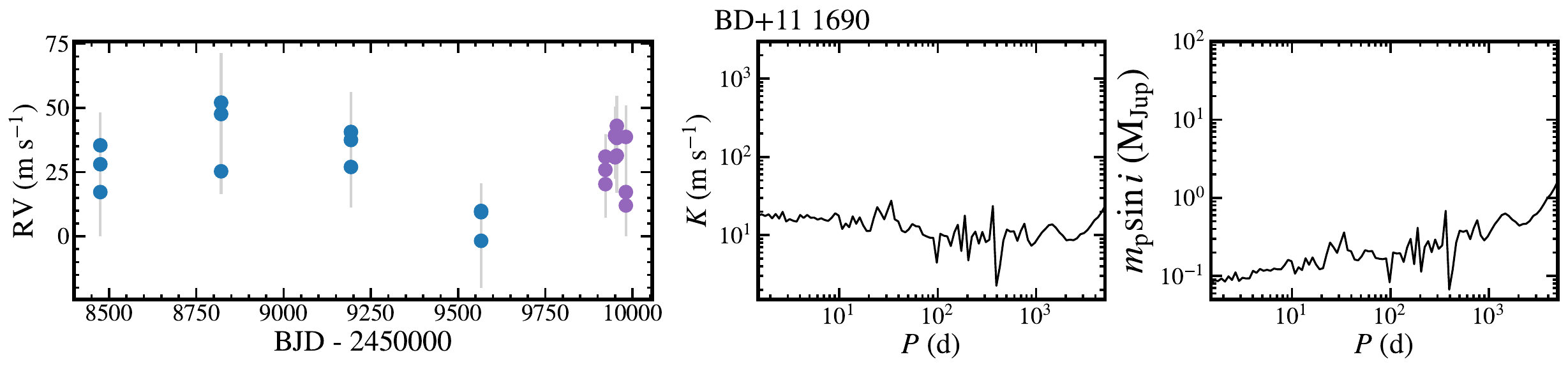}
     \includegraphics[width=\linewidth]{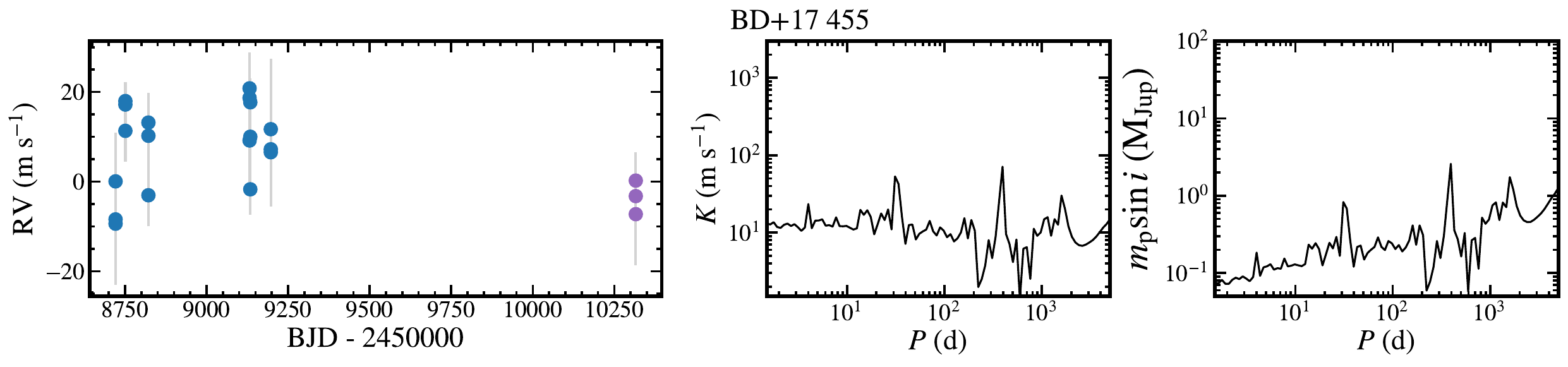}
    \caption{HPF time series and corresponding completeness functions for targets 2MASS J03515733+2320219, 2MASS J05234246+0651581, ASAS J232157+0721.3, BD+11 1690, and BD+17 455.}
    \label{fig:rv_comp1}
\end{figure}

\begin{figure}[p]
     \centering
     \includegraphics[width=\linewidth]{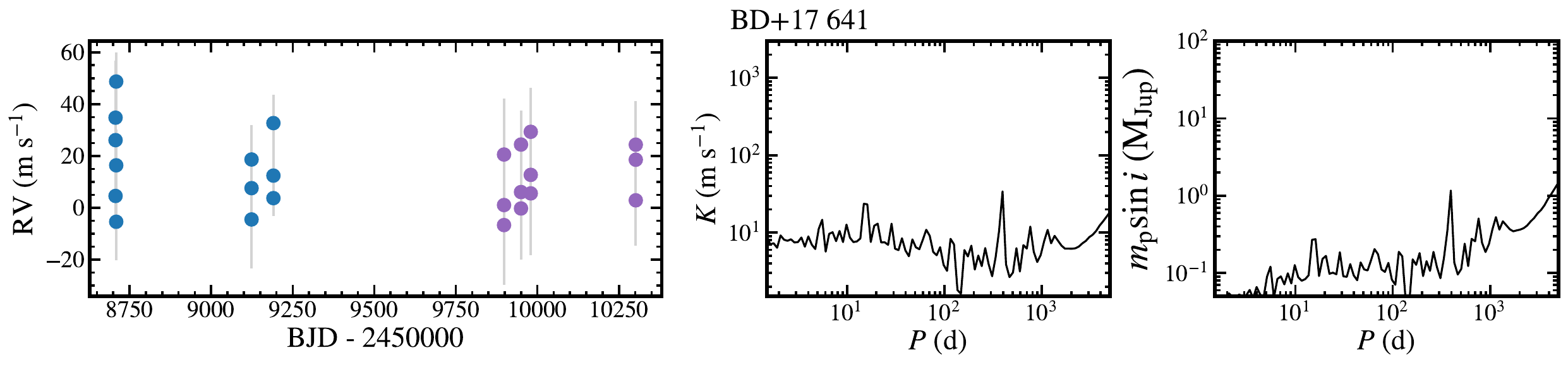}
     \includegraphics[width=\linewidth]{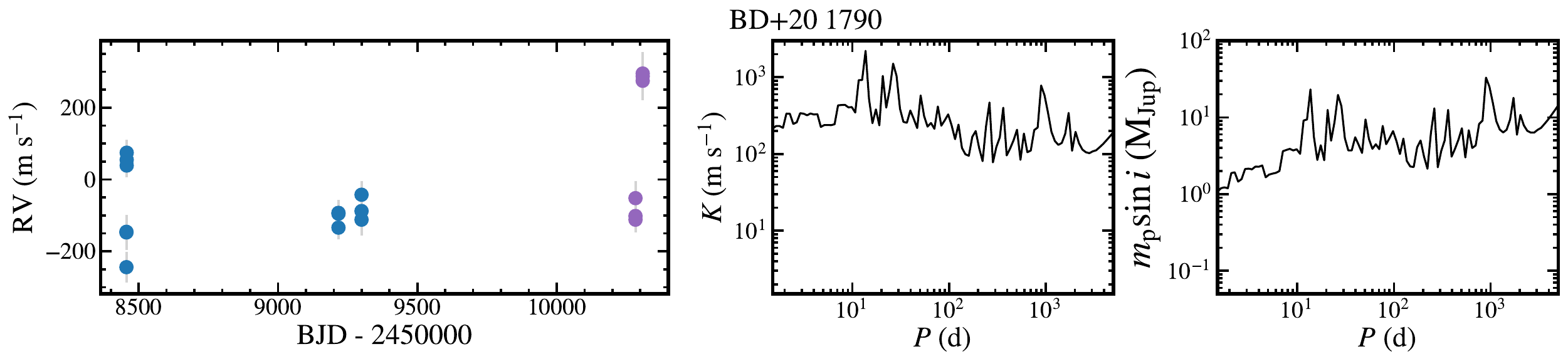}
     \includegraphics[width=\linewidth]{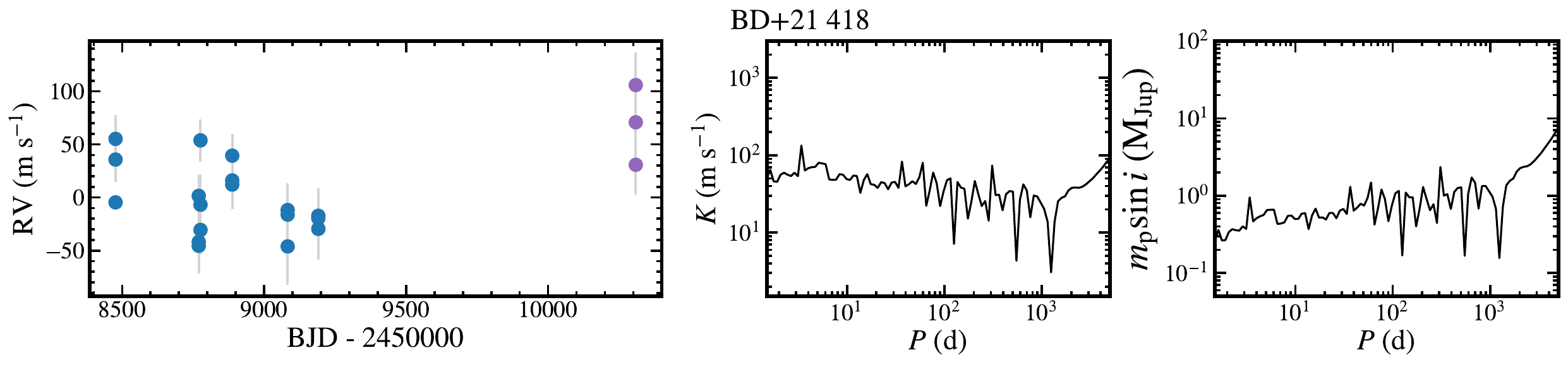}
     \includegraphics[width=\linewidth]{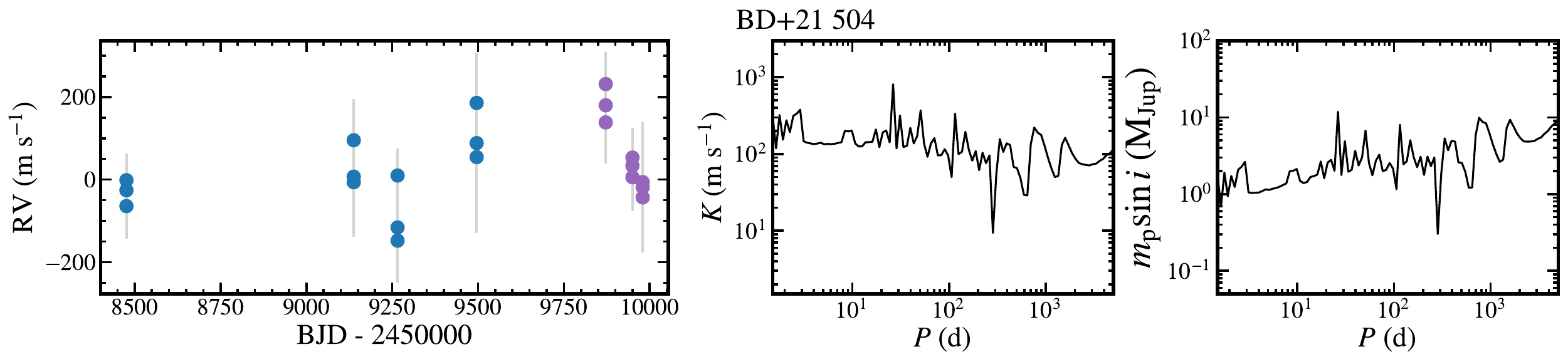}
     \includegraphics[width=\linewidth]{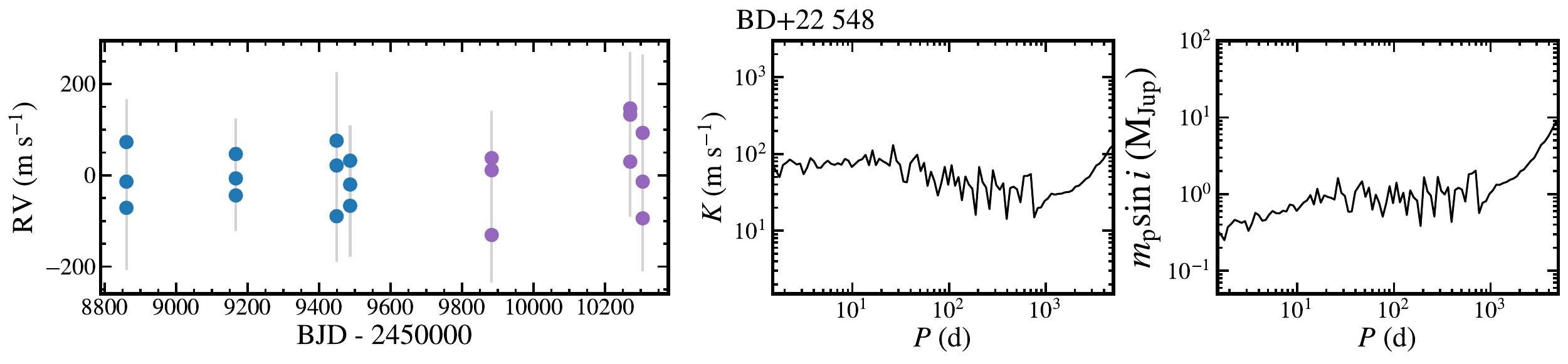}
    \caption{HPF time series and corresponding completeness functions for targets BD+17 641, BD+20 1790, BD+21 418, BD+21 504, and BD+22 548.}
    \label{fig:rv_comp2}
\end{figure}

\begin{figure}[p]
     \centering
     \includegraphics[width=\linewidth]{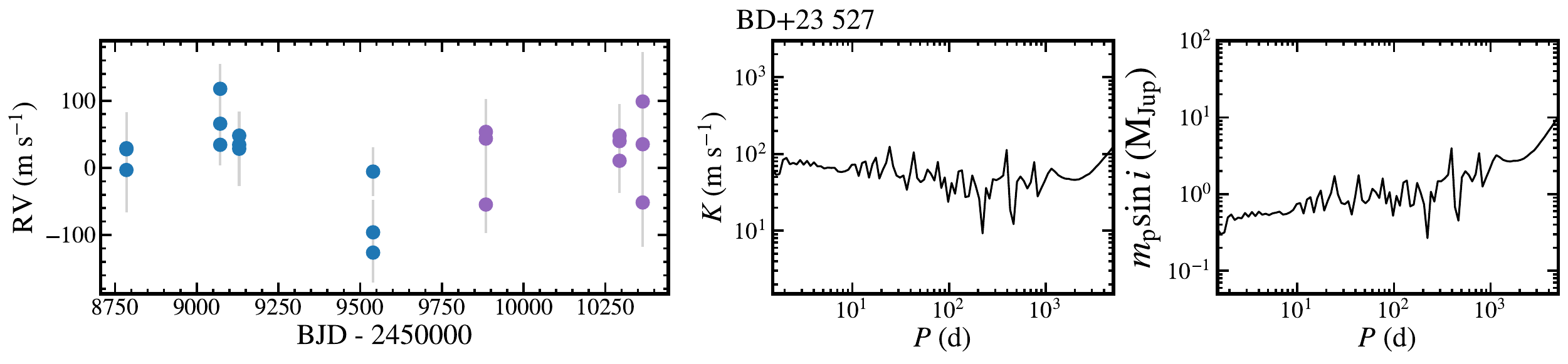}
     \includegraphics[width=\linewidth]{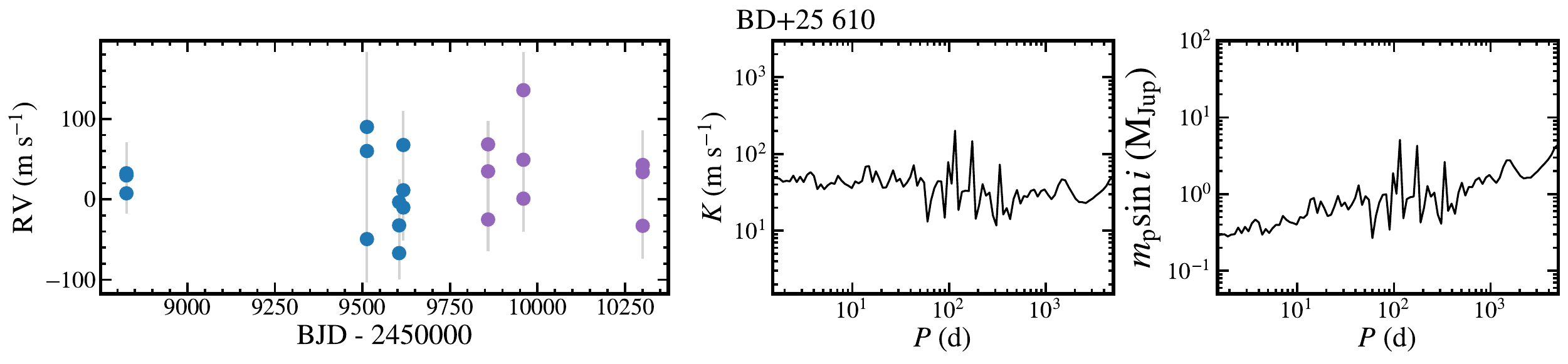}
     \includegraphics[width=\linewidth]{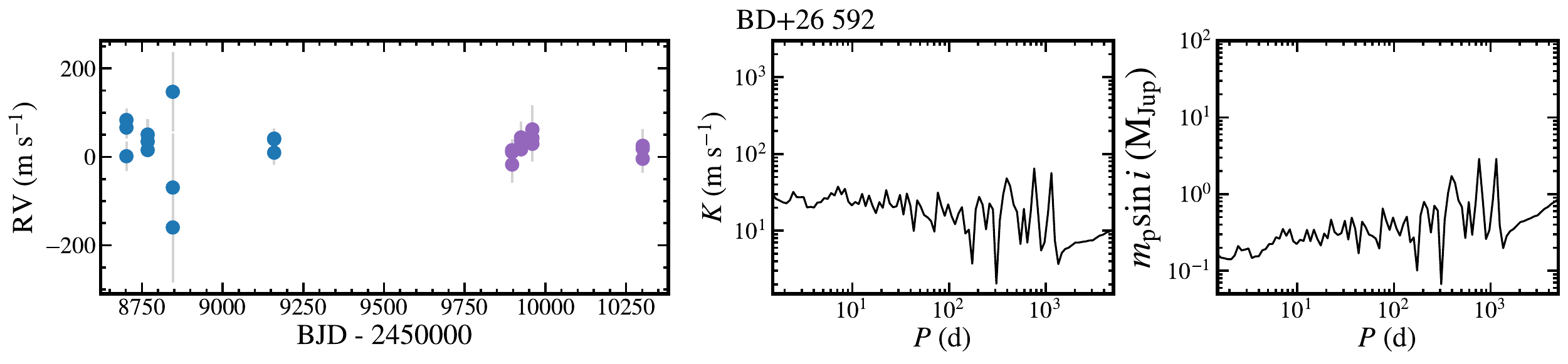}
     \includegraphics[width=\linewidth]{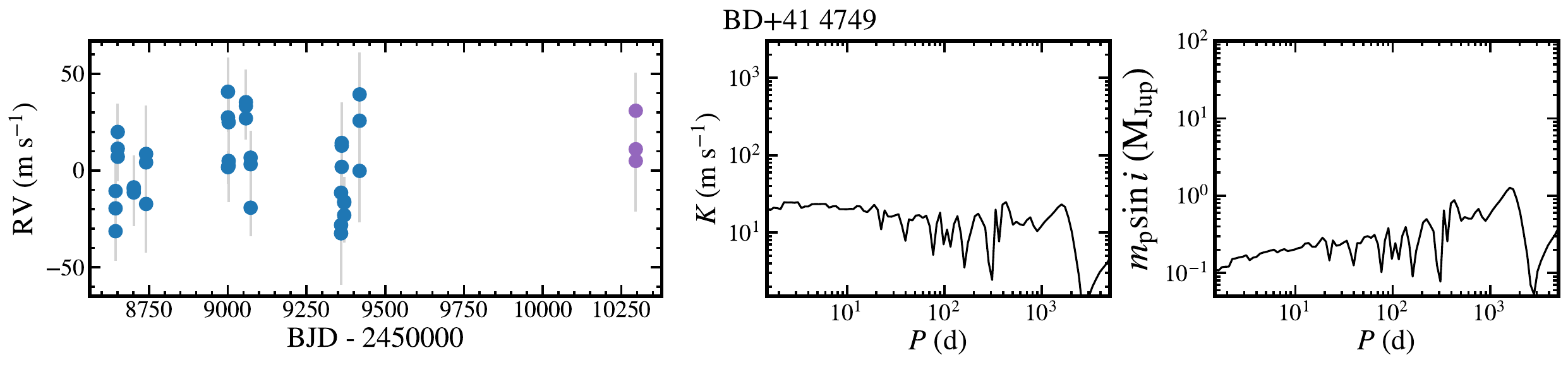}
     \includegraphics[width=\linewidth]{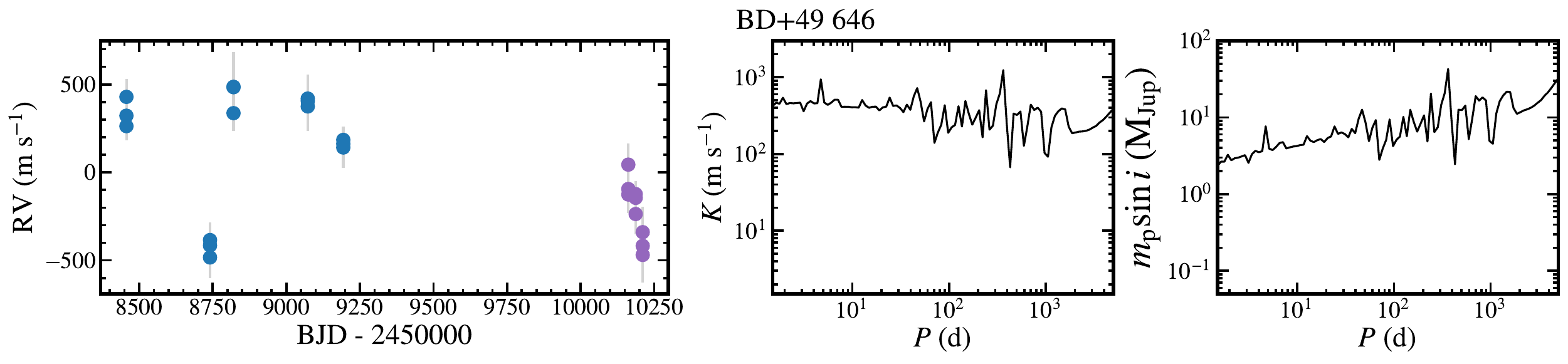}
    \caption{HPF time series and corresponding completeness functions for targets BD+23 527, BD+25 610, BD+26 592, BD+41 4749, and BD+49 646.}
    \label{fig:rv_comp3}
\end{figure}

\begin{figure}[p]
     \centering
     \includegraphics[width=\linewidth]{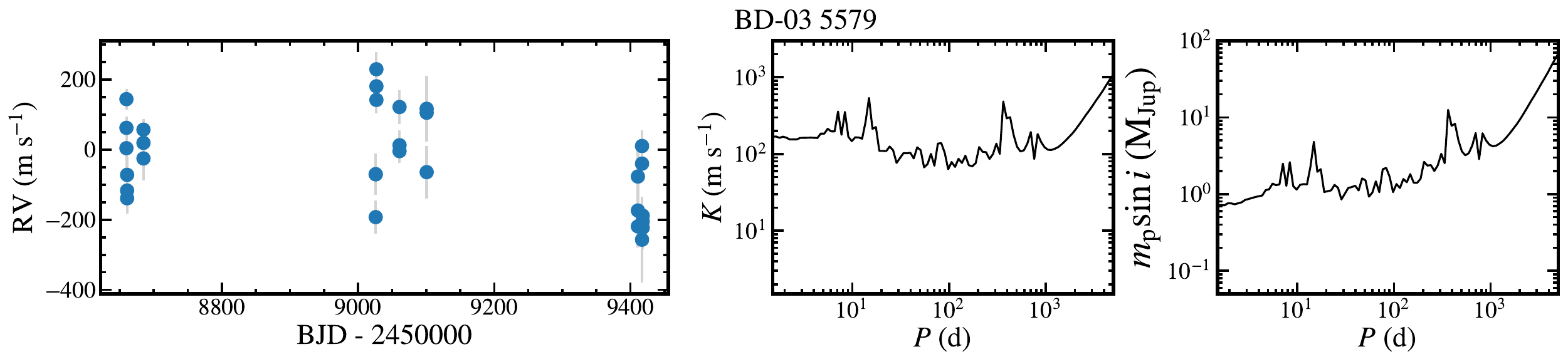}
     \includegraphics[width=\linewidth]{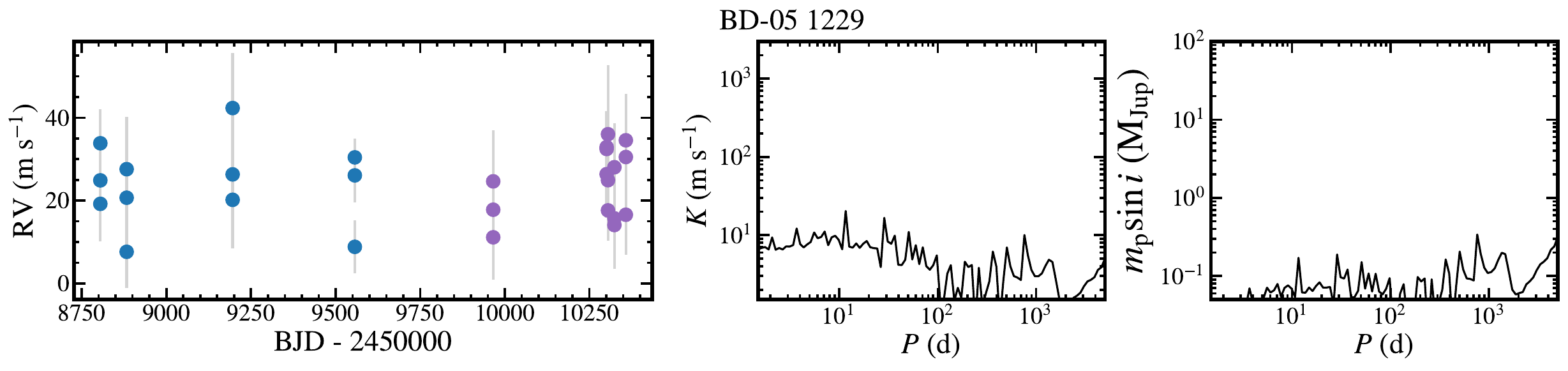}
     \includegraphics[width=\linewidth]{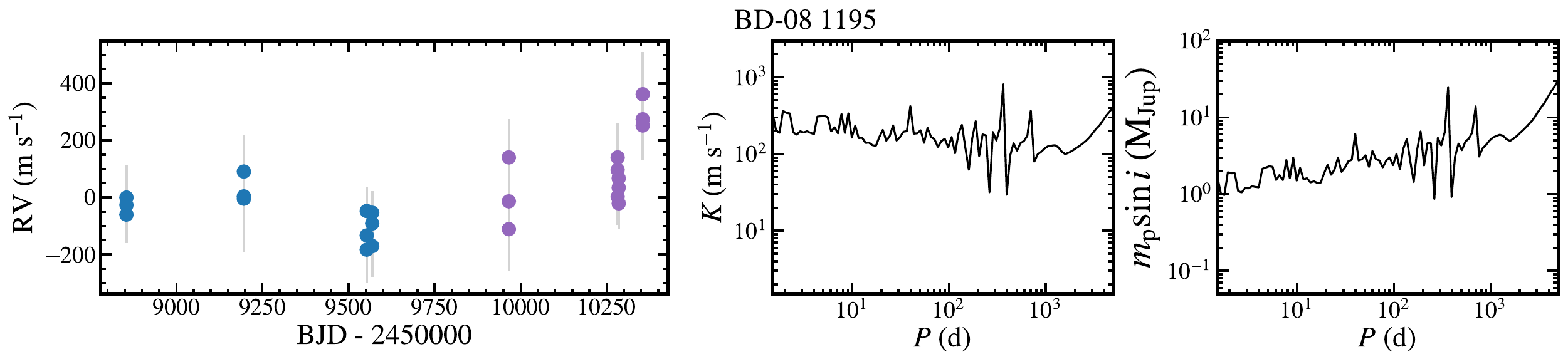}
     \includegraphics[width=\linewidth]{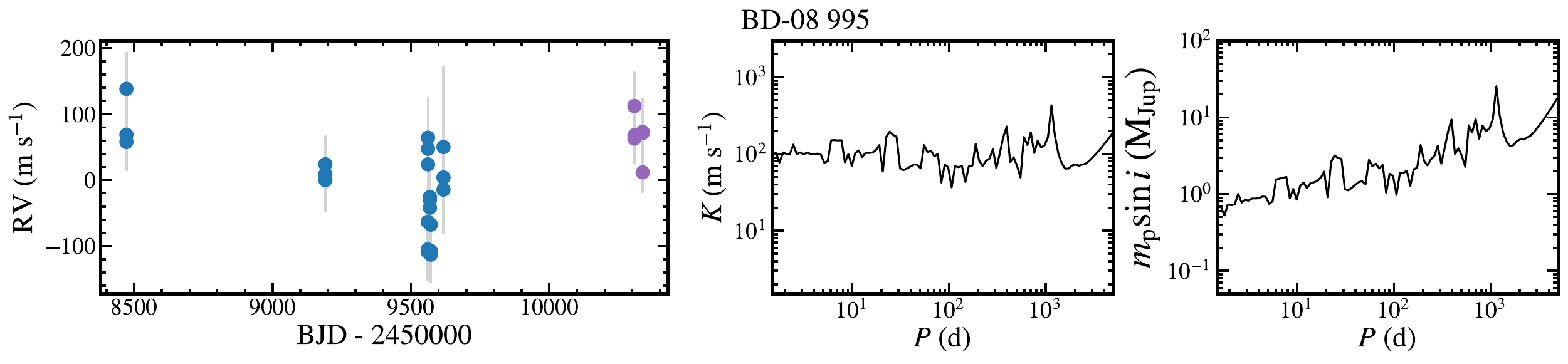}
     \includegraphics[width=\linewidth]{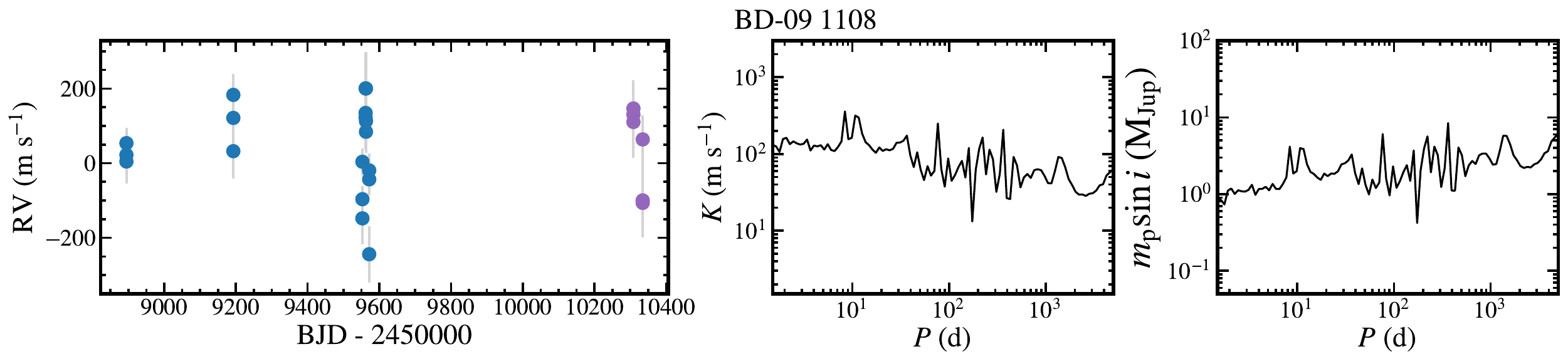}
    \caption{HPF time series and corresponding completeness functions for targets BD-03 5579, BD-05 1229, BD-08 1195, BD-08 995, and BD-09 1108.}
    \label{fig:rv_comp4}
\end{figure}

\begin{figure}[p]
     \centering
     \includegraphics[width=\linewidth]{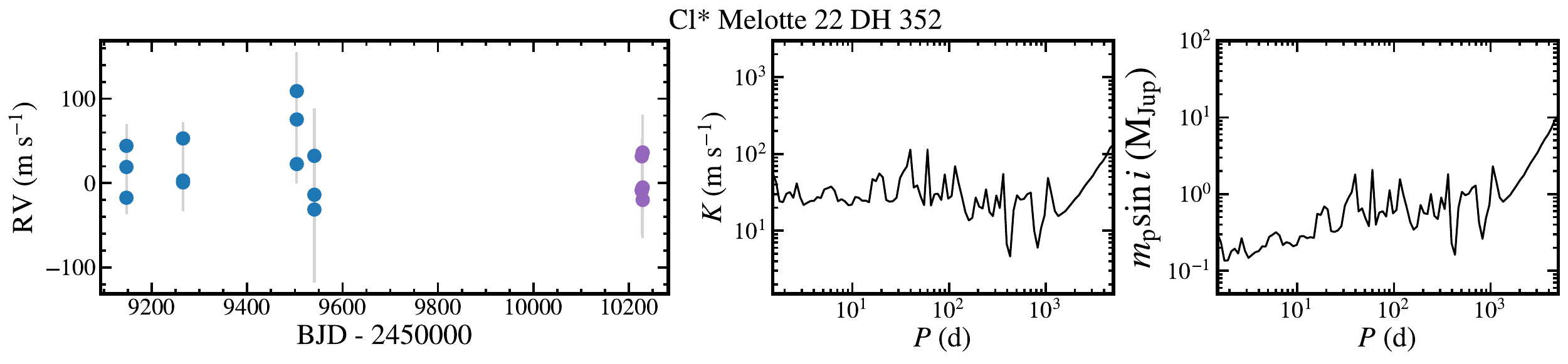}
     \includegraphics[width=\linewidth]{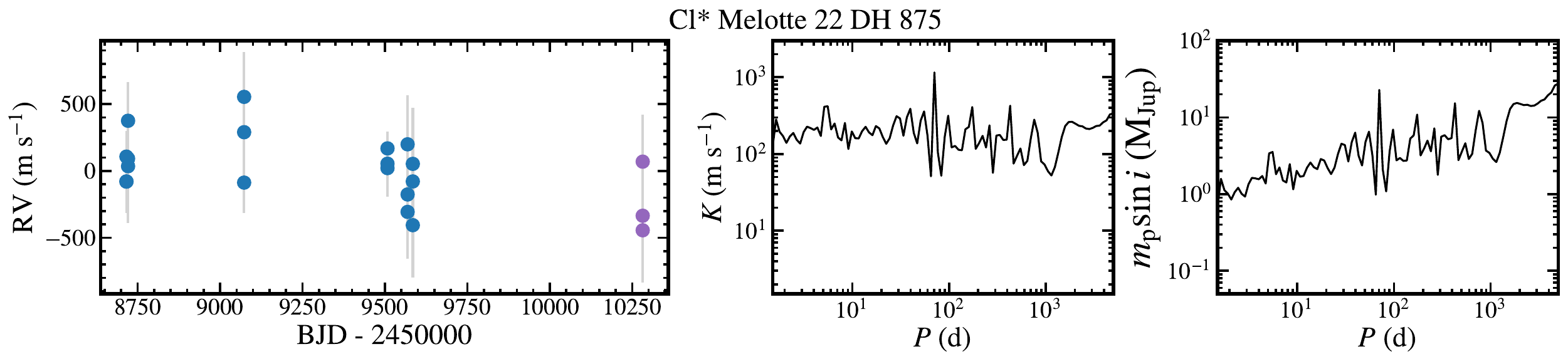}
     \includegraphics[width=\linewidth]{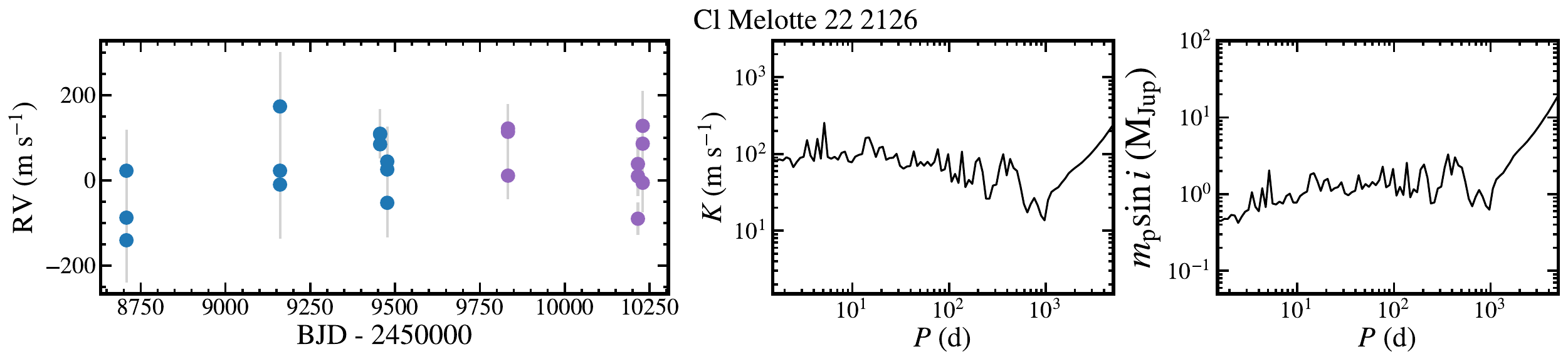}
     \includegraphics[width=\linewidth]{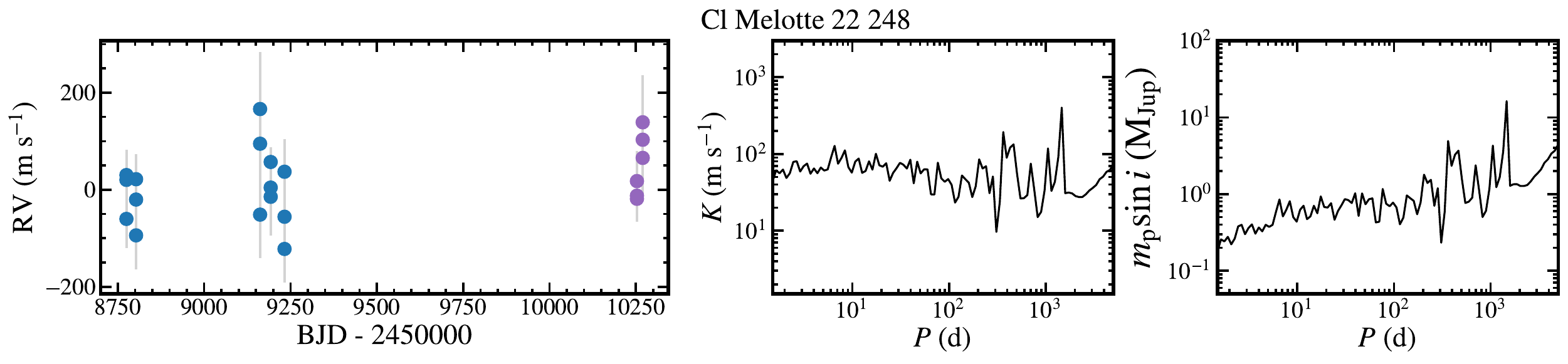}
     \includegraphics[width=\linewidth]{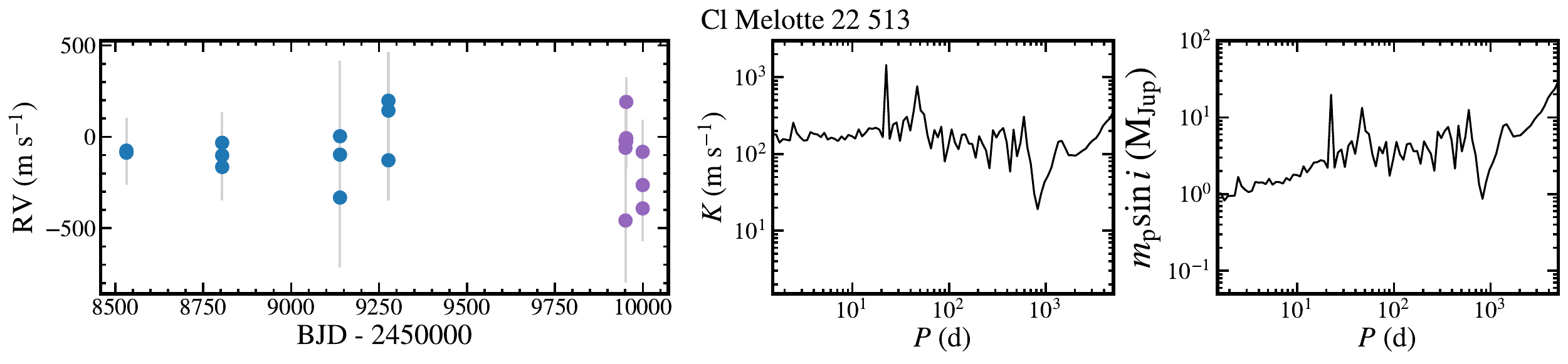}
    \caption{HPF time series and corresponding completeness functions for targets Cl* Melotte 22 DH 352, Cl* Melotte 22 DH 875, Cl Melotte 22 2126, Cl Melotte 22 248, and Cl Melotte 22 513.}
    \label{fig:rv_comp5}
\end{figure}

\begin{figure}[p]
     \centering
     \includegraphics[width=\linewidth]{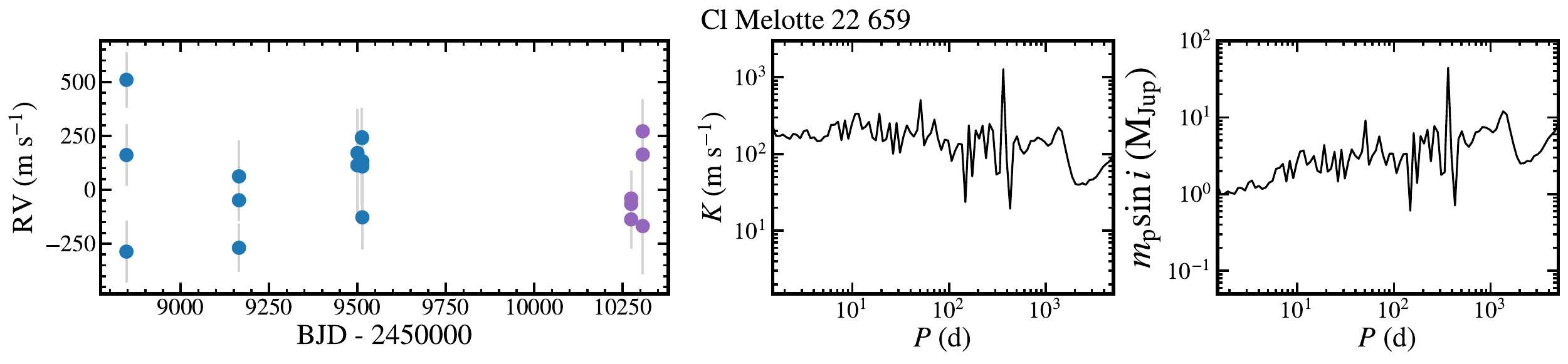}
     \includegraphics[width=\linewidth]{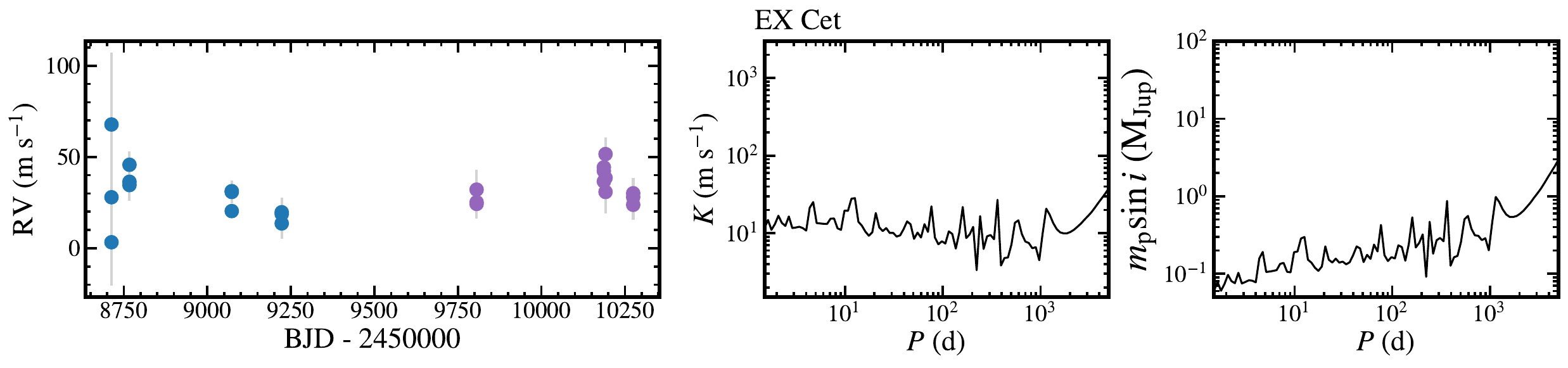}
     \includegraphics[width=\linewidth]{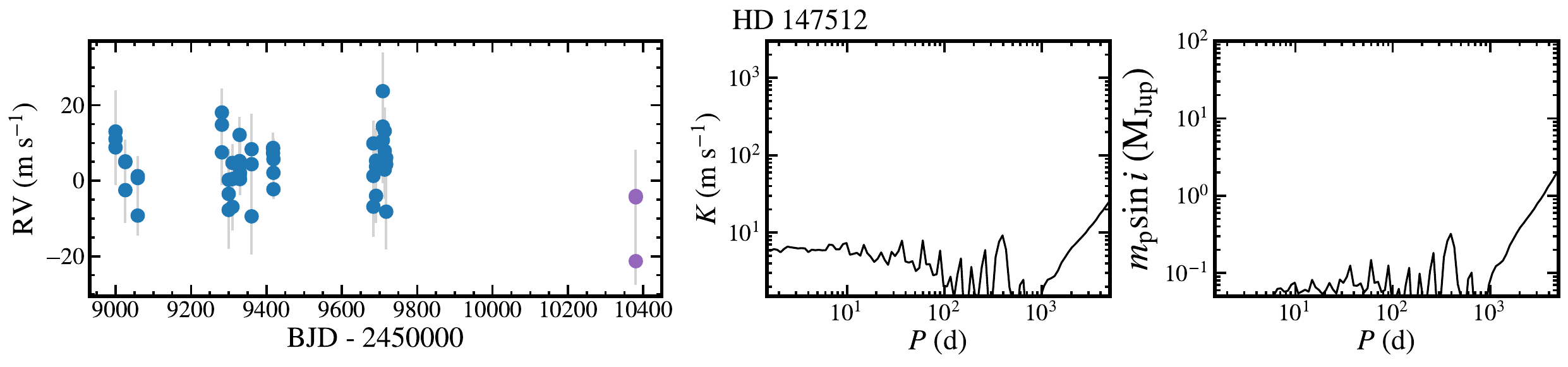}
     \includegraphics[width=\linewidth]{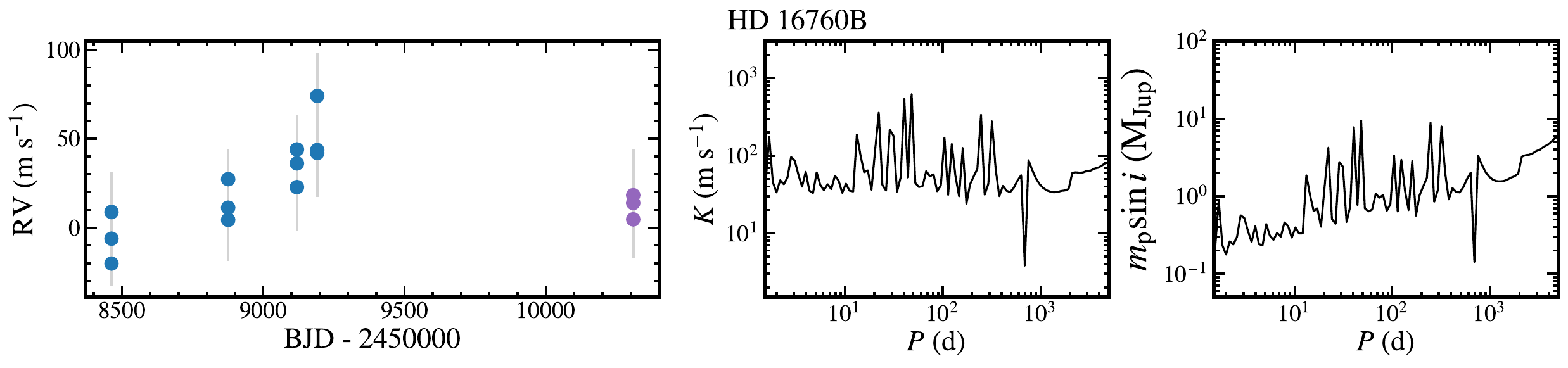}
     \includegraphics[width=\linewidth]{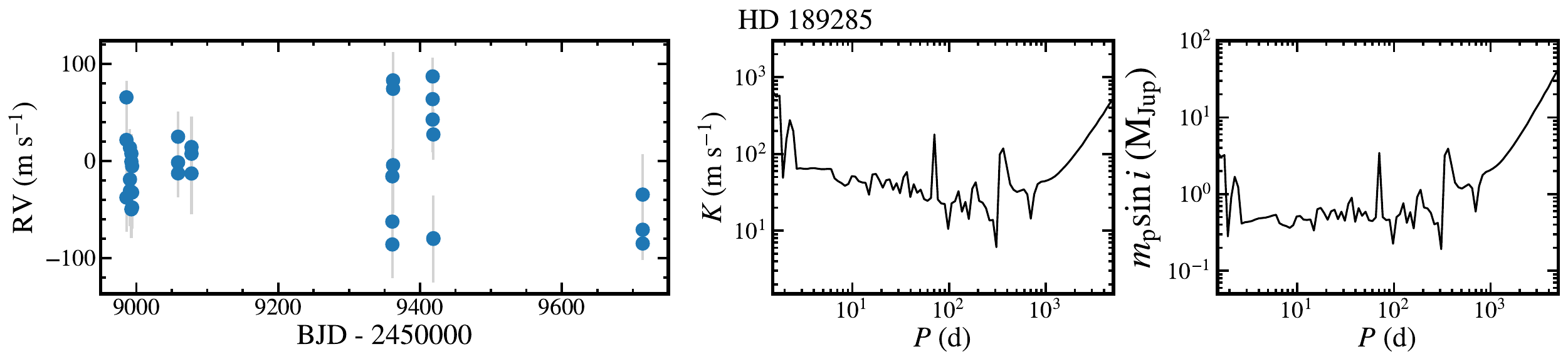}
    \caption{HPF time series and corresponding completeness functions for targets Cl Melotte 22 659, EX Cet, HD 147512, HD 16760B, and HD 189285.}
    \label{fig:rv_comp6}
\end{figure}

\begin{figure}[p]
     \centering
     \includegraphics[width=\linewidth]{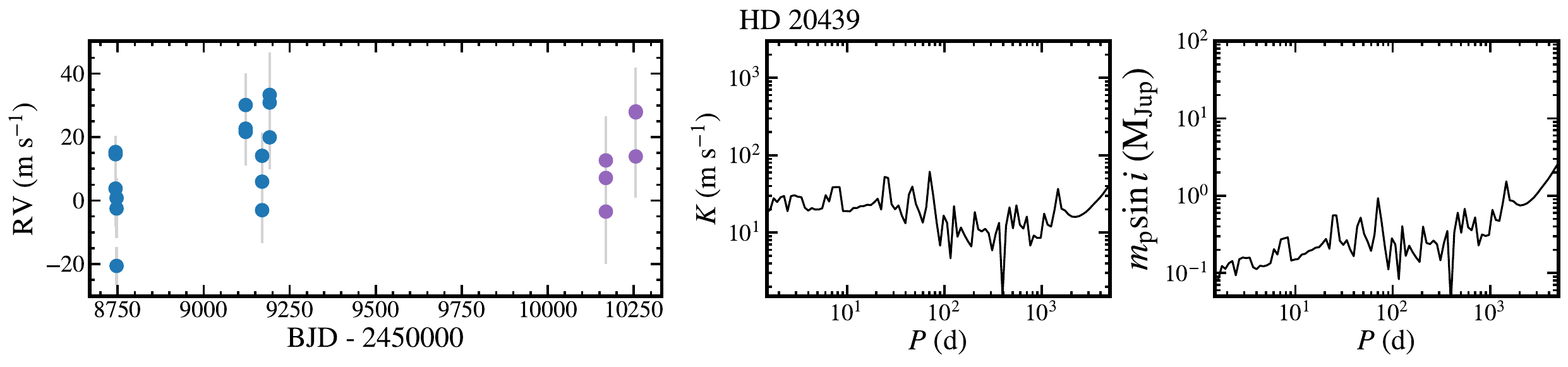}
     \includegraphics[width=\linewidth]{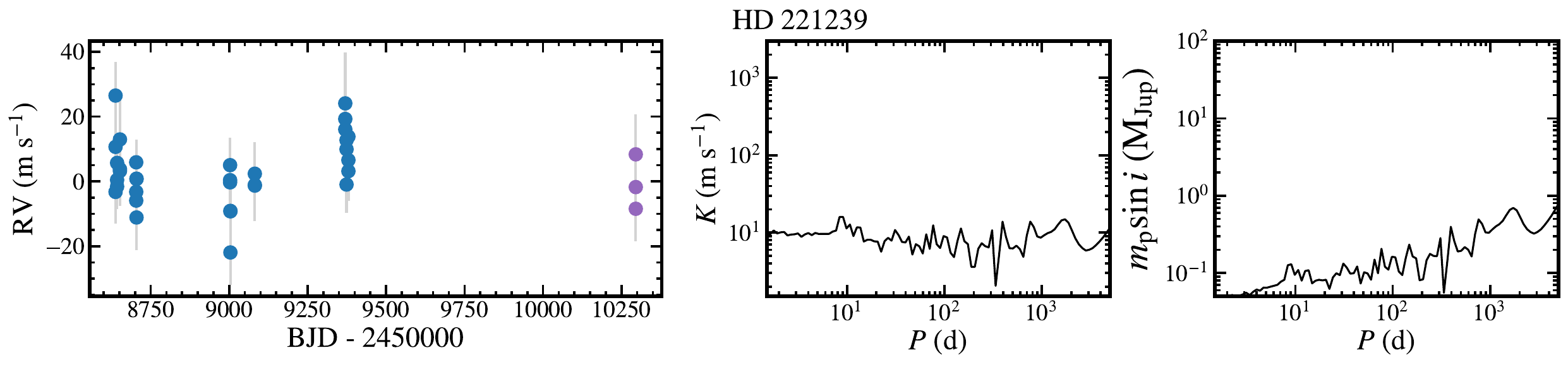}
     \includegraphics[width=\linewidth]{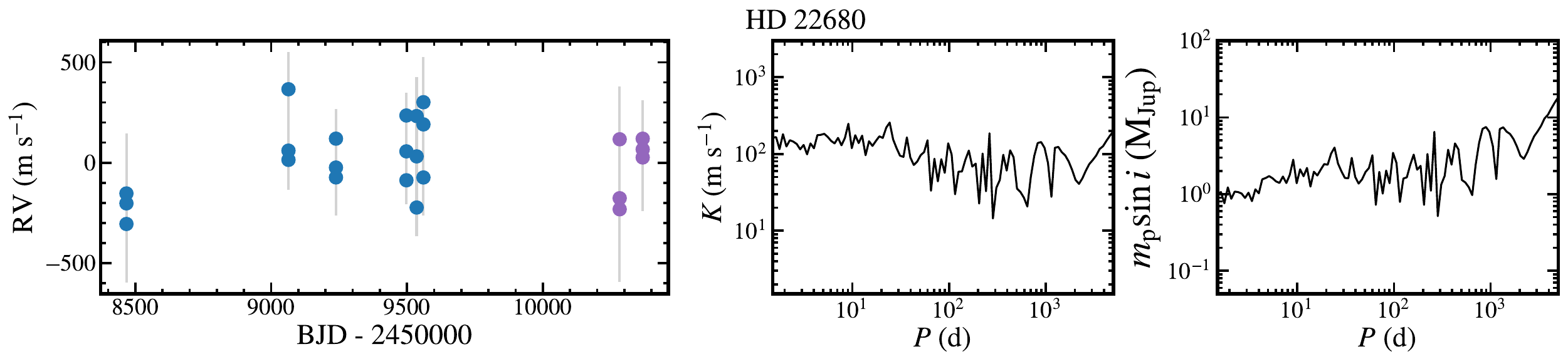}
     \includegraphics[width=\linewidth]{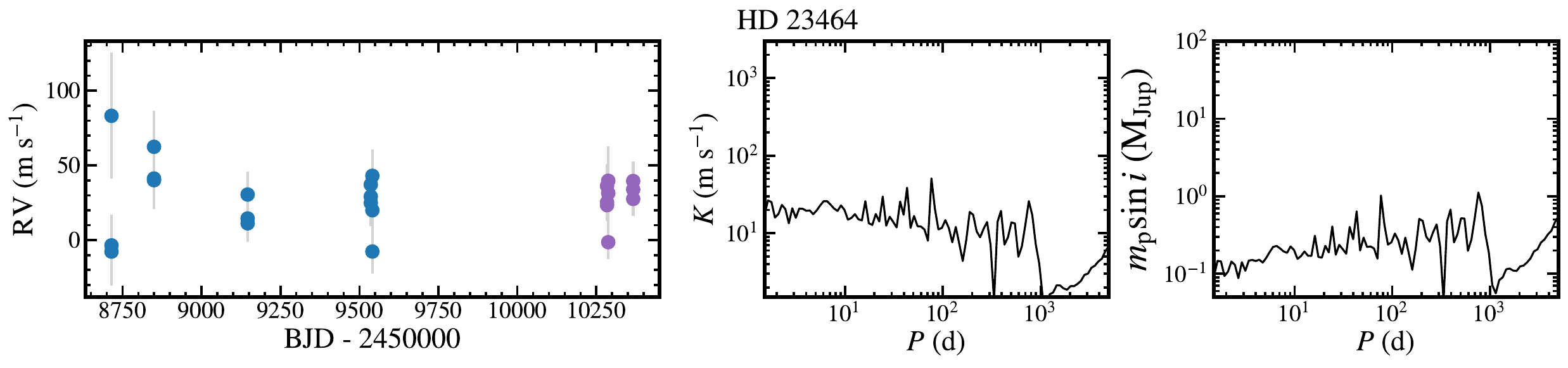}
     \includegraphics[width=\linewidth]{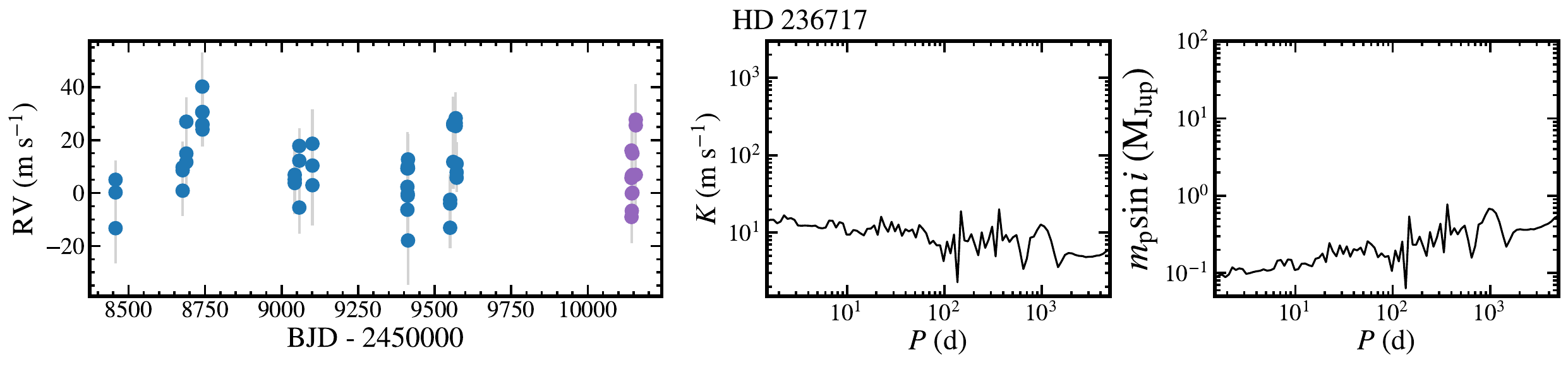}
    \caption{HPF time series and corresponding completeness functions for targets HD 20439, HD 221239, HD 22680, HD 23464, and HD 236717.}
    \label{fig:rv_comp7}
\end{figure}

\begin{figure}[p]
     \centering
     \includegraphics[width=\linewidth]{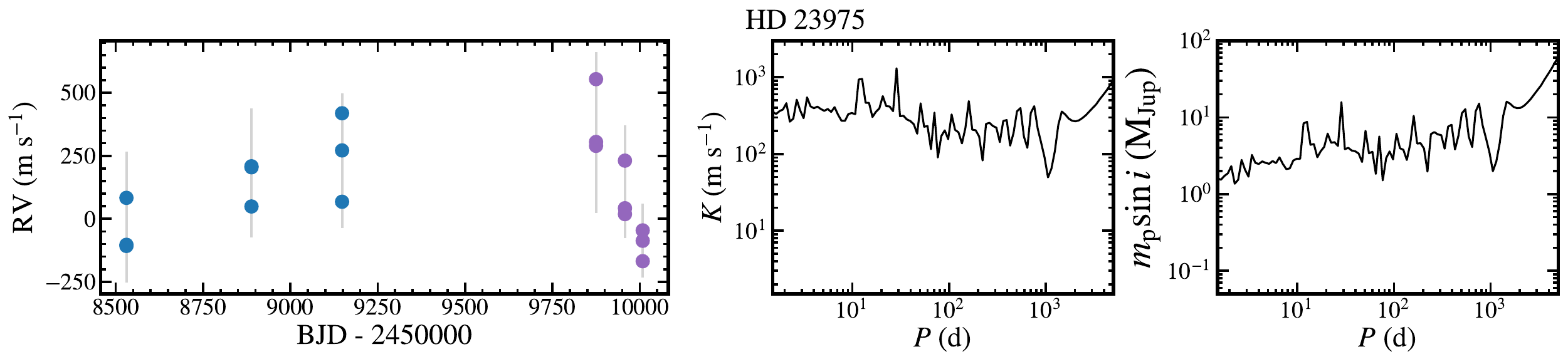}
     \includegraphics[width=\linewidth]{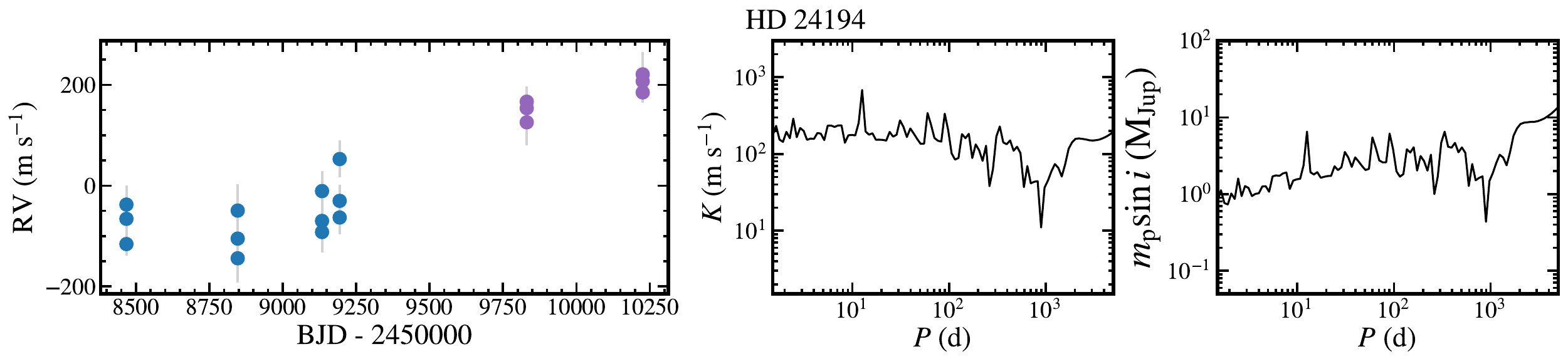}
     \includegraphics[width=\linewidth]{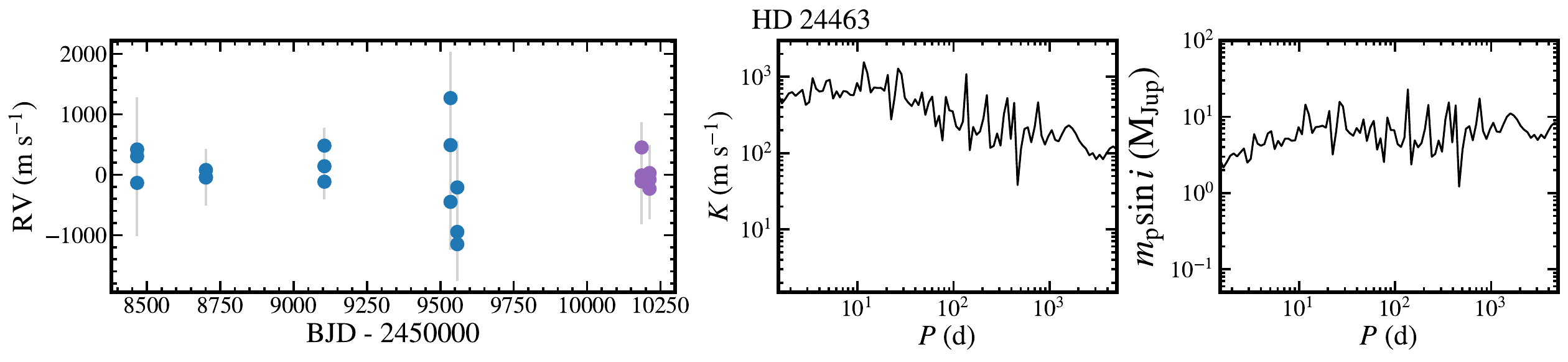}
     \includegraphics[width=\linewidth]{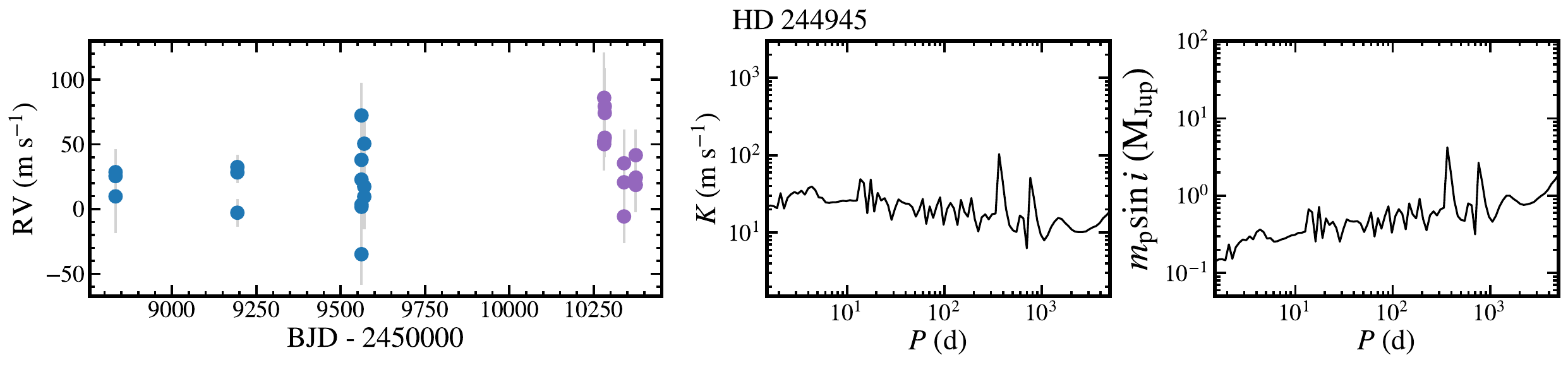}
     \includegraphics[width=\linewidth]{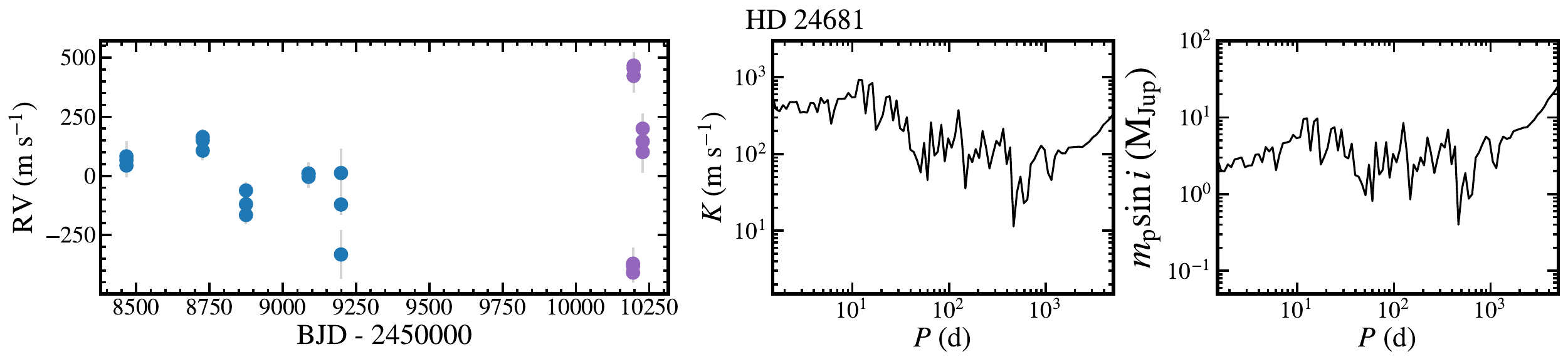}
    \caption{HPF time series and corresponding completeness functions for targets HD 23975, HD 24194, HD 24463, HD 244945, and HD 24681.}
    \label{fig:rv_comp8}
\end{figure}

\begin{figure}[p]
     \centering
     \includegraphics[width=\linewidth]{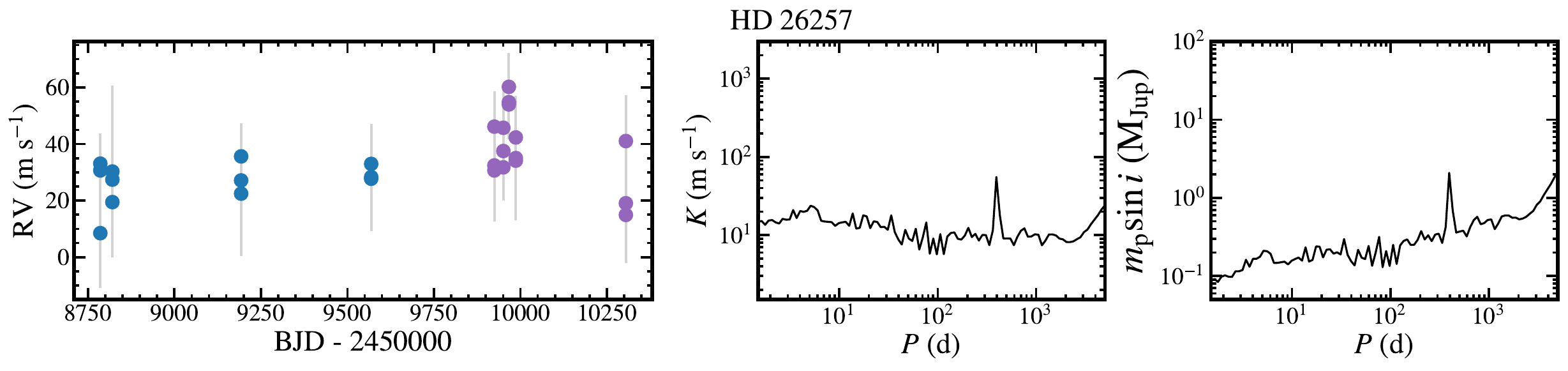}
     \includegraphics[width=\linewidth]{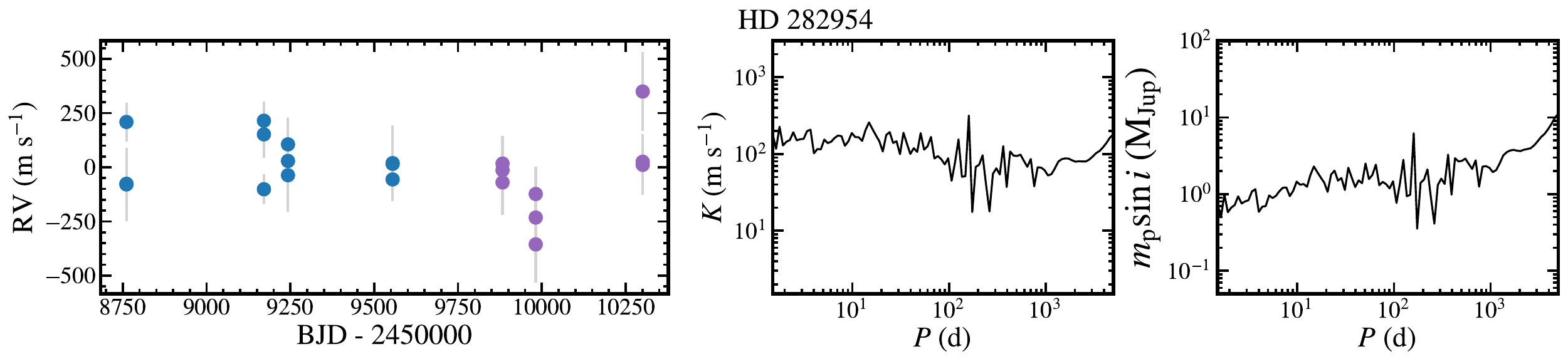}
     \includegraphics[width=\linewidth]{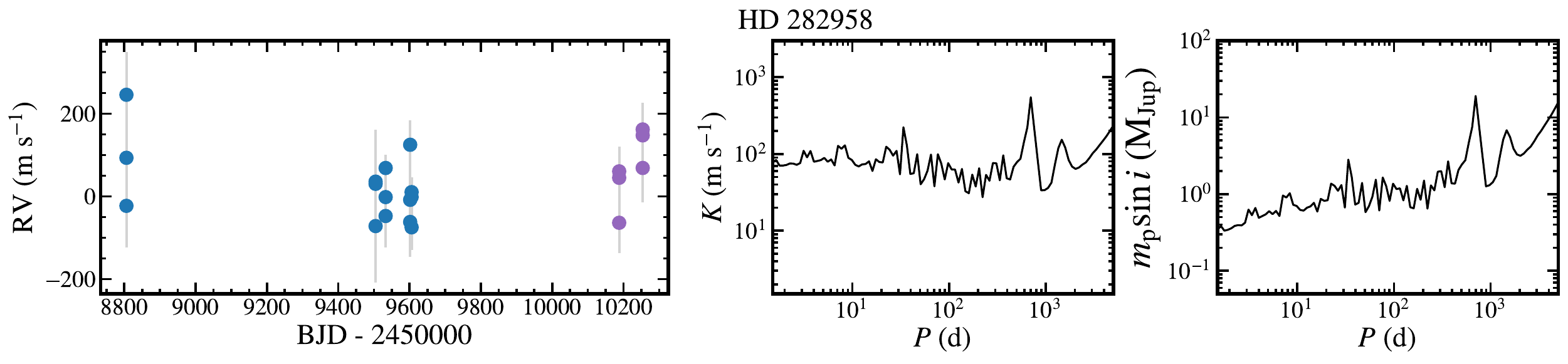}
     \includegraphics[width=\linewidth]{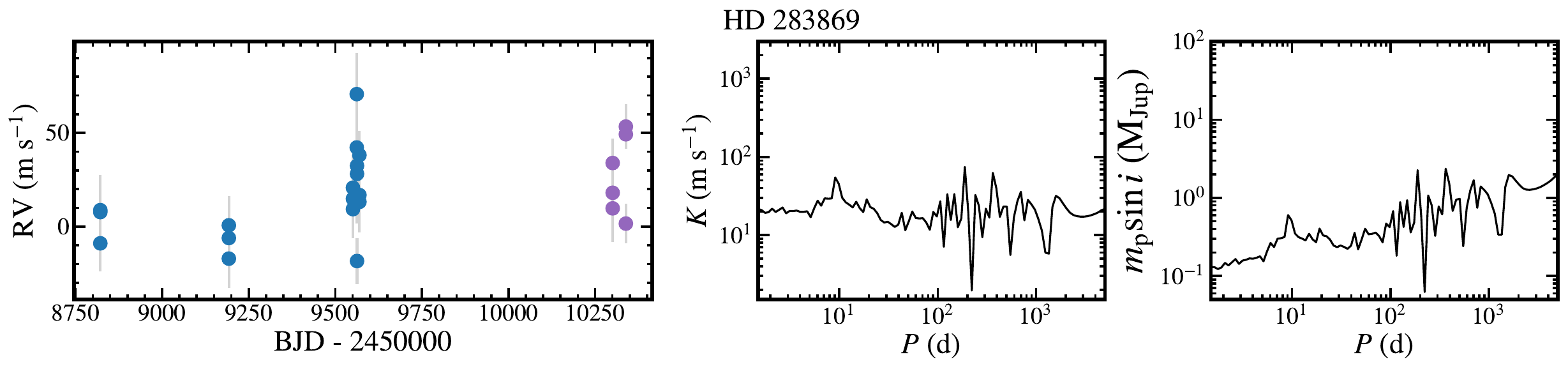}
     \includegraphics[width=\linewidth]{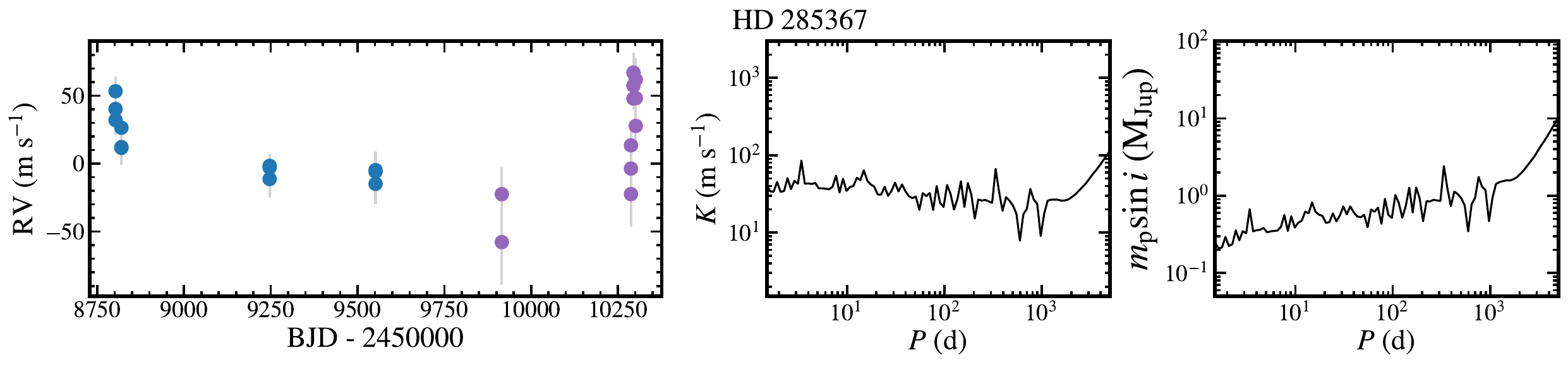}
    \caption{HPF time series and corresponding completeness functions for targets HD 26257, HD 282954, HD 282958, HD 283869, and HD 285367.}
    \label{fig:rv_comp9}
\end{figure}

\begin{figure}[p]
     \centering
     \includegraphics[width=\linewidth]{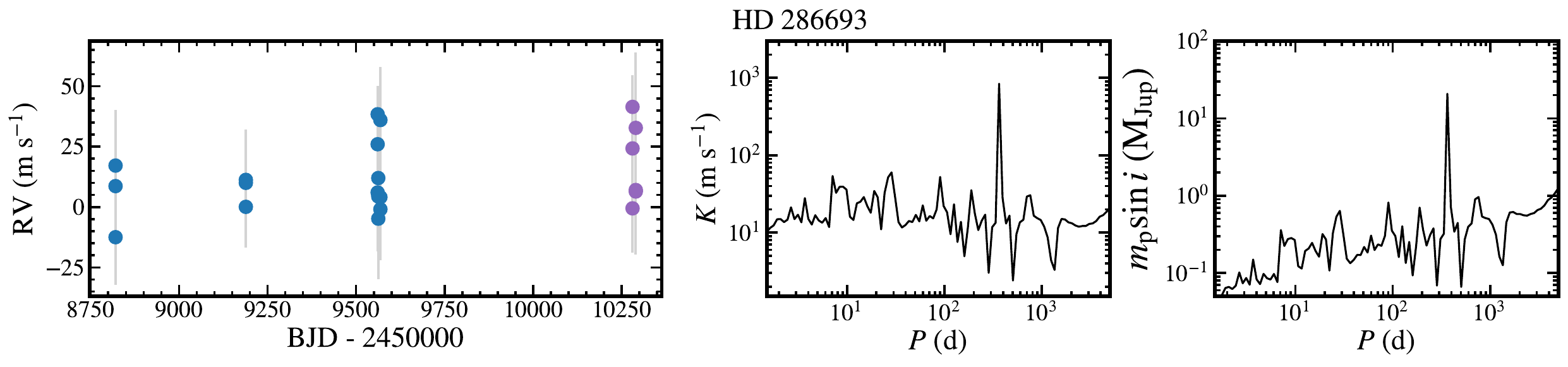}
     \includegraphics[width=\linewidth]{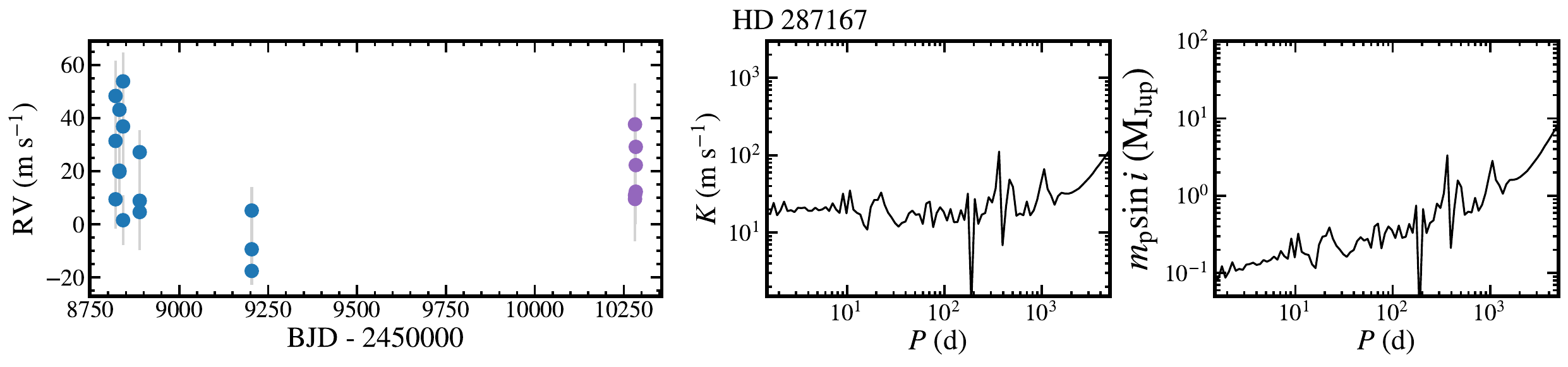}
     \includegraphics[width=\linewidth]{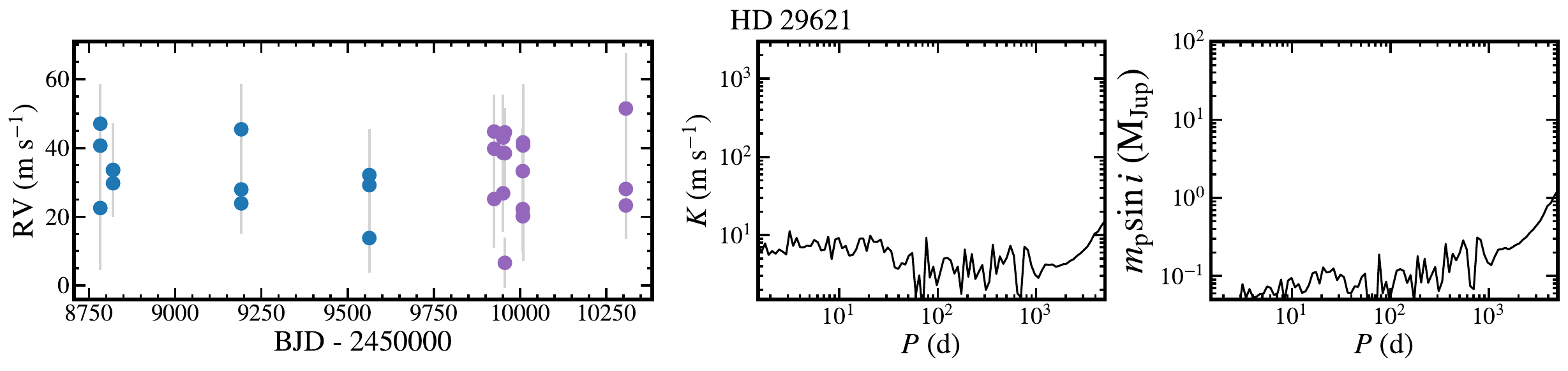}
     \includegraphics[width=\linewidth]{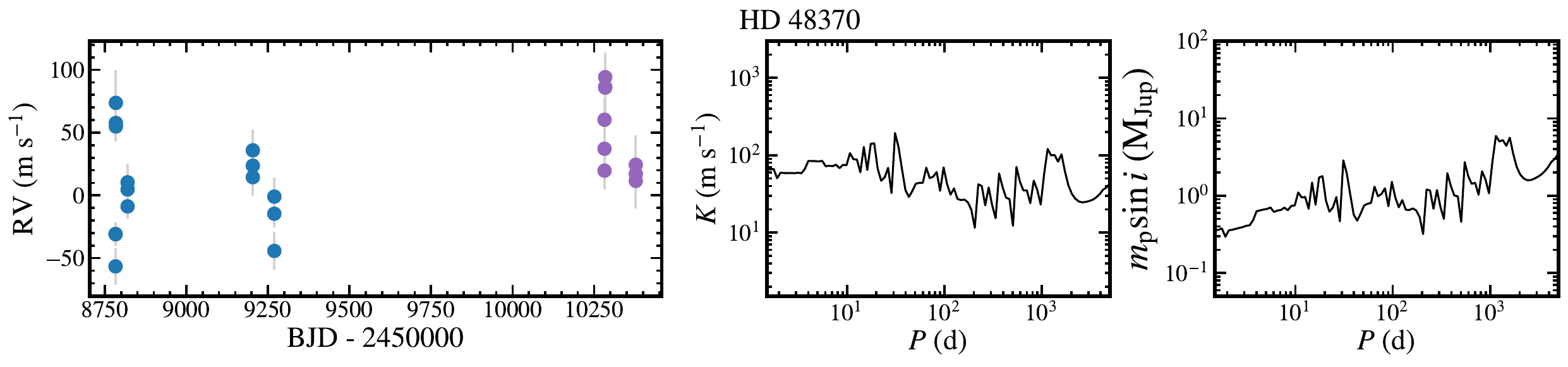}
     \includegraphics[width=\linewidth]{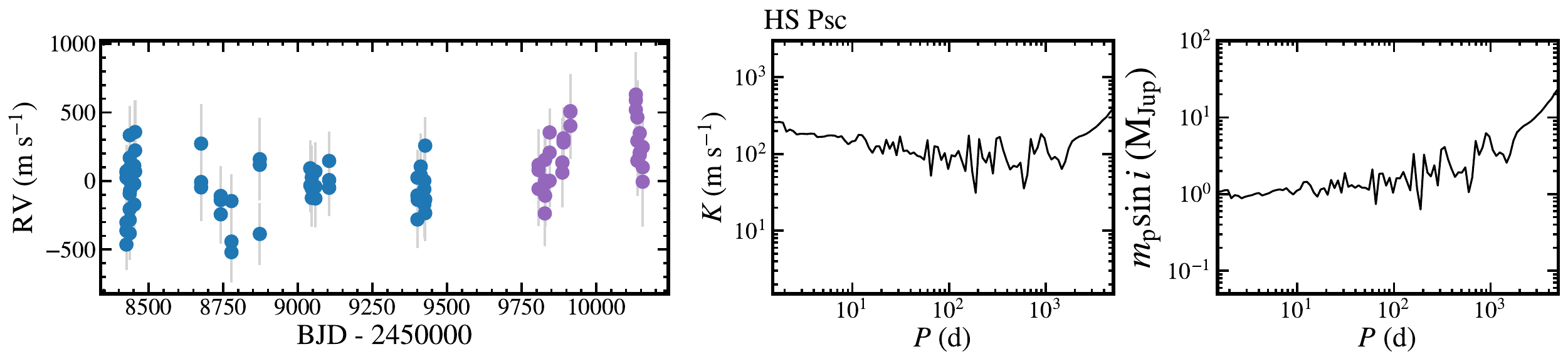}
    \caption{HPF time series and corresponding completeness functions for targets HD 286693, HD 287167, HD 29621, HD 48370, and HS Psc.}
    \label{fig:rv_comp10}
\end{figure}

\begin{figure}[p]
     \centering
     \includegraphics[width=\linewidth]{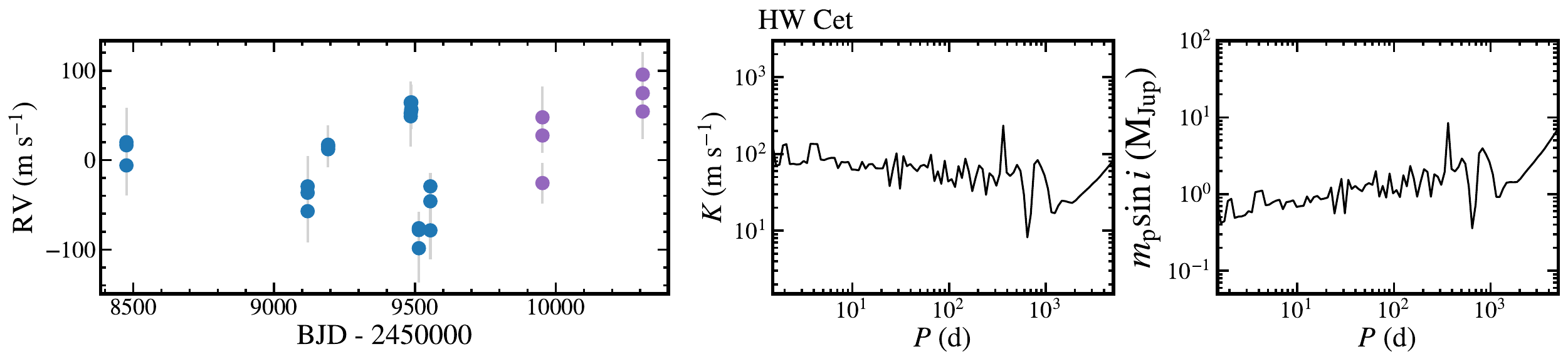}
     \includegraphics[width=\linewidth]{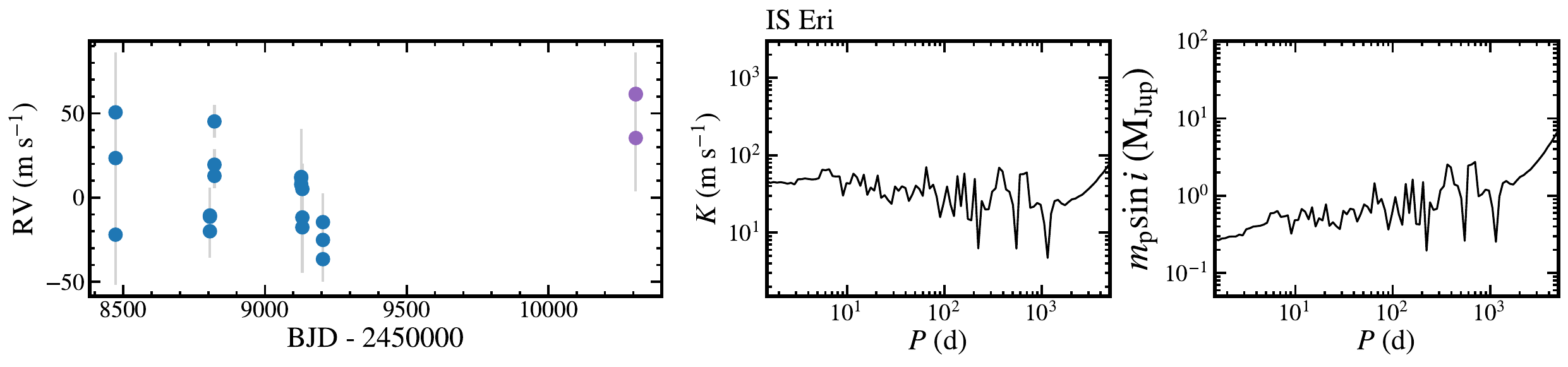}
     \includegraphics[width=\linewidth]{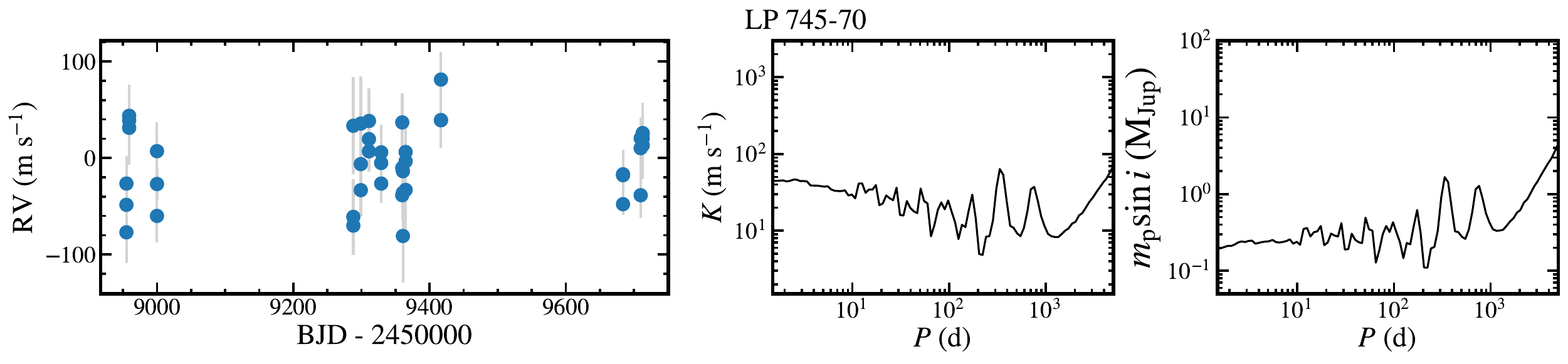}
     \includegraphics[width=\linewidth]{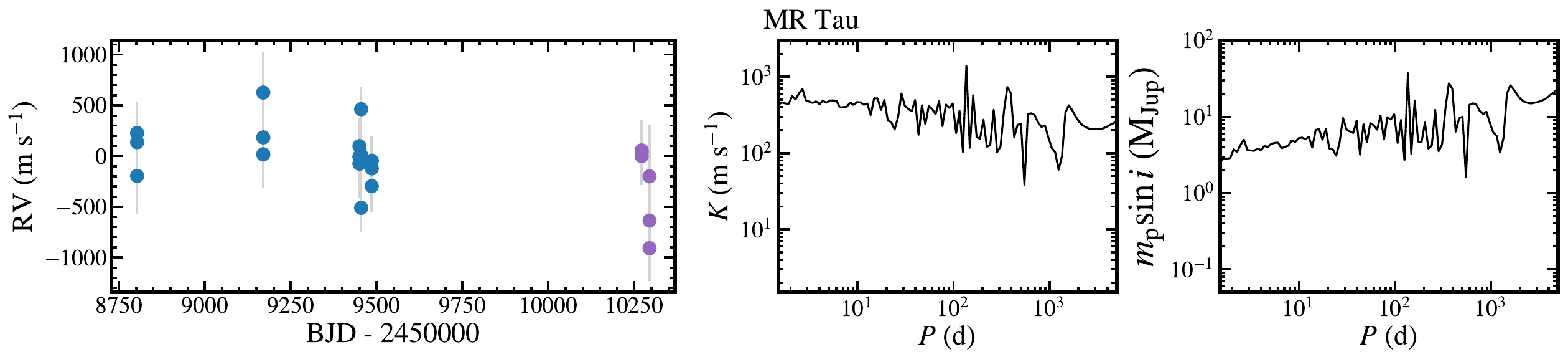}
     \includegraphics[width=\linewidth]{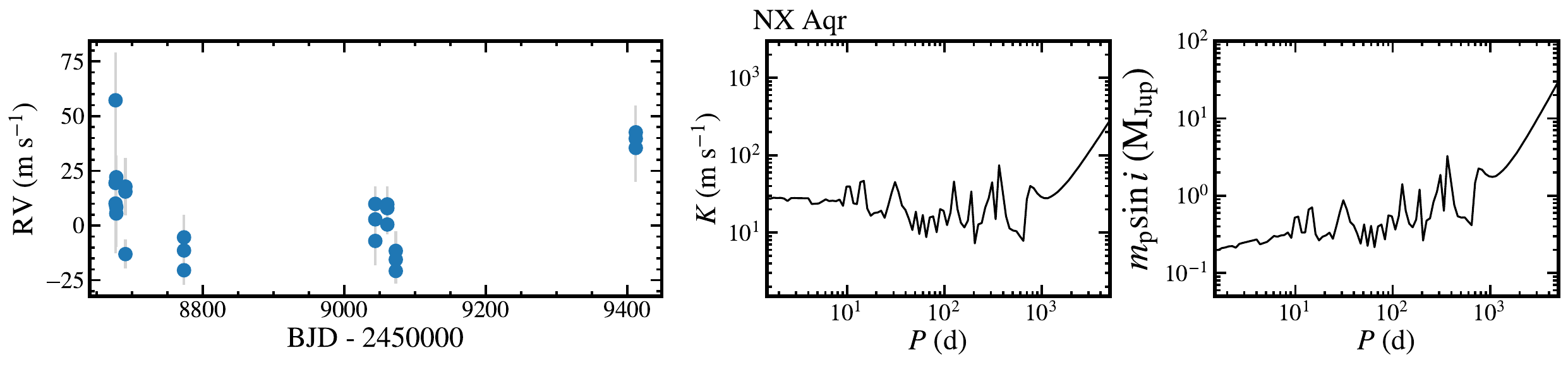}
    \caption{HPF time series and corresponding completeness functions for targets HW Cet, IS Eri, LP 745-70, MR Tau, and NX Aqr.}
    \label{fig:rv_comp11}
\end{figure}

\begin{figure}[p]
     \centering
     \includegraphics[width=\linewidth]{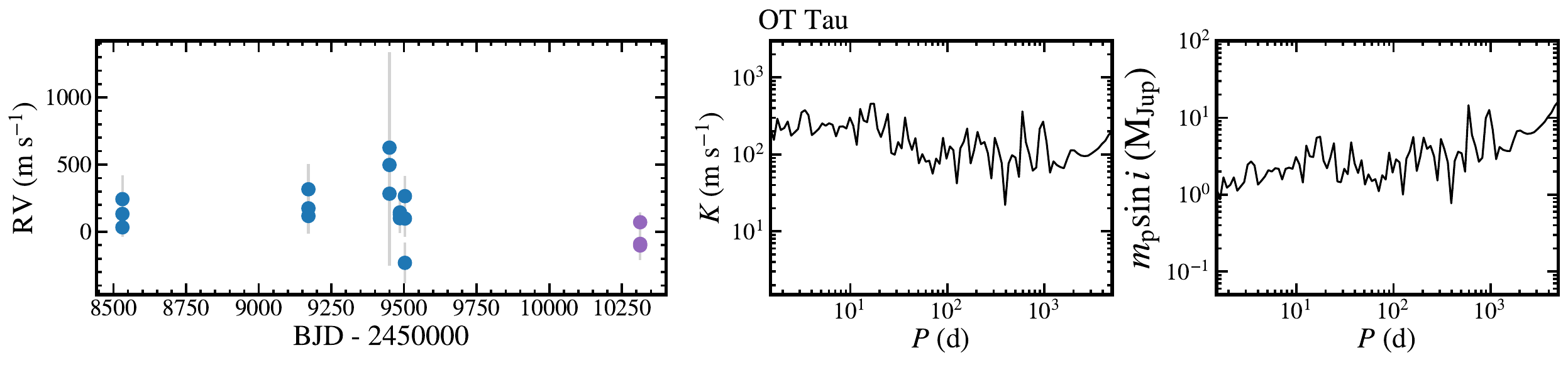}
     \includegraphics[width=\linewidth]{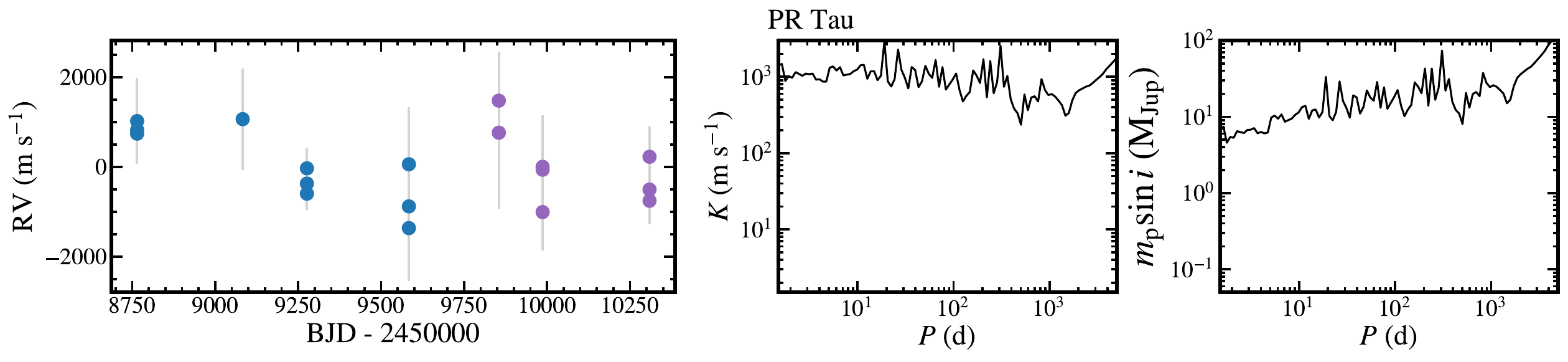}
     \includegraphics[width=\linewidth]{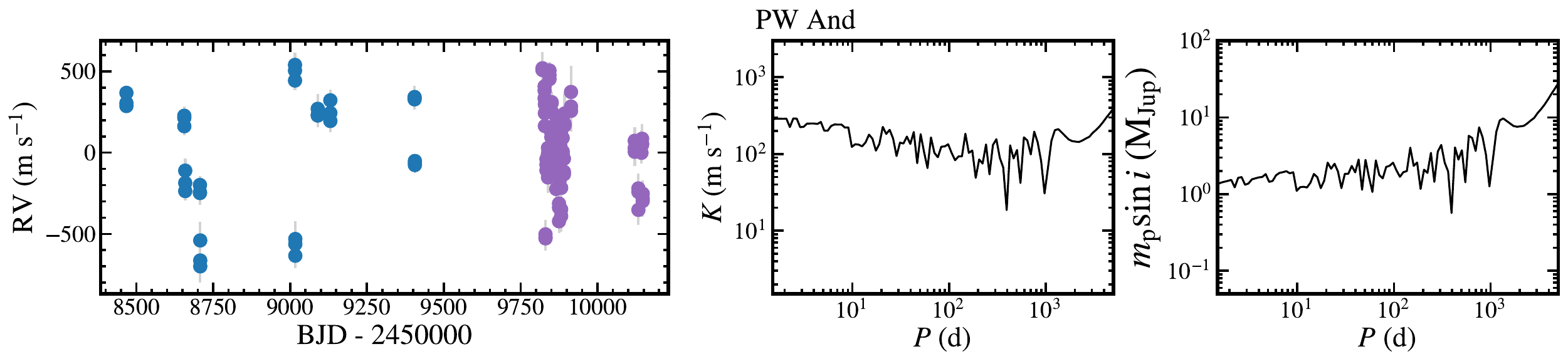}
     \includegraphics[width=\linewidth]{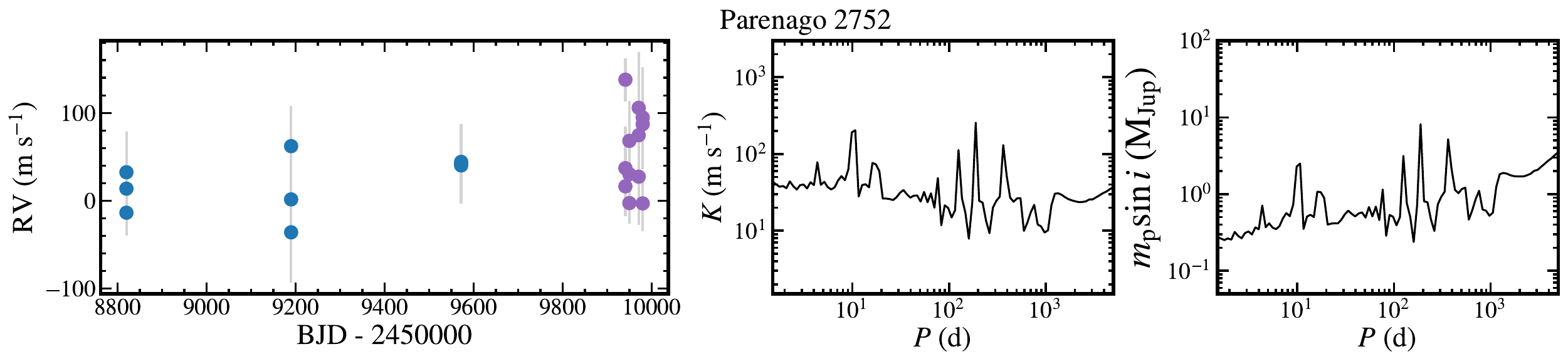}
     \includegraphics[width=\linewidth]{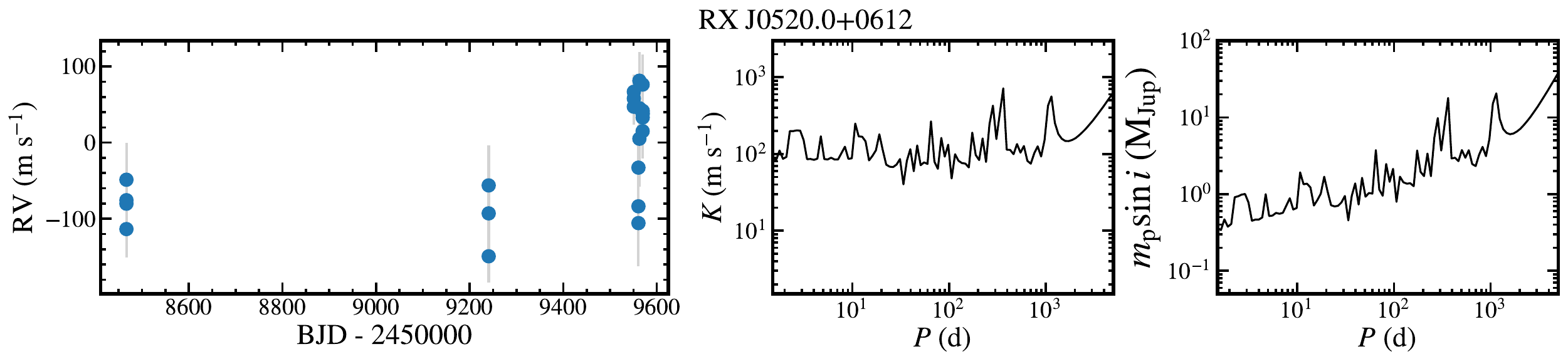}
    \caption{HPF time series and corresponding completeness functions for targets OT Tau, PR Tau, PW And, Parenago 2752, and RX J0520.0+0612.}
    \label{fig:rv_comp12}
\end{figure}

\begin{figure}[p]
     \centering
     \includegraphics[width=\linewidth]{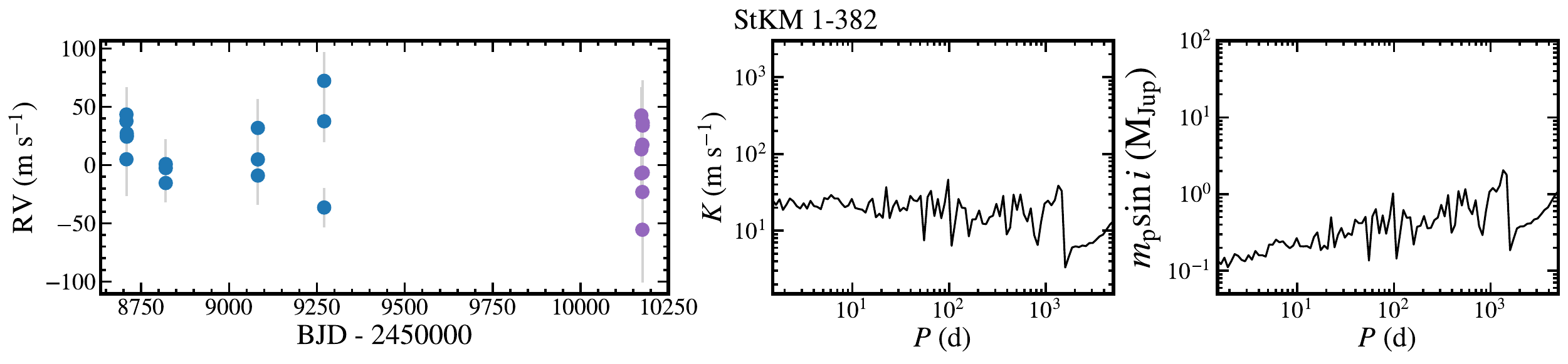}
     \includegraphics[width=\linewidth]{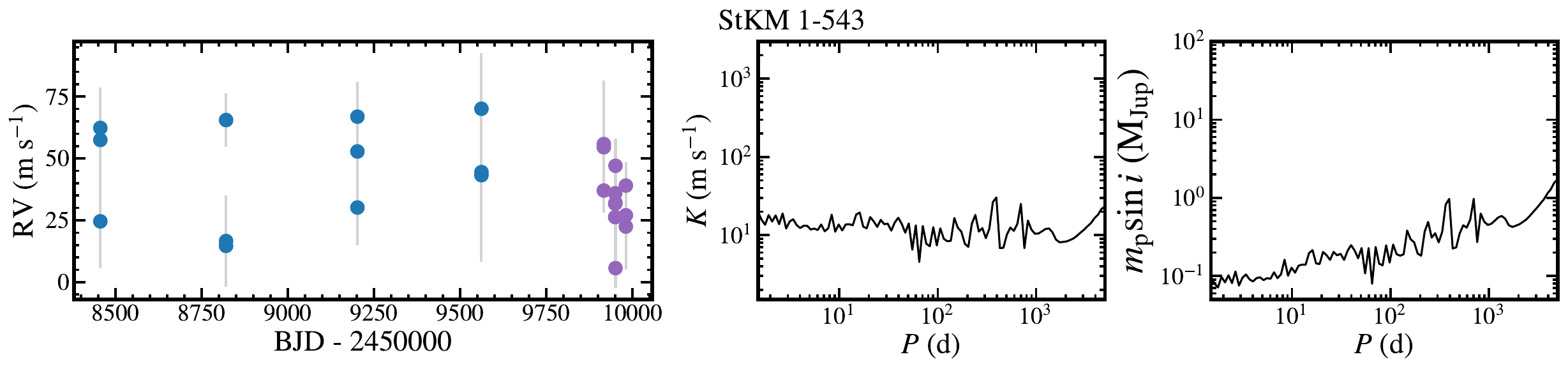}
     \includegraphics[width=\linewidth]{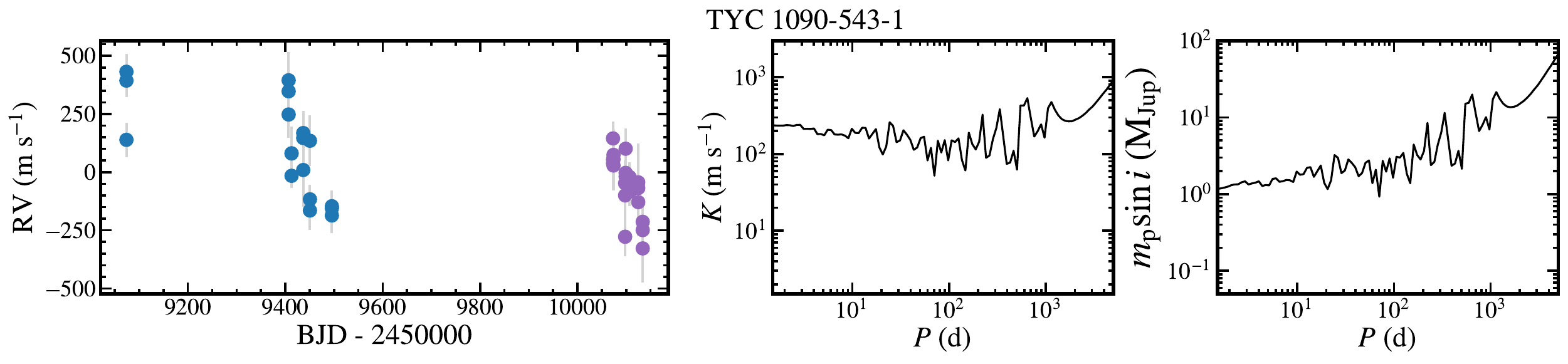}
     \includegraphics[width=\linewidth]{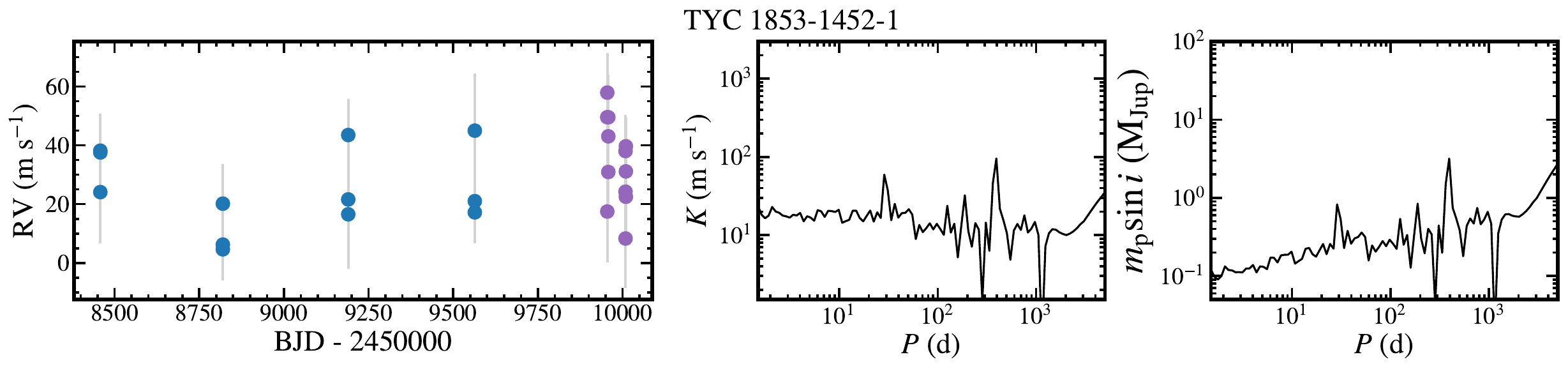}
     \includegraphics[width=\linewidth]{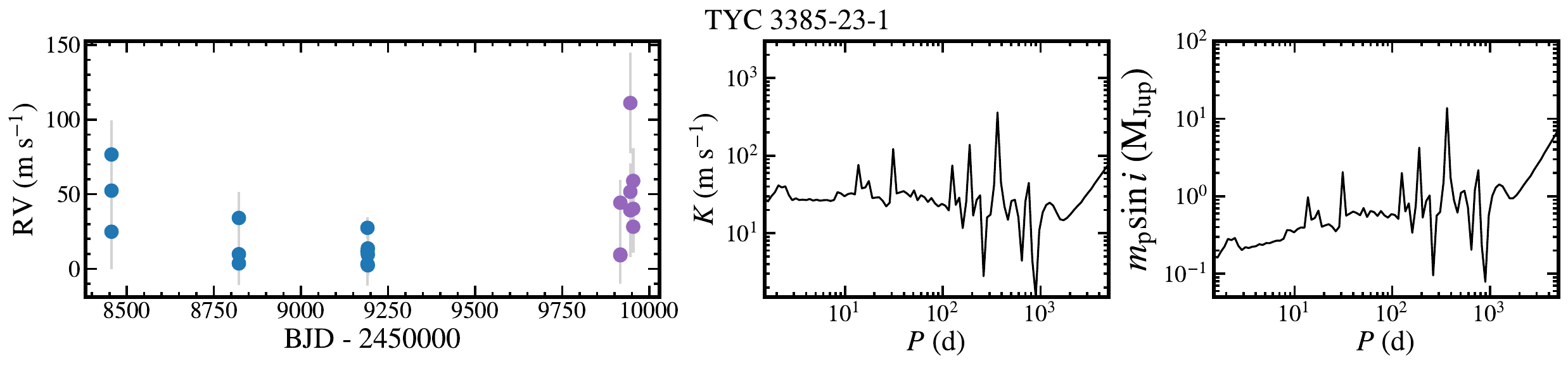}
    \caption{HPF time series and corresponding completeness functions for targets StKM 1-382, StKM 1-543, TYC 1090-543-1, TYC 1853-1452-1, and TYC 3385-23-1.}
    \label{fig:rv_comp13}
\end{figure}

\begin{figure}[p]
     \centering
     \includegraphics[width=\linewidth]{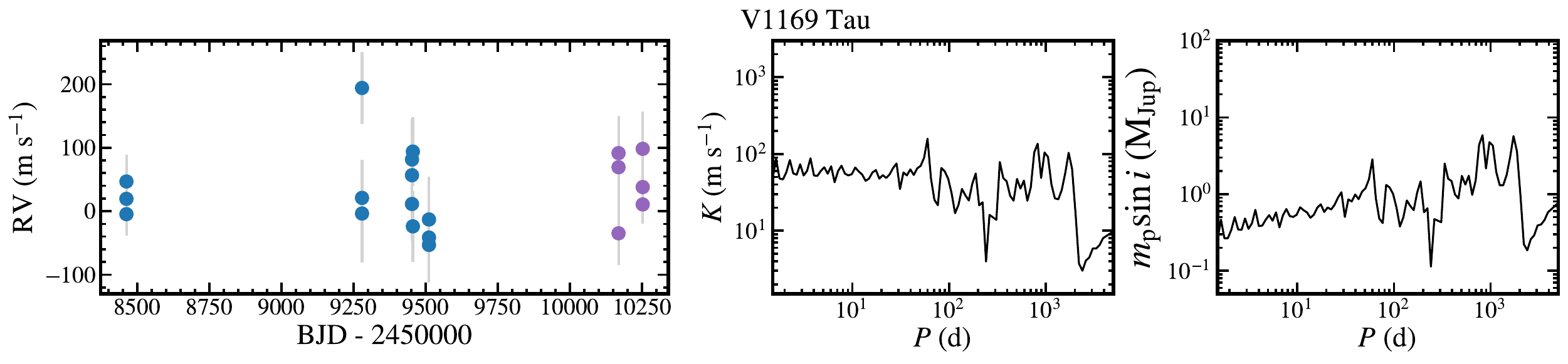}
     \includegraphics[width=\linewidth]{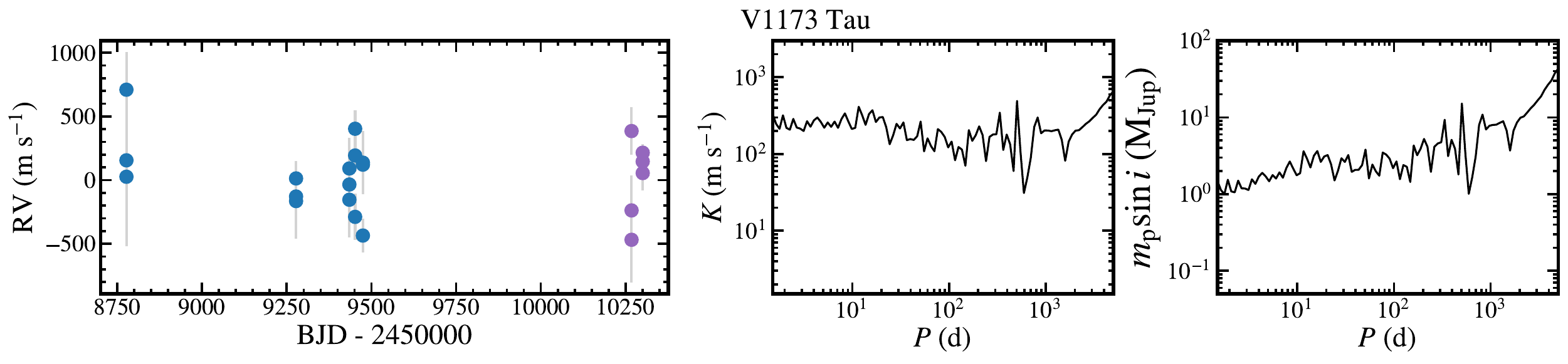}
     \includegraphics[width=\linewidth]{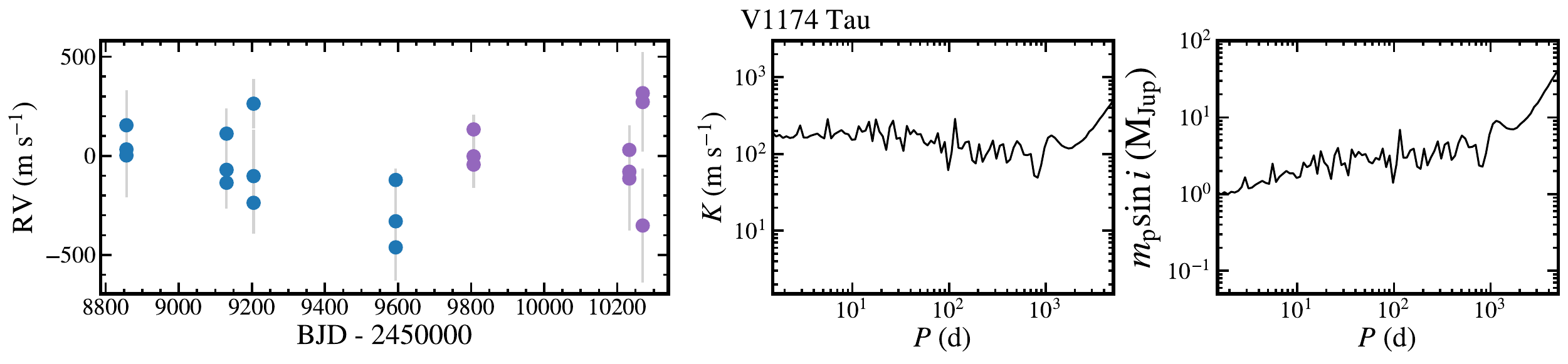}
     \includegraphics[width=\linewidth]{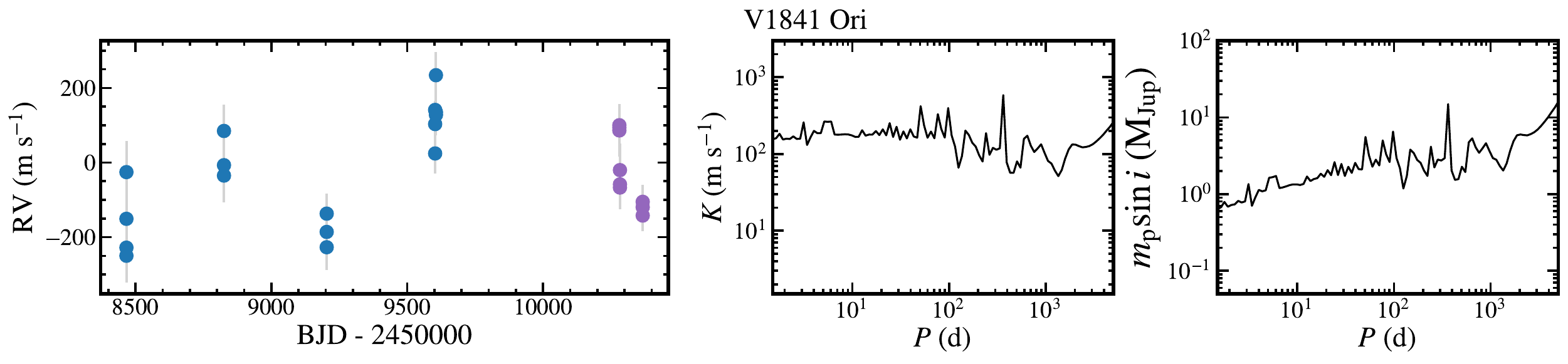}
     \includegraphics[width=\linewidth]{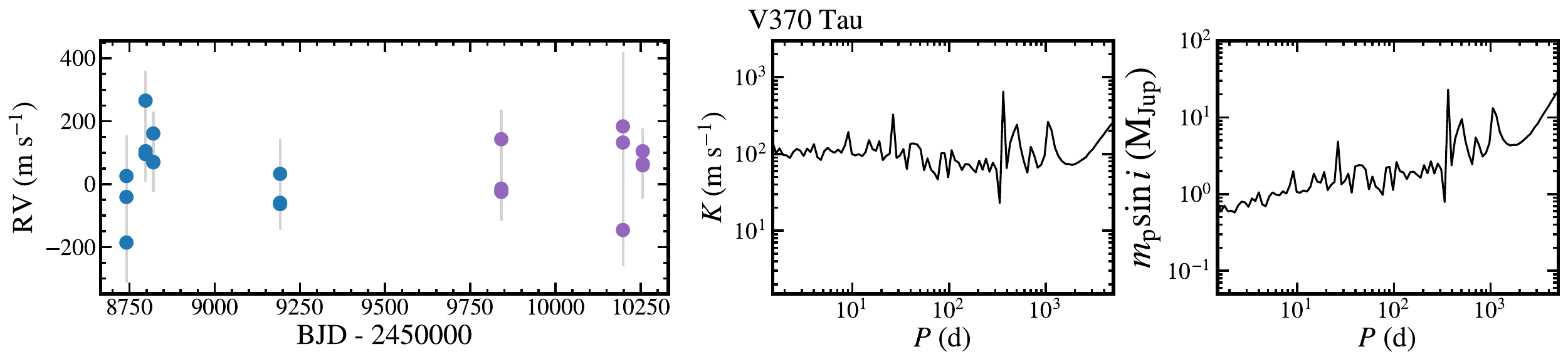}
    \caption{HPF time series and corresponding completeness functions for targets V1169 Tau, V1173 Tau, V1174 Tau, V1841 Ori, and V370 Tau.}
    \label{fig:rv_comp14}
\end{figure}

\begin{figure}[p]
     \centering
     \includegraphics[width=\linewidth]{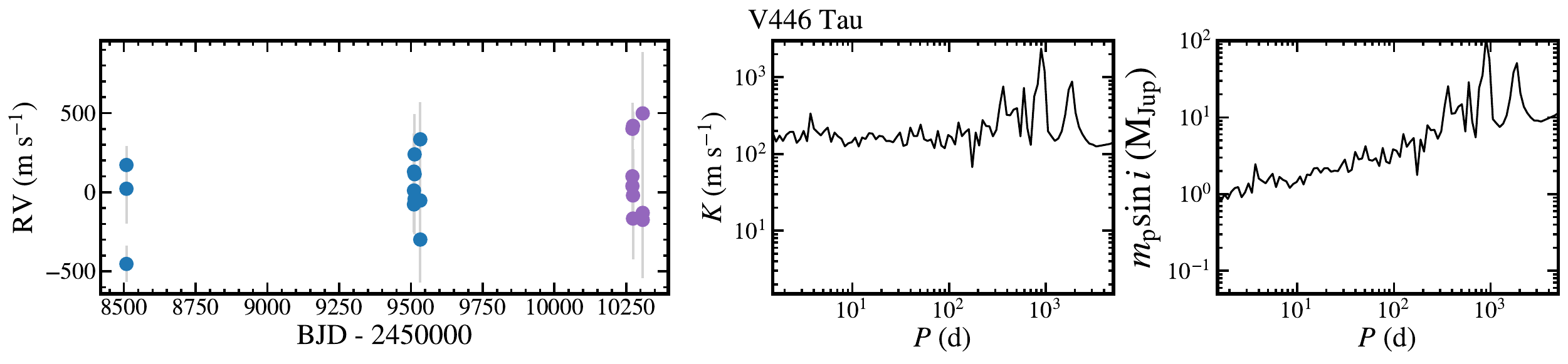}
     \includegraphics[width=\linewidth]{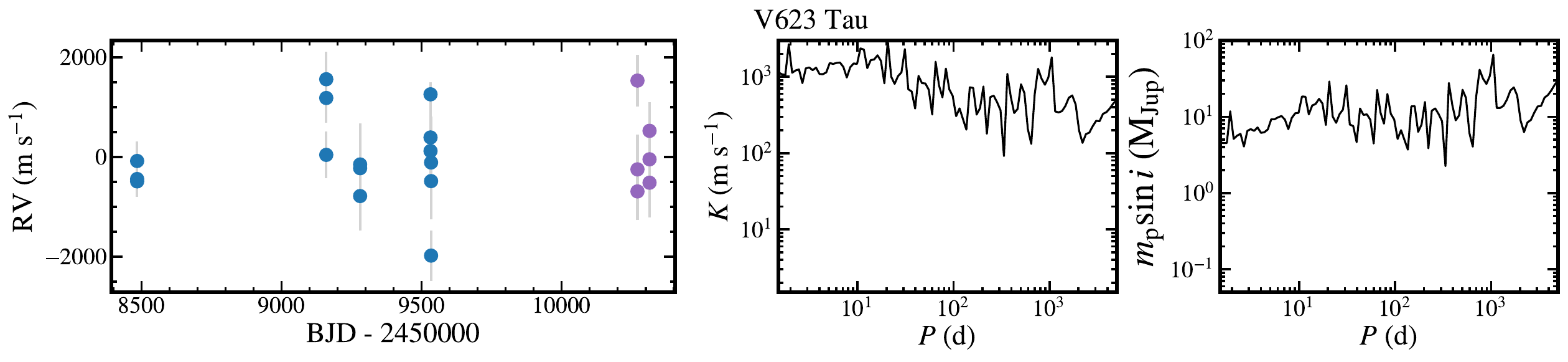}
     \includegraphics[width=\linewidth]{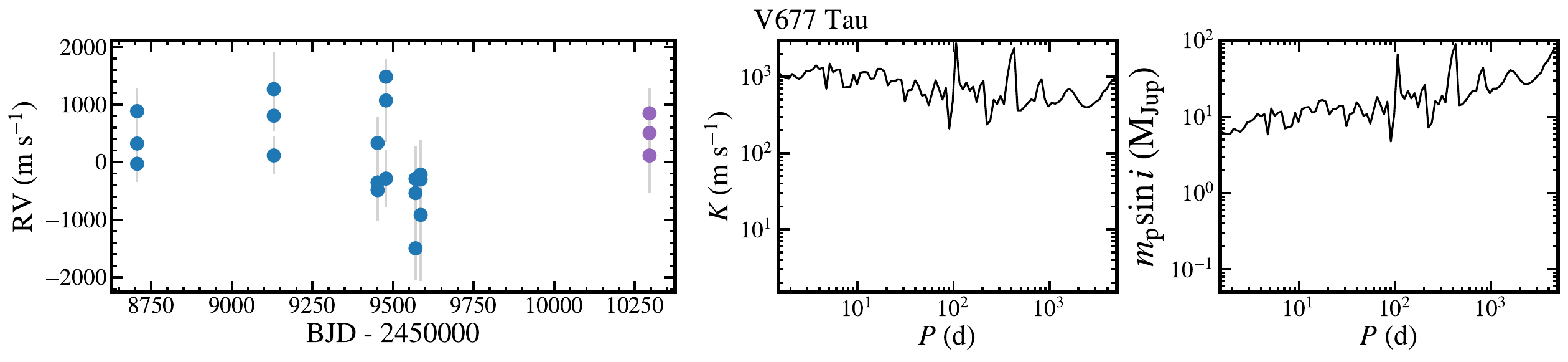}
     \includegraphics[width=\linewidth]{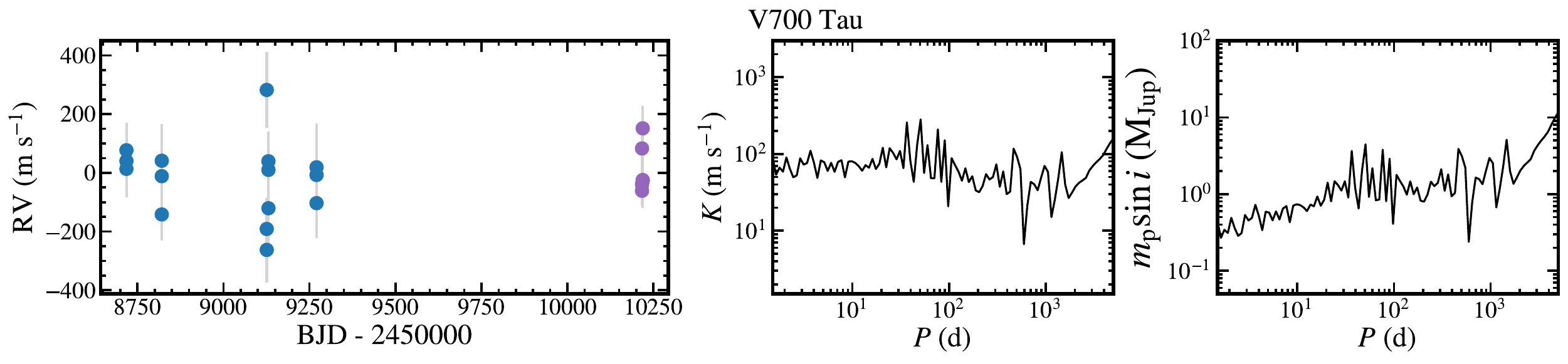}
     \includegraphics[width=\linewidth]{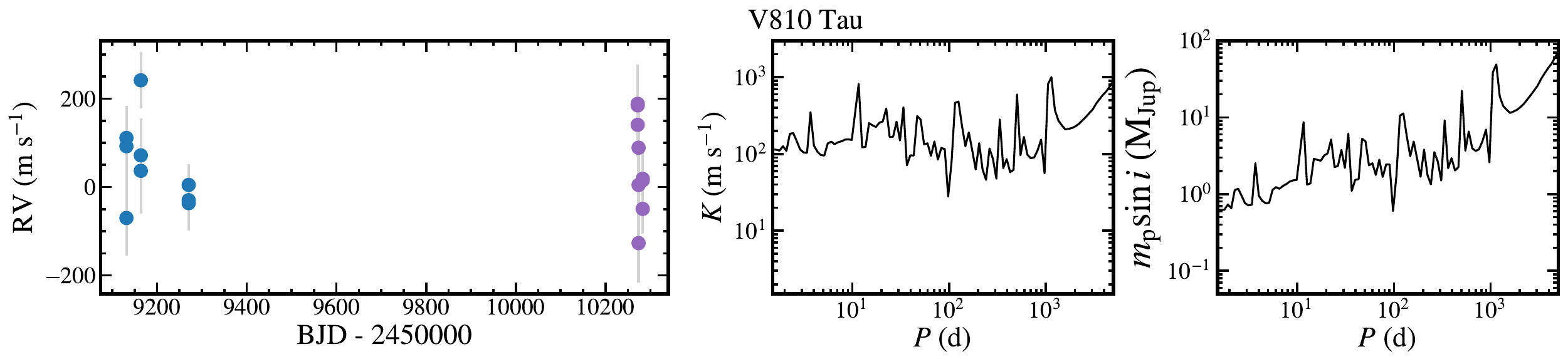}
    \caption{HPF time series and corresponding completeness functions for targets V446 Tau, V623 Tau, V677 Tau, V700 Tau, and V810 Tau.}
    \label{fig:rv_comp15}
\end{figure}

\begin{figure}[p]
     \centering
     \includegraphics[width=\linewidth]{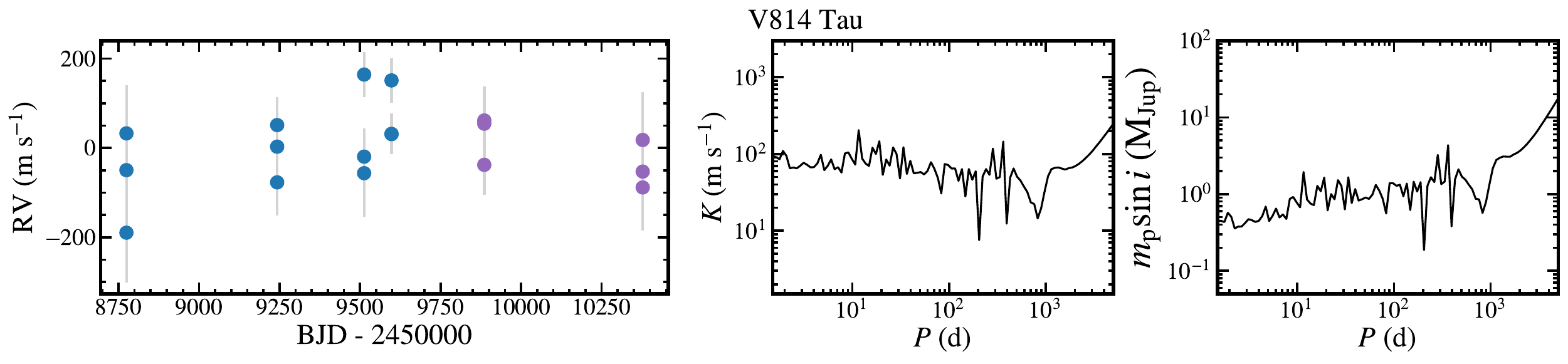}
     \includegraphics[width=\linewidth]{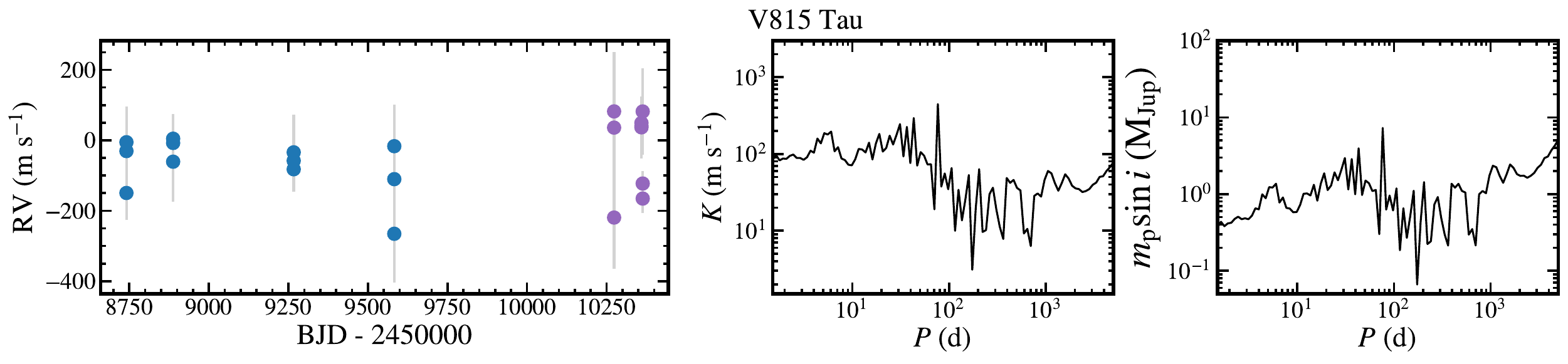}
     \includegraphics[width=\linewidth]{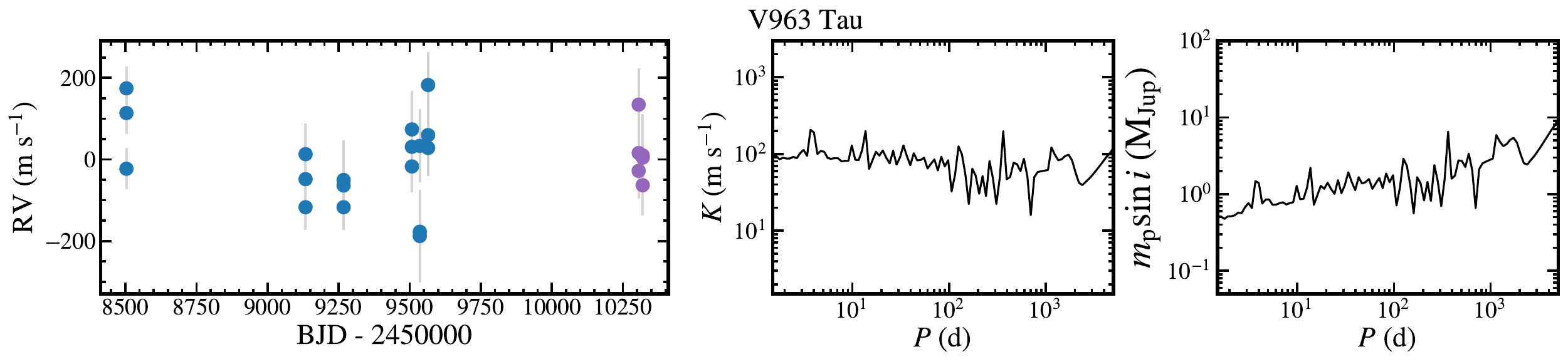}
     \includegraphics[width=\linewidth]{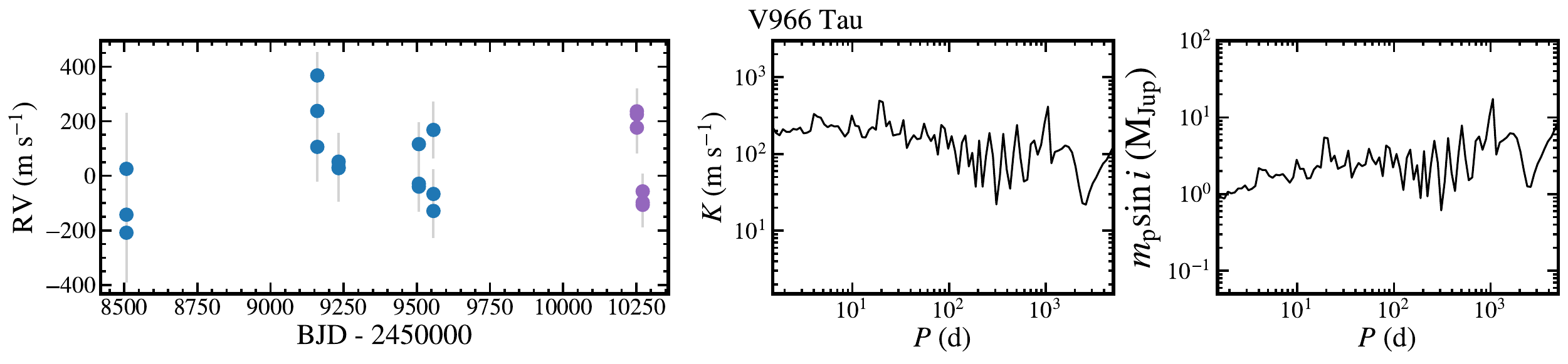}
     \includegraphics[width=\linewidth]{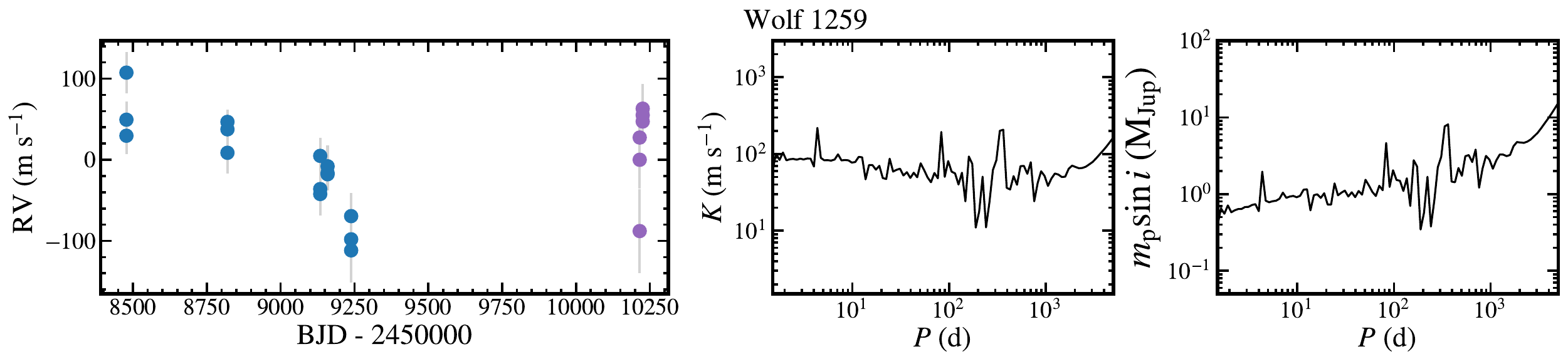}
    \caption{HPF time series and corresponding completeness functions for targets V814 Tau, V815 Tau, V963 Tau, V966 Tau, and Wolf 1259.}
    \label{fig:rv_comp16}
\end{figure}

\Crefrange{fig:rv_comp0}{fig:rv_comp16} show the HPF RV time series for stars in our statistical sample. The center and right panels display the corresponding completeness functions in the $P$--$K$ and the $m \sin i$--$P$ domains, respectively. A description of these HPF RVs is provided in \autoref{sec:hpf_obs}. The methodology for calculating these completeness functions is described in \autoref{sec:completeness}. For all measurements, see \autoref{tab:hpf_measurements} in \autoref{sec:appendix_measurements}.

 \clearpage

 \clearpage

 \clearpage

 \clearpage

 \clearpage

 \clearpage

 \clearpage

 \clearpage

 \clearpage

 \clearpage

 \clearpage

 \clearpage

 \clearpage

 \clearpage

 \clearpage

 \clearpage

 \clearpage

\clearpage
\FloatBarrier

\bibliography{survey}{}

\begin{thebibliography}{}
\expandafter\ifx\csname natexlab\endcsname\relax\def\natexlab#1{#1}\fi
\providecommand{\url}[1]{\href{#1}{#1}}
\providecommand{\dodoi}[1]{doi:~\href{http://doi.org/#1}{\nolinkurl{#1}}}
\providecommand{\doeprint}[1]{\href{http://ascl.net/#1}{\nolinkurl{http://ascl.net/#1}}}
\providecommand{\doarXiv}[1]{\href{https://arxiv.org/abs/#1}{\nolinkurl{https://arxiv.org/abs/#1}}}

\bibitem[{{Abt}(1988)}]{Abt1988}
{Abt}, H.~A. 1988, \apj, 331, 922, \dodoi{10.1086/166609}

\bibitem[{{Adolfsson}(1954)}]{Adolfsson1954}
{Adolfsson}, T. 1954, Arkiv for Astronomi, 1, 425

\bibitem[{{Aguilera-G{\'o}mez} {et~al.}(2018){Aguilera-G{\'o}mez}, {Ram{\'\i}rez}, \& {Chanam{\'e}}}]{Aguilera-Gomez2018}
{Aguilera-G{\'o}mez}, C., {Ram{\'\i}rez}, I., \& {Chanam{\'e}}, J. 2018, \aap, 614, A55, \dodoi{10.1051/0004-6361/201732209}

\bibitem[{{Aigrain} {et~al.}(2012){Aigrain}, {Pont}, \& {Zucker}}]{Aigrain2012}
{Aigrain}, S., {Pont}, F., \& {Zucker}, S. 2012, \mnras, 419, 3147, \dodoi{10.1111/j.1365-2966.2011.19960.x}

\bibitem[{{Albrecht} {et~al.}(2012){Albrecht}, {Winn}, {Johnson}, {Howard}, {Marcy}, {Butler}, {Arriagada}, {Crane}, {Shectman}, {Thompson}, {Hirano}, {Bakos}, \& {Hartman}}]{Albrecht2012}
{Albrecht}, S., {Winn}, J.~N., {Johnson}, J.~A., {et~al.} 2012, \apj, 757, 18, \dodoi{10.1088/0004-637X/757/1/18}

\bibitem[{{Alcala} {et~al.}(1996){Alcala}, {Terranegra}, {Wichmann}, {Chavarria-K.}, {Krautter}, {Schmitt}, {Moreno-Corral}, {de Lara}, \& {Wagner}}]{Alcala1996}
{Alcala}, J.~M., {Terranegra}, L., {Wichmann}, R., {et~al.} 1996, \aaps, 119, 7

\bibitem[{{Anderson} \& {Lai}(2017)}]{Anderson2017}
{Anderson}, K.~R., \& {Lai}, D. 2017, \mnras, 472, 3692, \dodoi{10.1093/mnras/stx2250}

\bibitem[{Artin(1964)}]{Artin1964}
Artin, E. 1964, The Gamma Function, Athena series (Holt, Rinehart and Winston).
\newblock \url{https://books.google.com/books?id=NNc-AAAAIAAJ}

\bibitem[{{Astropy Collaboration} {et~al.}(2013){Astropy Collaboration}, {Robitaille}, {Tollerud}, {Greenfield}, {Droettboom}, {Bray}, {Aldcroft}, {Davis}, {Ginsburg}, {Price-Whelan}, {Kerzendorf}, {Conley}, {Crighton}, {Barbary}, {Muna}, {Ferguson}, {Grollier}, {Parikh}, {Nair}, {Unther}, {Deil}, {Woillez}, {Conseil}, {Kramer}, {Turner}, {Singer}, {Fox}, {Weaver}, {Zabalza}, {Edwards}, {Azalee Bostroem}, {Burke}, {Casey}, {Crawford}, {Dencheva}, {Ely}, {Jenness}, {Labrie}, {Lim}, {Pierfederici}, {Pontzen}, {Ptak}, {Refsdal}, {Servillat}, \& {Streicher}}]{Astropy2013}
{Astropy Collaboration}, {Robitaille}, T.~P., {Tollerud}, E.~J., {et~al.} 2013, \aap, 558, A33, \dodoi{10.1051/0004-6361/201322068}

\bibitem[{{Astropy Collaboration} {et~al.}(2018){Astropy Collaboration}, {Price-Whelan}, {Sip{\H{o}}cz}, {G{\"u}nther}, {Lim}, {Crawford}, {Conseil}, {Shupe}, {Craig}, {Dencheva}, {Ginsburg}, {Vand erPlas}, {Bradley}, {P{\'e}rez-Su{\'a}rez}, {de Val-Borro}, {Aldcroft}, {Cruz}, {Robitaille}, {Tollerud}, {Ardelean}, {Babej}, {Bach}, {Bachetti}, {Bakanov}, {Bamford}, {Barentsen}, {Barmby}, {Baumbach}, {Berry}, {Biscani}, {Boquien}, {Bostroem}, {Bouma}, {Brammer}, {Bray}, {Breytenbach}, {Buddelmeijer}, {Burke}, {Calderone}, {Cano Rodr{\'\i}guez}, {Cara}, {Cardoso}, {Cheedella}, {Copin}, {Corrales}, {Crichton}, {D'Avella}, {Deil}, {Depagne}, {Dietrich}, {Donath}, {Droettboom}, {Earl}, {Erben}, {Fabbro}, {Ferreira}, {Finethy}, {Fox}, {Garrison}, {Gibbons}, {Goldstein}, {Gommers}, {Greco}, {Greenfield}, {Groener}, {Grollier}, {Hagen}, {Hirst}, {Homeier}, {Horton}, {Hosseinzadeh}, {Hu}, {Hunkeler}, {Ivezi{\'c}}, {Jain}, {Jenness}, {Kanarek}, {Kendrew}, {Kern}, {Kerzendorf}, {Khvalko}, {King}, {Kirkby}, {Kulkarni},
  {Kumar}, {Lee}, {Lenz}, {Littlefair}, {Ma}, {Macleod}, {Mastropietro}, {McCully}, {Montagnac}, {Morris}, {Mueller}, {Mumford}, {Muna}, {Murphy}, {Nelson}, {Nguyen}, {Ninan}, {N{\"o}the}, {Ogaz}, {Oh}, {Parejko}, {Parley}, {Pascual}, {Patil}, {Patil}, {Plunkett}, {Prochaska}, {Rastogi}, {Reddy Janga}, {Sabater}, {Sakurikar}, {Seifert}, {Sherbert}, {Sherwood-Taylor}, {Shih}, {Sick}, {Silbiger}, {Singanamalla}, {Singer}, {Sladen}, {Sooley}, {Sornarajah}, {Streicher}, {Teuben}, {Thomas}, {Tremblay}, {Turner}, {Terr{\'o}n}, {van Kerkwijk}, {de la Vega}, {Watkins}, {Weaver}, {Whitmore}, {Woillez}, {Zabalza}, \& {Astropy Contributors}}]{Astropy2018}
{Astropy Collaboration}, {Price-Whelan}, A.~M., {Sip{\H{o}}cz}, B.~M., {et~al.} 2018, \aj, 156, 123, \dodoi{10.3847/1538-3881/aabc4f}

\bibitem[{{Astropy Collaboration} {et~al.}(2022){Astropy Collaboration}, {Price-Whelan}, {Lim}, {Earl}, {Starkman}, {Bradley}, {Shupe}, {Patil}, {Corrales}, {Brasseur}, {N{\"o}the}, {Donath}, {Tollerud}, {Morris}, {Ginsburg}, {Vaher}, {Weaver}, {Tocknell}, {Jamieson}, {van Kerkwijk}, {Robitaille}, {Merry}, {Bachetti}, {G{\"u}nther}, {Aldcroft}, {Alvarado-Montes}, {Archibald}, {B{\'o}di}, {Bapat}, {Barentsen}, {Baz{\'a}n}, {Biswas}, {Boquien}, {Burke}, {Cara}, {Cara}, {Conroy}, {Conseil}, {Craig}, {Cross}, {Cruz}, {D'Eugenio}, {Dencheva}, {Devillepoix}, {Dietrich}, {Eigenbrot}, {Erben}, {Ferreira}, {Foreman-Mackey}, {Fox}, {Freij}, {Garg}, {Geda}, {Glattly}, {Gondhalekar}, {Gordon}, {Grant}, {Greenfield}, {Groener}, {Guest}, {Gurovich}, {Handberg}, {Hart}, {Hatfield-Dodds}, {Homeier}, {Hosseinzadeh}, {Jenness}, {Jones}, {Joseph}, {Kalmbach}, {Karamehmetoglu}, {Ka{\l}uszy{\'n}ski}, {Kelley}, {Kern}, {Kerzendorf}, {Koch}, {Kulumani}, {Lee}, {Ly}, {Ma}, {MacBride}, {Maljaars}, {Muna}, {Murphy}, {Norman},
  {O'Steen}, {Oman}, {Pacifici}, {Pascual}, {Pascual-Granado}, {Patil}, {Perren}, {Pickering}, {Rastogi}, {Roulston}, {Ryan}, {Rykoff}, {Sabater}, {Sakurikar}, {Salgado}, {Sanghi}, {Saunders}, {Savchenko}, {Schwardt}, {Seifert-Eckert}, {Shih}, {Jain}, {Shukla}, {Sick}, {Simpson}, {Singanamalla}, {Singer}, {Singhal}, {Sinha}, {Sip{\H{o}}cz}, {Spitler}, {Stansby}, {Streicher}, {{\v{S}}umak}, {Swinbank}, {Taranu}, {Tewary}, {Tremblay}, {de Val-Borro}, {Van Kooten}, {Vasovi{\'c}}, {Verma}, {de Miranda Cardoso}, {Williams}, {Wilson}, {Winkel}, {Wood-Vasey}, {Xue}, {Yoachim}, {Zhang}, {Zonca}, \& {Astropy Project Contributors}}]{Astropy2022}
{Astropy Collaboration}, {Price-Whelan}, A.~M., {Lim}, P.~L., {et~al.} 2022, \apj, 935, 167, \dodoi{10.3847/1538-4357/ac7c74}

\bibitem[{{Bailer-Jones} {et~al.}(2021){Bailer-Jones}, {Rybizki}, {Fouesneau}, {Demleitner}, \& {Andrae}}]{Bailer-Jones2021}
{Bailer-Jones}, C.~A.~L., {Rybizki}, J., {Fouesneau}, M., {Demleitner}, M., \& {Andrae}, R. 2021, \aj, 161, 147, \dodoi{10.3847/1538-3881/abd806}

\bibitem[{{Bailey} {et~al.}(2012){Bailey}, {White}, {Blake}, {Charbonneau}, {Barman}, {Tanner}, \& {Torres}}]{Bailey2012}
{Bailey}, John~I., I., {White}, R.~J., {Blake}, C.~H., {et~al.} 2012, \apj, 749, 16, \dodoi{10.1088/0004-637X/749/1/16}

\bibitem[{{Bailey} {et~al.}(2018){Bailey}, {Mateo}, {White}, {Shectman}, \& {Crane}}]{Bailey2018}
{Bailey}, J.~I., {Mateo}, M., {White}, R.~J., {Shectman}, S.~A., \& {Crane}, J.~D. 2018, \mnras, 475, 1609, \dodoi{10.1093/mnras/stx3266}

\bibitem[{{Baraffe} {et~al.}(1998){Baraffe}, {Chabrier}, {Allard}, \& {Hauschildt}}]{Baraffe1998}
{Baraffe}, I., {Chabrier}, G., {Allard}, F., \& {Hauschildt}, P.~H. 1998, \aap, 337, 403, \dodoi{10.48550/arXiv.astro-ph/9805009}

\bibitem[{{Barrag{\'a}n} {et~al.}(2022){Barrag{\'a}n}, {Aigrain}, {Rajpaul}, \& {Zicher}}]{Barragan2022}
{Barrag{\'a}n}, O., {Aigrain}, S., {Rajpaul}, V.~M., \& {Zicher}, N. 2022, \mnras, 509, 866, \dodoi{10.1093/mnras/stab2889}

\bibitem[{{Barrag{\'a}n} {et~al.}(2019){Barrag{\'a}n}, {Gandolfi}, \& {Antoniciello}}]{Barragan2019}
{Barrag{\'a}n}, O., {Gandolfi}, D., \& {Antoniciello}, G. 2019, \mnras, 482, 1017, \dodoi{10.1093/mnras/sty2472}

\bibitem[{{Bell} {et~al.}(2015){Bell}, {Mamajek}, \& {Naylor}}]{Bell2015}
{Bell}, C. P.~M., {Mamajek}, E.~E., \& {Naylor}, T. 2015, \mnras, 454, 593, \dodoi{10.1093/mnras/stv1981}

\bibitem[{{Bell} {et~al.}(2017){Bell}, {Murphy}, \& {Mamajek}}]{Bell2017}
{Bell}, C. P.~M., {Murphy}, S.~J., \& {Mamajek}, E.~E. 2017, \mnras, 468, 1198, \dodoi{10.1093/mnras/stx535}

\bibitem[{{Belokurov} {et~al.}(2020){Belokurov}, {Penoyre}, {Oh}, {Iorio}, {Hodgkin}, {Evans}, {Everall}, {Koposov}, {Tout}, {Izzard}, {Clarke}, \& {Brown}}]{Belokurov2020}
{Belokurov}, V., {Penoyre}, Z., {Oh}, S., {et~al.} 2020, \mnras, 496, 1922, \dodoi{10.1093/mnras/staa1522}

\bibitem[{{Binks} {et~al.}(2018){Binks}, {Jeffries}, \& {Ward}}]{Binks2018}
{Binks}, A.~S., {Jeffries}, R.~D., \& {Ward}, J.~L. 2018, \mnras, 473, 2465, \dodoi{10.1093/mnras/stx2252}

\bibitem[{{Boisse} {et~al.}(2012){Boisse}, {Bonfils}, \& {Santos}}]{Boisse2012}
{Boisse}, I., {Bonfils}, X., \& {Santos}, N.~C. 2012, \aap, 545, A109, \dodoi{10.1051/0004-6361/201219115}

\bibitem[{{Boisse} {et~al.}(2011){Boisse}, {Bouchy}, {H{\'e}brard}, {Bonfils}, {Santos}, \& {Vauclair}}]{Boisse2011}
{Boisse}, I., {Bouchy}, F., {H{\'e}brard}, G., {et~al.} 2011, \aap, 528, A4, \dodoi{10.1051/0004-6361/201014354}

\bibitem[{{Bouchy} {et~al.}(2013){Bouchy}, {D{\'\i}az}, {H{\'e}brard}, {Arnold}, {Boisse}, {Delfosse}, {Perruchot}, \& {Santerne}}]{Bouchy2013}
{Bouchy}, F., {D{\'\i}az}, R.~F., {H{\'e}brard}, G., {et~al.} 2013, \aap, 549, A49, \dodoi{10.1051/0004-6361/201219979}

\bibitem[{{Bowler} {et~al.}(2015){Bowler}, {Liu}, {Shkolnik}, \& {Tamura}}]{Bowler2015}
{Bowler}, B.~P., {Liu}, M.~C., {Shkolnik}, E.~L., \& {Tamura}, M. 2015, \apjs, 216, 7, \dodoi{10.1088/0067-0049/216/1/7}

\bibitem[{{Bowler} {et~al.}(2019){Bowler}, {Hinkley}, {Ziegler}, {Baranec}, {Gizis}, {Law}, {Liu}, {Shah}, {Shkolnik}, {Riaz}, \& {Riddle}}]{Bowler2019}
{Bowler}, B.~P., {Hinkley}, S., {Ziegler}, C., {et~al.} 2019, \apj, 877, 60, \dodoi{10.3847/1538-4357/ab1018}

\bibitem[{{Bowler} {et~al.}(2021){Bowler}, {Cochran}, {Endl}, {Franson}, {Brandt}, {Dupuy}, {MacQueen}, {Kratter}, {Mawet}, \& {Ruane}}]{Bowler2021}
{Bowler}, B.~P., {Cochran}, W.~D., {Endl}, M., {et~al.} 2021, \aj, 161, 106, \dodoi{10.3847/1538-3881/abd243}

\bibitem[{{Breger}(1984)}]{Breger1984}
{Breger}, M. 1984, \aaps, 57, 217

\bibitem[{{Brewer} {et~al.}(2016){Brewer}, {Fischer}, {Valenti}, \& {Piskunov}}]{Brewer2016}
{Brewer}, J.~M., {Fischer}, D.~A., {Valenti}, J.~A., \& {Piskunov}, N. 2016, \apjs, 225, 32, \dodoi{10.3847/0067-0049/225/2/32}

\bibitem[{{Buchhave} {et~al.}(2012){Buchhave}, {Latham}, {Johansen}, {Bizzarro}, {Torres}, {Rowe}, {Batalha}, {Borucki}, {Brugamyer}, {Caldwell}, {Bryson}, {Ciardi}, {Cochran}, {Endl}, {Esquerdo}, {Ford}, {Geary}, {Gilliland}, {Hansen}, {Isaacson}, {Laird}, {Lucas}, {Marcy}, {Morse}, {Robertson}, {Shporer}, {Stefanik}, {Still}, \& {Quinn}}]{Buchhave2012}
{Buchhave}, L.~A., {Latham}, D.~W., {Johansen}, A., {et~al.} 2012, \nat, 486, 375, \dodoi{10.1038/nature11121}

\bibitem[{{Buchner} {et~al.}(2014){Buchner}, {Georgakakis}, {Nandra}, {Hsu}, {Rangel}, {Brightman}, {Merloni}, {Salvato}, {Donley}, \& {Kocevski}}]{Buchner2014}
{Buchner}, J., {Georgakakis}, A., {Nandra}, K., {et~al.} 2014, \aap, 564, A125, \dodoi{10.1051/0004-6361/201322971}

\bibitem[{{Butler} {et~al.}(2006){Butler}, {Wright}, {Marcy}, {Fischer}, {Vogt}, {Tinney}, {Jones}, {Carter}, {Johnson}, {McCarthy}, \& {Penny}}]{Butler2006}
{Butler}, R.~P., {Wright}, J.~T., {Marcy}, G.~W., {et~al.} 2006, \apj, 646, 505, \dodoi{10.1086/504701}

\bibitem[{{Butler} {et~al.}(2017){Butler}, {Vogt}, {Laughlin}, {Burt}, {Rivera}, {Tuomi}, {Teske}, {Arriagada}, {Diaz}, {Holden}, \& {Keiser}}]{Butler2017}
{Butler}, R.~P., {Vogt}, S.~S., {Laughlin}, G., {et~al.} 2017, \aj, 153, 208, \dodoi{10.3847/1538-3881/aa66ca}

\bibitem[{{Cale} {et~al.}(2021){Cale}, {Reefe}, {Plavchan}, {Tanner}, {Gaidos}, {Gagn{\'e}}, {Gao}, {Kane}, {B{\'e}jar}, {Lodieu}, {Anglada-Escud{\'e}}, {Ribas}, {Pall{\'e}}, {Quirrenbach}, {Amado}, {Reiners}, {Caballero}, {Rosa Zapatero Osorio}, {Dreizler}, {Howard}, {Fulton}, {Xuesong Wang}, {Collins}, {El Mufti}, {Wittrock}, {Gilbert}, {Barclay}, {Klein}, {Martioli}, {Wittenmyer}, {Wright}, {Addison}, {Hirano}, {Tamura}, {Kotani}, {Narita}, {Vermilion}, {Lee}, {Geneser}, {Teske}, {Quinn}, {Latham}, {Esquerdo}, {Calkins}, {Berlind}, {Zohrabi}, {Stibbards}, {Kotnana}, {Jenkins}, {Twicken}, {Henze}, {Kidwell}, {Burke}, {Villase{\~n}or}, \& {Boyd}}]{Cale2021}
{Cale}, B.~L., {Reefe}, M., {Plavchan}, P., {et~al.} 2021, \aj, 162, 295, \dodoi{10.3847/1538-3881/ac2c80}

\bibitem[{{Cannon} \& {Pickering}(1993)}]{Cannon1993}
{Cannon}, A.~J., \& {Pickering}, E.~C. 1993, {VizieR Online Data Catalog: Henry Draper Catalogue and Extension (Cannon+ 1918-1924; ADC 1989)}, VizieR On-line Data Catalog: III/135A. Originally published in: Harv. Ann. 91-100 (1918-1924)

\bibitem[{{Carleo} {et~al.}(2020){Carleo}, {Malavolta}, {Lanza}, {Damasso}, {Desidera}, {Borsa}, {Mallonn}, {Pinamonti}, {Gratton}, {Alei}, {Benatti}, {Mancini}, {Maldonado}, {Biazzo}, {Esposito}, {Frustagli}, {Gonz{\'a}lez-{\'A}lvarez}, {Micela}, {Scandariato}, {Sozzetti}, {Affer}, {Bignamini}, {Bonomo}, {Claudi}, {Cosentino}, {Covino}, {Fiorenzano}, {Giacobbe}, {Harutyunyan}, {Leto}, {Maggio}, {Molinari}, {Nascimbeni}, {Pagano}, {Pedani}, {Piotto}, {Poretti}, {Rainer}, {Redfield}, {Baffa}, {Baruffolo}, {Buchschacher}, {Billotti}, {Cecconi}, {Falcini}, {Fantinel}, {Fini}, {Galli}, {Ghedina}, {Ghinassi}, {Giani}, {Gonzalez}, {Gonzalez}, {Guerra}, {Hernandez Diaz}, {Hernandez}, {Iuzzolino}, {Lodi}, {Oliva}, {Origlia}, {Perez Ventura}, {Puglisi}, {Riverol}, {Riverol}, {San Juan}, {Sanna}, {Scuderi}, {Seemann}, {Sozzi}, \& {Tozzi}}]{Carleo2020}
{Carleo}, I., {Malavolta}, L., {Lanza}, A.~F., {et~al.} 2020, \aap, 638, A5, \dodoi{10.1051/0004-6361/201937369}

\bibitem[{{Carmona} {et~al.}(2023){Carmona}, {Delfosse}, {Bellotti}, {Cort{\'e}s-Zuleta}, {Ould-Elhkim}, {Heidari}, {Mignon}, {Donati}, {Moutou}, {Cook}, {Artigau}, {Fouqu{\'e}}, {Martioli}, {Cadieux}, {Morin}, {Forveille}, {Boisse}, {H{\'e}brard}, {D{\'\i}az}, {Lafreni{\`e}re}, {Kiefer}, {Petit}, {Doyon}, {Acu{\~n}a}, {Arnold}, {Bonfils}, {Bouchy}, {Bourrier}, {Dalal}, {Deleuil}, {Demangeon}, {Dumusque}, {Hara}, {Hoyer}, {Mousis}, {Santerne}, {S{\'e}grasan}, {Stalport}, \& {Udry}}]{Carmona2023}
{Carmona}, A., {Delfosse}, X., {Bellotti}, S., {et~al.} 2023, arXiv e-prints, arXiv:2303.16712, \dodoi{10.48550/arXiv.2303.16712}

\bibitem[{{Cayrel de Strobel} {et~al.}(2001){Cayrel de Strobel}, {Soubiran}, \& {Ralite}}]{CayreldeStrobel2001}
{Cayrel de Strobel}, G., {Soubiran}, C., \& {Ralite}, N. 2001, \aap, 373, 159, \dodoi{10.1051/0004-6361:20010525}

\bibitem[{{Chatterjee} {et~al.}(2008){Chatterjee}, {Ford}, {Matsumura}, \& {Rasio}}]{Chatterjee2008}
{Chatterjee}, S., {Ford}, E.~B., {Matsumura}, S., \& {Rasio}, F.~A. 2008, \apj, 686, 580, \dodoi{10.1086/590227}

\bibitem[{{Chubak} {et~al.}(2012){Chubak}, {Marcy}, {Fischer}, {Howard}, {Isaacson}, {Johnson}, \& {Wright}}]{Chubak2012}
{Chubak}, C., {Marcy}, G., {Fischer}, D.~A., {et~al.} 2012, arXiv e-prints, arXiv:1207.6212, \dodoi{10.48550/arXiv.1207.6212}

\bibitem[{{Cochran} {et~al.}(2002){Cochran}, {Hatzes}, \& {Paulson}}]{Cochran2002}
{Cochran}, W.~D., {Hatzes}, A.~P., \& {Paulson}, D.~B. 2002, \aj, 124, 565, \dodoi{10.1086/341170}

\bibitem[{{Crockett} {et~al.}(2012){Crockett}, {Mahmud}, {Prato}, {Johns-Krull}, {Jaffe}, {Hartigan}, \& {Beichman}}]{Crockett2012}
{Crockett}, C.~J., {Mahmud}, N.~I., {Prato}, L., {et~al.} 2012, \apj, 761, 164, \dodoi{10.1088/0004-637X/761/2/164}

\bibitem[{{Cutri} {et~al.}(2003){Cutri}, {Skrutskie}, {van Dyk}, {Beichman}, {Carpenter}, {Chester}, {Cambresy}, {Evans}, {Fowler}, {Gizis}, {Howard}, {Huchra}, {Jarrett}, {Kopan}, {Kirkpatrick}, {Light}, {Marsh}, {McCallon}, {Schneider}, {Stiening}, {Sykes}, {Weinberg}, {Wheaton}, {Wheelock}, \& {Zacarias}}]{Cutri2003}
{Cutri}, R.~M., {Skrutskie}, M.~F., {van Dyk}, S., {et~al.} 2003, VizieR Online Data Catalog, II/246

\bibitem[{{David} {et~al.}(2019){David}, {Petigura}, {Luger}, {Foreman-Mackey}, {Livingston}, {Mamajek}, \& {Hillenbrand}}]{David2019}
{David}, T.~J., {Petigura}, E.~A., {Luger}, R., {et~al.} 2019, \apjl, 885, L12, \dodoi{10.3847/2041-8213/ab4c99}

\bibitem[{{Donahue}(1993)}]{Donahue1993}
{Donahue}, R.~A. 1993, PhD thesis, New Mexico State University

\bibitem[{{Donati} {et~al.}(1992){Donati}, {Brown}, {Semel}, {Rees}, {Dempsey}, {Matthews}, {Henry}, \& {Hall}}]{Donati1992}
{Donati}, J.~F., {Brown}, S.~F., {Semel}, M., {et~al.} 1992, \aap, 265, 682

\bibitem[{{Donati} {et~al.}(2014){Donati}, {H{\'e}brard}, {Hussain}, {Moutou}, {Grankin}, {Boisse}, {Morin}, {Gregory}, {Vidotto}, {Bouvier}, {Alencar}, {Delfosse}, {Doyon}, {Takami}, {Jardine}, {Fares}, {Cameron}, {M{\'e}nard}, {Dougados}, {Herczeg}, \& {Matysse Collaboration}}]{Donati2014}
{Donati}, J.~F., {H{\'e}brard}, E., {Hussain}, G., {et~al.} 2014, \mnras, 444, 3220, \dodoi{10.1093/mnras/stu1679}

\bibitem[{{Dumusque} {et~al.}(2011){Dumusque}, {Santos}, {Udry}, {Lovis}, \& {Bonfils}}]{Dumusque2011b}
{Dumusque}, X., {Santos}, N.~C., {Udry}, S., {Lovis}, C., \& {Bonfils}, X. 2011, \aap, 527, A82, \dodoi{10.1051/0004-6361/201015877}

\bibitem[{{Elliott} {et~al.}(2016){Elliott}, {Bayo}, {Melo}, {Torres}, {Sterzik}, {Quast}, {Montes}, \& {Brahm}}]{Elliott2016}
{Elliott}, P., {Bayo}, A., {Melo}, C.~H.~F., {et~al.} 2016, \aap, 590, A13, \dodoi{10.1051/0004-6361/201628253}

\bibitem[{{Fabrycky} \& {Tremaine}(2007)}]{Fabrycky2007}
{Fabrycky}, D., \& {Tremaine}, S. 2007, \apj, 669, 1298, \dodoi{10.1086/521702}

\bibitem[{{Feroz} {et~al.}(2009){Feroz}, {Hobson}, \& {Bridges}}]{Feroz2009}
{Feroz}, F., {Hobson}, M.~P., \& {Bridges}, M. 2009, \mnras, 398, 1601, \dodoi{10.1111/j.1365-2966.2009.14548.x}

\bibitem[{{Feroz} {et~al.}(2019){Feroz}, {Hobson}, {Cameron}, \& {Pettitt}}]{Feroz2019}
{Feroz}, F., {Hobson}, M.~P., {Cameron}, E., \& {Pettitt}, A.~N. 2019, The Open Journal of Astrophysics, 2, 10, \dodoi{10.21105/astro.1306.2144}

\bibitem[{{Findeisen} \& {Hillenbrand}(2010)}]{Findeisen2010}
{Findeisen}, K., \& {Hillenbrand}, L. 2010, \aj, 139, 1338, \dodoi{10.1088/0004-6256/139/4/1338}

\bibitem[{{Fischer} \& {Valenti}(2005)}]{Fischer2005}
{Fischer}, D.~A., \& {Valenti}, J. 2005, \apj, 622, 1102, \dodoi{10.1086/428383}

\bibitem[{{Foreman-Mackey} {et~al.}(2013){Foreman-Mackey}, {Hogg}, {Lang}, \& {Goodman}}]{Foreman-Mackey2013}
{Foreman-Mackey}, D., {Hogg}, D.~W., {Lang}, D., \& {Goodman}, J. 2013, \pasp, 125, 306, \dodoi{10.1086/670067}

\bibitem[{{Foreman-Mackey} {et~al.}(2019){Foreman-Mackey}, {Farr}, {Sinha}, {Archibald}, {Hogg}, {Sanders}, {Zuntz}, {Williams}, {Nelson}, {de Val-Borro}, {Erhardt}, {Pashchenko}, \& {Pla}}]{Foreman-Mackey2019}
{Foreman-Mackey}, D., {Farr}, W., {Sinha}, M., {et~al.} 2019, The Journal of Open Source Software, 4, 1864, \dodoi{10.21105/joss.01864}

\bibitem[{{Frelikh} {et~al.}(2019){Frelikh}, {Jang}, {Murray-Clay}, \& {Petrovich}}]{Frelikh2019}
{Frelikh}, R., {Jang}, H., {Murray-Clay}, R.~A., \& {Petrovich}, C. 2019, \apjl, 884, L47, \dodoi{10.3847/2041-8213/ab4a7b}

\bibitem[{{Fressin} {et~al.}(2013){Fressin}, {Torres}, {Charbonneau}, {Bryson}, {Christiansen}, {Dressing}, {Jenkins}, {Walkowicz}, \& {Batalha}}]{Fressin2013}
{Fressin}, F., {Torres}, G., {Charbonneau}, D., {et~al.} 2013, \apj, 766, 81, \dodoi{10.1088/0004-637X/766/2/81}

\bibitem[{{Fulton} {et~al.}(2018){Fulton}, {Petigura}, {Blunt}, \& {Sinukoff}}]{Fulton2018}
{Fulton}, B.~J., {Petigura}, E.~A., {Blunt}, S., \& {Sinukoff}, E. 2018, \pasp, 130, 044504, \dodoi{10.1088/1538-3873/aaaaa8}

\bibitem[{{Gagn{\'e}} {et~al.}(2018{\natexlab{a}}){Gagn{\'e}}, {Fontaine}, {Simon}, \& {Faherty}}]{Gagne2018c}
{Gagn{\'e}}, J., {Fontaine}, G., {Simon}, A., \& {Faherty}, J.~K. 2018{\natexlab{a}}, \apjl, 861, L13, \dodoi{10.3847/2041-8213/aacdff}

\bibitem[{{Gagn{\'e}} {et~al.}(2018{\natexlab{b}}){Gagn{\'e}}, {Roy-Loubier}, {Faherty}, {Doyon}, \& {Malo}}]{Gagne2018b}
{Gagn{\'e}}, J., {Roy-Loubier}, O., {Faherty}, J.~K., {Doyon}, R., \& {Malo}, L. 2018{\natexlab{b}}, \apj, 860, 43, \dodoi{10.3847/1538-4357/aac2b8}

\bibitem[{{Gagn{\'e}} {et~al.}(2016){Gagn{\'e}}, {Plavchan}, {Gao}, {Anglada-Escude}, {Furlan}, {Davison}, {Tanner}, {Henry}, {Riedel}, {Brinkworth}, {Latham}, {Bottom}, {White}, {Mills}, {Beichman}, {Johnson}, {Ciardi}, {Wallace}, {Mennesson}, {von Braun}, {Vasisht}, {Prato}, {Kane}, {Mamajek}, {Walp}, {Crawford}, {Rougeot}, {Geneser}, \& {Catanzarite}}]{Gagne2016}
{Gagn{\'e}}, J., {Plavchan}, P., {Gao}, P., {et~al.} 2016, \apj, 822, 40, \dodoi{10.3847/0004-637X/822/1/40}

\bibitem[{{Gagn{\'e}} {et~al.}(2018{\natexlab{c}}){Gagn{\'e}}, {Mamajek}, {Malo}, {Riedel}, {Rodriguez}, {Lafreni{\`e}re}, {Faherty}, {Roy-Loubier}, {Pueyo}, {Robin}, \& {Doyon}}]{Gagne2018a}
{Gagn{\'e}}, J., {Mamajek}, E.~E., {Malo}, L., {et~al.} 2018{\natexlab{c}}, \apj, 856, 23, \dodoi{10.3847/1538-4357/aaae09}

\bibitem[{{Gaia Collaboration}(2022)}]{Gaia2022}
{Gaia Collaboration}. 2022, {VizieR Online Data Catalog: Gaia DR3 Part 3. Non-single stars (Gaia Collaboration, 2022)}, VizieR On-line Data Catalog: I/357. Originally published in: Astron. Astrophys., in prep. (2022)

\bibitem[{{Gaia Collaboration} {et~al.}(2018){Gaia Collaboration}, {Babusiaux}, {van Leeuwen}, {Barstow}, {Jordi}, {Vallenari}, {Bossini}, {Bressan}, {Cantat-Gaudin}, {van Leeuwen}, {Brown}, {Prusti}, {de Bruijne}, {Bailer-Jones}, {Biermann}, {Evans}, {Eyer}, {Jansen}, {Klioner}, {Lammers}, {Lindegren}, {Luri}, {Mignard}, {Panem}, {Pourbaix}, {Randich}, {Sartoretti}, {Siddiqui}, {Soubiran}, {Walton}, {Arenou}, {Bastian}, {Cropper}, {Drimmel}, {Katz}, {Lattanzi}, {Bakker}, {Cacciari}, {Casta{\~n}eda}, {Chaoul}, {Cheek}, {De Angeli}, {Fabricius}, {Guerra}, {Holl}, {Masana}, {Messineo}, {Mowlavi}, {Nienartowicz}, {Panuzzo}, {Portell}, {Riello}, {Seabroke}, {Tanga}, {Th{\'e}venin}, {Gracia-Abril}, {Comoretto}, {Garcia-Reinaldos}, {Teyssier}, {Altmann}, {Andrae}, {Audard}, {Bellas-Velidis}, {Benson}, {Berthier}, {Blomme}, {Burgess}, {Busso}, {Carry}, {Cellino}, {Clementini}, {Clotet}, {Creevey}, {Davidson}, {De Ridder}, {Delchambre}, {Dell'Oro}, {Ducourant}, {Fern{\'a}ndez-Hern{\'a}ndez}, {Fouesneau},
  {Fr{\'e}mat}, {Galluccio}, {Garc{\'\i}a-Torres}, {Gonz{\'a}lez-N{\'u}{\~n}ez}, {Gonz{\'a}lez-Vidal}, {Gosset}, {Guy}, {Halbwachs}, {Hambly}, {Harrison}, {Hern{\'a}ndez}, {Hestroffer}, {Hodgkin}, {Hutton}, {Jasniewicz}, {Jean-Antoine-Piccolo}, {Jordan}, {Korn}, {Krone-Martins}, {Lanzafame}, {Lebzelter}, {L{\"o}ffler}, {Manteiga}, {Marrese}, {Mart{\'\i}n-Fleitas}, {Moitinho}, {Mora}, {Muinonen}, {Osinde}, {Pancino}, {Pauwels}, {Petit}, {Recio-Blanco}, {Richards}, {Rimoldini}, {Robin}, {Sarro}, {Siopis}, {Smith}, {Sozzetti}, {S{\"u}veges}, {Torra}, {van Reeven}, {Abbas}, {Abreu Aramburu}, {Accart}, {Aerts}, {Altavilla}, {{\'A}lvarez}, {Alvarez}, {Alves}, {Anderson}, {Andrei}, {Anglada Varela}, {Antiche}, {Antoja}, {Arcay}, {Astraatmadja}, {Bach}, {Baker}, {Balaguer-N{\'u}{\~n}ez}, {Balm}, {Barache}, {Barata}, {Barbato}, {Barblan}, {Barklem}, {Barrado}, {Barros}, {Bartholom{\'e} Mu{\~n}oz}, {Bassilana}, {Becciani}, {Bellazzini}, {Berihuete}, {Bertone}, {Bianchi}, {Bienaym{\'e}}, {Blanco-Cuaresma}, {Boch},
  {Boeche}, {Bombrun}, {Borrachero}, {Bouquillon}, {Bourda}, {Bragaglia}, {Bramante}, {Breddels}, {Brouillet}, {Br{\"u}semeister}, {Brugaletta}, {Bucciarelli}, {Burlacu}, {Busonero}, {Butkevich}, {Buzzi}, {Caffau}, {Cancelliere}, {Cannizzaro}, {Carballo}, {Carlucci}, {Carrasco}, {Casamiquela}, {Castellani}, {Castro-Ginard}, {Charlot}, {Chemin}, {Chiavassa}, {Cocozza}, {Costigan}, {Cowell}, {Crifo}, {Crosta}, {Crowley}, {Cuypers}, {Dafonte}, {Damerdji}, {Dapergolas}, {David}, {David}, {de Laverny}, {De Luise}, {De March}, {de Martino}, {de Souza}, {de Torres}, {Debosscher}, {del Pozo}, {Delbo}, {Delgado}, {Delgado}, {Diakite}, {Diener}, {Distefano}, {Dolding}, {Drazinos}, {Dur{\'a}n}, {Edvardsson}, {Enke}, {Eriksson}, {Esquej}, {Eynard Bontemps}, {Fabre}, {Fabrizio}, {Faigler}, {Falc{\~a}o}, {Farr{\`a}s Casas}, {Federici}, {Fedorets}, {Fernique}, {Figueras}, {Filippi}, {Findeisen}, {Fonti}, {Fraile}, {Fraser}, {Fr{\'e}zouls}, {Gai}, {Galleti}, {Garabato}, {Garc{\'\i}a-Sedano}, {Garofalo}, {Garralda}, {Gavel},
  {Gavras}, {Gerssen}, {Geyer}, {Giacobbe}, {Gilmore}, {Girona}, {Giuffrida}, {Glass}, {Gomes}, {Granvik}, {Gueguen}, {Guerrier}, {Guiraud}, {Guti{\'e}}, {Haigron}, {Hatzidimitriou}, {Hauser}, {Haywood}, {Heiter}, {Helmi}, {Heu}, {Hilger}, {Hobbs}, {Hofmann}, {Holland}, {Huckle}, {Hypki}, {Icardi}, {Jan{\ss}en}, {Jevardat de Fombelle}, {Jonker}, {Juh{\'a}sz}, {Julbe}, {Karampelas}, {Kewley}, {Klar}, {Kochoska}, {Kohley}, {Kolenberg}, {Kontizas}, {Kontizas}, {Koposov}, {Kordopatis}, {Kostrzewa-Rutkowska}, {Koubsky}, {Lambert}, {Lanza}, {Lasne}, {Lavigne}, {Le Fustec}, {Le Poncin-Lafitte}, {Lebreton}, {Leccia}, {Leclerc}, {Lecoeur-Taibi}, {Lenhardt}, {Leroux}, {Liao}, {Licata}, {Lindstr{\o}m}, {Lister}, {Livanou}, {Lobel}, {L{\'o}pez}, {Managau}, {Mann}, {Mantelet}, {Marchal}, {Marchant}, {Marconi}, {Marinoni}, {Marschalk{\'o}}, {Marshall}, {Martino}, {Marton}, {Mary}, {Massari}, {Matijevi{\v{c}}}, {Mazeh}, {McMillan}, {Messina}, {Michalik}, {Millar}, {Molina}, {Molinaro}, {Moln{\'a}r}, {Montegriffo}, {Mor},
  {Morbidelli}, {Morel}, {Morris}, {Mulone}, {Muraveva}, {Musella}, {Nelemans}, {Nicastro}, {Noval}, {O'Mullane}, {Ord{\'e}novic}, {Ord{\'o}{\~n}ez-Blanco}, {Osborne}, {Pagani}, {Pagano}, {Pailler}, {Palacin}, {Palaversa}, {Panahi}, {Pawlak}, {Piersimoni}, {Pineau}, {Plachy}, {Plum}, {Poggio}, {Poujoulet}, {Pr{\v{s}}a}, {Pulone}, {Racero}, {Ragaini}, {Rambaux}, {Ramos-Lerate}, {Regibo}, {Reyl{\'e}}, {Riclet}, {Ripepi}, {Riva}, {Rivard}, {Rixon}, {Roegiers}, {Roelens}, {Romero-G{\'o}mez}, {Rowell}, {Royer}, {Ruiz-Dern}, {Sadowski}, {Sagrist{\`a} Sell{\'e}s}, {Sahlmann}, {Salgado}, {Salguero}, {Sanna}, {Santana-Ros}, {Sarasso}, {Savietto}, {Schultheis}, {Sciacca}, {Segol}, {Segovia}, {S{\'e}gransan}, {Shih}, {Siltala}, {Silva}, {Smart}, {Smith}, {Solano}, {Solitro}, {Sordo}, {Soria Nieto}, {Souchay}, {Spagna}, {Spoto}, {Stampa}, {Steele}, {Steidelm{\"u}ller}, {Stephenson}, {Stoev}, {Suess}, {Surdej}, {Szabados}, {Szegedi-Elek}, {Tapiador}, {Taris}, {Tauran}, {Taylor}, {Teixeira}, {Terrett}, {Teyssandier},
  {Thuillot}, {Titarenko}, {Torra Clotet}, {Turon}, {Ulla}, {Utrilla}, {Uzzi}, {Vaillant}, {Valentini}, {Valette}, {van Elteren}, {Van Hemelryck}, {Vaschetto}, {Vecchiato}, {Veljanoski}, {Viala}, {Vicente}, {Vogt}, {von Essen}, {Voss}, {Votruba}, {Voutsinas}, {Walmsley}, {Weiler}, {Wertz}, {Wevers}, {Wyrzykowski}, {Yoldas}, {{\v{Z}}erjal}, {Ziaeepour}, {Zorec}, {Zschocke}, {Zucker}, {Zurbach}, \& {Zwitter}}]{Gaia2018}
{Gaia Collaboration}, {Babusiaux}, C., {van Leeuwen}, F., {et~al.} 2018, \aap, 616, A10, \dodoi{10.1051/0004-6361/201832843}

\bibitem[{{Gaia Collaboration} {et~al.}(2022){Gaia Collaboration}, {Vallenari}, {Brown}, {Prusti}, {de Bruijne}, {Arenou}, {Babusiaux}, {Biermann}, {Creevey}, {Ducourant}, {Evans}, {Eyer}, {Guerra}, {Hutton}, {Jordi}, {Klioner}, {Lammers}, {Lindegren}, {Luri}, {Mignard}, {Panem}, {Pourbaix}, {Randich}, {Sartoretti}, {Soubiran}, {Tanga}, {Walton}, {Bailer-Jones}, {Bastian}, {Drimmel}, {Jansen}, {Katz}, {Lattanzi}, {van Leeuwen}, {Bakker}, {Cacciari}, {Casta{\~n}eda}, {De Angeli}, {Fabricius}, {Fouesneau}, {Fr{\'e}mat}, {Galluccio}, {Guerrier}, {Heiter}, {Masana}, {Messineo}, {Mowlavi}, {Nicolas}, {Nienartowicz}, {Pailler}, {Panuzzo}, {Riclet}, {Roux}, {Seabroke}, {Sordo{\o}rcit}, {Th{\'e}venin}, {Gracia-Abril}, {Portell}, {Teyssier}, {Altmann}, {Andrae}, {Audard}, {Bellas-Velidis}, {Benson}, {Berthier}, {Blomme}, {Burgess}, {Busonero}, {Busso}, {C{\'a}novas}, {Carry}, {Cellino}, {Cheek}, {Clementini}, {Damerdji}, {Davidson}, {de Teodoro}, {Nu{\~n}ez Campos}, {Delchambre}, {Dell'Oro}, {Esquej},
  {Fern{\'a}ndez-Hern{\'a}ndez}, {Fraile}, {Garabato}, {Garc{\'\i}a-Lario}, {Gosset}, {Haigron}, {Halbwachs}, {Hambly}, {Harrison}, {Hern{\'a}ndez}, {Hestroffer}, {Hodgkin}, {Holl}, {Jan{\ss}en}, {Jevardat de Fombelle}, {Jordan}, {Krone-Martins}, {Lanzafame}, {L{\"o}ffler}, {Marchal}, {Marrese}, {Moitinho}, {Muinonen}, {Osborne}, {Pancino}, {Pauwels}, {Recio-Blanco}, {Reyl{\'e}}, {Riello}, {Rimoldini}, {Roegiers}, {Rybizki}, {Sarro}, {Siopis}, {Smith}, {Sozzetti}, {Utrilla}, {van Leeuwen}, {Abbas}, {{\'A}brah{\'a}m}, {Abreu Aramburu}, {Aerts}, {Aguado}, {Ajaj}, {Aldea-Montero}, {Altavilla}, {{\'A}lvarez}, {Alves}, {Anders}, {Anderson}, {Anglada Varela}, {Antoja}, {Baines}, {Baker}, {Balaguer-N{\'u}{\~n}ez}, {Balbinot}, {Balog}, {Barache}, {Barbato}, {Barros}, {Barstow}, {Bartolom{\'e}}, {Bassilana}, {Bauchet}, {Becciani}, {Bellazzini}, {Berihuete}, {Bernet}, {Bertone}, {Bianchi}, {Binnenfeld}, {Blanco-Cuaresma}, {Blazere}, {Boch}, {Bombrun}, {Bossini}, {Bouquillon}, {Bragaglia}, {Bramante}, {Breedt},
  {Bressan}, {Brouillet}, {Brugaletta}, {Bucciarelli}, {Burlacu}, {Butkevich}, {Buzzi}, {Caffau}, {Cancelliere}, {Cantat-Gaudin}, {Carballo}, {Carlucci}, {Carnerero}, {Carrasco}, {Casamiquela}, {Castellani}, {Castro-Ginard}, {Chaoul}, {Charlot}, {Chemin}, {Chiaramida}, {Chiavassa}, {Chornay}, {Comoretto}, {Contursi}, {Cooper}, {Cornez}, {Cowell}, {Crifo}, {Cropper}, {Crosta}, {Crowley}, {Dafonte}, {Dapergolas}, {David}, {David}, {de Laverny}, {De Luise}, {De March}, {De Ridder}, {de Souza}, {de Torres}, {del Peloso}, {del Pozo}, {Delbo}, {Delgado}, {Delisle}, {Demouchy}, {Dharmawardena}, {Di Matteo}, {Diakite}, {Diener}, {Distefano}, {Dolding}, {Edvardsson}, {Enke}, {Fabre}, {Fabrizio}, {Faigler}, {Fedorets}, {Fernique}, {Fienga}, {Figueras}, {Fournier}, {Fouron}, {Fragkoudi}, {Gai}, {Garcia-Gutierrez}, {Garcia-Reinaldos}, {Garc{\'\i}a-Torres}, {Garofalo}, {Gavel}, {Gavras}, {Gerlach}, {Geyer}, {Giacobbe}, {Gilmore}, {Girona}, {Giuffrida}, {Gomel}, {Gomez}, {Gonz{\'a}lez-N{\'u}{\~n}ez},
  {Gonz{\'a}lez-Santamar{\'\i}a}, {Gonz{\'a}lez-Vidal}, {Granvik}, {Guillout}, {Guiraud}, {Guti{\'e}rrez-S{\'a}nchez}, {Guy}, {Hatzidimitriou}, {Hauser}, {Haywood}, {Helmer}, {Helmi}, {Sarmiento}, {Hidalgo}, {Hilger}, {H{\l}adczuk}, {Hobbs}, {Holland}, {Huckle}, {Jardine}, {Jasniewicz}, {Jean-Antoine Piccolo}, {Jim{\'e}nez-Arranz}, {Jorissen}, {Juaristi Campillo}, {Julbe}, {Karbevska}, {Kervella}, {Khanna}, {Kontizas}, {Kordopatis}, {Korn}, {K{\'o}sp{\'a}l}, {Kostrzewa-Rutkowska}, {Kruszy{\'n}ska}, {Kun}, {Laizeau}, {Lambert}, {Lanza}, {Lasne}, {Le Campion}, {Lebreton}, {Lebzelter}, {Leccia}, {Leclerc}, {Lecoeur-Taibi}, {Liao}, {Licata}, {Lindstr{\o}m}, {Lister}, {Livanou}, {Lobel}, {Lorca}, {Loup}, {Madrero Pardo}, {Magdaleno Romeo}, {Managau}, {Mann}, {Manteiga}, {Marchant}, {Marconi}, {Marcos}, {Marcos Santos}, {Mar{\'\i}n Pina}, {Marinoni}, {Marocco}, {Marshall}, {Polo}, {Mart{\'\i}n-Fleitas}, {Marton}, {Mary}, {Masip}, {Massari}, {Mastrobuono-Battisti}, {Mazeh}, {McMillan}, {Messina}, {Michalik},
  {Millar}, {Mints}, {Molina}, {Molinaro}, {Moln{\'a}r}, {Monari}, {Mongui{\'o}}, {Montegriffo}, {Montero}, {Mor}, {Mora}, {Morbidelli}, {Morel}, {Morris}, {Muraveva}, {Murphy}, {Musella}, {Nagy}, {Noval}, {Oca{\~n}a}, {Ogden}, {Ordenovic}, {Osinde}, {Pagani}, {Pagano}, {Palaversa}, {Palicio}, {Pallas-Quintela}, {Panahi}, {Payne-Wardenaar}, {Pe{\~n}alosa Esteller}, {Penttil{\"a}}, {Pichon}, {Piersimoni}, {Pineau}, {Plachy}, {Plum}, {Poggio}, {Pr{\v{s}}a}, {Pulone}, {Racero}, {Ragaini}, {Rainer}, {Raiteri}, {Rambaux}, {Ramos}, {Ramos-Lerate}, {Re Fiorentin}, {Regibo}, {Richards}, {Rios Diaz}, {Ripepi}, {Riva}, {Rix}, {Rixon}, {Robichon}, {Robin}, {Robin}, {Roelens}, {Rogues}, {Rohrbasser}, {Romero-G{\'o}mez}, {Rowell}, {Royer}, {Ruz Mieres}, {Rybicki}, {Sadowski}, {S{\'a}ez N{\'u}{\~n}ez}, {Sagrist{\`a} Sell{\'e}s}, {Sahlmann}, {Salguero}, {Samaras}, {Sanchez Gimenez}, {Sanna}, {Santove{\~n}a}, {Sarasso}, {Schultheis}, {Sciacca}, {Segol}, {Segovia}, {S{\'e}gransan}, {Semeux}, {Shahaf}, {Siddiqui}, {Siebert},
  {Siltala}, {Silvelo}, {Slezak}, {Slezak}, {Smart}, {Snaith}, {Solano}, {Solitro}, {Souami}, {Souchay}, {Spagna}, {Spina}, {Spoto}, {Steele}, {Steidelm{\"u}ller}, {Stephenson}, {S{\"u}veges}, {Surdej}, {Szabados}, {Szegedi-Elek}, {Taris}, {Taylo}, {Teixeira}, {Tolomei}, {Tonello}, {Torra}, {Torra}, {Torralba Elipe}, {Trabucchi}, {Tsounis}, {Turon}, {Ulla}, {Unger}, {Vaillant}, {van Dillen}, {van Reeven}, {Vanel}, {Vecchiato}, {Viala}, {Vicente}, {Voutsinas}, {Weiler}, {Wevers}, {Wyrzykowski}, {Yoldas}, {Yvard}, {Zhao}, {Zorec}, {Zucker}, \& {Zwitter}}]{GaiaDR3_2022}
{Gaia Collaboration}, {Vallenari}, A., {Brown}, A.~G.~A., {et~al.} 2022, arXiv e-prints, arXiv:2208.00211, \dodoi{10.48550/arXiv.2208.00211}

\bibitem[{{Garcia Lopez} {et~al.}(2001){Garcia Lopez}, {Rebolo}, \& {Zapaterio Osorio}}]{Garcia2001}
{Garcia Lopez}, R.~J., {Rebolo}, R., \& {Zapaterio Osorio}, M.~R., eds. 2001, Astronomical Society of the Pacific Conference Series, Vol. 223, {11th Cambridge Workshop on Cool Stars, Stellar Systems and the Sun}

\bibitem[{{Gelman} \& {Rubin}(1992)}]{Gelman1992}
{Gelman}, A., \& {Rubin}, D.~B. 1992, Statistical Science, 7, 457, \dodoi{10.1214/ss/1177011136}

\bibitem[{{Ginsburg} {et~al.}(2019){Ginsburg}, {Sip{\H{o}}cz}, {Brasseur}, {Cowperthwaite}, {Craig}, {Deil}, {Guillochon}, {Guzman}, {Liedtke}, {Lian Lim}, {Lockhart}, {Mommert}, {Morris}, {Norman}, {Parikh}, {Persson}, {Robitaille}, {Segovia}, {Singer}, {Tollerud}, {de Val-Borro}, {Valtchanov}, {Woillez}, {Astroquery Collaboration}, \& {a subset of astropy Collaboration}}]{Ginsburg2019}
{Ginsburg}, A., {Sip{\H{o}}cz}, B.~M., {Brasseur}, C.~E., {et~al.} 2019, \aj, 157, 98, \dodoi{10.3847/1538-3881/aafc33}

\bibitem[{{Goodman} \& {Weare}(2010)}]{Goodman2010}
{Goodman}, J., \& {Weare}, J. 2010, Communications in Applied Mathematics and Computational Science, 5, 65, \dodoi{10.2140/camcos.2010.5.65}

\bibitem[{{Grandjean} {et~al.}(2021){Grandjean}, {Lagrange}, {Meunier}, {Rubini}, {Desidera}, {Galland}, {Borgniet}, {Zicher}, {Messina}, {Chauvin}, {Sterzik}, \& {Pantoja}}]{Grandjean2021}
{Grandjean}, A., {Lagrange}, A.~M., {Meunier}, N., {et~al.} 2021, \aap, 650, A39, \dodoi{10.1051/0004-6361/202039672}

\bibitem[{{Grandjean} {et~al.}(2023){Grandjean}, {Lagrange}, {Meunier}, {Chauvin}, {Borgniet}, {Desidera}, {Galland}, {Kiefer}, {Messina}, {Iglesias}, {Nicholson}, {Pantoja}, {Rubini}, {Sedaghati}, {Sterzik}, \& {Zicher}}]{Grandjean2023}
---. 2023, \aap, 669, A12, \dodoi{10.1051/0004-6361/202141235}

\bibitem[{{Gray}(1992)}]{Gray1992}
{Gray}, D.~F. 1992, {The observation and analysis of stellar photospheres.}, Vol.~20 (Cambridge University Press)

\bibitem[{{Gray}(1999)}]{Gray1999}
---. 1999, in Astronomical Society of the Pacific Conference Series, Vol. 185, IAU Colloq. 170: Precise Stellar Radial Velocities, ed. J.~B. {Hearnshaw} \& C.~D. {Scarfe} (Astronomical Society of the Pacific), 243

\bibitem[{{Gray} {et~al.}(2006){Gray}, {Corbally}, {Garrison}, {McFadden}, {Bubar}, {McGahee}, {O'Donoghue}, \& {Knox}}]{Gray2006}
{Gray}, R.~O., {Corbally}, C.~J., {Garrison}, R.~F., {et~al.} 2006, \aj, 132, 161, \dodoi{10.1086/504637}

\bibitem[{{Gray} {et~al.}(2003){Gray}, {Corbally}, {Garrison}, {McFadden}, \& {Robinson}}]{Gray2003}
{Gray}, R.~O., {Corbally}, C.~J., {Garrison}, R.~F., {McFadden}, M.~T., \& {Robinson}, P.~E. 2003, \aj, 126, 2048, \dodoi{10.1086/378365}

\bibitem[{{Gregory}(2005)}]{Gregory2005}
{Gregory}, P.~C. 2005, \apj, 631, 1198, \dodoi{10.1086/432594}

\bibitem[{{Griffin}(2001)}]{Griffin2001}
{Griffin}, R.~F. 2001, The Observatory, 121, 244

\bibitem[{{Haisch} {et~al.}(2001){Haisch}, {Lada}, \& {Lada}}]{Haisch2001}
{Haisch}, Karl~E., J., {Lada}, E.~A., \& {Lada}, C.~J. 2001, \apjl, 553, L153, \dodoi{10.1086/320685}

\bibitem[{{Haro} {et~al.}(1982){Haro}, {Chavira}, \& {Gonzalez}}]{Haro1982}
{Haro}, G., {Chavira}, E., \& {Gonzalez}, G. 1982, Boletin del Instituto de Tonantzintla, 3, 3

\bibitem[{{Haywood} {et~al.}(2014){Haywood}, {Collier Cameron}, {Queloz}, {Barros}, {Deleuil}, {Fares}, {Gillon}, {Lanza}, {Lovis}, {Moutou}, {Pepe}, {Pollacco}, {Santerne}, {S{\'e}gransan}, \& {Unruh}}]{Haywood2014}
{Haywood}, R.~D., {Collier Cameron}, A., {Queloz}, D., {et~al.} 2014, \mnras, 443, 2517, \dodoi{10.1093/mnras/stu1320}

\bibitem[{{Hester} \& {Desch}(2005)}]{Hester2005}
{Hester}, J.~J., \& {Desch}, S.~J. 2005, in Astronomical Society of the Pacific Conference Series, Vol. 341, Chondrites and the Protoplanetary Disk, ed. A.~N. {Krot}, E.~R.~D. {Scott}, \& B.~{Reipurth}, 107, \dodoi{10.48550/arXiv.astro-ph/0506190}

\bibitem[{{Hill} {et~al.}(2021){Hill}, {Lee}, {MacQueen}, {Kelz}, {Drory}, {Vattiat}, {Good}, {Ramsey}, {Kriel}, {Peterson}, {DePoy}, {Gebhardt}, {Marshall}, {Tuttle}, {Bauer}, {Chonis}, {Fabricius}, {Froning}, {H{\"a}user}, {Indahl}, {Jahn}, {Landriau}, {Leck}, {Montesano}, {Prochaska}, {Snigula}, {Zeimann}, {Bryant}, {Damm}, {Fowler}, {Janowiecki}, {Martin}, {Mrozinski}, {Odewahn}, {Rostopchin}, {Shetrone}, {Spencer}, {Mentuch Cooper}, {Armandroff}, {Bender}, {Dalton}, {Hopp}, {Komatsu}, {Nicklas}, {Ramsey}, {Roth}, {Schneider}, {Sneden}, \& {Steinmetz}}]{Hill2021}
{Hill}, G.~J., {Lee}, H., {MacQueen}, P.~J., {et~al.} 2021, \aj, 162, 298, \dodoi{10.3847/1538-3881/ac2c02}

\bibitem[{{Hill} \& {Schilt}(1952)}]{Hill1952}
{Hill}, S.~J., \& {Schilt}, J. 1952, Contributions from the Rutherford Observatory of Columbia University New York, 32, 1

\bibitem[{{Hinkel} {et~al.}(2015){Hinkel}, {Kane}, {Henry}, {Feng}, {Boyajian}, {Wright}, {Fischer}, \& {Howard}}]{Hinkel2015}
{Hinkel}, N.~R., {Kane}, S.~R., {Henry}, G.~W., {et~al.} 2015, \apj, 803, 8, \dodoi{10.1088/0004-637X/803/1/8}

\bibitem[{{Houk} \& {Swift}(1999)}]{Houk1999}
{Houk}, N., \& {Swift}, C. 1999, {Michigan catalogue of two-dimensional spectral types for the HD Stars ; vol. 5}, Vol.~5 (University of Michigan)

\bibitem[{{Howard} {et~al.}(2010){Howard}, {Marcy}, {Johnson}, {Fischer}, {Wright}, {Isaacson}, {Valenti}, {Anderson}, {Lin}, \& {Ida}}]{Howard2010}
{Howard}, A.~W., {Marcy}, G.~W., {Johnson}, J.~A., {et~al.} 2010, Science, 330, 653, \dodoi{10.1126/science.1194854}

\bibitem[{{Hunter}(2007)}]{Hunter4160265}
{Hunter}, J.~D. 2007, Computing in Science Engineering, 9, 90

\bibitem[{{Jackson} {et~al.}(2009){Jackson}, {Jeffries}, \& {Maxted}}]{JacksonRJ2009}
{Jackson}, R.~J., {Jeffries}, R.~D., \& {Maxted}, P.~F.~L. 2009, \mnras, 399, L89, \dodoi{10.1111/j.1745-3933.2009.00729.x}

\bibitem[{{Jahn} \& {Stepien}(1984)}]{Jahn1984}
{Jahn}, K., \& {Stepien}, K. 1984, \actaa, 34, 1

\bibitem[{{Johns-Krull} {et~al.}(2016){Johns-Krull}, {McLane}, {Prato}, {Crockett}, {Jaffe}, {Hartigan}, {Beichman}, {Mahmud}, {Chen}, {Skiff}, {Cauley}, {Jones}, \& {Mace}}]{Johns-Krull2016}
{Johns-Krull}, C.~M., {McLane}, J.~N., {Prato}, L., {et~al.} 2016, \apj, 826, 206, \dodoi{10.3847/0004-637X/826/2/206}

\bibitem[{{Johnson} {et~al.}(2010){Johnson}, {Aller}, {Howard}, \& {Crepp}}]{Johnson2010}
{Johnson}, J.~A., {Aller}, K.~M., {Howard}, A.~W., \& {Crepp}, J.~R. 2010, \pasp, 122, 905, \dodoi{10.1086/655775}

\bibitem[{{Kanodia} \& {Wright}(2018)}]{Kanodia2018a}
{Kanodia}, S., \& {Wright}, J. 2018, Research Notes of the American Astronomical Society, 2, 4, \dodoi{10.3847/2515-5172/aaa4b7}

\bibitem[{{Kanodia} {et~al.}(2018){Kanodia}, {Mahadevan}, {Ramsey}, {Stefansson}, {Monson}, {Hearty}, {Blakeslee}, {Lubar}, {Bender}, {Ninan}, {Sterner}, {Roy}, {Halverson}, \& {Robertson}}]{Kanodia2018b}
{Kanodia}, S., {Mahadevan}, S., {Ramsey}, L.~W., {et~al.} 2018, in Society of Photo-Optical Instrumentation Engineers (SPIE) Conference Series, Vol. 10702, Ground-based and Airborne Instrumentation for Astronomy VII, ed. C.~J. {Evans}, L.~{Simard}, \& H.~{Takami}, 107026Q, \dodoi{10.1117/12.2313491}

\bibitem[{{Kaplan} {et~al.}(2019){Kaplan}, {Bender}, {Terrien}, {Ninan}, {Roy}, \& {Mahadevan}}]{Kaplan2019}
{Kaplan}, K.~F., {Bender}, C.~F., {Terrien}, R.~C., {et~al.} 2019, in Astronomical Society of the Pacific Conference Series, Vol. 523, Astronomical Data Analysis Software and Systems XXVII, ed. P.~J. {Teuben}, M.~W. {Pound}, B.~A. {Thomas}, \& E.~M. {Warner}, 567

\bibitem[{{Keenan} \& {McNeil}(1989)}]{Keenan1989}
{Keenan}, P.~C., \& {McNeil}, R.~C. 1989, \apjs, 71, 245, \dodoi{10.1086/191373}

\bibitem[{{Kley} \& {Nelson}(2012)}]{Kley2012}
{Kley}, W., \& {Nelson}, R.~P. 2012, \araa, 50, 211, \dodoi{10.1146/annurev-astro-081811-125523}

\bibitem[{{Kraft}(1967)}]{Kraft1967}
{Kraft}, R.~P. 1967, \apj, 150, 551, \dodoi{10.1086/149359}

\bibitem[{{Lada} \& {Lada}(2003)}]{Lada2003}
{Lada}, C.~J., \& {Lada}, E.~A. 2003, \araa, 41, 57, \dodoi{10.1146/annurev.astro.41.011802.094844}

\bibitem[{{Lagrange} {et~al.}(2010){Lagrange}, {Desort}, \& {Meunier}}]{Lagrange2010}
{Lagrange}, A.~M., {Desort}, M., \& {Meunier}, N. 2010, \aap, 512, A38, \dodoi{10.1051/0004-6361/200913071}

\bibitem[{{Lagrange} {et~al.}(2013){Lagrange}, {Meunier}, {Chauvin}, {Sterzik}, {Galland}, {Lo Curto}, {Rameau}, \& {Sosnowska}}]{Lagrange2013}
{Lagrange}, A.~M., {Meunier}, N., {Chauvin}, G., {et~al.} 2013, \aap, 559, A83, \dodoi{10.1051/0004-6361/201220770}

\bibitem[{{Lightkurve Collaboration} {et~al.}(2018){Lightkurve Collaboration}, {Cardoso}, {Hedges}, {Gully-Santiago}, {Saunders}, {Cody}, {Barclay}, {Hall}, {Sagear}, {Turtelboom}, {Zhang}, {Tzanidakis}, {Mighell}, {Coughlin}, {Bell}, {Berta-Thompson}, {Williams}, {Dotson}, \& {Barentsen}}]{Lightkurve2018}
{Lightkurve Collaboration}, {Cardoso}, J.~V.~d.~M., {Hedges}, C., {et~al.} 2018, {Lightkurve: Kepler and TESS time series analysis in Python}, Astrophysics Source Code Library.
\newblock \doeprint{1812.013}

\bibitem[{Lindegren(2018)}]{Lindegren2018}
Lindegren, L. 2018.
\newblock \url{http://www.rssd.esa.int/doc_fetch.php?id=3757412}

\bibitem[{{Llorente de Andr{\'e}s} {et~al.}(2021){Llorente de Andr{\'e}s}, {Chavero}, {de la Reza}, {Roca-F{\`a}brega}, \& {Cifuentes}}]{LlorentedeAndres2021}
{Llorente de Andr{\'e}s}, F., {Chavero}, C., {de la Reza}, R., {Roca-F{\`a}brega}, S., \& {Cifuentes}, C. 2021, \aap, 654, A137, \dodoi{10.1051/0004-6361/202141339}

\bibitem[{{Lopez} \& {Fortney}(2013)}]{Lopez2013}
{Lopez}, E.~D., \& {Fortney}, J.~J. 2013, \apj, 776, 2, \dodoi{10.1088/0004-637X/776/1/2}

\bibitem[{{Lovis} {et~al.}(2011){Lovis}, {Dumusque}, {Santos}, {Bouchy}, {Mayor}, {Pepe}, {Queloz}, {S{\'e}gransan}, \& {Udry}}]{Lovis2011}
{Lovis}, C., {Dumusque}, X., {Santos}, N.~C., {et~al.} 2011, arXiv e-prints, arXiv:1107.5325, \dodoi{10.48550/arXiv.1107.5325}

\bibitem[{{Luhn} {et~al.}(2020){Luhn}, {Wright}, {Howard}, \& {Isaacson}}]{Luhn2020}
{Luhn}, J.~K., {Wright}, J.~T., {Howard}, A.~W., \& {Isaacson}, H. 2020, \aj, 159, 235, \dodoi{10.3847/1538-3881/ab855a}

\bibitem[{{Lynden-Bell} \& {Pringle}(1974)}]{Lynden-Bell1974}
{Lynden-Bell}, D., \& {Pringle}, J.~E. 1974, \mnras, 168, 603, \dodoi{10.1093/mnras/168.3.603}

\bibitem[{{Mahadevan} {et~al.}(2012){Mahadevan}, {Ramsey}, {Bender}, {Terrien}, {Wright}, {Halverson}, {Hearty}, {Nelson}, {Burton}, {Redman}, {Osterman}, {Diddams}, {Kasting}, {Endl}, \& {Deshpande}}]{Mahadevan2012}
{Mahadevan}, S., {Ramsey}, L., {Bender}, C., {et~al.} 2012, in Society of Photo-Optical Instrumentation Engineers (SPIE) Conference Series, Vol. 8446, Ground-based and Airborne Instrumentation for Astronomy IV, ed. I.~S. {McLean}, S.~K. {Ramsay}, \& H.~{Takami}, 84461S, \dodoi{10.1117/12.926102}

\bibitem[{{Mahadevan} {et~al.}(2014){Mahadevan}, {Ramsey}, {Terrien}, {Halverson}, {Roy}, {Hearty}, {Levi}, {Stefansson}, {Robertson}, {Bender}, {Schwab}, \& {Nelson}}]{Mahadevan2014}
{Mahadevan}, S., {Ramsey}, L.~W., {Terrien}, R., {et~al.} 2014, in Society of Photo-Optical Instrumentation Engineers (SPIE) Conference Series, Vol. 9147, Ground-based and Airborne Instrumentation for Astronomy V, ed. S.~K. {Ramsay}, I.~S. {McLean}, \& H.~{Takami}, 91471G, \dodoi{10.1117/12.2056417}

\bibitem[{Mahadevan {et~al.}(2018)Mahadevan, Anderson, Balderrama, Bender, Bevins, Blakeslee, Cole, Conran, Diddams, Dykhouse, Darling, Fredrick, Halverson, Hearty, Jennings, Kaplan, Kanodia, Levi, Lubar, Metcalf, Monson, Ninan, Nitroy, Ramsey, Robertson, Roy, Schwab, Shetrone, Spencer, Stefansson, Terrien, \& Wright}]{Mahadevan2018}
Mahadevan, S., Anderson, T., Balderrama, E., {et~al.} 2018, in Ground-based and Airborne Instrumentation for Astronomy VII, ed. C.~J. Evans, L.~Simard, \& H.~Takami, Vol. 10702, International Society for Optics and Photonics (SPIE), 1070214, \dodoi{10.1117/12.2313835}

\bibitem[{{Malo} {et~al.}(2014){Malo}, {Artigau}, {Doyon}, {Lafreni{\`e}re}, {Albert}, \& {Gagn{\'e}}}]{Malo2014}
{Malo}, L., {Artigau}, {\'E}., {Doyon}, R., {et~al.} 2014, \apj, 788, 81, \dodoi{10.1088/0004-637X/788/1/81}

\bibitem[{{Malo} {et~al.}(2013){Malo}, {Doyon}, {Lafreni{\`e}re}, {Artigau}, {Gagn{\'e}}, {Baron}, \& {Riedel}}]{Malo2013}
{Malo}, L., {Doyon}, R., {Lafreni{\`e}re}, D., {et~al.} 2013, \apj, 762, 88, \dodoi{10.1088/0004-637X/762/2/88}

\bibitem[{{Mann} {et~al.}(2013){Mann}, {Brewer}, {Gaidos}, {L{\'e}pine}, \& {Hilton}}]{Mann2013}
{Mann}, A.~W., {Brewer}, J.~M., {Gaidos}, E., {L{\'e}pine}, S., \& {Hilton}, E.~J. 2013, \aj, 145, 52, \dodoi{10.1088/0004-6256/145/2/52}

\bibitem[{{Martin} \& {Livio}(2012)}]{Martin2012}
{Martin}, R.~G., \& {Livio}, M. 2012, \mnras, 425, L6, \dodoi{10.1111/j.1745-3933.2012.01290.x}

\bibitem[{{Marzari} \& {Nagasawa}(2019)}]{Marzari2019}
{Marzari}, F., \& {Nagasawa}, M. 2019, \aap, 625, A121, \dodoi{10.1051/0004-6361/201935065}

\bibitem[{{Mason} {et~al.}(2001){Mason}, {Wycoff}, {Hartkopf}, {Douglass}, \& {Worley}}]{Mason2001}
{Mason}, B.~D., {Wycoff}, G.~L., {Hartkopf}, W.~I., {Douglass}, G.~G., \& {Worley}, C.~E. 2001, \aj, 122, 3466, \dodoi{10.1086/323920}

\bibitem[{{MAST Team}(2021)}]{doi.org10.17909/t9-nmc8-f686}
{MAST Team}. 2021, TESS Light Curves - All Sectors,  STScI/MAST, \dodoi{10.17909/T9-NMC8-F686}

\bibitem[{{M}c{K}inney(2010)}]{mckinney-proc-scipy-2010}
{M}c{K}inney, W. 2010, in {P}roceedings of the 9th {P}ython in {S}cience {C}onference, ed. {S}t\'efan van~der {W}alt \& {J}arrod {M}illman, 56 -- 61, \dodoi{10.25080/Majora-92bf1922-00a}

\bibitem[{{Mendoza V.}(1956)}]{Mendoza1956}
{Mendoza V.}, E.~E. 1956, \apj, 123, 54, \dodoi{10.1086/146129}

\bibitem[{{Metcalf} {et~al.}(2019){Metcalf}, {Anderson}, {Bender}, {Blakeslee}, {Brand}, {Carlson}, {Cochran}, {Diddams}, {Endl}, {Fredrick}, {Halverson}, {Hickstein}, {Hearty}, {Jennings}, {Kanodia}, {Kaplan}, {Levi}, {Lubar}, {Mahadevan}, {Monson}, {Ninan}, {Nitroy}, {Osterman}, {Papp}, {Quinlan}, {Ramsey}, {Robertson}, {Roy}, {Schwab}, {Sigurdsson}, {Srinivasan}, {Stefansson}, {Sterner}, {Terrien}, {Wolszczan}, {Wright}, \& {Ycas}}]{Metcalf2019a}
{Metcalf}, A.~J., {Anderson}, T., {Bender}, C.~F., {et~al.} 2019, Optica, 6, 233, \dodoi{10.1364/OPTICA.6.000233}

\bibitem[{{Meunier} \& {Lagrange}(2013)}]{Meunier2013}
{Meunier}, N., \& {Lagrange}, A.~M. 2013, \aap, 551, A101, \dodoi{10.1051/0004-6361/201219917}

\bibitem[{{Meunier} {et~al.}(2017){Meunier}, {Lagrange}, {Mbemba Kabuiku}, {Alex}, {Mignon}, \& {Borgniet}}]{Meunier2017}
{Meunier}, N., {Lagrange}, A.~M., {Mbemba Kabuiku}, L., {et~al.} 2017, \aap, 597, A52, \dodoi{10.1051/0004-6361/201629052}

\bibitem[{{Montes} {et~al.}(2001{\natexlab{a}}){Montes}, {L{\'o}pez-Santiago}, {Fern{\'a}ndez-Figueroa}, \& {G{\'a}lvez}}]{Montes2001b}
{Montes}, D., {L{\'o}pez-Santiago}, J., {Fern{\'a}ndez-Figueroa}, M.~J., \& {G{\'a}lvez}, M.~C. 2001{\natexlab{a}}, \aap, 379, 976, \dodoi{10.1051/0004-6361:20011385}

\bibitem[{{Montes} {et~al.}(2001{\natexlab{b}}){Montes}, {L{\'o}pez-Santiago}, {G{\'a}lvez}, {Fern{\'a}ndez-Figueroa}, {De Castro}, \& {Cornide}}]{Montes2001a}
{Montes}, D., {L{\'o}pez-Santiago}, J., {G{\'a}lvez}, M.~C., {et~al.} 2001{\natexlab{b}}, \mnras, 328, 45, \dodoi{10.1046/j.1365-8711.2001.04781.x}

\bibitem[{{Morris} {et~al.}(2019){Morris}, {Curtis}, {Sakari}, {Hawley}, \& {Agol}}]{Morris2019}
{Morris}, B.~M., {Curtis}, J.~L., {Sakari}, C., {Hawley}, S.~L., \& {Agol}, E. 2019, \aj, 158, 101, \dodoi{10.3847/1538-3881/ab2e04}

\bibitem[{{Mustill} {et~al.}(2017){Mustill}, {Davies}, \& {Johansen}}]{Mustill2017}
{Mustill}, A.~J., {Davies}, M.~B., \& {Johansen}, A. 2017, \mnras, 468, 3000, \dodoi{10.1093/mnras/stx693}

\bibitem[{{Nesterov} {et~al.}(1995){Nesterov}, {Kuzmin}, {Ashimbaeva}, {Volchkov}, {R{\"o}ser}, \& {Bastian}}]{Nesterov1995}
{Nesterov}, V.~V., {Kuzmin}, A.~V., {Ashimbaeva}, N.~T., {et~al.} 1995, \aaps, 110, 367

\bibitem[{{Nielsen} {et~al.}(2008){Nielsen}, {Close}, {Biller}, {Masciadri}, \& {Lenzen}}]{Nielsen2008}
{Nielsen}, E.~L., {Close}, L.~M., {Biller}, B.~A., {Masciadri}, E., \& {Lenzen}, R. 2008, \apj, 674, 466, \dodoi{10.1086/524344}

\bibitem[{{Nielsen} {et~al.}(2019){Nielsen}, {De Rosa}, {Macintosh}, {Wang}, {Ruffio}, {Chiang}, {Marley}, {Saumon}, {Savransky}, {Ammons}, {Bailey}, {Barman}, {Blain}, {Bulger}, {Burrows}, {Chilcote}, {Cotten}, {Czekala}, {Doyon}, {Duch{\^e}ne}, {Esposito}, {Fabrycky}, {Fitzgerald}, {Follette}, {Fortney}, {Gerard}, {Goodsell}, {Graham}, {Greenbaum}, {Hibon}, {Hinkley}, {Hirsch}, {Hom}, {Hung}, {Dawson}, {Ingraham}, {Kalas}, {Konopacky}, {Larkin}, {Lee}, {Lin}, {Maire}, {Marchis}, {Marois}, {Metchev}, {Millar-Blanchaer}, {Morzinski}, {Oppenheimer}, {Palmer}, {Patience}, {Perrin}, {Poyneer}, {Pueyo}, {Rafikov}, {Rajan}, {Rameau}, {Rantakyr{\"o}}, {Ren}, {Schneider}, {Sivaramakrishnan}, {Song}, {Soummer}, {Tallis}, {Thomas}, {Ward-Duong}, \& {Wolff}}]{Nielsen2019}
{Nielsen}, E.~L., {De Rosa}, R.~J., {Macintosh}, B., {et~al.} 2019, \aj, 158, 13, \dodoi{10.3847/1538-3881/ab16e9}

\bibitem[{{Ninan} {et~al.}(2018){Ninan}, {Bender}, {Mahadevan}, {Ford}, {Monson}, {Kaplan}, {Terrien}, {Roy}, {Robertson}, {Kanodia}, \& {Stefansson}}]{Ninan2018}
{Ninan}, J.~P., {Bender}, C.~F., {Mahadevan}, S., {et~al.} 2018, in Society of Photo-Optical Instrumentation Engineers (SPIE) Conference Series, Vol. 10709, High Energy, Optical, and Infrared Detectors for Astronomy VIII, 107092U, \dodoi{10.1117/12.2312787}

\bibitem[{{Ninan} {et~al.}(2019){Ninan}, {Mahadevan}, {Stefansson}, {Bender}, {Roy}, {Kaplan}, {Fredrick}, {Metcalf}, {Monson}, {Terrien}, {Ramsey}, \& {Diddams}}]{Ninan2019}
{Ninan}, J.~P., {Mahadevan}, S., {Stefansson}, G., {et~al.} 2019, Journal of Astronomical Telescopes, Instruments, and Systems, 5, 041511, \dodoi{10.1117/1.JATIS.5.4.041511}

\bibitem[{{Owen} \& {Wu}(2017)}]{Owen2017}
{Owen}, J.~E., \& {Wu}, Y. 2017, \apj, 847, 29, \dodoi{10.3847/1538-4357/aa890a}

\bibitem[{{Paulson} {et~al.}(2004){Paulson}, {Saar}, {Cochran}, \& {Henry}}]{Paulson2004}
{Paulson}, D.~B., {Saar}, S.~H., {Cochran}, W.~D., \& {Henry}, G.~W. 2004, \aj, 127, 1644, \dodoi{10.1086/381948}

\bibitem[{{Paulson} \& {Yelda}(2006)}]{Paulson2006}
{Paulson}, D.~B., \& {Yelda}, S. 2006, \pasp, 118, 706, \dodoi{10.1086/504115}

\bibitem[{{Pecaut} \& {Mamajek}(2016)}]{Pecaut2016}
{Pecaut}, M.~J., \& {Mamajek}, E.~E. 2016, \mnras, 461, 794, \dodoi{10.1093/mnras/stw1300}

\bibitem[{{Perryman} {et~al.}(2014){Perryman}, {Hartman}, {Bakos}, \& {Lindegren}}]{Perryman2014}
{Perryman}, M., {Hartman}, J., {Bakos}, G.~{\'A}., \& {Lindegren}, L. 2014, \apj, 797, 14, \dodoi{10.1088/0004-637X/797/1/14}

\bibitem[{{Portegies Zwart}(2016)}]{Portegies2016}
{Portegies Zwart}, S.~F. 2016, \mnras, 457, 313, \dodoi{10.1093/mnras/stv2831}

\bibitem[{{Prato} {et~al.}(2008){Prato}, {Huerta}, {Johns-Krull}, {Mahmud}, {Jaffe}, \& {Hartigan}}]{Prato2008}
{Prato}, L., {Huerta}, M., {Johns-Krull}, C.~M., {et~al.} 2008, \apjl, 687, L103, \dodoi{10.1086/593201}

\bibitem[{{Prosser} {et~al.}(1991){Prosser}, {Stauffer}, \& {Kraft}}]{Prosser1991}
{Prosser}, C.~F., {Stauffer}, J., \& {Kraft}, R.~P. 1991, \aj, 101, 1361, \dodoi{10.1086/115772}

\bibitem[{{Queloz} {et~al.}(2001){Queloz}, {Henry}, {Sivan}, {Baliunas}, {Beuzit}, {Donahue}, {Mayor}, {Naef}, {Perrier}, \& {Udry}}]{Queloz2001}
{Queloz}, D., {Henry}, G.~W., {Sivan}, J.~P., {et~al.} 2001, \aap, 379, 279, \dodoi{10.1051/0004-6361:20011308}

\bibitem[{{Quinn}(2016)}]{Quinn2016}
{Quinn}, S.~N. 2016, PhD thesis, Georgia State University

\bibitem[{{Ram{\'\i}rez} {et~al.}(2012){Ram{\'\i}rez}, {Fish}, {Lambert}, \& {Allende Prieto}}]{Ramirez2012}
{Ram{\'\i}rez}, I., {Fish}, J.~R., {Lambert}, D.~L., \& {Allende Prieto}, C. 2012, \apj, 756, 46, \dodoi{10.1088/0004-637X/756/1/46}

\bibitem[{{Ramsey} {et~al.}(1998){Ramsey}, {Adams}, {Barnes}, {Booth}, {Cornell}, {Fowler}, {Gaffney}, {Glaspey}, {Good}, {Hill}, {Kelton}, {Krabbendam}, {Long}, {MacQueen}, {Ray}, {Ricklefs}, {Sage}, {Sebring}, {Spiesman}, \& {Steiner}}]{Ramsey1998}
{Ramsey}, L.~W., {Adams}, M.~T., {Barnes}, T.~G., {et~al.} 1998, in Society of Photo-Optical Instrumentation Engineers (SPIE) Conference Series, Vol. 3352, Advanced Technology Optical/IR Telescopes VI, ed. L.~M. {Stepp}, 34--42, \dodoi{10.1117/12.319287}

\bibitem[{{Reid} {et~al.}(2004){Reid}, {Cruz}, {Allen}, {Mungall}, {Kilkenny}, {Liebert}, {Hawley}, {Fraser}, {Covey}, {Lowrance}, {Kirkpatrick}, \& {Burgasser}}]{Reid2004}
{Reid}, I.~N., {Cruz}, K.~L., {Allen}, P., {et~al.} 2004, \aj, 128, 463, \dodoi{10.1086/421374}

\bibitem[{{Robertson} {et~al.}(2014){Robertson}, {Mahadevan}, {Endl}, \& {Roy}}]{Robertson2014}
{Robertson}, P., {Mahadevan}, S., {Endl}, M., \& {Roy}, A. 2014, Science, 345, 440, \dodoi{10.1126/science.1253253}

\bibitem[{{Roeser} \& {Bastian}(1988)}]{Roeser1988}
{Roeser}, S., \& {Bastian}, U. 1988, \aaps, 74, 449

\bibitem[{{Rosenthal} {et~al.}(2021){Rosenthal}, {Fulton}, {Hirsch}, {Isaacson}, {Howard}, {Dedrick}, {Sherstyuk}, {Blunt}, {Petigura}, {Knutson}, {Behmard}, {Chontos}, {Crepp}, {Crossfield}, {Dalba}, {Fischer}, {Henry}, {Kane}, {Kosiarek}, {Marcy}, {Rubenzahl}, {Weiss}, \& {Wright}}]{Rosenthal2021}
{Rosenthal}, L.~J., {Fulton}, B.~J., {Hirsch}, L.~A., {et~al.} 2021, \apjs, 255, 8, \dodoi{10.3847/1538-4365/abe23c}

\bibitem[{{Saar} {et~al.}(1998){Saar}, {Butler}, \& {Marcy}}]{Saar1998}
{Saar}, S.~H., {Butler}, R.~P., \& {Marcy}, G.~W. 1998, \apjl, 498, L153, \dodoi{10.1086/311325}

\bibitem[{{Saar} \& {Donahue}(1997)}]{Saar1997}
{Saar}, S.~H., \& {Donahue}, R.~A. 1997, \apj, 485, 319, \dodoi{10.1086/304392}

\bibitem[{{Santos} {et~al.}(2000){Santos}, {Mayor}, {Naef}, {Pepe}, {Queloz}, {Udry}, \& {Blecha}}]{Santos2000}
{Santos}, N.~C., {Mayor}, M., {Naef}, D., {et~al.} 2000, \aap, 361, 265

\bibitem[{{Savitzky} \& {Golay}(1964)}]{Savitzky1964}
{Savitzky}, A., \& {Golay}, M.~J.~E. 1964, Analytical Chemistry, 36, 1627

\bibitem[{{Schlieder} {et~al.}(2010){Schlieder}, {L{\'e}pine}, \& {Simon}}]{Schlieder2010}
{Schlieder}, J.~E., {L{\'e}pine}, S., \& {Simon}, M. 2010, \aj, 140, 119, \dodoi{10.1088/0004-6256/140/1/119}

\bibitem[{{Schuessler} {et~al.}(1996){Schuessler}, {Caligari}, {Ferriz-Mas}, {Solanki}, \& {Stix}}]{Schuessler1996}
{Schuessler}, M., {Caligari}, P., {Ferriz-Mas}, A., {Solanki}, S.~K., \& {Stix}, M. 1996, \aap, 314, 503

\bibitem[{{Sharma}(2017)}]{Sharma2017}
{Sharma}, S. 2017, \araa, 55, 213, \dodoi{10.1146/annurev-astro-082214-122339}

\bibitem[{{Shetrone} {et~al.}(2007){Shetrone}, {Cornell}, {Fowler}, {Gaffney}, {Laws}, {Mader}, {Mason}, {Odewahn}, {Roman}, {Rostopchin}, {Schneider}, {Umbarger}, \& {Westfall}}]{Shetrone2007}
{Shetrone}, M., {Cornell}, M.~E., {Fowler}, J.~R., {et~al.} 2007, \pasp, 119, 556, \dodoi{10.1086/519291}

\bibitem[{{Shvonski} {et~al.}(2016){Shvonski}, {Mamajek}, {Kim}, {Meyer}, \& {Pecaut}}]{Shvonski2016}
{Shvonski}, A.~J., {Mamajek}, E.~E., {Kim}, J.~S., {Meyer}, M.~R., \& {Pecaut}, M.~J. 2016, arXiv e-prints, arXiv:1612.06924, \dodoi{10.48550/arXiv.1612.06924}

\bibitem[{{Skiff}(2014)}]{Skiff2014}
{Skiff}, B.~A. 2014, {VizieR Online Data Catalog: Catalogue of Stellar Spectral Classifications (Skiff, 2009- )}, VizieR On-line Data Catalog: B/mk. Originally published in: Lowell Observatory (October 2014)

\bibitem[{{Smith} {et~al.}(2012){Smith}, {Stumpe}, {Van Cleve}, {Jenkins}, {Barclay}, {Fanelli}, {Girouard}, {Kolodziejczak}, {McCauliff}, {Morris}, \& {Twicken}}]{Smith2012}
{Smith}, J.~C., {Stumpe}, M.~C., {Van Cleve}, J.~E., {et~al.} 2012, \pasp, 124, 1000, \dodoi{10.1086/667697}

\bibitem[{{Soderblom} {et~al.}(2005){Soderblom}, {Nelan}, {Benedict}, {McArthur}, {Ramirez}, {Spiesman}, \& {Jones}}]{Soderblom2005}
{Soderblom}, D.~R., {Nelan}, E., {Benedict}, G.~F., {et~al.} 2005, \aj, 129, 1616, \dodoi{10.1086/427860}

\bibitem[{{Soderblom} {et~al.}(1993){Soderblom}, {Stauffer}, {Hudon}, \& {Jones}}]{Soderblom1993}
{Soderblom}, D.~R., {Stauffer}, J.~R., {Hudon}, J.~D., \& {Jones}, B.~F. 1993, \apjs, 85, 315, \dodoi{10.1086/191767}

\bibitem[{{Stanford-Moore} {et~al.}(2020){Stanford-Moore}, {Nielsen}, {De Rosa}, {Macintosh}, \& {Czekala}}]{Stanford-Moore2020}
{Stanford-Moore}, S.~A., {Nielsen}, E.~L., {De Rosa}, R.~J., {Macintosh}, B., \& {Czekala}, I. 2020, \apj, 898, 27, \dodoi{10.3847/1538-4357/ab9a35}

\bibitem[{{Stauffer} {et~al.}(2007){Stauffer}, {Hartmann}, {Fazio}, {Allen}, {Patten}, {Lowrance}, {Hurt}, {Rebull}, {Cutri}, {Ramirez}, {Young}, {Rieke}, {Gorlova}, {Muzerolle}, {Slesnick}, \& {Skrutskie}}]{Stauffer2007}
{Stauffer}, J.~R., {Hartmann}, L.~W., {Fazio}, G.~G., {et~al.} 2007, \apjs, 172, 663, \dodoi{10.1086/518961}

\bibitem[{{Stefansson} {et~al.}(2016){Stefansson}, {Hearty}, {Robertson}, {Mahadevan}, {Anderson}, {Levi}, {Bender}, {Nelson}, {Monson}, {Blank}, {Halverson}, {Henderson}, {Ramsey}, {Roy}, {Schwab}, \& {Terrien}}]{Stefansson2016}
{Stefansson}, G., {Hearty}, F., {Robertson}, P., {et~al.} 2016, \apj, 833, 175, \dodoi{10.3847/1538-4357/833/2/175}

\bibitem[{{Stefansson} {et~al.}(2020{\natexlab{a}}){Stefansson}, {Mahadevan}, {Maney}, {Ninan}, {Robertson}, {Rajagopal}, {Haase}, {Allen}, {Ford}, {Winn}, {Wolfgang}, {Dawson}, {Wisniewski}, {Bender}, {Ca{\~n}as}, {Cochran}, {Diddams}, {Fredrick}, {Halverson}, {Hearty}, {Hebb}, {Kanodia}, {Levi}, {Metcalf}, {Monson}, {Ramsey}, {Roy}, {Schwab}, {Terrien}, \& {Wright}}]{Stefansson2020b}
{Stefansson}, G., {Mahadevan}, S., {Maney}, M., {et~al.} 2020{\natexlab{a}}, \aj, 160, 192, \dodoi{10.3847/1538-3881/abb13a}

\bibitem[{{Stefansson} {et~al.}(2020{\natexlab{b}}){Stefansson}, {Ca{\~n}as}, {Wisniewski}, {Robertson}, {Mahadevan}, {Maney}, {Kanodia}, {Beard}, {Bender}, {Brunt}, {Clemens}, {Cochran}, {Diddams}, {Endl}, {Ford}, {Fredrick}, {Halverson}, {Hearty}, {Hebb}, {Huehnerhoff}, {Jennings}, {Kaplan}, {Levi}, {Lubar}, {Metcalf}, {Monson}, {Morris}, {Ninan}, {Nitroy}, {Ramsey}, {Roy}, {Schwab}, {Sigurdsson}, {Terrien}, \& {Wright}}]{Stefansson2020a}
{Stefansson}, G., {Ca{\~n}as}, C., {Wisniewski}, J., {et~al.} 2020{\natexlab{b}}, \aj, 159, 100, \dodoi{10.3847/1538-3881/ab5f15}

\bibitem[{{Stephenson}(1986)}]{Stephenson1986}
{Stephenson}, C.~B. 1986, \aj, 91, 144, \dodoi{10.1086/113994}

\bibitem[{{Stumpe} {et~al.}(2014){Stumpe}, {Smith}, {Catanzarite}, {Van Cleve}, {Jenkins}, {Twicken}, \& {Girouard}}]{Stumpe2014}
{Stumpe}, M.~C., {Smith}, J.~C., {Catanzarite}, J.~H., {et~al.} 2014, \pasp, 126, 100, \dodoi{10.1086/674989}

\bibitem[{{Stumpe} {et~al.}(2012){Stumpe}, {Smith}, {Van Cleve}, {Twicken}, {Barclay}, {Fanelli}, {Girouard}, {Jenkins}, {Kolodziejczak}, {McCauliff}, \& {Morris}}]{Stumpe2012}
{Stumpe}, M.~C., {Smith}, J.~C., {Van Cleve}, J.~E., {et~al.} 2012, \pasp, 124, 985, \dodoi{10.1086/667698}

\bibitem[{{Su{\'a}rez Mascare{\~n}o} {et~al.}(2017){Su{\'a}rez Mascare{\~n}o}, {Rebolo}, {Gonz{\'a}lez Hern{\'a}ndez}, \& {Esposito}}]{Suarez2017}
{Su{\'a}rez Mascare{\~n}o}, A., {Rebolo}, R., {Gonz{\'a}lez Hern{\'a}ndez}, J.~I., \& {Esposito}, M. 2017, \mnras, 468, 4772, \dodoi{10.1093/mnras/stx771}

\bibitem[{{Su{\'a}rez Mascare{\~n}o} {et~al.}(2021){Su{\'a}rez Mascare{\~n}o}, {Damasso}, {Lodieu}, {Sozzetti}, {B{\'e}jar}, {Benatti}, {Zapatero Osorio}, {Micela}, {Rebolo}, {Desidera}, {Murgas}, {Claudi}, {Gonz{\'a}lez Hern{\'a}ndez}, {Malavolta}, {del Burgo}, {D'Orazi}, {Amado}, {Locci}, {Tabernero}, {Marzari}, {Aguado}, {Turrini}, {Cardona Guill{\'e}n}, {Toledo-Padr{\'o}n}, {Maggio}, {Aceituno}, {Bauer}, {Caballero}, {Chinchilla}, {Esparza-Borges}, {Gonz{\'a}lez-{\'A}lvarez}, {Granzer}, {Luque}, {Mart{\'\i}n}, {Nowak}, {Oshagh}, {Pall{\'e}}, {Parviainen}, {Quirrenbach}, {Reiners}, {Ribas}, {Strassmeier}, {Weber}, \& {Mallonn}}]{Suarez2022}
{Su{\'a}rez Mascare{\~n}o}, A., {Damasso}, M., {Lodieu}, N., {et~al.} 2021, Nature Astronomy, 6, 232, \dodoi{10.1038/s41550-021-01533-7}

\bibitem[{{Takarada} {et~al.}(2020){Takarada}, {Sato}, {Omiya}, {Hori}, \& {Fujii}}]{Takarada2020}
{Takarada}, T., {Sato}, B., {Omiya}, M., {Hori}, Y., \& {Fujii}, M.~S. 2020, \pasj, 72, 104, \dodoi{10.1093/pasj/psaa105}

\bibitem[{{Tang} {et~al.}(2023){Tang}, {Stahl}, {Prato}, {Schaefer}, {Johns-Krull}, {Skiff}, {Beichman}, \& {Uyama}}]{Tang2023}
{Tang}, S.-Y., {Stahl}, A.~G., {Prato}, L., {et~al.} 2023, \apj, 950, 92, \dodoi{10.3847/1538-4357/acc58b}

\bibitem[{{Teske} {et~al.}(2021){Teske}, {Wang}, {Wolfgang}, {Gan}, {Plotnykov}, {Armstrong}, {Butler}, {Cale}, {Crane}, {Howard}, {Jensen}, {Law}, {Shectman}, {Plavchan}, {Valencia}, {Vanderburg}, {Ricker}, {Vanderspek}, {Latham}, {Seager}, {Winn}, {Jenkins}, {Adibekyan}, {Barrado}, {Barros}, {Benkhaldoun}, {Brown}, {Bryant}, {Burt}, {Caldwell}, {Charbonneau}, {Cloutier}, {Collins}, {Collins}, {Colon}, {Conti}, {Demangeon}, {Eastman}, {Elmufti}, {Feng}, {Flowers}, {Guerrero}, {Hojjatpanah}, {Irwin}, {Isopi}, {Lillo-Box}, {Mallia}, {Massey}, {Mori}, {Mullally}, {Narita}, {Nishiumi}, {Osborn}, {Paegert}, {de Leon}, {Quinn}, {Reefe}, {Schwarz}, {Shporer}, {Soubkiou}, {Sousa}, {Stockdale}, {Str{\o}m}, {Tan}, {Tang}, {Tenenbaum}, {Wheatley}, {Wittrock}, {Yahalomi}, \& {Zohrabi}}]{Teske2021}
{Teske}, J., {Wang}, S.~X., {Wolfgang}, A., {et~al.} 2021, \apjs, 256, 33, \dodoi{10.3847/1538-4365/ac0f0a}

\bibitem[{{Torres} {et~al.}(2006){Torres}, {Quast}, {da Silva}, {de La Reza}, {Melo}, \& {Sterzik}}]{Torres2006}
{Torres}, C.~A.~O., {Quast}, G.~R., {da Silva}, L., {et~al.} 2006, \aap, 460, 695, \dodoi{10.1051/0004-6361:20065602}

\bibitem[{{Torres} {et~al.}(2008){Torres}, {Quast}, {Melo}, \& {Sterzik}}]{Torres2008}
{Torres}, C.~A.~O., {Quast}, G.~R., {Melo}, C.~H.~F., \& {Sterzik}, M.~F. 2008, in Handbook of Star Forming Regions, Volume II, ed. B.~{Reipurth}, Vol.~5 (Astronomical Society of the Pacific), 757, \dodoi{10.48550/arXiv.0808.3362}

\bibitem[{{Torres}(1999)}]{Torres1999}
{Torres}, G. 1999, \pasp, 111, 169, \dodoi{10.1086/316313}

\bibitem[{{Torres} {et~al.}(2021){Torres}, {Latham}, \& {Quinn}}]{Torres2021}
{Torres}, G., {Latham}, D.~W., \& {Quinn}, S.~N. 2021, \apj, 921, 117, \dodoi{10.3847/1538-4357/ac1585}

\bibitem[{{Torres} {et~al.}(2020){Torres}, {Melis}, {Kraus}, {Dupuy}, {Chilcote}, \& {Crepp}}]{Torres2020}
{Torres}, G., {Melis}, C., {Kraus}, A.~L., {et~al.} 2020, \apj, 898, 2, \dodoi{10.3847/1538-4357/ab9c20}

\bibitem[{{Torres} {et~al.}(2002){Torres}, {Neuh{\"a}user}, \& {Guenther}}]{Torres2002}
{Torres}, G., {Neuh{\"a}user}, R., \& {Guenther}, E.~W. 2002, \aj, 123, 1701, \dodoi{10.1086/339178}

\bibitem[{{Tran} {et~al.}(2021){Tran}, {Bowler}, {Cochran}, {Endl}, {Stef{\'a}nsson}, {Mahadevan}, {Ninan}, {Bender}, {Halverson}, {Roy}, \& {Terrien}}]{Tran2021}
{Tran}, Q.~H., {Bowler}, B.~P., {Cochran}, W.~D., {et~al.} 2021, \aj, 161, 173, \dodoi{10.3847/1538-3881/abe041}

\bibitem[{{Tran} {et~al.}(2024){Tran}, {Bowler}, {Cochran}, {Halverson}, {Mahadevan}, {Ninan}, {Robertson}, {Stef{\'a}nsson}, \& {Terrien}}]{Tran2024}
---. 2024, \aj, 167, 193, \dodoi{10.3847/1538-3881/ad2eaf}

\bibitem[{{Triaud} {et~al.}(2010){Triaud}, {Collier Cameron}, {Queloz}, {Anderson}, {Gillon}, {Hebb}, {Hellier}, {Loeillet}, {Maxted}, {Mayor}, {Pepe}, {Pollacco}, {S{\'e}gransan}, {Smalley}, {Udry}, {West}, \& {Wheatley}}]{Triaud2010}
{Triaud}, A.~H.~M.~J., {Collier Cameron}, A., {Queloz}, D., {et~al.} 2010, \aap, 524, A25, \dodoi{10.1051/0004-6361/201014525}

\bibitem[{{Udry} {et~al.}(2000){Udry}, {Mayor}, {Naef}, {Pepe}, {Queloz}, {Santos}, {Burnet}, {Confino}, \& {Melo}}]{Udry2000}
{Udry}, S., {Mayor}, M., {Naef}, D., {et~al.} 2000, \aap, 356, 590

\bibitem[{{van der Walt} {et~al.}(2011){van der Walt}, {Colbert}, \& {Varoquaux}}]{vanderWalt2011}
{van der Walt}, S., {Colbert}, S.~C., \& {Varoquaux}, G. 2011, Computing in Science and Engineering, 13, 22, \dodoi{10.1109/MCSE.2011.37}

\bibitem[{{Vanderburg} {et~al.}(2016){Vanderburg}, {Plavchan}, {Johnson}, {Ciardi}, {Swift}, \& {Kane}}]{Vanderburg2016}
{Vanderburg}, A., {Plavchan}, P., {Johnson}, J.~A., {et~al.} 2016, \mnras, 459, 3565, \dodoi{10.1093/mnras/stw863}

\bibitem[{{VanderPlas}(2018)}]{VanderPlas2018}
{VanderPlas}, J.~T. 2018, \apjs, 236, 16, \dodoi{10.3847/1538-4365/aab766}

\bibitem[{{Virtanen} {et~al.}(2020){Virtanen}, {Gommers}, {Oliphant}, {Haberland}, {Reddy}, {Cournapeau}, {Burovski}, {Peterson}, {Weckesser}, {Bright}, {van der Walt}, {Brett}, {Wilson}, {Jarrod Millman}, {Mayorov}, {Nelson}, {Jones}, {Kern}, {Larson}, {Carey}, {Polat}, {Feng}, {Moore}, {Vand erPlas}, {Laxalde}, {Perktold}, {Cimrman}, {Henriksen}, {Quintero}, {Harris}, {Archibald}, {Ribeiro}, {Pedregosa}, {van Mulbregt}, \& {Contributors}}]{Virtanen2020}
{Virtanen}, P., {Gommers}, R., {Oliphant}, T.~E., {et~al.} 2020, Nature Methods, \dodoi{https://doi.org/10.1038/s41592-019-0686-2}

\bibitem[{{Ward}(1997)}]{Ward1997}
{Ward}, W.~R. 1997, \icarus, 126, 261, \dodoi{10.1006/icar.1996.5647}

\bibitem[{{White} {et~al.}(2007){White}, {Gabor}, \& {Hillenbrand}}]{White2007}
{White}, R.~J., {Gabor}, J.~M., \& {Hillenbrand}, L.~A. 2007, \aj, 133, 2524, \dodoi{10.1086/514336}

\bibitem[{{Wilson}(1963)}]{Wilson1963}
{Wilson}, O.~C. 1963, \apj, 138, 832, \dodoi{10.1086/147689}

\bibitem[{{Wittenmyer} {et~al.}(2009){Wittenmyer}, {Endl}, {Cochran}, {Levison}, \& {Henry}}]{Wittenmyer2009b}
{Wittenmyer}, R.~A., {Endl}, M., {Cochran}, W.~D., {Levison}, H.~F., \& {Henry}, G.~W. 2009, \apjs, 182, 97, \dodoi{10.1088/0067-0049/182/1/97}

\bibitem[{{Wright} {et~al.}(2004){Wright}, {Marcy}, {Butler}, \& {Vogt}}]{Wright2004}
{Wright}, J.~T., {Marcy}, G.~W., {Butler}, R.~P., \& {Vogt}, S.~S. 2004, \apjs, 152, 261, \dodoi{10.1086/386283}

\bibitem[{{Wright} {et~al.}(2012){Wright}, {Marcy}, {Howard}, {Johnson}, {Morton}, \& {Fischer}}]{Wright2012}
{Wright}, J.~T., {Marcy}, G.~W., {Howard}, A.~W., {et~al.} 2012, \apj, 753, 160, \dodoi{10.1088/0004-637X/753/2/160}

\bibitem[{{Wu} \& {Murray}(2003)}]{Wu2003}
{Wu}, Y., \& {Murray}, N. 2003, \apj, 589, 605, \dodoi{10.1086/374598}

\bibitem[{{Yee} {et~al.}(2017){Yee}, {Petigura}, \& {von Braun}}]{Yee2017}
{Yee}, S.~W., {Petigura}, E.~A., \& {von Braun}, K. 2017, \apj, 836, 77, \dodoi{10.3847/1538-4357/836/1/77}

\bibitem[{{Yoss}(1961)}]{Yoss1961}
{Yoss}, K.~M. 1961, \apj, 134, 809, \dodoi{10.1086/147209}

\bibitem[{{Yu} {et~al.}(2017){Yu}, {Donati}, {H{\'e}brard}, {Moutou}, {Malo}, {Grankin}, {Hussain}, {Collier Cameron}, {Vidotto}, {Baruteau}, {Alencar}, {Bouvier}, {Petit}, {Takami}, {Herczeg}, {Gregory}, {Jardine}, {Morin}, {M{\'e}nard}, \& {Matysse Collaboration}}]{Yu2017}
{Yu}, L., {Donati}, J.~F., {H{\'e}brard}, E.~M., {et~al.} 2017, \mnras, 467, 1342, \dodoi{10.1093/mnras/stx009}

\bibitem[{{Zakhozhay} {et~al.}(2022){Zakhozhay}, {Launhardt}, {M{\"u}ller}, {Brems}, {Eigenthaler}, {Gennaro}, {Hempel}, {Hempel}, {Henning}, {Kennedy}, {Kim}, {K{\"u}rster}, {Lachaume}, {Manerikar}, {Patel}, {Pavlov}, {Reffert}, \& {Trifonov}}]{Zakhozhay2022}
{Zakhozhay}, O.~V., {Launhardt}, R., {M{\"u}ller}, A., {et~al.} 2022, \aap, 667, A63, \dodoi{10.1051/0004-6361/202244213}

\bibitem[{{Zechmeister} \& {K{\"u}rster}(2009)}]{Zechmeister2009}
{Zechmeister}, M., \& {K{\"u}rster}, M. 2009, \aap, 496, 577, \dodoi{10.1051/0004-6361:200811296}

\bibitem[{{Zechmeister} {et~al.}(2018){Zechmeister}, {Reiners}, {Amado}, {Azzaro}, {Bauer}, {B{\'e}jar}, {Caballero}, {Guenther}, {Hagen}, {Jeffers}, {Kaminski}, {K{\"u}rster}, {Launhardt}, {Montes}, {Morales}, {Quirrenbach}, {Reffert}, {Ribas}, {Seifert}, {Tal-Or}, \& {Wolthoff}}]{Zechmeister2018}
{Zechmeister}, M., {Reiners}, A., {Amado}, P.~J., {et~al.} 2018, \aap, 609, A12, \dodoi{10.1051/0004-6361/201731483}

\bibitem[{{Zuckerman}(2019)}]{Zuckerman2019}
{Zuckerman}, B. 2019, \apj, 870, 27, \dodoi{10.3847/1538-4357/aaee66}

\bibitem[{{Zuckerman} {et~al.}(2006){Zuckerman}, {Bessell}, {Song}, \& {Kim}}]{Zuckerman2006}
{Zuckerman}, B., {Bessell}, M.~S., {Song}, I., \& {Kim}, S. 2006, \apjl, 649, L115, \dodoi{10.1086/508060}

\bibitem[{{Zuckerman} {et~al.}(2011){Zuckerman}, {Rhee}, {Song}, \& {Bessell}}]{Zuckerman2011}
{Zuckerman}, B., {Rhee}, J.~H., {Song}, I., \& {Bessell}, M.~S. 2011, \apj, 732, 61, \dodoi{10.1088/0004-637X/732/2/61}

\bibitem[{{Zuckerman} {et~al.}(2004){Zuckerman}, {Song}, \& {Bessell}}]{Zuckerman2004}
{Zuckerman}, B., {Song}, I., \& {Bessell}, M.~S. 2004, \apjl, 613, L65, \dodoi{10.1086/425036}

\bibitem[{{Zuckerman} {et~al.}(2013){Zuckerman}, {Vican}, {Song}, \& {Schneider}}]{Zuckerman2013}
{Zuckerman}, B., {Vican}, L., {Song}, I., \& {Schneider}, A. 2013, \apj, 778, 5, \dodoi{10.1088/0004-637X/778/1/5}

\end{thebibliography}
\bibliographystyle{aasjournal}

\end{document}